\documentclass[%
 reprint,superscriptaddress,
 amsmath,amssymb,
 aps,
]{revtex4-2} 
\usepackage{simplewick}
\usepackage{algorithmic}
\usepackage{multirow}
\usepackage[table,xcdraw]{xcolor}
\usepackage{graphicx}
\graphicspath{ {./figures/} }
\usepackage{float}
\usepackage{caption}
\usepackage{dcolumn}
\usepackage{bm}
\usepackage{amssymb}
\usepackage{tabularx}
\usepackage{mathtools,halloweenmath}
\usepackage{physics}
\usepackage{graphicx}
\usepackage{amsmath}
\usepackage[version=4]{mhchem}
\usepackage{siunitx}
\usepackage{tabularx}
\usepackage{braket}
\usepackage{wick}
\usepackage{xcolor}
\usepackage{ulem}
\usepackage{comment
}
\usepackage{hyperref} 
\usepackage{multirow}
\usepackage{lipsum}
\usepackage{array}
\usepackage{microtype}
\usepackage{float}
\usepackage{appendix}
\usepackage{amsmath}
\usepackage{comment}
\usepackage{amsthm}
\usepackage{gensymb}

\usepackage[color=yellow]{todonotes} 

\usepackage[ruled,lined]{algorithm2e}

\usepackage{mathtools}

\begin{document}
\preprint{APS/123-QED} 
\title[]{Enhancing the Harrow-Hassidim-Lloyd (HHL) algorithm in systems with large condition numbers}
\author{Peniel Bertrand Tsemo}
\affiliation{Centre for Quantum Engineering, Research and Education, TCG CREST, Sector V, Salt Lake, Kolkata 700091, India}
\affiliation{Department of Physics, IIT Tirupati, Chindepalle, Andhra Pradesh 517619, India}
\author{Akshaya Jayashankar}
\affiliation{Centre for Quantum Engineering, Research and Education, TCG CREST, Sector V, Salt Lake, Kolkata 700091, India}
\author{K. Sugisaki}
\affiliation{Centre for Quantum Engineering, Research and Education, TCG CREST, Sector V, Salt Lake, Kolkata 700091, India}
\affiliation{Graduate School of Science and Technology, Keio University, 7-1 Shinkawasaki, Saiwai-ku, Kawasaki, Kanagawa 212-0032, Japan}
\affiliation{Quantum Computing Center, Keio University, 3-14-1 Hiyoshi, Kohoku-ku, Yokohama, Kanagawa 223-8522, Japan}
\affiliation{Keio University Sustainable Quantum Artificial Intelligence Center (KSQAIC), Keio University, 2-15-45 Mita, Minato-ku, Tokyo 108-8345, Japan}
\author{Nishanth Baskaran}
\affiliation{Department of Electrical and Computer Engineering, The University of British Columbia, Vancouver, British Columbia V6T 1Z4, Canada}
\author{Sayan Chakraborty}
\affiliation{Institute for Advancing Intelligence, TCG CREST, Sector V, Salt Lake, Kolkata 700091, India}
\affiliation{Academy of Scientific and Innovative Research (AcSIR), Ghaziabad- 201002, India}
\author{V. S. Prasannaa}
\email{srinivasaprasannaa@gmail.com}
\affiliation{Centre for Quantum Engineering, Research and Education, TCG CREST, Sector V, Salt Lake, Kolkata 700091, India}
\affiliation{Academy of Scientific and Innovative Research (AcSIR), Ghaziabad- 201002, India}

\begin{abstract}
Although the Harrow-Hassidim-Lloyd (HHL) algorithm offers an exponential speedup in system size for treating linear equations of the form $A\vec{x}=\vec{b}$ on quantum computers when compared to their traditional counterparts, it faces a challenge related to the condition number ($\mathcal{\kappa}$) scaling of the $A$ matrix. In this work, we address the issue by introducing the post-selection-improved HHL (Psi-HHL) framework that operates on a simple yet effective premise: subtracting mixed and wrong signals to extract correct signals while providing the benefit of optimal scaling in the condition number of $A$ (denoted as $\mathcal{\kappa}$) for large $\mathcal{\kappa}$ scenarios. This approach, which leads to minimal increase in circuit depth, has the important practical implication of having to use substantially fewer shots relative to the traditional HHL algorithm. The term `signal' refers to a feature of $|x\rangle$. We design circuits for overlap and expectation value estimation in the Psi-HHL framework. We demonstrate performance of Psi-HHL via numerical simulations. We carry out two sets of computations, where we go up to 26-qubit calculations, to demonstrate the ability of Psi-HHL to handle situations involving large $\mathcal{\kappa}$ matrices via: (a) a set of toy matrices, for which we go up to size $64 \times 64$ and $\mathcal{\kappa}$ values of up to $\approx$ 1 million, and (b) application to quantum chemistry, where we consider matrices up to size $256 \times 256$ that reach $\mathcal{\kappa}$ of about 393. The molecular systems that we consider are Li$_{\mathrm{2}}$, KH, RbH, and CsH. 
\end{abstract}

\maketitle

\tableofcontents

\section{Introduction}\label{Sec:Intro}

The Harrow-Hassidim-Lloyd (HHL) algorithm was proposed to solve equations of the form $A\vec{x}=\vec{b}$, where $A$ is a known and sparse $2^N \times 2^N$ matrix and $\vec{b}$ is 
a $2^N \times 1$ vector, on quantum computers with a time complexity that is exponentially lower in system size than those of the best known classical methods~\cite{Harrow2009QuantumEquations}. Further, via amplitude encoding, $\vec{b}$, which is a $2^N \times 1$ vector, is encoded into a log$_2{2^N}$-qubit state (for brevity, we shall hereafter denote $2^N$ as $n$). Thus, when the HHL approach is applied to `killer applications' of quantum computing such as quantum chemistry, there is an exponential suppression in qubit number relative to the other known approaches such as the quantum phase estimation (QPE) and the variational quantum eigensolver (VQE) algorithms~\cite{AHL2023}, thus opening new avenues to apply HHL for calculating properties of large molecules. \\ 

Despite possessing the aforementioned advantages, the HHL algorithm relies upon the probability of successfully post-selecting $1$ on the HHL ancilla qubit, that is, $P(1)$~\cite{Harrow2009QuantumEquations}. It is known that there can be a substantial decrease in $P(1)$ with increase in the condition number of the problem matrix $A$, which we shall denote by $\mathcal{\kappa}$. In other words, the fraction of usable shots in the HHL algorithm becomes very low when large $\mathcal{\kappa}$ input matrices are involved. Unfortunately, it is not uncommon to encounter the issue while employing HHL for some interesting problems (for example, see Ref.~\cite{Golden2022} for application of the algorithm to hydrological systems), including quantum chemistry, which we shall discuss in detail in our Results section. Several works in literature have addressed the issue of handling large $\mathcal{\kappa}$ matrices either directly (where $\mathcal{\kappa}$ of the $A$ matrix itself is altered via approaches such as preconditioning) or indirectly \cite{ampamp,vtaa,Clader,Shao,Tong,Golden2022,Babukhin}. For a thorough survey, we direct the reader to Section 1.2 of Ref. \cite{Orsucci2021}. \\ 

In this work, we propose a new framework to address the issue of handling $A$ matrices with large condition numbers, by introducing the post-selection-improved HHL (Psi-HHL) algorithm, which relies upon the intuition of extracting the correct signal by subtracting a wrong signal from a mixed signal (that contains both the correct and wrong signals). A signal refers to a scalar that is a feature of $\ket{x}$. This is discussed in Section \ref{psihhl}, after we introduce briefly the HHL algorithm in Section \ref{hhl-algo}. A simple modification involving execution of two (three) HHL routines for overlap computation (expectation value calculation) enables one to carry out calculations for treating systems of linear equations involving large-$\mathcal{\kappa}$ matrices without having to contend with a large number of shots. In fact, we argue that the condition number scales linearly with system size with our approach for large $\mathcal{\kappa}$ situations (see Section \ref{Subsec:Complexity}), while still staying in the realm of a typical HHL computation. This compares favourably with several other notable works in literature, including but not limited to the original HHL work: $\mathcal{O}(\mathcal{\kappa}^3)$ and $\mathcal{O}(\mathcal{\kappa}^2)$ without and with amplitude amplification respectively~\cite{Harrow2009QuantumEquations}, $\mathcal{O}(\mathcal{\kappa}\mathrm{log} ^3\mathcal{\kappa})$ with variable time amplitude amplification~\cite{vtaa}, and works by Childs \textit{et al}: $\mathcal{O}(\mathcal{\kappa} \mathrm{polylog} (\mathcal{\kappa}))$~\cite{Childs2017}, Wossnig \textit{et al}: $\mathcal{O}(\mathcal{\kappa}^2)$, and Costa \textit{et al}'s discrete quantum adiabatic approach based quantum linear solver algorithm \cite{Costa}: $\mathcal{O}(\mathcal{\kappa})$. In particular, the last approach also achieves linear scaling in condition number, but we do so without invoking any additional algorithmic machinery  beyond the scope of the traditional HHL algorithm. In order to provide a very qualitative perspective, we note that for condition number of about 1 million, $\mathcal{\kappa}^3 : \mathcal{\kappa}^2 : \mathcal{\kappa} \mathrm{log}^3\mathcal{\kappa} : \mathcal{\kappa}$ is $\sim 10^{18}: 10^{12}: 7.9 \times 10^9: 10^6$. For completeness, we add that the best known classical approach, conjugate gradient, scales as $\mathcal{O}(\sqrt{\mathcal{\kappa}})$~\cite{JRS1994}, but its scaling with respect to system size is, of course, exponentially more expensive than that for HHL. We also address the ability of the Psi-HHL algorithm to handle instances where $A$ is singular in Section \ref{Subsec:Singular}. \\ 

For the purposes of this work, we pick the overlap, $\mathfrak{o}_{x,b}$, between the input ($|b\rangle$) and output ($|x\rangle$) states of HHL (see Section \ref{overlap-psihhl}). In the case of molecules, the quantity corresponds to molecular correlation energies (up to some normalization constants), $E_c$, as our signal. It is to be noted that Psi-HHL itself is applicable to other features such as expectation values (we explore this option in Section \ref{Subsec:Transition}). We also explore a simple representative case where the Psi-HHL framework for overlap calculation can be made a primitive in a larger quantum circuit, in Section \ref{Subsec:Primitive}. \\ 

Section \ref{Sec:Results} presents our results for overlap calculations from numerical simulations for Psi-HHL applied to (a) suitably chosen sets of $4 \times 4$, $8 \times 8$, $16 \times 16$, $32 \times 32$, and $64 \times 64$ toy matrices, as well as for (b) molecular calculations ($64 \times 64$ and $256 \times 256$ $A$ matrix sizes) in the linearized coupled cluster (LCC) HHL and Psi-HHL frameworks. We conclude in Section \ref{Sec:Conclusion}. \\ 

\begin{figure*}[t] 
\includegraphics[scale=0.20,width=1.0\textwidth]{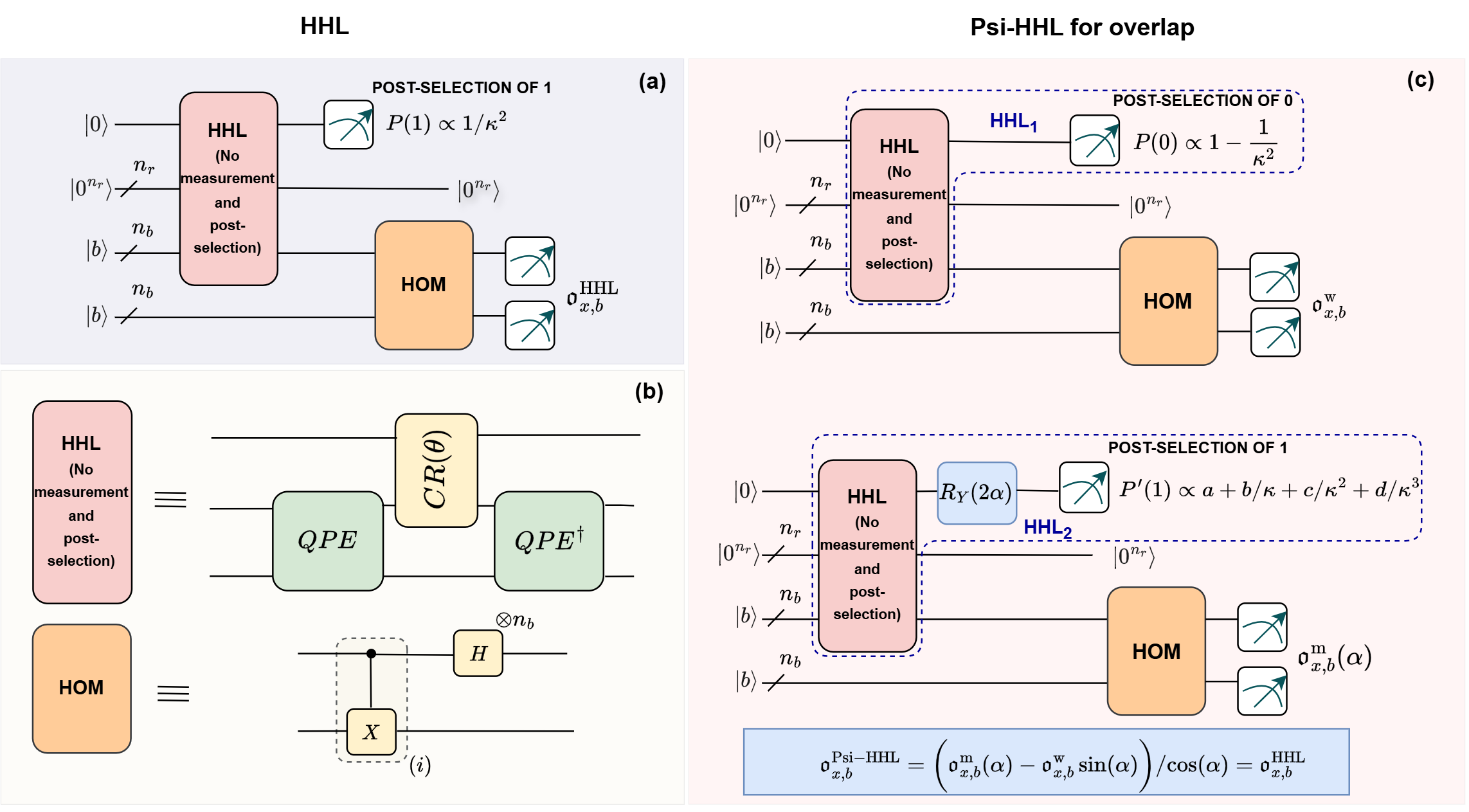} 
\caption{(a) Schematic of the HHL algorithm, where $1$ is post-selected and for which $P(1)$ would be low for large $\mathcal{\kappa}$. $\ket{b}$ is the input state register with $n_b$ qubits, while $n_r$ refers to the number of clock register qubits. The Hong-Ou-Mandel (HOM) module is a destructive version of the controlled-SWAP test without an additional ancilla qubit, and we use it to extract $|\langle b | x \rangle|$, using which we evaluate $\mathfrak{o}_{x,b}^{\mathrm{HHL}} = - \|\ket{b}_{\rm un}\|^2 \ \|\ket{x}_{\rm un}\|\ \ |\langle b | x \rangle|$ (interpreted as the correlation energy, $E_{\mathrm{c}}^{\mathrm{HHL}}$, when HHL is applied to molecular calculations), which is the feature of $\ket{x}$ that we are interested in for our numerical simulations. The subscript `un' stands for un-normalized. Sub-figure (b) provides a block-diagram description of the HHL and the HOM modules. $QPE$ refers to the quantum phase estimation while $CR(\theta)$ refers to the controlled-rotation module that inverts the eigenvalues of $A$. The $CR(\theta)$ circuit is built out of uniformly controlled rotations involving a set of angles $\{\theta_i\}$, denoted compactly as just $\theta$ in the figure. The dashed cambered rectangle around the CNOT gate in the HOM circuit module, and which is accompanied by $(i)$, indicates that the gate is controlled on the i$^{\mathrm{th}}$ qubit of the $\ket{b}$ register. Sub-figure (c) shows HHL$_{1}$ employed inside an overlap calculation circuit in the top panel, where $P(0)$ is post-selected, and also shows HHL$_{2}$ used inside an overlap calculation circuit in the bottom panel. In HHL$_2$, the traditional HHL module is appended with an additional mixing unitary, $R_Y(2\alpha)$, and where $P'(1)$ is post-selected. Psi-HHL relies on subtracting the \textit{final} outcomes of these two circuits, and as the expression at the bottom of panel (c) indicates, we recover $\mathfrak{o}_{x,b}$ (or the correlation energy when applied to molecules) via the procedure to within a simple factor. In our notation, $\mathfrak{o}_{x,b}^{\mathrm{HHL}}$ and $\mathfrak{o}_{x,b}^{\mathrm{Psi-HHL}}$ are obtained from HHL and Psi-HHL respectively. Further, $a, b, c, d \in \mathbb{R}$ are constants. } 
\label{fig:HHL}
\end{figure*} 

\section{Theory and Methodology}\label{Sec:Theory}

\subsection{The HHL algorithm}\label{hhl-algo}

We briefly sketch the relevant details of the HHL algorithm (we refer the readers to Refs. \cite{HHLreview0,Morrell:2021wvm, HHLreview1,HHLreview2} for detailed descriptions of the algorithm) in Figures~\ref{fig:HHL}(a) and \ref{fig:HHL}(b), where an amplitude-encoded input $\arrowvert b\rangle = \sum_i b_i \arrowvert \nu_i \rangle$, such that $\sum_i |b_i|^2 = 1$ where $\{\arrowvert \nu_i \rangle\}$ refers to the eigenbasis of $A$, is fed into the HHL algorithm along with $n_r$ number of clock register qubits and one HHL ancillary qubit. The HHL algorithm itself involves QPE, followed by a non-trivial controlled-rotation module, and then a QPE$^\dag$ module. At the end of the circuit, we have the following state vector,

\begin{eqnarray} \label{eq:one}
\sum_i b_i \left(\sqrt{1-\frac{C^2}{\tilde{\lambda}^2_i}} \ket{0} + \frac{C}{\tilde{\lambda}_i} \ket{1} \right) \otimes \ket{0^{n_r}} \otimes \ket{\nu_i}.
\end{eqnarray} 

\noindent In the above expression, $\{\tilde{\lambda}_{i}\}$ refers to the estimates of the actual eigenvalues of $A$ ($1 \geq \tilde{\lambda}_i \geq 1/\mathcal{\kappa}$) obtained via the QPE module of HHL. Typically, $C$ is chosen to be the minimum eigenvalue of $A$, but it is worth noting that incurring the large classical overhead associated with computing it can always be circumvented via approaches such as AdaptHHL~\cite{AHL2023}. When one post-selects the outcome `$1$' on the HHL ancilla qubit, one can think of it as choosing the `correct' signal. \\ 

The solution vector $|x\rangle \propto A^{-1} |b \rangle$, and the input vector $|b\rangle$ can be written as,
  \begin{equation}\label{eq:unnormalized}
|x\rangle=|x\rangle_{\rm un}/\||x\rangle_{\rm un}\| \ \mathrm{and}\ \ket{b}=\ket{b}_{\rm un}/\|\ket{b}_{\rm un}\|\
\end{equation} 
\noindent  respectively, where subscript `$\mathrm{un}$' refers to \textit{unnormalized} states--- $|b\rangle_{\rm un}$ is the unnormalized input state after amplitude encoding and $|x\rangle_{\rm un}=\sum_i \frac{b_i C}{\tilde{\lambda}_i}|\nu_i \rangle$, with $\sum_i |b_i|^2 = 1$. The probability of successfully post-selecting outcome `$1$' is given as $P(1)=\||x\rangle_{\rm un}\|^2$.
Finally, when one wants to predict a feature of the solution vector, say, the overlap between $\ket{x}$ and $\ket{b}$, it can be done by appending an additional circuit (the Hong-Ou-Mandel module \cite{HOM}), abbreviated as HOM in Figure~\ref{fig:HHL}(b) that inputs the output, $\arrowvert x \rangle$, of the HHL algorithm and $\arrowvert b \rangle$, to output the overlap between the two quantities. The overlap is then calculated as 

\begin{eqnarray}\label{eqn:Ecorr}
\mathfrak{o}_{x,b}^{\rm HHL} &=& - \|\ket{b}_{\rm un}\|^2 \ \|\ket{x}_{\rm un}\|\ \ |\langle b | x \rangle|  \\ \nonumber
&=& - \|\ket{b}_{\rm{un}}\|^2 \sum_i |b_i|^2 \frac{C}{\tilde{\lambda}_i}. 
\end{eqnarray} 

\noindent 

For the sake of simplicity, we set this quantity hereafter as the feature of $\ket{x}$ that is of interest to us. \\ 

The number of trials needed to successfully post-select outcome $`1$' is $\mathcal{O}(\mathcal{\kappa}^2)$. Thus, in the event of large $\mathcal{\kappa}$, the success probability falls quickly, making it practically difficult to carry computations using large $\mathcal{\kappa}$ matrices. In this work, we demonstrate a new approach--- `Psi-HHL', towards handling large $\mathcal{\kappa}$ matrices by leveraging the probability of successfully post-selecting the `wrong' outcome $`0$', $P(0)$, in HHL. 

\subsection{The Psi-HHL framework} \label{psihhl} 

In this section, we introduce the two essential ingredients--- HHL$_1$ and HHL$_2$, which make up the Psi-HHL framework. For different features (signals) of $|x\rangle$ to be extracted, one employs slightly different sets of circuits that involve HHL$_1$ and HHL$_2$ modules. In the HHL$_1$ module, one executes the HHL circuit as it is, except that the `wrong' signal is post-selected, that is, `$0$' is post-selected. The HHL$_2$ entails the traditional HHL module appended with a mixing $2 \times 2$ unitary ($R_Y(2\alpha)$) on the HHL ancillary qubit just before measurement, and post-selecting $`1$' from the `mixed' signal. The role of this unitary is to mix the coefficients of $|0\rangle$ and $|1\rangle$ in the traditional HHL (see Eq. \ref{eq:one}), thereby outputting a `mixed' signal. 

\subsubsection{HHL$_1$: Computing the wrong signal}

The first HHL module, HHL$_1$, serves to post-select the `wrong' signal of the HHL ancilla qubit, that is, the state attached to $\ket{0}$ in Eq.\ref{eq:one}. 
We re-write the state, $|\psi_{\rm out}\rangle$, at the output of all the registers except the clock qubits register, as follows, \begin{eqnarray} 
|\psi_{\rm out}\rangle&=&\sum_i b_i \left(\sqrt{1-\frac{C^2}{\tilde{\lambda}^2_i}} \ket{0} \otimes \ket{\nu_i} + \frac{C}{\tilde{\lambda}_i} \ket{1} \otimes   \ket{\nu_i}\right)\nonumber \\
&=&( |0\rangle\otimes  |x_ \mathrm{w}\rangle+ |1\rangle \otimes |x_\mathrm{r}  \rangle ). 
\end{eqnarray} 
Note that here $|x_\mathrm{w}\rangle$ and $|x_\mathrm{r}\rangle$ are the unnormalised vectors in the superposition. The subscripts $\rm w$ and $\rm r$ refer to `wrong' and `correct' respectively, where for the latter, we use $\rm r$ since $\rm c$ is used already to denote `correlation'. The normalised solution vector $|x_0\rangle$ which appears at the output register after successfully post-selecting outcome `$0$' is given by,
\begin{eqnarray}
\ket{x_{\rm 0}} &=&|x_\mathrm{w} \rangle/\||x_{\mathrm{w}} \rangle\| \nonumber \\
&=&\sum_i b_i \sqrt{1 - \frac{C^2}{\tilde{\lambda}_i^2} }\ket{\nu_i}/ \||x_{\mathrm{w}} \rangle\|, 
\end{eqnarray}
where,
\begin{equation}\label{eq:postselect0}
\||x_{\rm w}\rangle\| = \sqrt{P(0)}= \left(\sum_i \left|b_i\sqrt{1- \frac{C^2}{\tilde{\lambda}_i^2}}\right|^2\right)^{1/2}. 
\end{equation} 
Here, $P(0)$ is the probability of successfully post-selecting outcome `$0$'. 

\subsubsection{HHL$_2$: Computing the mixed signal}

The second HHL execution involves attaching the rotation gate $R_Y(2\alpha)$ on the HHL ancillary qubit before the measurement as illustrated in the lower panel of Figure~\ref{fig:HHL}(c). This seemingly trivial addition to the algorithm introduces a new degree of freedom via the angle $\alpha$, enabling one to boost the probability of successful matrix inversion in the case of large $\mathcal{\kappa}$. We recall that $R_Y(2\alpha)$ is a rotation about the $Y$ axis on the Bloch sphere as given below
\begin{equation} \label{eq:unitary}
R_Y(2\alpha) = \begin{pmatrix} \cos({\alpha}) & -\sin({\alpha}) \\ \sin({\alpha}) & \cos({\alpha})\end{pmatrix}.  \end{equation} 
The state vector at the output of HHL$_2$ module $|\psi'_{\rm out}\rangle$, ignoring the clock qubits register, after applying the unitary  in Eq.~\ref{eq:unitary} is given below,

\begin{eqnarray}\label{eq:hhl2}
|\psi'_{\rm out}\rangle &=& \sum_i b_i \left(\sqrt{1-\frac{C^{2}}{\tilde{\lambda}_i^2}} \cos({\alpha}) - \frac{C}{\tilde{\lambda}_i} \sin({\alpha}) \right) \ket{0} \otimes \ket{\nu_i}  \nonumber \\ 
&+&  b_i \left( \sqrt{1-\frac{C^2}{\tilde{\lambda}_i^2}} \sin({\alpha}) + \frac{C}{\tilde{\lambda}_i} \cos({\alpha})\right) \ket{1} \otimes \ket{\nu_i} \nonumber \\ 
&=&(|0\rangle \otimes|x_{\rm m}\rangle+|1\rangle \otimes|x'_{\rm m}\rangle ). 
\end{eqnarray}

\noindent Note that in Eq.~\ref{eq:hhl2}, $\ket{x_\mathrm{m}}$ and $\ket{x'_\mathrm{m}}$ are unnormalized vectors. The subscript `m' denotes `mixed', and while the unprimed vector accompanies $\ket{0}$, the primed one accompanies $\ket{1}$. The solution vector at the output after successfully post-selecting outcome `$1$' is given below

\begin{eqnarray}
    \ket{x'} &=& \frac{\ket{x'_{\rm m}}}{\|\ket{x'_{\rm m}}\|} \nonumber \\ \nonumber
    &=& \sum_i b_i \left( \sqrt{1-\frac{C^2}{\tilde{\lambda}_i^2} }\sin({\alpha}) + \frac{C}{\tilde{\lambda}_i}\cos({\alpha})\right) \ket{\nu_i}/ \|\ket{x'_{\rm m}}\|, 
\end{eqnarray}

\noindent where 

\begin{eqnarray}\label{eq:postselect1}
\|\ket{x'_{\rm m}}\| &=&\sqrt{P'(1)} \nonumber \\  &=&\left(\sum_i \left| b_i \left(\sqrt{1-\frac{C^2}{\tilde{\lambda}_i^2}} \sin({\alpha}) + \frac{ C}{\tilde{\lambda}_i} \cos({\alpha})\right) \right|^2\right)^{1/2}. 
\end{eqnarray}

\noindent Here, $P'(1)$ is the probability of successfully obtaining outcome `1' in HHL$_2$. 

\subsection{Complexity aspects} \label{Subsec:Complexity}

We now demonstrate the complexity (in $\mathcal{\kappa}$) of the Psi-HHL algorithm. 
We proceed henceforth in this sub-section by assuming the condition that the eigenvalues $\lambda_i$ are precisely captured with adequate number of QPE clock register qubits, $n_r$. 

\noindent a) \textbf{Success probability of the post-selection process in HHL$_1$}:  The probability of successfully post-selecting the outcome `0', $P(0)$, in HHL$_1$ as given in Eq.~\ref{eq:postselect0} can be expanded as~\cite{Linlin} 

\begin{eqnarray}\label{eqn:p0}
    P(0)&=&\sum_i \left|b_i\sqrt{1- \frac{C^2}{{\lambda}_i^2}}\right|^2 \nonumber \\ 
    &=& \|\sqrt{(\mathbb{I}-C^2 A^{-2})}|b\rangle\|^2, 
\end{eqnarray}
where $\mathbb{I}$ is the identity operator written in the eigenbasis of operator $A$, constant $C= \mathcal{O}(1/\mathcal{\kappa})$, where the condition number $\mathcal{\kappa}$ is taken as $\lambda_{\rm max}/\lambda_{\rm min}$  and $|b\rangle$ is the input to the HHL circuit. The set of values $\bigg\{\sqrt{1-\frac{C^2}{\lambda_i^2} }\bigg\}$ are the eigenvalues of the operator $\sqrt{\mathbb{I}-C^2 A^{-2}}$, with the largest eigenvalue being $\sqrt{1-\frac{C^2}{\lambda_{\rm max}^2}}$, and therefore 

\begin{eqnarray}\label{eq:prob0}
P(0)&\leq& \|\sqrt{(\mathbb{I}-C^2 A^{-2})}\|_{\rm op}^2\||b\rangle\| ^2 \nonumber \\ \nonumber
P(0)& \leq & 1-\frac{C^2}{\lambda_{\rm max}^2} \nonumber \\ 
&\leq & 1-\frac{1}{\mathcal{\kappa}^2}. 
\end{eqnarray}
 
\begin{figure*}[t] 
\centering
\hspace*{-0.55cm}
\begin{tabular}{cc}     \includegraphics[scale=0.68,keepaspectratio]{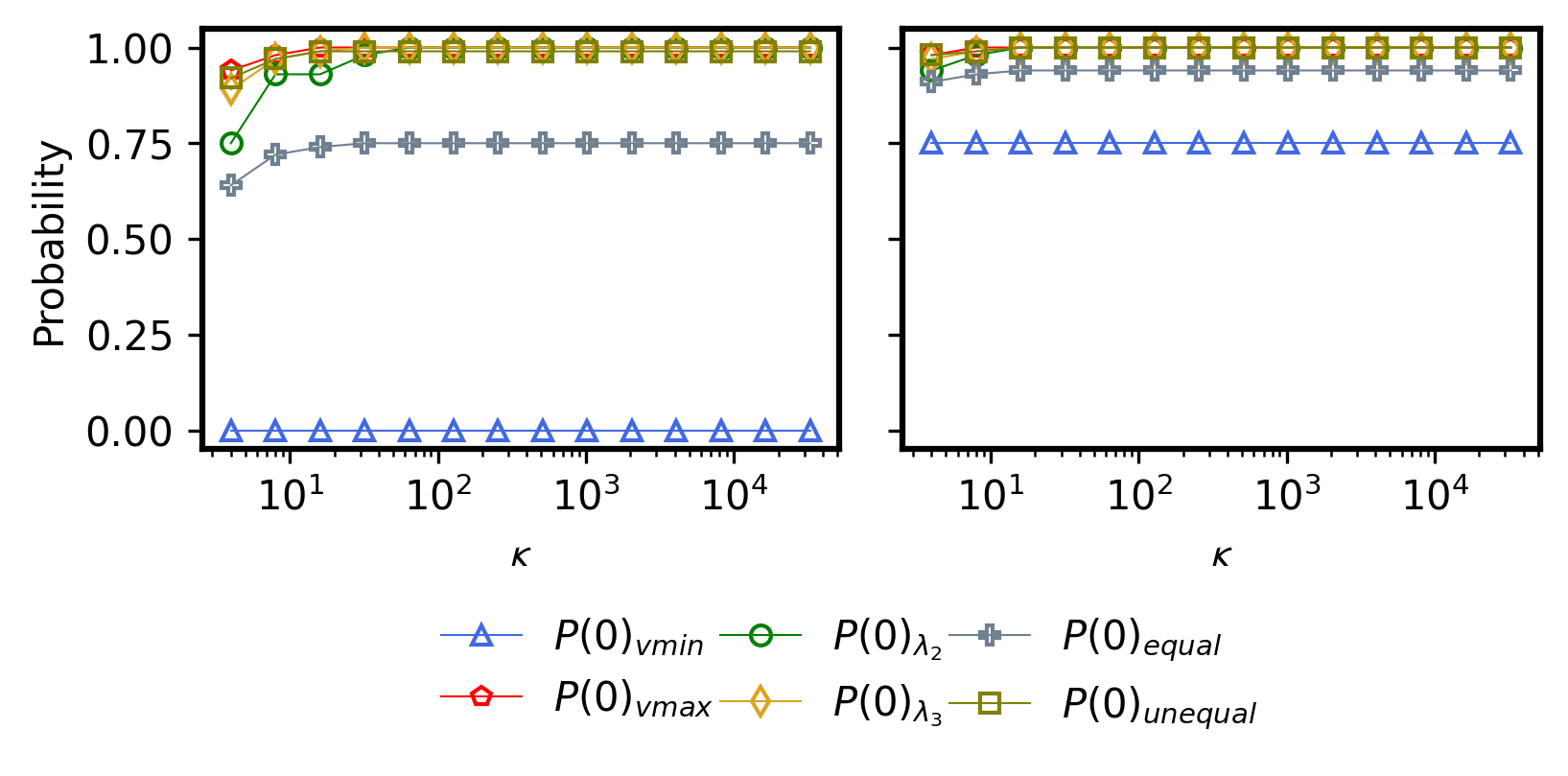}&
\includegraphics[scale=0.68,keepaspectratio]{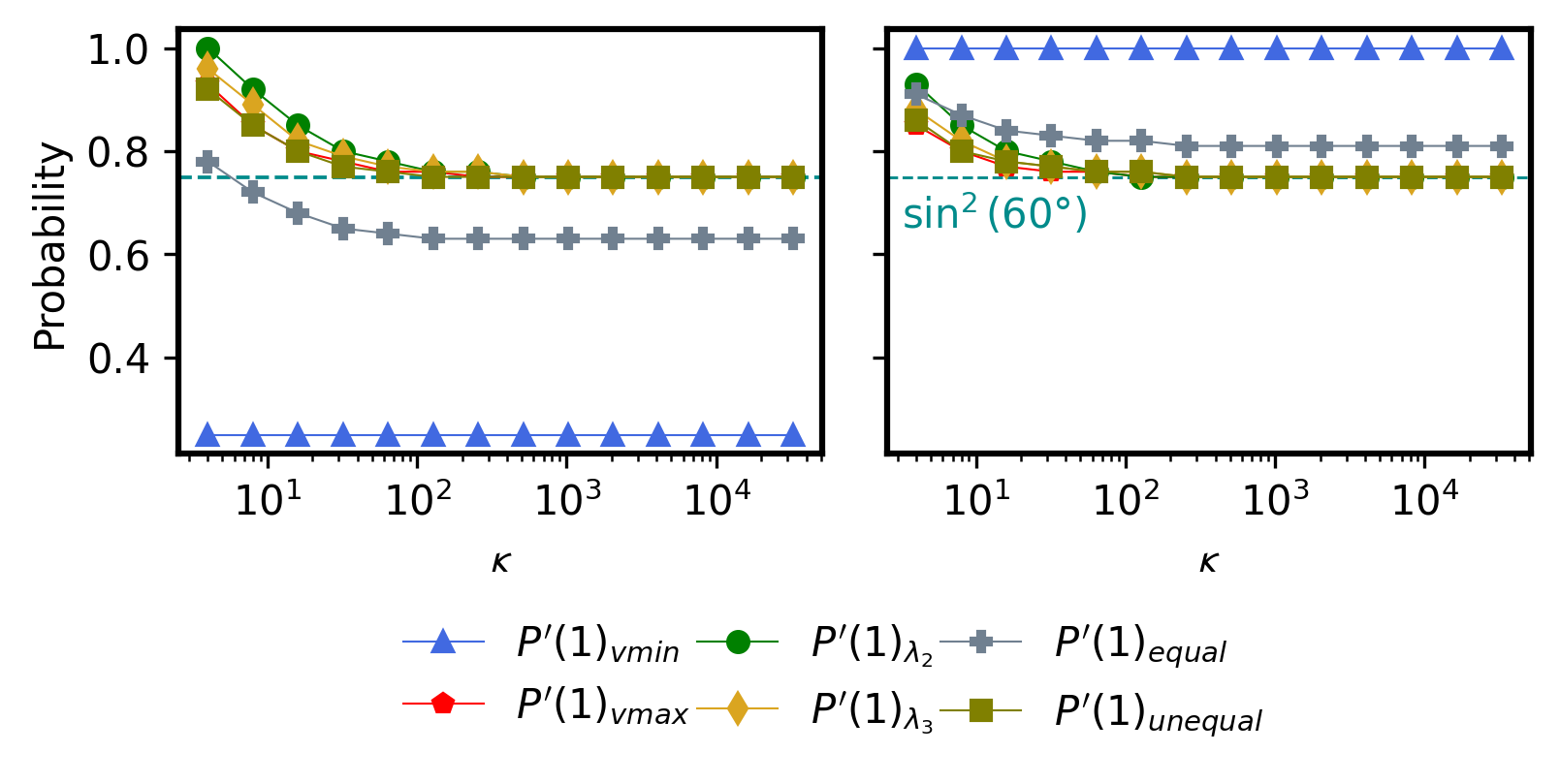}\\
(a)&(b)
\end{tabular}
\caption{Figure showing probabilities of post-selecting various outcomes in Psi-HHL. Sub-figures (a) and (b) show the behaviour of $P(0)$ and $P'(1)$ respectively with $\mathcal{\kappa}$, for $A$ being a $4 \times 4$ diagonal matrix and with six different choices for the vector $\vec{b}$: $P(0)_{\mathrm{vmin}}$ corresponds to the case where $\vec{b}$ is the eigenvector associated with the minimum eigenvalue of $A$, $P(0)_{\mathrm{vmax}}$ is the case where $\vec{b}$ is the eigenvector associated with the maximum eigenvalue of $A$, $P(0)_{\lambda_2}$ and $P(0)_{\lambda_3}$ correspond to the situations where $\vec{b}$ is the eigenvector associated with two other eigenvalues of $A$, $P(0)_{\mathrm{equal}}$ is the case where $\vec{b}$ is in equal superposition of all eigenstates of $A$ and finally $P(0)_{\mathrm{unequal}}$ considers the scenario where $\vec{b}$ is in an arbitrary superposition of eigenstates of $A$. In the left-most figure of sub-figure (a), we choose the parameter $C$ as $\lambda_{\mathrm{min}}$, that is, the minimum eigenvalue of $A$, whereas in right-most figure of sub-figure (a), $C= 0.5\lambda_{\mathrm{min}}$, that is, $\gamma$ is set to 0.5. Figures (b) represent $P'(1)$ where the legend's labels have a similar interpretation to those of $P(0)$. The HHL and Psi-HHL calculations were done using 1 million shots, and $\alpha$ was set to $60^\circ$ for Psi-HHL. }  
\label{fig:scaling}
\end{figure*}

\noindent The first line uses the property of sub-multiplicativity of matrix norms, that is, $\|A B\|\leq\|A\|\| B\|$. In  Eq.~\ref{eq:prob0}, we recall that we set $C=\lambda_{\rm min}$, $\|.\|_{\rm op}$ is the operator norm, and $\||b\rangle \|=1$. Note that the smallest eigenvalue of this operator is $\sqrt{1-\frac{C^2}{\lambda_{\rm min}^2}}=0$, and corresponds to $P(0)=0$, which leads to expending a large number of shots in HHL$_1$. One could circumvent this issue by appropriately choosing a scaled $C = \gamma \lambda_{\min}$, where $\gamma \in 
( 0,1)$, leading to $P(0)\geq 1-\gamma^2$. This idea has been illustrated via an example as shown in Figure~\ref{fig:scaling}, where we choose the scaling $\gamma=0.5$. \\ 

\noindent b) \textbf {Success probability of the post-selection process in HHL$_2$}: A similar expression for the probability of successful post-selection in HHL$_2$,  $P'(1)$, can be obtained by further expanding Eq.~\ref{eq:postselect1} as follows, 

\begin{eqnarray}\label{eq:p1}
P'(1)&=& \| [\sqrt{(\mathbb{I}-C^2A^{-2}}) \mathrm{sin}(\alpha) \nonumber \\ 
&+& (C A^{-1}) \mathrm{cos}(\alpha) ]|b\rangle  \|^{2}. 
\end{eqnarray}
The operator $B \coloneq \sqrt{\mathbb{I}-C^2A^{-2}} \sin(\alpha) + (C A^{-1}) \cos(\alpha)$ in its eigenbasis $\{|\nu_i\rangle\}$ could be given in the diagonal form as 

\begin{eqnarray} \label{eq:matrix1}
\mathrm{diag}( \cdots, \sqrt{1-C^2/\lambda^2_{\mathrm{max}}}\sin(\alpha)+(C/\lambda_{\mathrm{max}})\cos(\alpha), \cdots, \nonumber \\ 
\sqrt{1- C^2/\lambda^2_{\mathrm{min}}}\sin(\alpha)+(C/\lambda_{\mathrm{min}})\cos(\alpha),  
\cdots ),  \nonumber \\
\end{eqnarray} 

\noindent where the entries that occur in the parentheses after `diag' are the diagonal elements of the matrix. Note that the largest and smallest eigenvalue of the matrix given in Eq.~\ref{eq:matrix1} depend on the choice of $\alpha$. When $\alpha=\pi/2$, the component of that eigenvalue attached to $\sin(\alpha)$ remains, and we do not extract the desired solution. However, when $\alpha$ is too small, say $0$, the component attached to $\cos(\alpha)$ remains and we get the same complexity as the traditional HHL approach. Therefore, $\alpha$ is chosen to satisfy the following properties: 

\begin{itemize}
\item The success probability, $P'(1)$ is large enough.
\item The solution component that matters to us does not vanish in the superpositions that occur along the diagonal entries in Eq.~\ref{eq:matrix1}.
\end{itemize}

We demonstrate the case where $\alpha$ is close to $\pi/2$ and $C=\lambda_{\min}$, in which case the operator in Eq.~\ref{eq:matrix1} could be written in the form where the largest eigenvalue is the first entry and the smallest eigenvalue is the last entry, as given below,

\begin{eqnarray}  \label{eq:matrix2}
\mathrm{diag}(\cdots, \sqrt{1-1/\mathcal{\kappa}^2}\sin(\alpha)+(1/\mathcal{\kappa})\cos(\alpha), \cdots, 
\cos(\alpha)). \nonumber \\ 
\end{eqnarray}

Using once again the sub-multiplicativity of matrix norms in Eq.~\ref{eq:p1}, we can write 

\begin{eqnarray}
P'(1)& \leq & \rm  \| [\sqrt{\mathbb{I}-C^2A^{-2}} \ sin(\alpha) \nonumber \\ 
&+& CA^{-1} \cos(\alpha) ] \|^{2}_{\rm op} \| |b\rangle \|^2 ,
\end{eqnarray}

\noindent where $\|.\|_{\rm op}$ is the operator norm yielding the largest eigenvalue of the operator, and $\||b\rangle\|=1$. Then $P'(1)$ is bounded the following way, 

\begin{eqnarray}\label{eq:bounds}
\cos^2(\alpha)\leq P'(1)&\leq& \rm \bigg|\bigg[\bigg(\sqrt{1- \frac{1}{\mathcal{\kappa}^2}}\bigg) sin(\alpha) + \frac{1}{\mathcal{\kappa}} cos(\alpha) \bigg]  \bigg|^2 \nonumber \\ \nonumber\\ 
&=& \rm \bigg(1- \frac{1}{\mathcal{\kappa}^2}\bigg) \sin^{2}(\alpha) + \frac{1}{\mathcal{\kappa}^2} cos^{2}(\alpha) \nonumber \\ 
&+& 2\left(\sqrt{1- \frac{1}{\mathcal{\kappa}^2}}\right)\frac{1}{\mathcal{\kappa}} \sin(\alpha) \cos(\alpha) \nonumber \\ 
&\leq& \rm sin^{2}(\alpha) + \frac{1}{\mathcal{\kappa}} \ sin(2\alpha) +  \frac{1}{\mathcal{\kappa}^2} \cos(2\alpha) \nonumber \\ 
&-& \frac{1}{\mathcal{\kappa}^3}\sin(\alpha)\cos(\alpha). 
\end{eqnarray}

In this bound, the left hand side is the square of the smallest eigenvalue of the matrix given in Eq.~\ref{eq:matrix2}, while the right hand side is the square of the  largest eigenvalue of the matrix in Eq.~\ref{eq:matrix2}. 

We note that the scaling analysis discussed in this section is independent of the feature extraction module, and is focused on the ease/difficulty of post-selection that happens before it. That is, as long as one can design circuits based on the HHL$_{\mathrm{1}}$ and the HHL$_{\mathrm{2}}$ modules such that mixed and wrong signals cancel to yield the correct signal, the linear scaling is preserved. 

For the choice of $C=\gamma \lambda_{\min}$, where $\gamma \in ( 0,1)$, one can push the lower bound for the choice of given $\alpha$. This implies that fewer measurements are required to invert matrices with large condition numbers, as compared to the traditional HHL technique illustrated in Figure~\ref{fig:HHL}(a). We obtain 

\begin{eqnarray}
 \rm \bigg|\bigg[\bigg(\sqrt{1- \frac{\gamma^2}{\mathcal{\kappa}^2}}\bigg) sin(\alpha) + \frac{\gamma}{\mathcal{\kappa}} cos(\alpha) \bigg]  \bigg|^2&\leq& P'(1) \nonumber \\ 
 &\leq& \bigg|\bigg[(\sqrt{1- {\gamma^2}}) \mathrm{sin}(\alpha) \nonumber \\ 
 &+& {\gamma} \mathrm{cos}(\alpha) \bigg]  \bigg|^2 . 
\end{eqnarray}

We show this in Figure~\ref{fig:scaling}, where we pick $\alpha=60^\circ$, the choice of $A$ matrix is $A =\begin{pmatrix} 0.25&0.00&0.00&0.00\\0.00&0.75&0.00&0.00\\0.00&0.00&0.50&0.00\\0.00&0.00&0.00&1.00\end{pmatrix}$, and with six different choices of $\vec{b}$ (the eigenvector associated with the minimum eigenvalue of $A$, the eigenvector associated with the maximum eigenvalue of $A$, eigenvector associated with two other eigenvalues of $A$, equal superposition of all eigenstates of $A$, and an arbitrary superposition of eigenstates of $A$). 

While the $A$  and $B$ matrices share the same basis $\{\ket{\nu_i}\}$, an eigenstate of $B$ matrix that achieves the lower bound for $P'(1)$ may not often correspond to the largest/smallest eigenvalue  of $A$ matrix, for a given choice of angle $\alpha$. 

\noindent Note that for an arbitrary input $|b\rangle$ which is a nontrivial superposition of all the eigenvectors $\{|\nu_i\rangle\}$ of $A$, we have the following condition,
\begin{eqnarray}\label{eq:convex}
{P'(1)} &=&\left(\sum_i \left| b_i \left(\sqrt{1-\frac{C^2}{{\lambda}_i^2}} \sin({\alpha}) + \frac{ C}{{\lambda}_i} \cos({\alpha})\right) \right|^2\right) \nonumber \\
&=&\left(\sum_i \mid b_i\mid^2{\lambda}_i'^2\right).
\end{eqnarray} 

\noindent Here, $\{\lambda_i'\}$ are the eigenvalues of the operator $\rm \sqrt{\mathbb{I}-C^2A^{-2}} \ sin(\alpha) + CA^{-1} \cos(\alpha)  $. Eq.~\ref{eq:convex} is a convex combination of the square of eigenvalues, $\lambda_i'^2$, therefore, an arbitrary input $|b\rangle$ will yield a $P'(1)$ satisfying the condition given in Eq.~\ref{eq:bounds}. We augment our findings with the simulations in the subsequent Sec.~\ref{Sec:Results}. 

\subsection{Singular matrices}\label{Subsec:Singular}

When the HHL algorithm is executed for a large system size for some target application, the user ideally should not possess knowledge of some crucial properties of $A$, including its condition number and invertibility. In the extreme event that $A$ happens to be singular and its condition number is not known \textit{a priori}, the algorithm always outputs $0$, that is, $P(1)=0$, and a user has to supply a rather large number of shots before perhaps drawing on the realization that the problem may not produce a $1$. However, Psi-HHL, by construction, circumvents this issue that HHL suffers from. For illustration, we pick two small but extreme examples, as follows, 

\begin{itemize}
\item A $2 \times 2$ matrix given by $ A =\begin{pmatrix} 0.25&0.00\\0.00&0.00\end{pmatrix}$ and the $\vec{b}$ is $\begin{pmatrix} 0.00\\1.00\end{pmatrix}$, 
\item A $4 \times 4$ matrix $A =\begin{pmatrix} 1.00&2.00&0.00&0.00\\2.00&4.00&0.00&0.00\\0.00&0.00&0.00&0.00\\0.00&0.00&0.00&2.00\end{pmatrix} $  and with $\vec{b}=\begin{pmatrix} 0.00 \\ 0.00\\1.00\\0.00\end{pmatrix}$.
\end{itemize} 

\begin{figure}[t] 
\centering
\hspace*{-0.55cm}
\begin{tabular}{c}     \includegraphics[width=8.5cm, height=3.7cm]{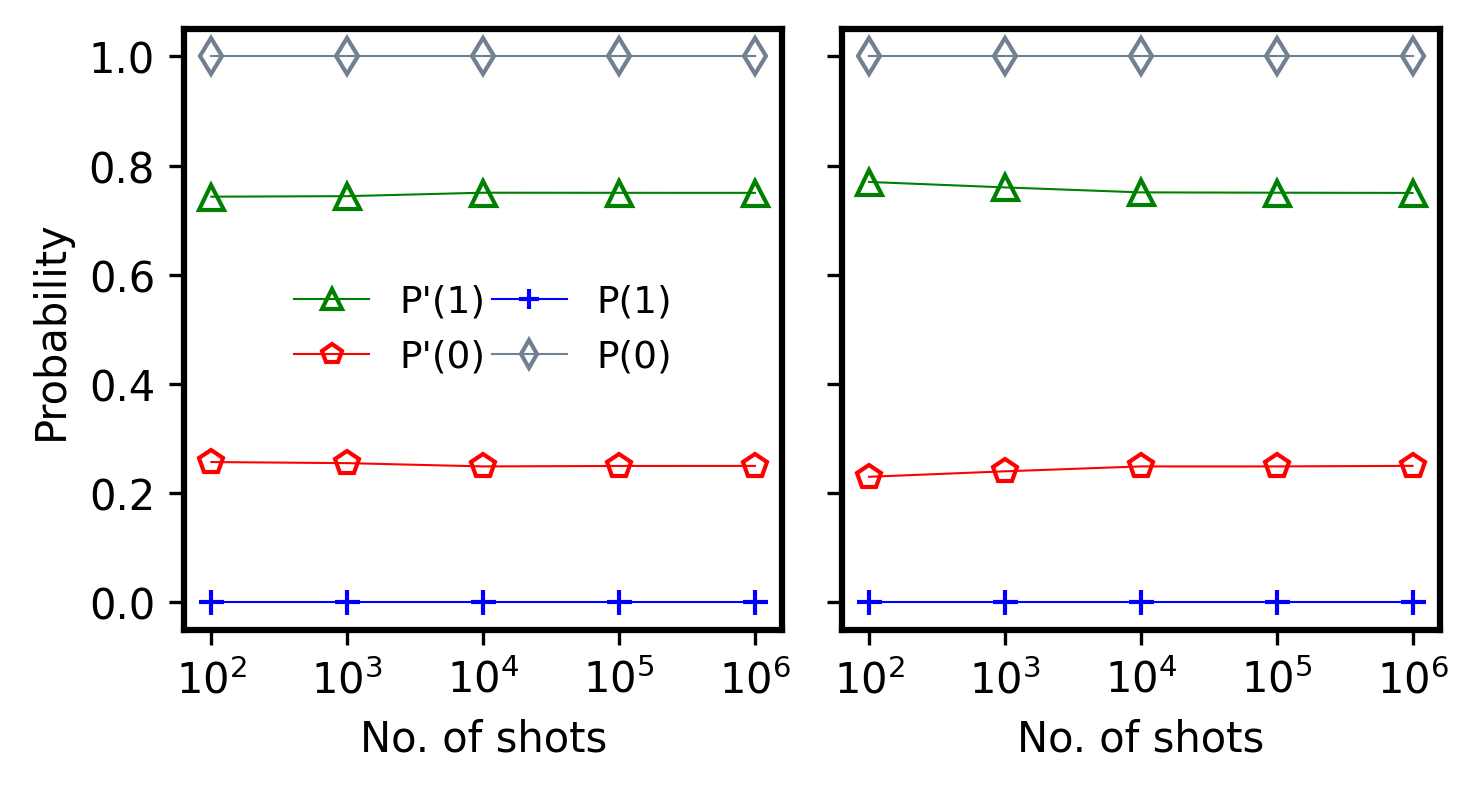}\\ 
\ \ \ \ \ \ \ \ \ \ (a)\ \ \ \ \ \ \ \  \ \ \ \ \ \ \ \ \ \ \ \ \ \ \ \ \ \ \ \ (b) 
\end{tabular}
\caption{Figure showing the performance of HHL and Psi-HHL for singular matrices of size $2 \times 2$ (sub-figure (a)) and $4 \times 4$ (sub-figure (b)). }
\label{fig:singular}
\end{figure}

In both the cases above $\vec{b} \in N(A)$, such that $N(A)$ is the null space of $A$ matrix. We execute HHL as well as Psi-HHL for these two cases by setting $C/\lambda_i$=$0$ whenever the eigenvalues of $A$ are zero, that is, $\lambda_i=0$. The traditional HHL algorithm would expend all the supplied shots without being able to post-select outcome $1$, since $P(1)=0$. The HHL$_\mathrm{1}$ of Psi-HHL successfully post-selects outcome $0$ with probability $P(0)=1$. On the other hand, the HHL$_\mathrm{2}$ module of Psi-HHL post-selects outcome $1$ with probability $\sin^2(\alpha)$ (see Eq.~\ref{eq:hhl2}). The data points presented in Figure~\ref{fig:singular} demonstrate these probabilities for the two examples considered. Both these HHLs yield the solution $|x\rangle=|b\rangle$. The overlap $|\langle x|b\rangle|$ from the HOM module in HHL$_\mathrm{1}$ and HHL$_\mathrm{2}$ both yield a value of 1, with all of the additional $\sin({\alpha})$ factors cancelling out (see Figure~\ref{fig:HHL}). Subtracting the two overlap values results in 0, as expected. Thus, the Psi-HHL approach proves powerful when $P(0) \gg P(1)$. We now consider the case when $\vec{b}$ is an eigenvector of $A$. Here too, one could still get the same result via Psi-HHL by choosing the evolution time, $t$, as a multiple of $2\pi$, which could be incorrect. Therefore, in order to reliably calculate the outputs from both the cases using Psi-HHL, one chooses a non-trivial $t$ in the QPE module, such that it is not an integral multiple of $2\pi$, in order to avoid the possibility of obtaining $\lambda_i=0$ for both the cases. Choosing a non-trivial $t$ will resolve this issue, in the following way: 

\begin{itemize}
\item  In the case where $\vec{b}$ is an eigenvector of $A$, one obtains an eigenvalue $\lambda_i\neq 0$, yielding an associated nontrivial controlled-rotation angle, $\theta_i$. 
\item However, when $\vec{b}$ is a null space vector, one would still obtain an eigenvalue $\lambda_i=0$ leading to a controlled-rotation angle of $0$.
\end{itemize} 

It is worth adding that a striking feature of Psi-HHL is that the approach, unlike amplitude amplification, for example, needs no additional circuit elements or primitives, and thus the requirements on the properties of $A$ to execute the HHL algorithm are the same those for Psi-HHL too. Psi-HHL additionally accommodates the possibility of handling singular matrices, but this comes about not due to requirements from the properties of $A$ but rather the act of subtracting signals. 

\subsection{The Psi-HHL approach for extracting features of the solution vector} 

\subsubsection{Overlap calculation using Psi-HHL}\label{overlap-psihhl} 

We recall that Figure~\ref{fig:HHL}(c) shows how HHL$_1$ and HHL$_2$ are employed as a sub-routines in larger circuits to extract the overlap, $\mathfrak{o}_{x,b}$. We subtract the $\mathfrak{o}_{x,b}$ values associated with these circuits involving HHL$_1$ and HHL$_2$, thus leaving us with the $\mathfrak{o}_{x,b}$ corresponding to the `correct' signal. The gain in taking this route is the efficiency that we achieve in the number of shots expended. We summarize this in the following proposition: 

\newtheorem*{proposition*}{Proposition}
\begin{proposition*}
Psi-HHL yields the same overlap value, $\mathfrak{o}_{x,b}$, as traditional HHL through two implementations of HHL, but is more efficient in the number of shots than HHL, especially when dealing with $A$ matrices with large condition numbers. 
\end{proposition*} 

The overlap obtained from the circuit presented in the top panel of Figure~\ref{fig:HHL}(c) is then given as 
\begin{equation}\label{eqn:wrong}
\mathfrak{o}_{x,b}^{\rm w} = -\||b\rangle_{\rm un}\|^2 \ \||x_{\rm w}\rangle\| \  |\langle b|x_{0}\rangle| . 
\end{equation}
In the expression given above, $|b\rangle_{\rm un} $ is the unnormalized input vector after amplitude encoding, and 
\begin{equation}\label{wrongg}
 |\langle b|x_0\rangle|= \frac{\sum_i  |b_i|^2 \sqrt{1- \frac{C^2}{\tilde{\lambda}_i^2}}}{\| |x_{\rm w}\rangle \|}.
\end{equation}

On the other hand, the circuit presented in the bottom panel of Figure~\ref{fig:HHL}(c) yields 
\begin{equation}\label{eqn:mixed}
  \mathfrak{o}_{x,b}^{\rm m} = -\||b\rangle_{\rm un}\|^2 \ \| |x_{\rm m}'\rangle\| \ |\langle b|x'\rangle|, 
\end{equation}
where
\begin{equation}\label{mixedd} |\langle b|x'\rangle| = \frac{\sum_i |b_i|^2 \left( \sqrt{1- \frac{C^2}{\tilde{\lambda}_i^2}} \sin (\alpha) +\frac{C}{\tilde{\lambda}_i} \cos(\alpha) \right)}{\||x'_{\rm m}\rangle\|}.
\end{equation}
We choose $\alpha$ such that $\sin({\alpha})$ and $\cos({\alpha})$ are  positive and $P'(1)\gg P'(0)$. $P'(0)$ refers to the probability of post-selecting outcome `0' in HHL$_2$. 

In order to obtain the correct overlap value, we subtract Eq.~\ref{eqn:mixed} and Eq.~\ref{eqn:wrong}, thus leading to  

\begin{eqnarray}\label{eqn:EcorrpsiHHL}
 \mathfrak{o}_{x,b}^{\mathrm{Psi-HHL}} \mathrm{cot(\alpha)} &=&  \frac{\mathfrak{o}_{x,b}^{\mathrm{m}}}{\sin (\alpha)} - \mathfrak{o}_{x,b}^{\mathrm{w}} \nonumber \\ &=&  - \|\ket{b}_{\rm un}\|^2 \sum_i |b_i|^2 \frac{C}{\tilde{\lambda}_i}\rm  cot(\alpha). 
\end{eqnarray}
By comparing the right hand sides of Eq. \ref{eqn:Ecorr} and Eq. \ref{eqn:EcorrpsiHHL}, we see that the expressions for $\mathfrak{o}_{x,b}^{\rm HHL}$ and $\mathfrak{o}_{x,b}^{\mathrm{Psi-HHL}}$ are the same as shown below: 

\begin{eqnarray}\label{eq:psihhl}
  \mathfrak{o}_{x,b}^\mathrm{Psi-HHL}  =   \frac{\mathfrak{o}_{x,b}^\mathrm{m} - \mathfrak{o}_{x,b}^\mathrm{w}{\rm sin(\alpha)}}{\rm cos(\alpha)} = \mathfrak{o}_{x,b}^{\rm HHL}. 
\end{eqnarray} 

Although theoretically, we obtain the same overlap value from Psi-HHL, we observe that this technique is superior to traditional HHL in the way it handles large $\mathcal{\kappa}$ matrices, due to the optimal probability of success that one achieves while post-selecting, as we explain in the subsequent section. Therefore, $\mathfrak{o}_{x,b}^{\mathrm{Psi-HHL}} $ can predict the true overlap value, $\mathfrak{o}_{x,b}$, to a reasonable level of precision with fewer shots even for values of $\mathcal{\kappa}$ where HHL procedure fails to do so, which we illustrate in the Results section with examples. We find through our examples that for up to fairly large values of $\mathcal{\kappa}$ that are well beyond those that HHL can handle, Psi-HHL predicts $\mathfrak{o}_{x,b}$ with fewer shots. 

We also add that in problems where it is not immediately obvious \textit{a priori} that the condition number is large, one cannot determine whether to employ HHL or Psi-HHL. However, this is not a problem as Psi-HHL intrinsically involves the regular HHL procedure, and hence, if $P(1)$ is significant, one can choose not to opt for the Psi-HHL procedure. 

\paragraph{Overlap calculation along with sign information} 

It is important to note that the HOM module discussed in Section \ref{overlap-psihhl} gives the absolute value of the overlap. We discuss the Psi-HHL circuits required to obtain the overlap value without losing the sign information. Figures \ref{fig:ohhl}(a) and (b) present the relevant circuits (we see that in order to extract overlap values with sign information, we need to incur controlled versions of $\mathrm{HHL_1}$ and $\mathrm{HHL_2}$ (without measurement and post-selection under control)) for obtaining the wrong and mixed signals respectively. The former gives $\langle b | x_{\rm w} \rangle = ((2\mathcal{P}_R(0) - 1) +i(2\mathcal{P}_I(0) - 1))\mathfrak{\Gamma}$ with $\mathfrak{\Gamma}= \frac{1}{2}(1+ \|\ket{x_{\rm w}}\|^2)$. We note that $\|\ket{x_{\rm w}}\|^2$ is obtained from the post-selection step in Figure \ref{fig:ohhl}(a). $\mathcal{P}$ is the probability obtained in the measurement step at the end of the circuit (as opposed to the usual $P$ or $P'$ that is used for denoting probability that occurs in the post-selection step) and the subscripts `R' an `I' are to denote the circuits used to obtain real and imaginary parts of the quantity of interest. The latter gives $\langle b| x'_{\rm m} \rangle = ((2\mathcal{P'}_R(0) - 1) +i(2\mathcal{P'}_I(0) - 1))\beta$ with $\beta = \frac{1}{2}(1+ \|\ket{x'_{\rm m}}\|^2)$. Here, $\|\ket{x'_{\rm m}}\|^2$ is obtained from the post-selection step in Figure \ref{fig:ohhl}(b). Thus, knowing that $\langle b| x'_{\rm m} \rangle = \sin(\alpha) \langle b |x_{\rm w}\rangle + \cos(\alpha) \langle b| x_{\rm r}\rangle$, we could compute the right signal as 
\begin{equation}
  \langle b | x_{\rm r}\rangle = \frac{1}{\cos(\alpha)}(\langle b| x'_{\rm m} \rangle- \sin(\alpha) \langle b| x_{\rm w} \rangle).
\end{equation}

A step-by-step evaluation of the proposed circuits is discussed in Appendix \ref{SM:ocal}. 

\subsubsection{Expectation value calculation using Psi-HHL}\label{Subsec:Transition} 

\begin{figure}[t] 
\hspace*{-0.55cm}
\begin{tabular}{c}     
\includegraphics[width=8.9cm, height=2.3cm]{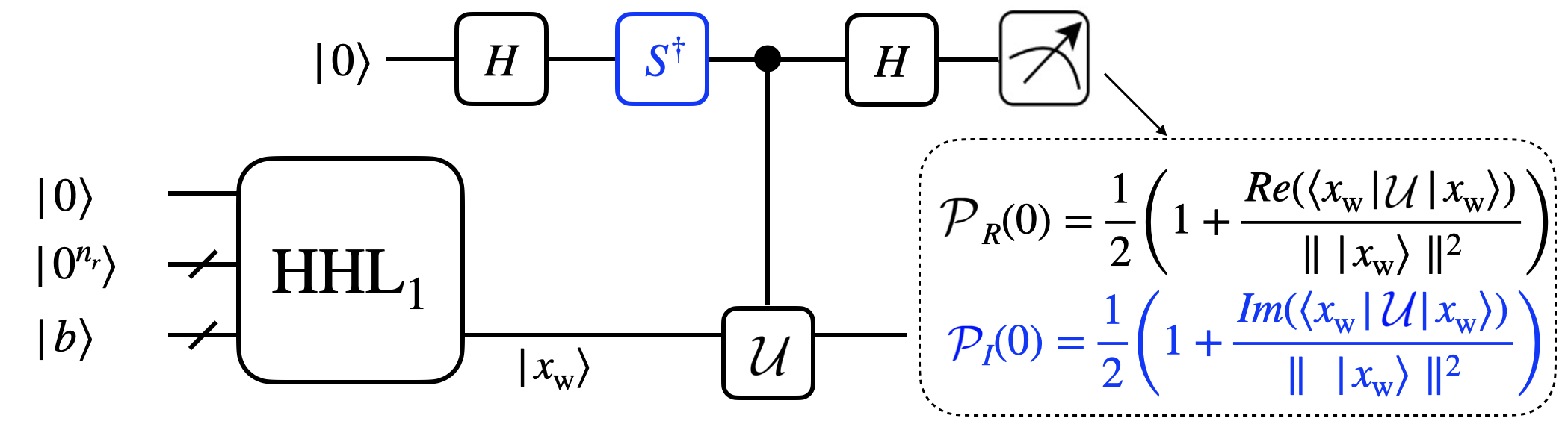} \\ 
(a) \\
\includegraphics[width=8.9cm, height=2.3cm]{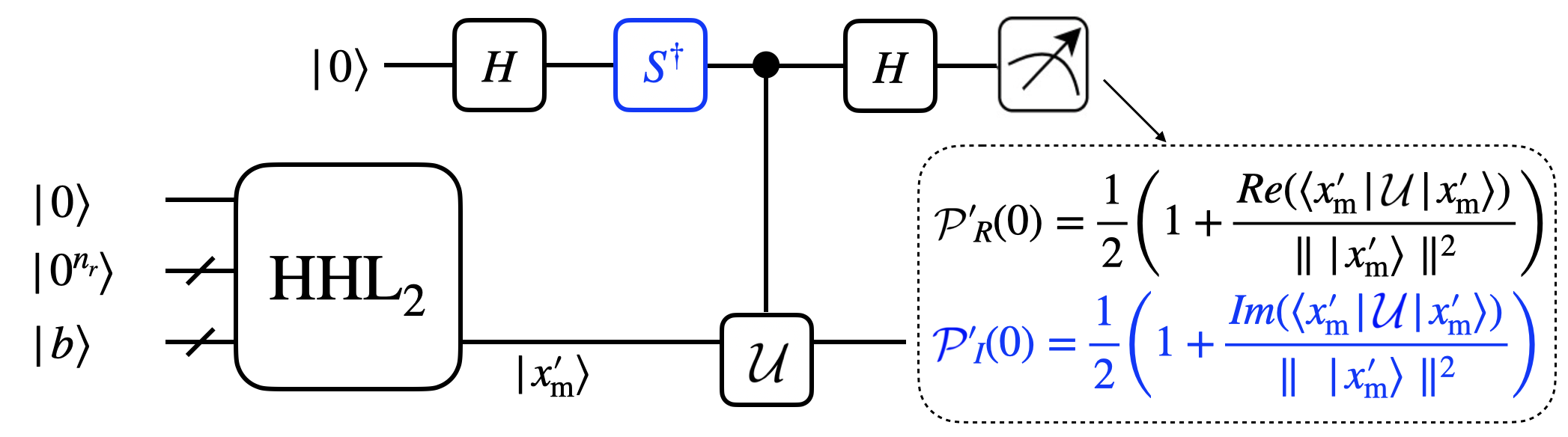} \\  
(b) \\ 
\includegraphics[width=9.5cm, height=2.5cm]{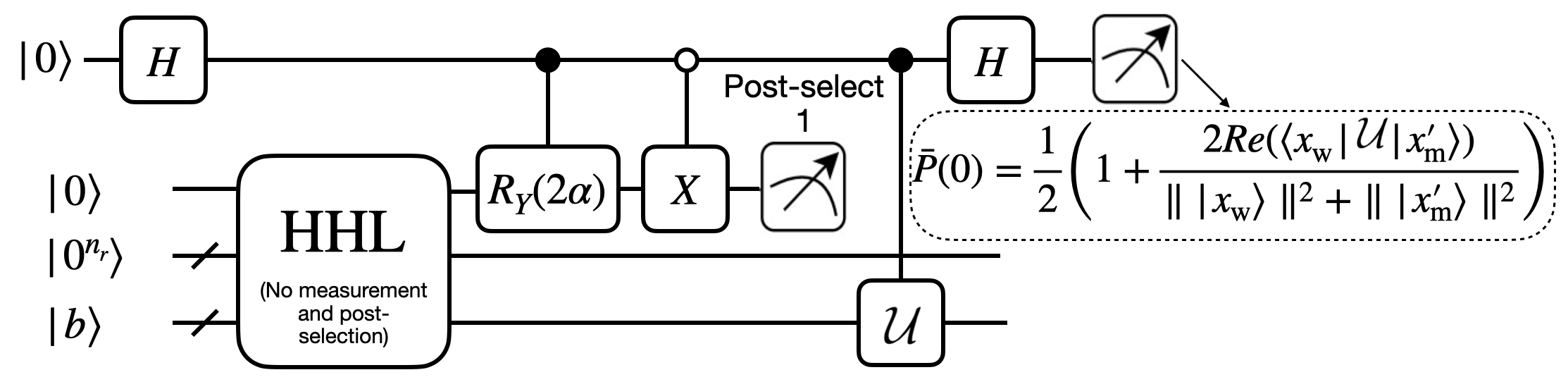} \\ 
(c) \\ 
\end{tabular}
\caption{Circuits for carrying out an expectation value calculation in the Psi-HHL framework. While $\mathrm{HHL_1}$ yields the wrong signal (sub-figure (a)), $\mathrm{HHL_2}$ has not only the mixed signal but also cross-terms (sub-figure (b)). This necessitates employing an additional circuit to cancel the latter (sub-figure (c)), thus recovering the correct signal, that is, the expectation value, $\langle x_{\rm r}|\mathcal{U}|x_{\rm r} \rangle$. In sub-figures (a) and (b), the blue parts are relevant when one evaluates the imaginary part. } 
\label{fig:ev}
\end{figure}

In this sub-section, we discuss the applicability of Psi-HHL to expectation value ($\langle x_{\mathrm{r}}|\mathcal{U}|x_{\mathrm{r}} \rangle$) calculations. It is critical to notice that Psi-HHL for overlap extraction works because of perfect cancellation between the mixed and wrong signals. However, when we extend the idea to expectation value evaluation via the Hadamard test module in place of the HOM module, as shown in Figures \ref{fig:ev}(a) and \ref{fig:ev}(b), a cross-term remains. This is because the circuit in Figure \ref{fig:ev}(b) yields $\bra{x'_{\rm m}} \mathcal{U} \ket{x'_{\rm m}} = \sin^2(\alpha) \bra{x_{\rm w}} \mathcal{U} \ket{x_{\rm w}} + 2\sin(\alpha) \cos(\alpha) Re\bra{x_{\rm w}} \mathcal{U} \ket{x_{\rm r}}  + \cos^2(\alpha) \bra{x_{\rm r}} \mathcal{U} \ket{x_{\rm r}}$. The issue of imperfect cancellation can be addressed by evaluating an additional circuit, as shown in Figure \ref{fig:ev}(c). The probability of measuring $0$ on the first qubit from the top is  $\bar{P}(0)
=  \frac{1}{2}\left( 1+ \frac{2Re(\bra{x_{\rm w}} \mathcal{U} \ket{x'_{\rm m}})}{\|\ket{x_{\rm w}}\|^2 + \|\ket{x'_{\rm m }}\|^2}\right).$ Through this equation, we calculate the term
$Re( \bra{x_{\rm w}}\mathcal{U} \ket{x'_{\rm m}} )$. Thus, the cross-term $Re(\bra{x_{\rm w}}  \mathcal{U} \ket{x_{\rm r}}) = \frac{\delta}{\cos(\alpha)} (2\bar{P}(0) - 1) - \frac{\sin(\alpha)}{\cos(\alpha)}(2\mathcal{P}_R(0)-1) \|\ket{x_{\rm w}}\|^2$, where $\delta=\frac{1}{2}(\|\ket{x_{\rm w}}\|^2 + \|\ket{x'_{\rm m }}\|^2)$. The final correct expectation value, $\langle x_{\mathrm{r}}|\mathcal{U}|x_{\mathrm{r}} \rangle$, is obtained by subtracting the quantities $\bra{x_{\rm w}}  \mathcal{U} \ket{x_{\rm w}} = ((2\mathcal{P}_R(0) - 1) + i(2\mathcal{P}_I(0) - 1))\|\ket{x_{\rm w}}\|^2$ obtained from Figure \ref{fig:ev}(a) and $Re(\bra{x_{\rm w}} \mathcal{U} \ket{x_{\rm r}})$ obtained from Figure \ref{fig:ev}(c), via the term $\bra{x'_{\rm m}} \mathcal{U} \ket{x'_{\rm m}} = ((2\mathcal{P'}_R(0) - 1) + i(2\mathcal{P'}_I(0) - 1))\|\ket{x'_{\rm m}}\|^2$. Its expression is 
\begin{eqnarray}
  \langle x_{\mathrm{r}}|\mathcal{U}|x_{\mathrm{r}} \rangle &=& \frac{1}{\cos^2(\alpha)}(\bra{x'_{\rm m}} \mathcal{U} \ket{x'_{\rm m}} - \sin^2(\alpha)\bra{x_{\rm w}} \mathcal{U} \ket{x_{\rm w}} \nonumber \\ &-&
  2\sin(\alpha)\cos(\alpha) Re\bra{x_{\rm w}} \mathcal{U}\ket{x_{\rm r}}),
\end{eqnarray}
where the values of $\|\ket{x_{\rm w}}\|^2,\ \|\ket{x'_{\rm m}}\|^2$ and $\delta$ are respectively the  probabilities of post-selection in Figure \ref{fig:ev}(a), \ref{fig:ev}(b) and \ref{fig:ev}(c).
Thus, the idea of Psi-HHL works for expectation value evaluation too, with the need to introduce an additional circuit to facilitate cancellation of the mixed and wrong signals to obtain the correct signal. An outline of the protocol is presented in Appendix \ref{SM:ev}. 

\subsubsection{Transition matrix element $(|\langle x_{\mathrm{r}}|W|\zeta\rangle|)$ calculation using Psi-HHL} 

We now briefly comment on the extension of Psi-HHL framework for overlap calculation to transition matrix element ($|\langle x_{\mathrm{r}}|W|\zeta\rangle|$) calculation. Psi-HHL only requires $|x_\mathrm{w} \rangle$ and $|x'_\mathrm{m}\rangle$ as the outputs from the $\mathrm{HHL_1}$ and the $\mathrm{HHL_2}$ steps respectively. We recall from Figure \ref{fig:HHL} that these two states are inputted into their respective HOM modules. Since Psi-HHL does not place any restriction on the \textit{second} state that is inputted to the HOM module, one can always consider an arbitrary state, $|w\rangle = W|\zeta\rangle$, in the place of $|b\rangle$. This naturally extends the applicability of Psi-HHL to transition matrix elements. 

\begin{figure}[t] 
\centering
\begin{tabular}{c}     
\includegraphics[width=7.5cm, height=2.1cm]{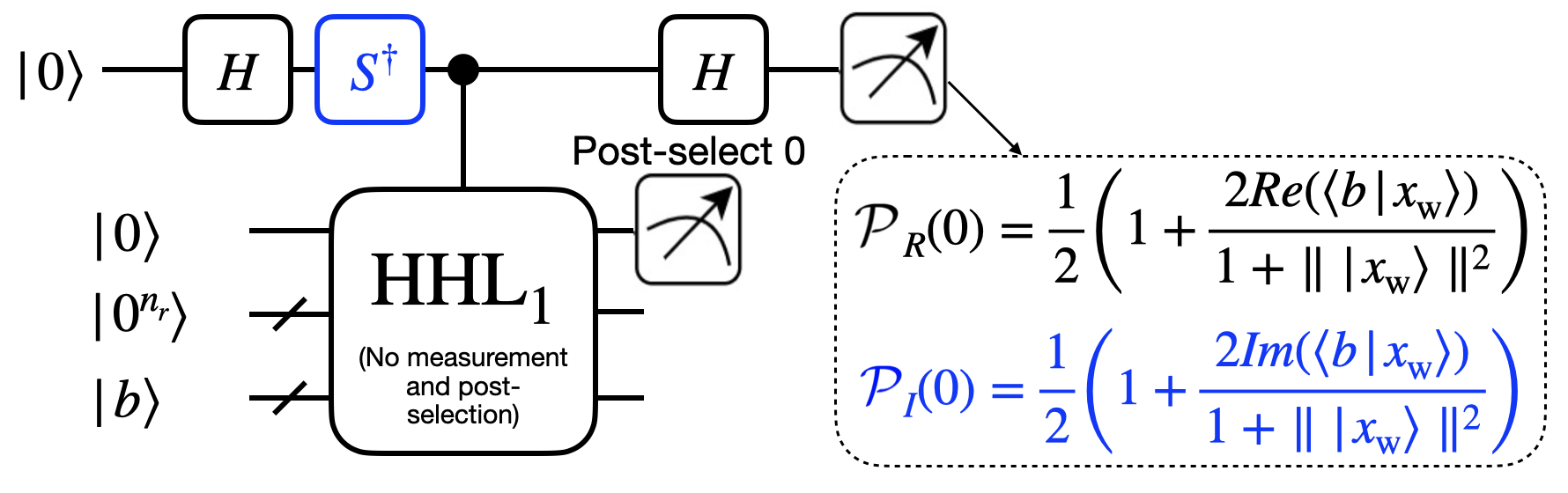} \\ 
(a) \\
\includegraphics[width=8.9cm, height=2.3cm]{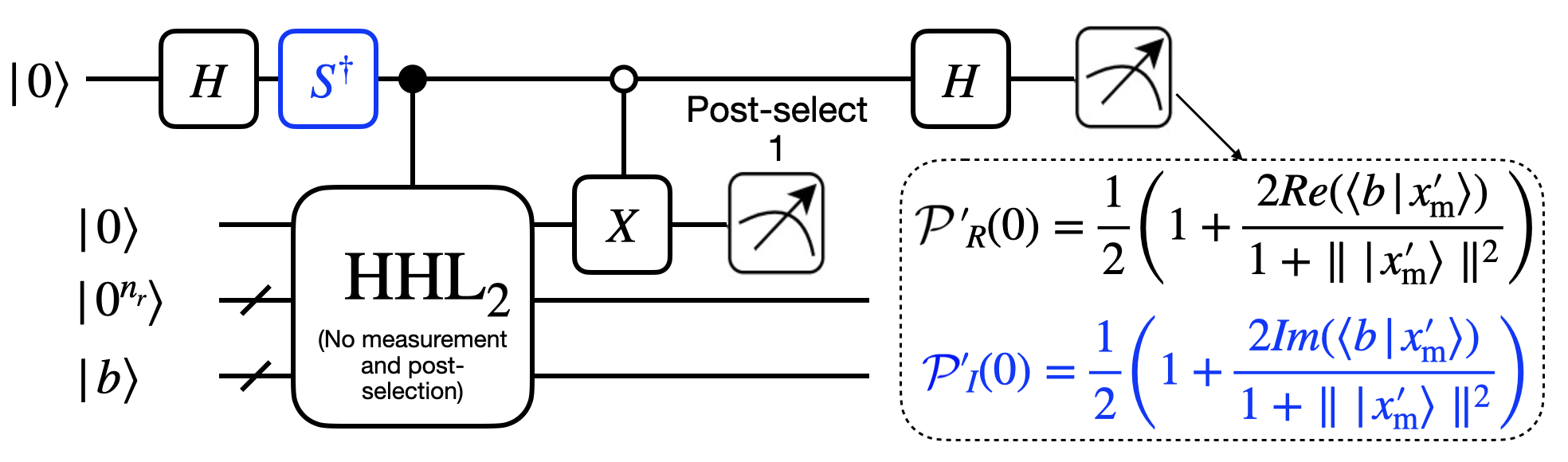} \\  
(b) \\ 
\end{tabular}
\caption{Circuits for obtaining the overlap values in the Psi-HHL framework. The blue parts are relevant when one evaluates the imaginary part. } 
\label{fig:ohhl}
\end{figure} 

\subsection{Psi-HHL as an algorithmic primitive}\label{Subsec:Primitive} 

When HHL is used as a primitive in larger algorithms, it can still be limited when $\mathcal{\kappa}$ is large. In such instances, Psi-HHL can become useful as a primitive. To showcase that the feature extraction still works as intended when Psi-HHL is a primitive in a larger algorithm, we pick a simple case of wanting to not extract a feature of $\ket{x}$, but rather a feature of a state of which $\ket{x}$ is a part. For that purpose, we pick a representative example where we evaluate the overlap $|\langle\Psi_{\rm w/m}|\Phi\rangle|$, where $\ket{\Phi}=V\ket{\chi}$ and $\ket{\Psi_{\rm w/m}}=\frac{1}{\sqrt{2}}(\ket{x_{\rm w/m} y}+\ket{yx_{\rm w/m}})$, where the indices $\rm w$ and $\rm m$ respectively denote the wrong and the mixed signals' solutions. $\ket{y}=\sum_j y_j\ket{\nu_j}$ and $\ket{\chi}$ are states obtained as outputs from either HHL or other algorithm(s). For simplicity, our example is such that state $\ket{y}$ has the same dimension as the state $\ket{x}$. We also make another simplifying assumption that the coefficients are real. We find that Psi-HHL indeed accommodates such scenarios, with no additional overheads. 

\begin{figure}[t] 
\centering
\begin{tabular}{c}     \includegraphics[width=8.5cm, height=3.4cm]{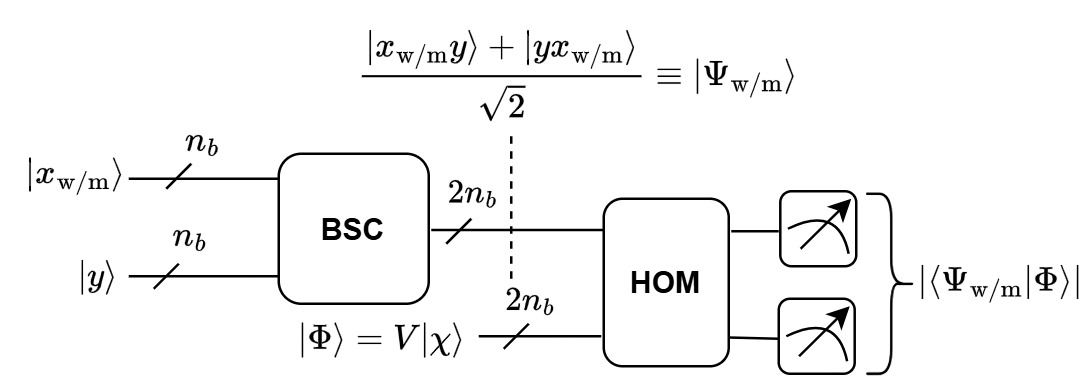}
\end{tabular}
\caption{A schematic of the scenario that we explore for the suitability of Psi-HHL. The feature of interest to us is an overlap, $|\langle\Psi_{\rm w/m}|\Phi\rangle|$, where $\ket{\Phi}=V\ket{\chi}$. `HOM' refers to the Hong-Ou-Mandel circuit module while `BSC' refers to Bell State Circuit which helps to prepare $\ket{\Psi_{\rm w/m}}=\frac{1}{\sqrt{2}}(\ket{x_{\rm w/m} y} + \ket{yx_{\rm  w/m}})$ given the inputs $\ket{x_{\rm w/m}}$ and $\ket{y}$.}
\label{fig:BSC}
\end{figure}

Figure \ref{fig:BSC} presents the representative situation of interest to us. We define the quantities involved in the eigenbasis of $A$, the correct state-vector: 

\begin{eqnarray} 
\ket{\Psi_{\rm r}}&=& \frac{1}{\sqrt{2}}(\ket{xy}+\ket{yx}),\\
&=& \frac{1}{\sqrt{2}}\sum_{i,j}\frac{C}{\lambda_i}b_iy_j\ket{\beta_{ij}}/ \||x_{\rm un}\rangle\|,
\end{eqnarray}

where $\ket{\beta_{ij}}=\ket{\nu_i \nu_j} + \ket{\nu_j \nu_i}$, and 

\begin{eqnarray} 
\ket{\Psi_{\rm w}}&=& \frac{1}{\sqrt{2}}(\ket{x_0y} + \ket{yx_0}),\\ 
&=& \frac{1}{\sqrt{2}}\sum_{i,j} b_i\sqrt{1-\frac{C^2}{\lambda_i^2}}y_j\ket{\beta_{ij}}/ \||x_{\rm w}\rangle\|,
\end{eqnarray}

and the mixed signal is 

\begin{eqnarray}
\ket{\Psi_{\rm m}} &=& \frac{1}{\sqrt{2}}(\ket{x'y}+\ket{yx'}),\\
&=& \frac{1}{\sqrt{2}}\sum_{i,j} b_i \bigg(\sqrt{1-\frac{C^2}{\lambda_i^2}} \sin(\alpha) \nonumber \\ &+& \frac{C}{\lambda_i} \cos(\alpha) \bigg) y_j\ket{\beta_{ij}}/\||x'_{\rm m}\rangle\| . 
\end{eqnarray} 

The state $\ket{\Phi}$ is written as 

\begin{eqnarray} 
\ket{\Phi} &=& \sum_{k,l} v_{kl} \ket{\nu_k \nu_l}.
\end{eqnarray}

Thus, the feature that we seek, that is, the correct signal, is given by 

\begin{eqnarray}
|\langle \Psi_{\rm r}|\Phi\rangle| = \frac{1}{\sqrt{2}}\sum_{i,j} \frac{C}{\lambda_i}b_i y_j (v_{ij} + v_{ji})/ \||x_{\rm un}\rangle\|. 
\end{eqnarray}\label{featfnR}

We now evaluate the expressions for the wrong and the mixed signals, given by 

\begin{eqnarray}
|\langle \Psi_{\rm w}|\Phi\rangle| = \frac{1}{\sqrt{2}}\sum_{i,j} b_i \sqrt{1-\frac{C^2}{\lambda_i^2}} y_j (v_{ij} + v_{ji})/ \||x_{\rm w}\rangle\|,
\end{eqnarray}

and 
 
\begin{eqnarray}
|\langle \Psi_{\rm m}|\Phi\rangle| &=& \frac{1}{\sqrt{2}}\left(\sum_{i,j}b_i\sqrt{1-\frac{C^2}{\lambda_i^2}}b_iy_j (v_{ij} + v_{ji}) \sin(\alpha) \right. \nonumber \\ &+& \left. \frac{C}{\lambda_i}b_iy_j (v_{ij} + v_{ji}) \cos(\alpha) \right) / \||x_{\rm m}\rangle\|, 
\end{eqnarray}\\
respectively, and we can now make use of the relation $\left(|\langle \Psi_{\rm m}|\Phi\rangle| \times \||x_{\rm m}\rangle\| - |\langle \Psi_{\rm w}|\Phi\rangle| \times \||x_{\rm w}\rangle\|\sin(\alpha)  \right) /\cos(\alpha)$ to recover the correct signal as $|\langle \Psi_{\rm r}|\Phi\rangle| \times \||x_{\rm un}\rangle\|= \frac{1}{\sqrt{2}}\sum_{i,j}\frac{C}{\lambda_i}b_iy_j\ket{\beta_{ij}}$. 
The state $\ket{\Phi}$, as Figure \ref{fig:BSC} shows, is constructed by starting with an input state, $\ket{\chi}$ and acting it upon by a unitary $V$. Thus, the feature of interest to us can also be viewed as a transition matrix element. 

\section{Results and discussions}\label{Sec:Results} 

In this section, we consider two sets of examples to compare the performance of Psi-HHL with that of HHL: 

\begin{itemize}
\item \textbf{Toy $A$ matrices of sizes $4 \times 4$, $8 \times 8$, $16 \times 16$, $32 \times 32$, and $64 \times 64$:} For practical reasons associated with (classical) computational resources, we do not choose matrices whose $\mathcal{\kappa}$ grows with the $A$ matrix size. Instead, we generate a sequence of $A$ matrices with systematically increasing condition numbers for every system size. For example, for $4 \times 4$ matrix size, we begin with a sparse and diagonally dominant matrix (the latter is not a requirement but simplifies matters for our illustrations) such that its smallest diagonal entry is $2^{-(n_r-1)}$ and the largest 1. Recall that $n_r$ is the number of QPE clock register qubits. We generate the sequence by varying $n_r$ from 2 to 20. In such a sequence, $\mathcal{\kappa}$ grows exponentially with $n_r$, thus enabling us to reach the failure point of HHL rather rapidly and with existing computational resources. We also note that changing the position of the minimum diagonal element of $A$ along the diagonal does not affect our analysis. 
\item \textbf{Quantum chemistry:} Since quantum chemistry is one of the main applications of interest with quantum computers, it is worth exploring this domain for our analysis \cite{AHL2023}. However, since we no longer are considering toy examples, we begin by \textit{qualitatively} assessing the applicability of HHL to the LCC problem by checking how the $A$ matrices grow with system size (and along the way, defining the system size), thereby providing a handle on the classical hardness of the problem. We also check the growth of $\mathcal{\kappa}$ itself with what we set to be the system size. Following this analysis, we move to our main results for this problem domain. 

\item We carry out the following analyses: 
\begin{enumerate}
\item For the toy matrices, we compare HHL and Psi-HHL, and demonstrate the ability of the latter in predicting $\mathfrak{o}_{x,b}$ across a range of condition number values. We quantify the quality of our results for the $\mathfrak{o}_{x,b}$ values through percentage fraction difference, abbreviated as PFD, given by $\frac{\mathfrak{o}_{x,b}-\mathfrak{o}_{x,b}^{\mathrm{HHL}}}{\mathfrak{o}_{x,b}}\times 100$ for HHL and $\frac{\mathfrak{o}_{x,b}-\mathfrak{o}_{x,b}^{\mathrm{Psi-HHL}}}{\mathfrak{o}_{x,b}}\times 100$ for Psi-HHL. We denote the overlap obtained using a suitable classical algorithm (we term it as `classical value' hereafter) by $\mathfrak{o}_{x,b}$, whereas $\mathfrak{o}_{x,b}^{\mathrm{HHL}}$ and $\mathfrak{o}_{x,b}^{\mathrm{Psi-HHL}}$ are the values for the quantity obtained from HHL and Psi-HHL, respectively. As we shall explain later in this section, we find that since $\mathcal{\kappa}$ grows polylogarithmically for the quantum chemistry case, it is not computationally feasible for us to simulate system sizes where HHL breaks down. 
\item For both the toy matrices as well as the quantum chemistry matrices, we study the PFDs of HHL and Psi-HHL approaches versus the number of shots. 
\item Additionally, for the chemistry case, we compare the performance of HHL and Psi-HHL across a limited range of geometries on a potential energy curve (PEC) for a suitably chosen molecule. 
\end{enumerate}
\end{itemize} 

All of our simulations are developed and executed on the Qiskit software development kit \cite{Qiskit2021}. 

\begin{figure*}[t]
\centering
    \begin{tabular}{cccc}
\includegraphics[scale=0.7]{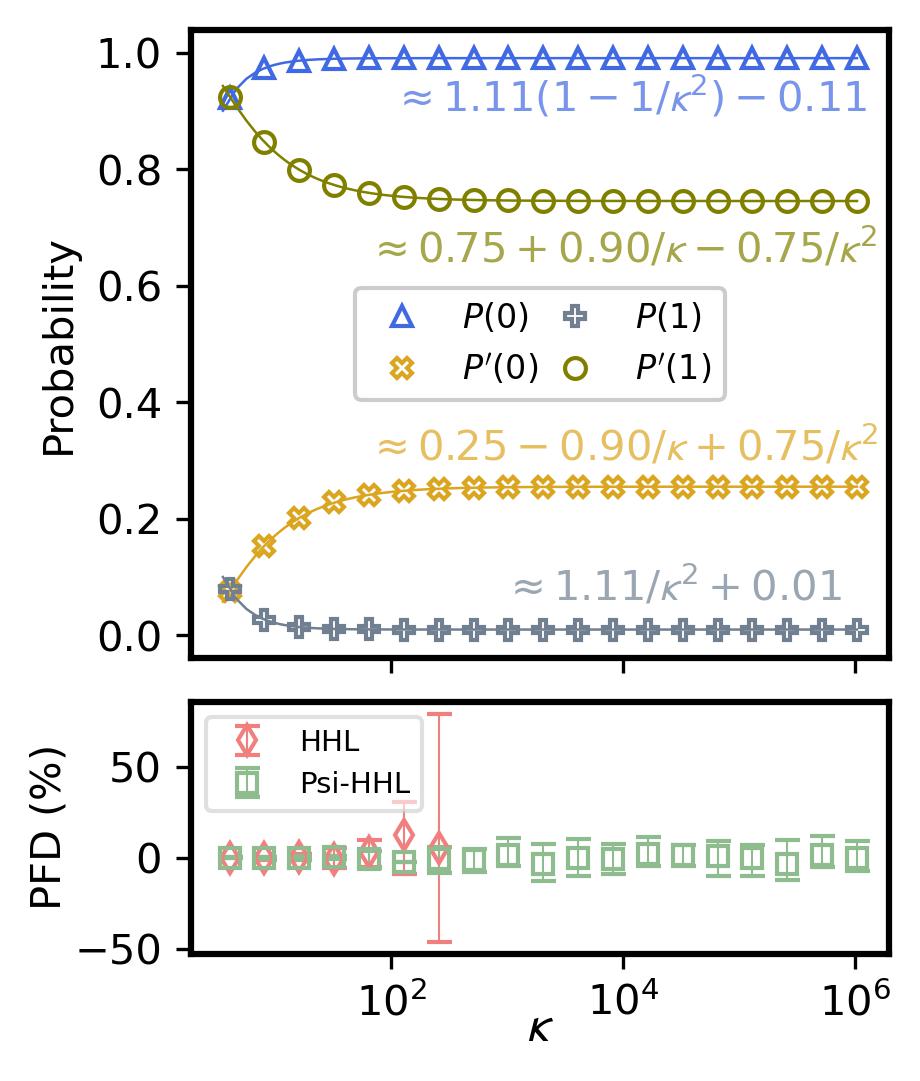}&
\includegraphics[scale=0.7]{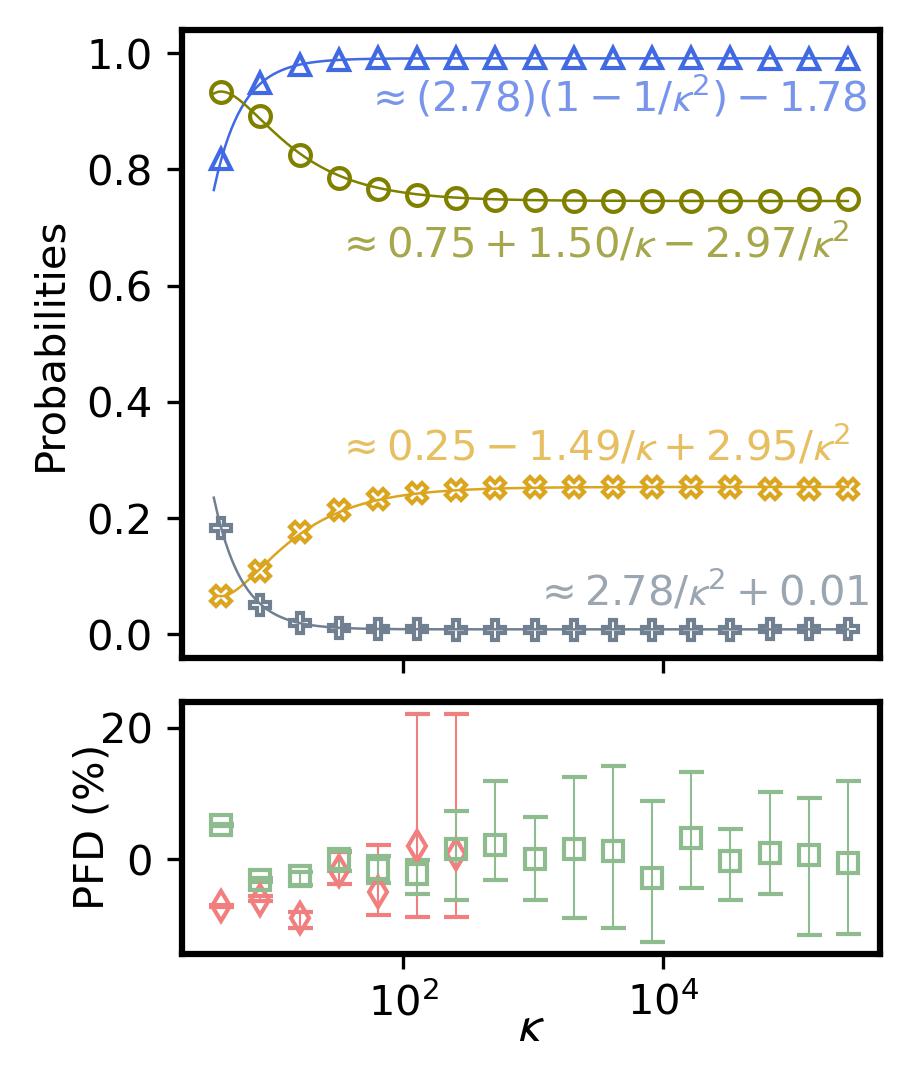}&\includegraphics[scale=0.7]{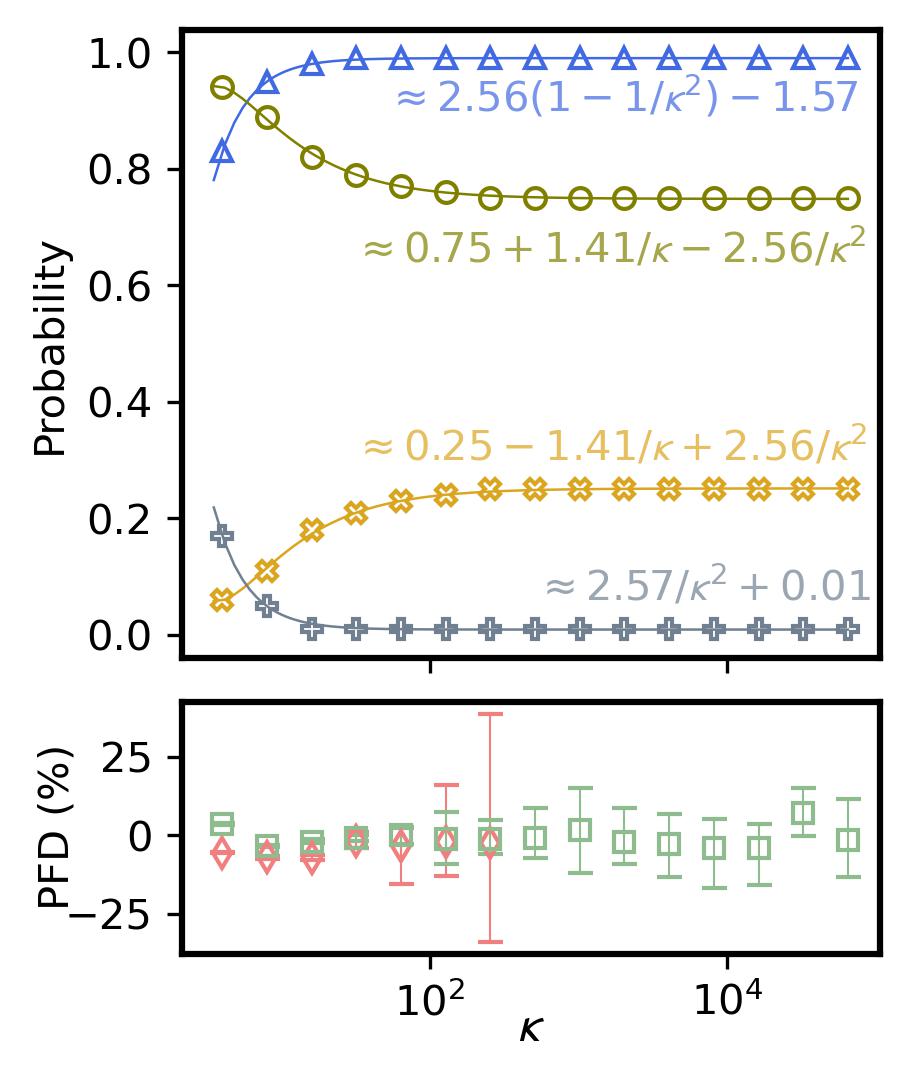}\\(a) &(b)&(c)  
\end{tabular}
\begin{tabular}{cccc}
&\includegraphics[scale=0.7]{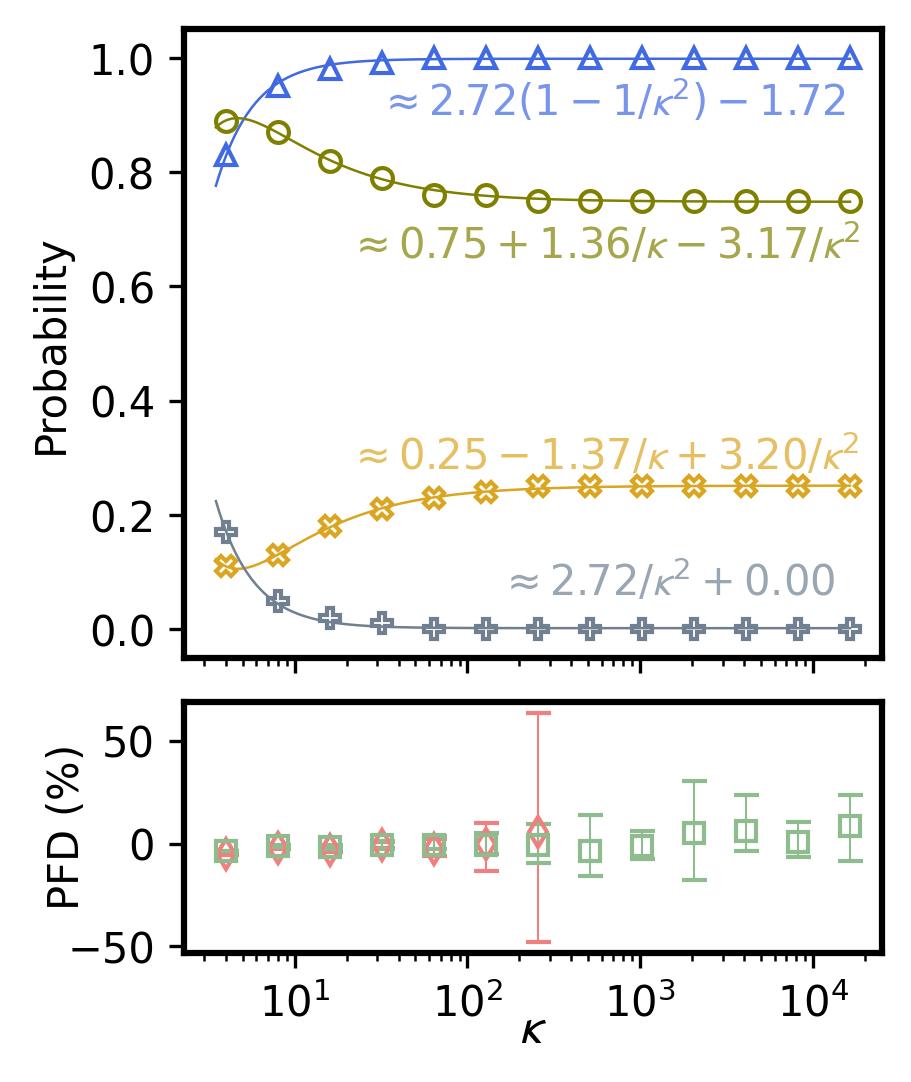}&
\includegraphics[scale=0.7]{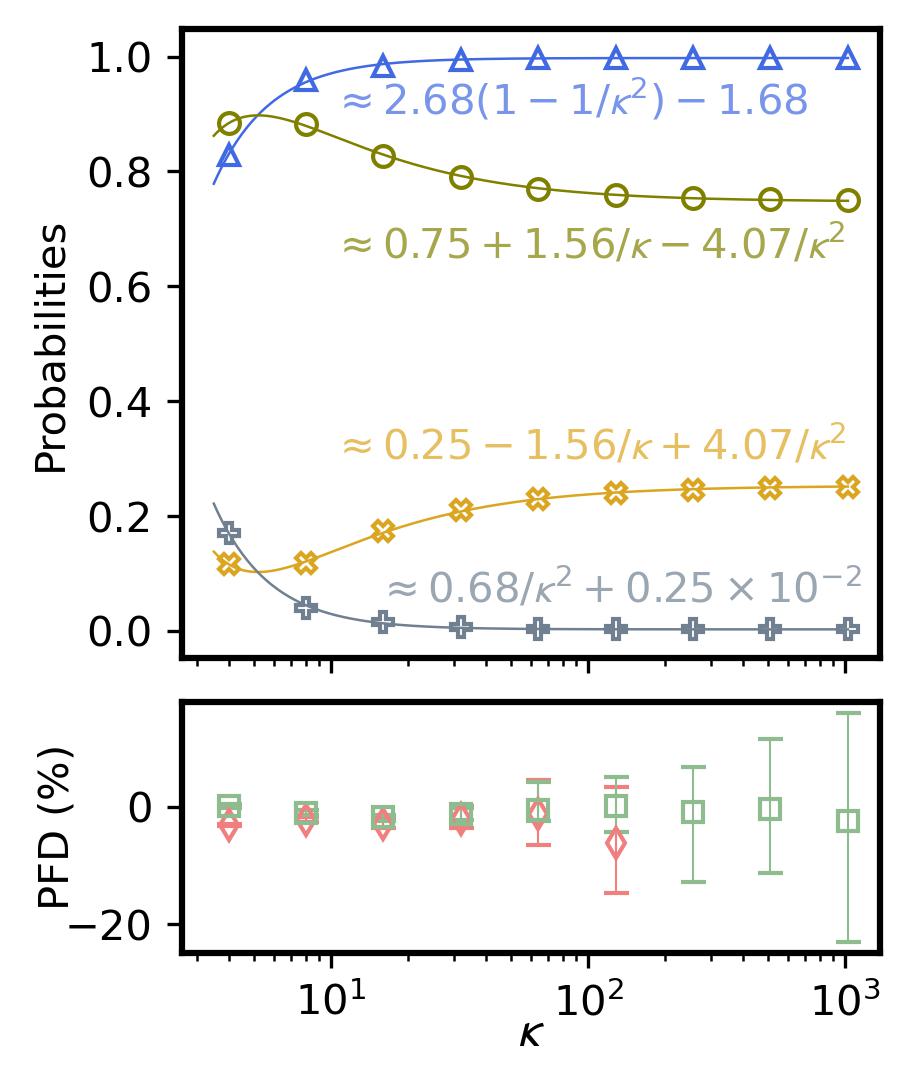}\\
&(d) &(e)  
\end{tabular}
\caption{Figure showing a comparison of the performance of HHL and Psi-HHL approaches for toy matrices. The top panels in each sub-figure present the average probabilities of HHL and Psi-HHL with varying values of $\mathcal{\kappa}$ for different $A$ matrix sizes: $4 \times 4$ (sub-figure (a)), $8 \times 8$ (sub-figure (b)), $16 \times 16$ (sub-figure (c)), $32 \times 32$ (sub-figure (d)), and $64\times 64$ (sub-figure (e)). The bottom panel of each sub-figure represents the average Percentage Fraction Differences (PFD) of the overlap, $E_c$, calculated with HHL and Psi-HHL with respect to the classical value versus $\mathcal{\kappa}$. Each data point results from 10 repeated runs of $10^6$ shots. Also, all the computations have been carried out with $A$ not diagonal and $\vec{b}$ in unequal superposition, as described in the main text. We also note here that the approximate equality is due to the fit coefficients in the figure being rounded off to two decimal places. For the Psi-HHL results, we set $\alpha$ to be 60 degrees. } 
\label{fig:mat-1e6shots}
\end{figure*} 

\subsection{Toy matrices: $4 \times 4$ through $64 \times 64$} \label{randdtoy}

We begin by discussing the details of our $A$ matrices and $\vec{b}$ vectors. We pick our $4 \times 4$ matrix to be non-diagonal, and is given by  

$$A=\begin{pmatrix}
2^{-(n_r-1)}&0.00& 0.00& 0.00\\ 0.00& 0.75& 0.10\times 10^{-3}& 0.00\\0.00& 0.10\times10^{-3}& 0.50& 0.00\\ 0.00& 0.00& 0.00& 1.00
\end{pmatrix},$$
and $\vec{b}=\begin{pmatrix}
0.10\\ 0.01\\ 0.20\\ 1.00
\end{pmatrix}$. The sparsity, $s$, of $A$ is defined as the number of non-zero entries in the row with the most number of non-zero elements, and is set to 2. Our choices for the other matrices, $8 \times 8$ through $64 \times 64$ can be found in Figure \ref{fig:mat-sizes} of the Appendix. We have set the following values for sparsity: 3 for $8 \times 8$, 4 for $16 \times 16$, 5 for $32\times 32$, and 7 for $64 \times 64$ matrices. Throughout, we have ensured that the eigenvalues of the $A$ matrix lies between $1/\mathcal{\kappa}$ and 1. We provide more examples for the $4 \times 4$ matrix size in Section \ref{SM: toy-matrices} of the Appendix. \\ 

Figures \ref{fig:mat-1e6shots}(a) through \ref{fig:mat-1e6shots}(e) provides our findings for $4 \times 4$ through the $64 \times 64$ examples considered. Each data point in the sub-figures is the average result obtained over 10 repetitions, with each repetition involving 1 million shots. For the Psi-HHL results, we set $\alpha$ to be 60 degrees. Each sub-figure provides two sets of data: one which compares the probabilities, $P(0)$, $P(1)$, $P'(0)$, and $P'(1)$, obtained by executing HHL and Psi-HHL, and the other providing the PFDs for the two methods. We choose condition numbers all the way till about 1 million. The figures show that our data fits for HHL align with the expected behaviour of $P(1) \propto \mathcal{\kappa}^{-2}$. We also find that $P'(1)$ scales as $\propto 1$ for large condition numbers (with the fit parameter $a_1=0.75$ matching with the Y-axis intercept for large condition number values). We also consider two more cases, where $A$ is diagonal but $\vec{b}$ is in an equal superposition, and $A$ diagonal and $\vec{b}$ in an unequal superposition. Furthermore, we also consider the situations where $\alpha$ is set to 70 and 80 degrees. The data for all of these cases are presented in Figures \ref{fig:4by4-1e5shots10reps} and \ref{fig:4by4-1e6shots10reps} of the Appendix, accompanied by Tables \ref{tab:A210reps} through \ref{tab:A410reps}. We also verify that the observed trends do not change with 50 repetitions, for which we pick the $4 \times 4$ matrix (with $\alpha = 60^\circ$ for the Psi-HHL calculations) from Figure \ref{fig:mat-1e6shots} as a representative example (see Figures \ref{fig:4by4-1e5shots} and \ref{fig:4by4-1e6shots} of the Appendix). The corresponding data is given in Table \ref{tab:A2} of the Appendix. Results obtained using other choices of angles, $70^\circ$ and $80^\circ$, are provided in Tables \ref{tab:A3} and \ref{tab:A4} respectively. For completeness, we also verify that the fit probabilities from Figure \ref{fig:mat-1e6shots} add to one (see Table \ref{tab:A5} of the Appendix). \\ 

\begin{figure}[t] \label{gHHL-limitation}
\centering
    \begin{tabular}{cc}
    \includegraphics[scale=0.60]{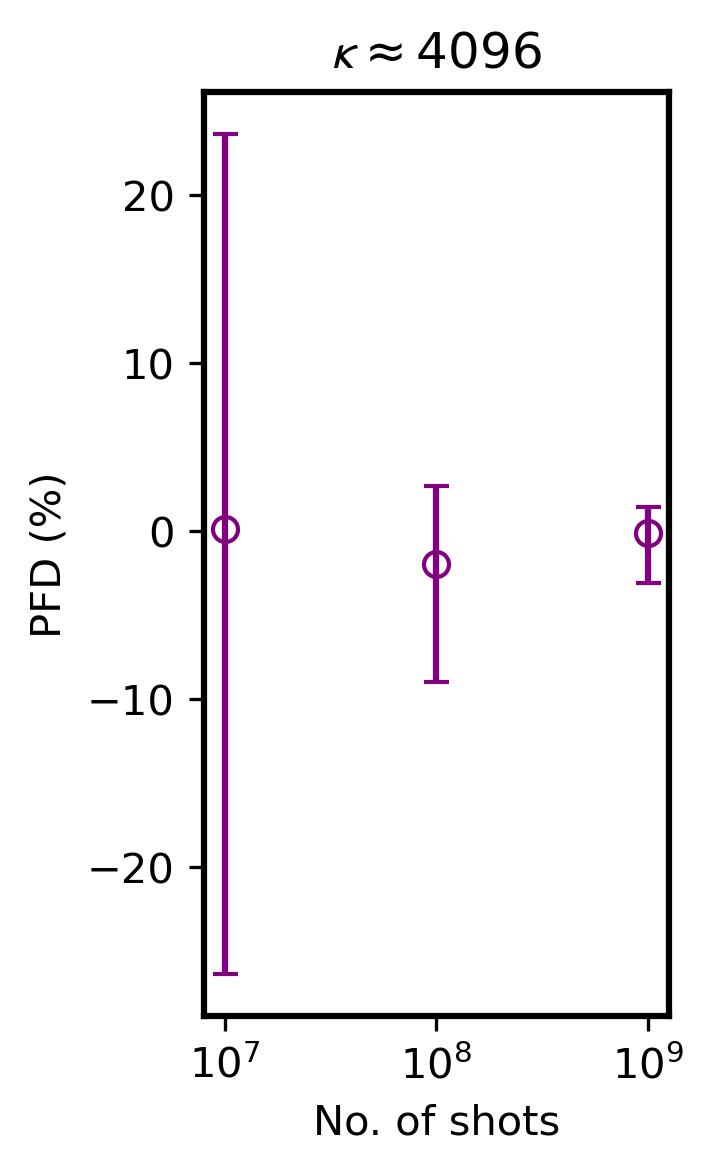} & 
    \includegraphics[scale=0.60]{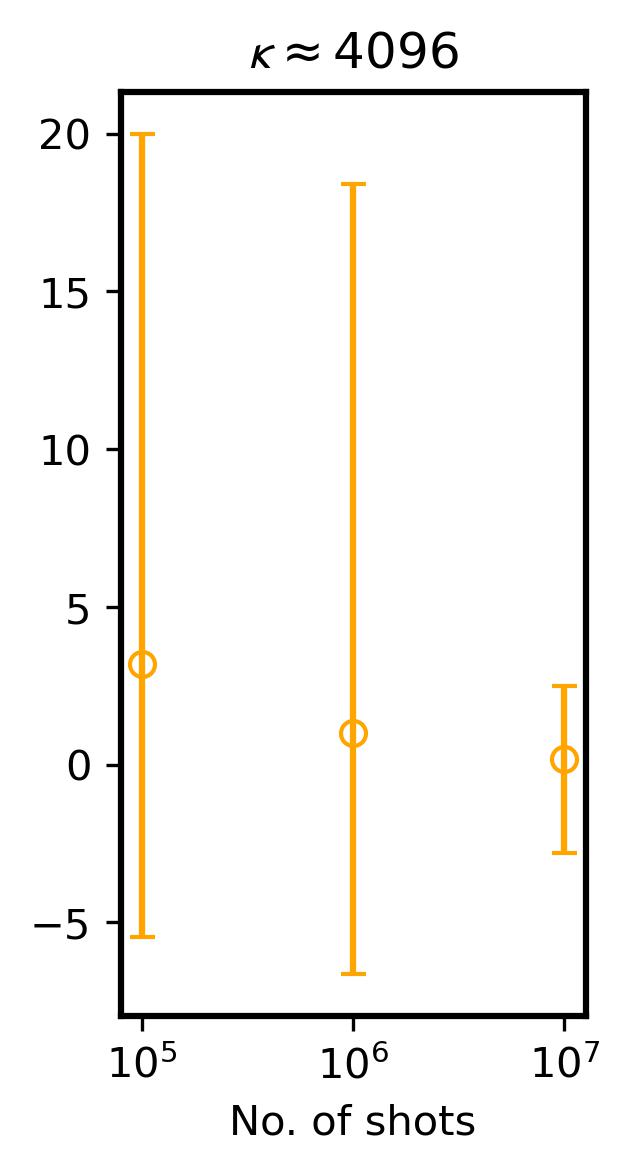}\\(a) & (b) \\ 
\end{tabular}
\caption{Figure showing a comparison of the spread in PFDs as well as the average PFD (over 10 repetitions), obtained using HHL (sub-figure (a)) and Psi-HHL (sub-figure (b)), against a range of three different numbers of shots and for the case where $A$ is of size $4 \times 4$, and for which $\mathcal{\kappa}\approx2^{12}=4096$. }
\label{fig:hhlvspsihhl}
\end{figure} 

We now comment on the observed PFDs from Figure \ref{fig:mat-1e6shots}. We immediately see that HHL fails at some value of $\mathcal{\kappa}$ for the chosen number of shots (indicated by `nan' values in Tables \ref{tab:A210reps} through \ref{tab:A4} of the Appendix). This is due to the fact that as $P(1)$ becomes very small, the fraction of the total number of shots that yield 1 upon measurement tend to very small values. Therefore, the number of shots available for the HOM module become too small to give any reasonable result beyond a certain value of $\mathcal{\kappa}$. Our data also show that the error bars, in which the upper and lower points of the bars reflect the largest and the smallest PFD values obtained using HHL, increases as the condition number increases, till it reaches a critical value of $\mathcal{\kappa}$, beyond which HHL fails. In contrast, we see that for Psi-HHL, even at $\mathcal{\kappa} \ \approx$ 1 million, the length of the error bar is about 17 percent. \\ 

We finally comment on our results from Figure \ref{fig:mat-1e6shots} on the larger size matrices ($8 \times 8$ through $64 \times 64$), where for Psi-HHL, we consider $\alpha=60^\circ$. Figures \ref{fig:8by8-1e6shots} through \ref{fig:64by64-1e6shots} provides the data for $70^\circ$ and $80^\circ$ cases, with Figure \ref{fig:8by8-1e6shots} also giving the information for $10^5$ shots. Tables \ref{tab:a8x8} through \ref{tab:a64x64} in the Appendix provide the accompanying data. We observe that broadly, the trends are similar to what we find for the smaller $4 \times 4$ cases. Figure \ref{fig:hhlvspsihhl1} of the Appendix summarizes the situations where HHL fails for all of the toy matrices considered, as a heat map. We find that for 1 million shots, HHL starts returning `nan' values between $\mathcal{\kappa}$ of $2^8$ and $2^{10}$. In view of the computational cost involved, we did not go all the way up to a condition number of $2^{20}$ for matrices above $4 \times 4$ size, and we stopped much earlier for the $64 \times 64$ case. \\ 

We now turn our attention to Figure \ref{fig:hhlvspsihhl}, where we see that HHL does not fail upon increasing the number of shots, but the number requires to be about 1 billion in order to get a comparable range of PFDs as Psi-HHL, while Psi-HHL only incurs 10 million shots. In particular, HHL gives a maximum PFD of 1.51 percent and a minimum of $-2.98$ percent with a billion shots, whereas Psi-HHL gives a maximum and minimum PFD of 2.33 and $-2.97$ percent respectively with only 10 million shots. \\ 

\begin{figure*}[t] 
    \centering
    \begin{tabular}{c}     \includegraphics[scale=0.65]{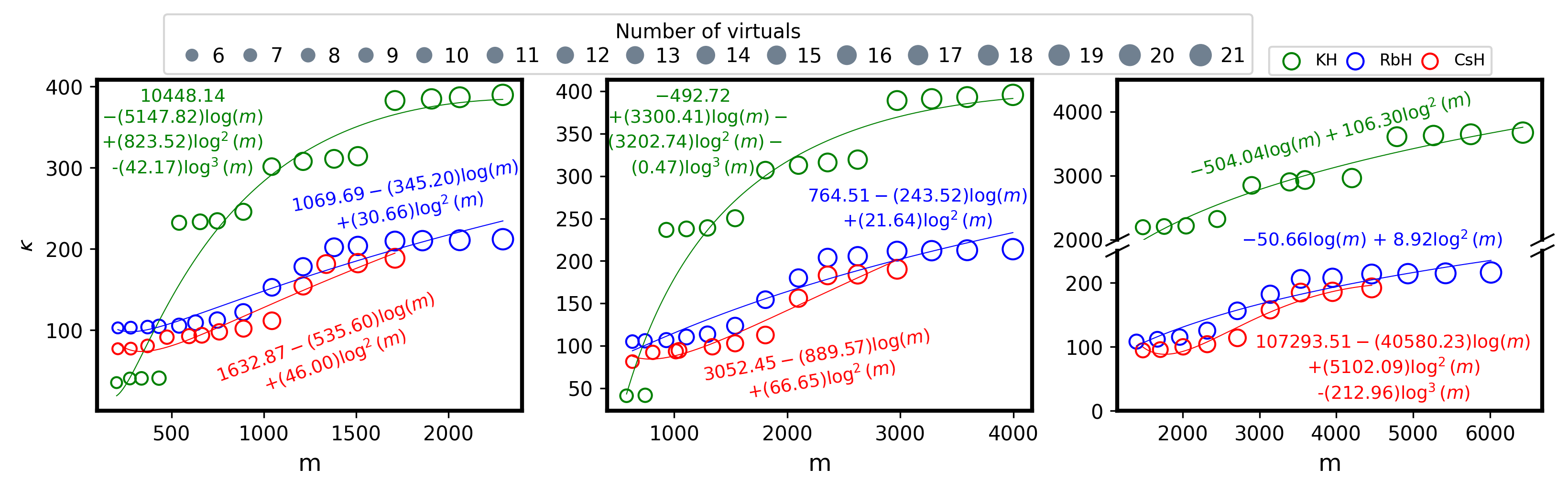} \\
     \ \ \ \ \  (a)\ \ \ \ \ \ \ \ \ \ \ \ \ \ \ \ \ \ \ \ \ \ \ \ \ \ \ \ \ \ \ \ \ \ \ \ \ \ \ \ \ \ \ \ \ \ \ (b) \ \ \ \ \ \ \ \ \ \ \ \ \ \ \ \ \ \ \ \ \ \ \ \ \ \ \ \ \ \ \ \ \ \ \ \ \ \ \ \ \ \ \ \ \ \ \ \ (c) 
    \end{tabular}
\caption{Plots showing the increase in $\mathcal{\kappa}$ with $A_{mol}$ matrix size, denoted as $m$, for the case of 6 (sub-figure (a)), 8 (sub-figure (b)) and 10 (sub-figure (c)) fixed occupied orbitals in the Sapporo-DKH3-DZP-2012 basis set. The legend to the left at the top of the figure shows solid circles whose areas increase proportionally to the number of virtual orbitals considered. } 
\label{fig:kappa}
\end{figure*} 

\subsection{Quantum chemistry matrices} 

Quantum chemistry is regarded as one of the killer applications of quantum computing~\cite{canden}. Therefore, it is interesting to consider the performance of Psi-HHL in this scenario. In particular, we consider quantum chemistry problems via the LCC approach~\cite{AHL2023}. \\ 

We start with preliminary \textit{qualitative} remarks on the suitability of HHL/Psi-HHL for carrying out LCC calculations, and to that end, discuss the growth of the $A$ matrix dimension and $\mathcal{\kappa}$ with system size. This in turn requires us to define system size. Since our focus is to demonstrate the performance of Psi-HHL and not a deep-dive into chemistry, we follow the following steps: a. choosing a few simple diatomic closed-shell molecules in their ground state and in their equilibrium bond lengths, b. picking a suitable single particle basis, c. deciding on the number of occupied orbitals, and d. systematically increasing the number of unoccupied orbitals (virtuals) for system size. We choose four representative molecules: the light Li$_\mathrm{2}$, the moderately heavy KH and RbH, and the heavy CsH, all in their equilibrium bond lengths (Li$_\mathrm{2}$: 5.0600 Bohr, KH: 4.2300 Bohr, RbH: 4.4726 Bohr, and CsH: 4.7135 Bohr)~\cite{Geum2001,Li2geo}. The adjective `light’ and `heavy’ refer to the atomic number of the heavier atom in a diatomic molecule. We pick the Sapporo double zeta basis sets for KH, RbH, and CsH (K, Rb, and Cs: Sapporo-DKH3-DZP-2012, and H: Sapporo-DZP)~\cite{Saporro}, while for Li$_\mathrm{2}$, we choose the STO-6G basis~\cite{BSE}. We employ the $C_{2v}$ point group symmetry for KH, RbH and CsH computations, while for the case of Li$_\mathrm{2}$, we pick the $D_{2h}$ symmetry. We do not include relativistic effects as our aim is to only compare HHL and Psi-HHL and not account for as many relevant physical effects as possible. We now proceed to discuss the dimension of the $A$ matrix and the condition number scaling in LCC for these chosen molecules. 

\subsubsection{Dimension of the $A$ matrix} 

Since the LCC matrix is not guaranteed to be of dimension $2^N \times 2^N \equiv n \times n$, we make a distinction between the dimension of the LCC matrix, $A_{mol}$, which is $m \times m$, and the dimension of the associated $A$ matrix, $n \times n$. We go from $A_{mol}$ to $A$ via padding, which we explain in Section \ref{SM: chemistry-matrices} of the Appendix. We note that $m$, which we shall hereafter think of as the system size, is the number of particle-hole excitations that one considers for an LCC calculation, which in turn is decided by choice of single particle basis and active space size. We denote the number of occupied orbitals as $n_o$ and virtuals by $n_v$. The number of rows (or columns) of the $A_{mol}$ matrix grows as $\sim n_o^\mathcal{E}n_v^\mathcal{E}$ (the number of orbitals are kept fixed), which is exponential. $\mathcal{E}$ refers to the excitation rank. For example, $\mathcal{E}=2$ is the LCCSD approximation, where S stands for single excitations and D for double excitations. For a given $\mathcal{E}$, as we increase the number of orbitals, the system size, $m$, grows \textit{further} polynomially (for example, see Figure \ref{fig:SM2} in the Appendix). For the purposes of this work, we fix $n_o$ and increase $n_v$ for a given molecule, all within the LCCSD approximation. The technical details involved in the generation of $A$ matrix as well as the $\vec{b}$ vector for our computations are discussed in Section \hyperref[SM: chemistry-matrices-A]{A2.A} of the Appendix. \\ 

\subsubsection{Condition number scaling in LCC} 

Although HHL offers an exponential advantage in terms of system size, it is important to check if for the LCCSD problem, the condition number itself scales favourably with system size, $m$, for the molecules that we consider. We point out towards the end of Section \hyperref[SM: chemistry-matrices-A]{A2.A} of the Appendix that it is sufficient to analyze $\mathcal{\kappa}$ versus $m$; this curve upper bounds $\mathcal{\kappa}$ versus $n$. We increase the size of the active space (via $n_v$), and thus $m$ in a systematic manner, by carrying out an orbital character analysis whose details we explain in Section \hyperref[SM: chemistry-matrices-B]{A2.B} of the Appendix and also accompanying Table \ref{tab:A300}. Furthermore, for this analysis, we consider KH, RbH and CsH, but not Li$_\mathrm{2}$, as the system is too small (in the STO-6G basis) to carry out a $\mathcal{\kappa}$ versus $m$ analysis. We start with $(n_o,n_v)$ of (6, 6), (8, 8), and (10, 10) for all the three molecules, and for each of these cases, we systematically increase $n_v$. Figure \ref{fig:kappa} presents our results. We see that the condition number varies as a polylogarithmic function of system size. This behaviour is also consistent with what one may expect from chemistry intuition: we begin by referring the reader to Table \ref{tab:A200} of the Appendix, where we list the largest and smallest eigenvalues of the $A$ matrix for each value of $m$, for KH, RbH, and CsH. The data shows that the increase in $\mathcal{\kappa}$ as we add virtual orbitals is driven by the decrease of the lowest eigenvalue of the $A$ matrix. We note that every matrix element of the $A$ matrix can be expressed as $\langle \Phi_p|H|\Phi_q \rangle$, where $H$ is the Hamiltonian and $\ket{\Phi_p}$ and $\ket{\Phi_q}$ are the excited determinants built from particle-hole excitations arising from the Hartree-Fock (HF) configuration. The smallest eigenvalue of the sparse and diagonally dominant $A$ matrix, which we saw as being important in determining the growth of $\mathcal{\kappa}$, 
can be qualitatively thought of as corresponding to the energy of the first excited state (assuming that the wave function of the first excited state has small contribution from the HF configuration). Thus, we expect that the decrease in the lowest eigenvalue of $A$ saturates as we keep adding virtual orbitals, and eventually, the lowest eigenvalue of the $A$ matrix will converge to a certain value. Thus, the increase of $\mathcal{\kappa}$ saturates too, thus leading one to expect that the condition number would vary as a polylogarithmic function of system size. \\ 

\begin{figure}[t]
    \begin{tabular}{c}
\includegraphics[scale=0.8]{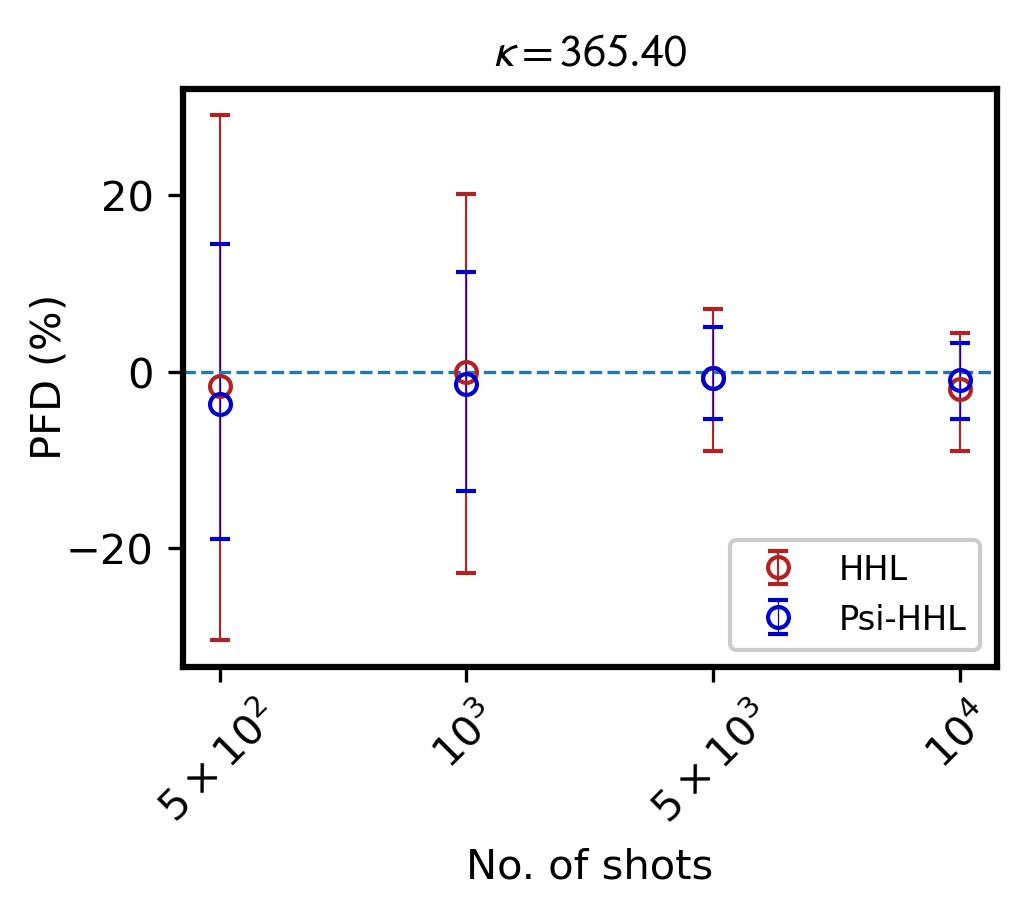}\\
    (a)\\
    \includegraphics[scale=0.8]{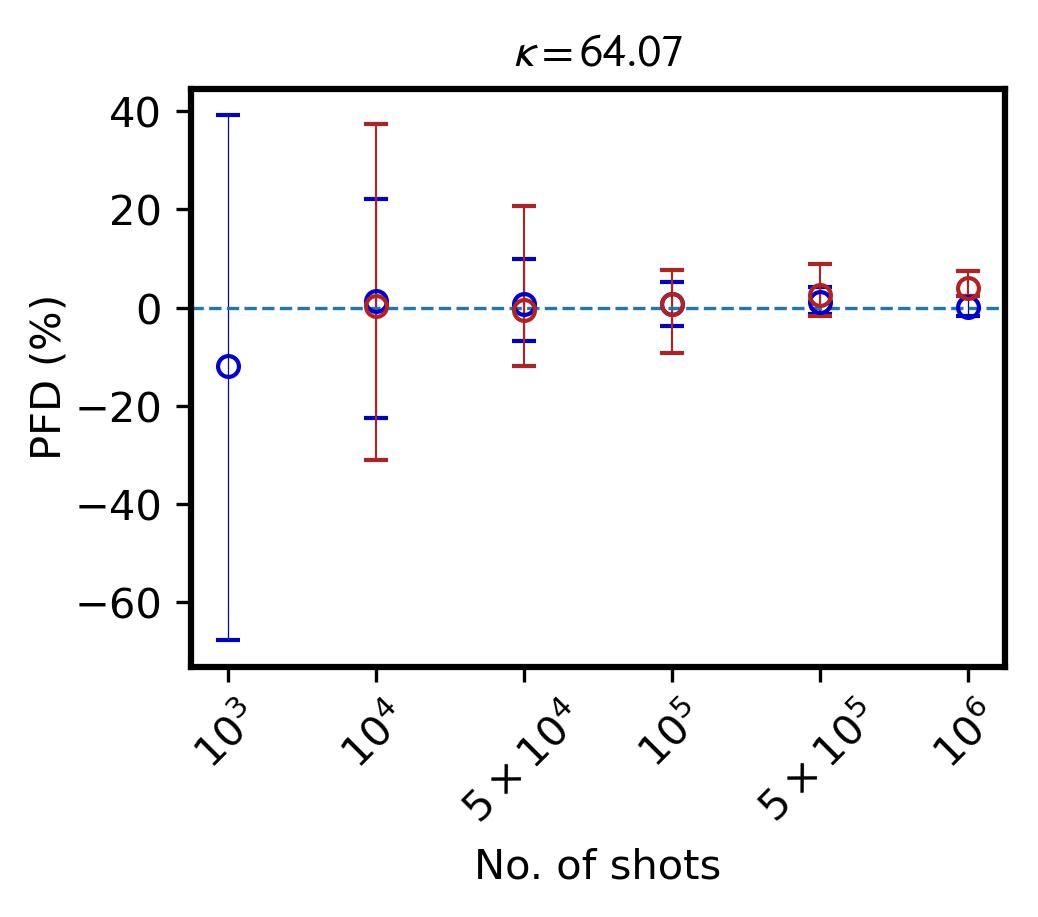}
\\(b) \\ 
\end{tabular}
\caption{Figure showing a comparison of the performance of HHL and Psi-HHL approaches, via PFDs versus number of shots (denoted in the figure as `No. of shots'), for two representative molecules, Li$_{\mathrm{2}}$ (sub-figure (a)) and KH (sub-figure (b)). The centre of each unfilled circle represents the average PFD over a range of 30 repetitions. Each Li$_{\mathrm{2}}$ calculation involves evaluating a 21-qubit circuit, with the $A$ matrix size being $64 \times 64$ ($\mathcal{\kappa}$ of 64.07), while for KH, we evaluate a 26-qubit circuit involving an $A$ matrix size of $256 \times 256$ ($\mathcal{\kappa}$ of 365.40). The zero PFD value is marked with a horizontal line for ease of visualization. }
\label{Qchem-rbh-csh}
\end{figure}

We also observe in each sub-figure (that is, for a fixed number of occupied orbitals) that for large $m$, the condition number of KH is the largest, followed by RbH, and then CsH. This trend can be explained qualitatively using chemical intuition, by looking at the considered orbitals for the (6, 6) case as a representative example. Recalling that each eigenvalue of $A$ can be thought of as an excited state energy, the highest excited state arises from electron excitations from the lowest occupied orbitals, which are $2p$ and $3d$ for KH and RbH cases respectively. It is clear that $2p$ of K has much lower orbital energy than the $3d$ orbital of Rb. Thus, we can expect that the $2p$ $\longrightarrow$ virtual excitation in KH has a higher excitation energy associated with it than the 3d $\longrightarrow$ virtual excitation in RbH. Similarly, we can expect that the $4d$ $\longrightarrow$ virtual excited state of CsH is lower in energy than the $3d$ $\longrightarrow$ virtual excited state of RbH. In summary, we can expect that the excitation energies decrease monotonically as we go from KH to CsH. We combine this observation with our data from Table \ref{tab:A200} where the maximum eigenvalue grows much faster than the decrease in minimum eigenvalue from KH to RbH to CsH for a given choice of $n_o$ and $n_v$, to infer that $\mathcal{\kappa}$ decreases monotonically from KH to CsH. \\ 

Furthermore, in Figure \ref{fig:kappa}(c), KH's condition number reaches a staggering 4000 with all the virtuals considered, whereas in Figures \ref{fig:kappa}(a) and (b), $\mathcal{\kappa}$ is substantially lower. This is not surprising, as for sub-figure \ref{fig:kappa}(c), we consider 10 occupied orbitals. Since KH has only 10 occupied orbitals, the active space contains the $1s$ orbital of K. We can invoke the fact that the orbital energy gap is larger between the inner s orbitals, for example, and the density of orbitals becomes higher in the vicinity of highest occupied molecular orbital and the lowest unoccupied molecular orbital. Thus, if we include inner core orbitals in our active space, we expect the excitation energy of the highest state, namely the largest eigenvalue of the $A$ matrix, to be larger. In contrast, for the 6, 8, and the 10 occupied orbitals' cases, RbH and CsH have comparable condition numbers. This can be attributed to the fact that the lowest orbital is always the $3d$ of Rb in all the three cases. Since the orbital character of the lowest orbital is the same among the three active spaces, they yield very similar condition numbers. \\ 

We add at this juncture that while considering only three molecules does not allow any quantitative inference on the behaviour of condition number with system size, it gives a qualitative indication that quantum chemistry using HHL-LCCSD may be a reasonable direction to consider. However, further analysis would be required to assess what one would expect for $\mathcal{\kappa}(m)$ in general for an HHL-LCC computation. 

\subsubsection{HHL and Psi-HHL results} 

We now discuss our results for the performance of the HHL and Psi-HHL algorithms for the molecules considered in this work.  \\ 

We begin by noting that it is not practical to carry out an analysis similar to that in Figure \ref{fig:mat-1e6shots}, since unlike the toy matrices case, the condition number grows much slower and the computations become too expensive too soon. We thus move to Figure \ref{Qchem-rbh-csh} (with the accompanying data given in Table \ref{tab:A100} of the Appendix), where we carry out a study on the same lines as Figure \ref{fig:hhlvspsihhl}. We vary the number of shots and compare the performance of HHL and Psi-HHL in predicting the PFD, as well as the range across which the PFDs are spread, across 30 repetitions. For this analysis, we consider Li$_{\mathrm{2}}$ and KH. The details are: 

\begin{itemize} 
\item Li$_{\mathrm{2}}$: 21-qubit circuit with an $A$ matrix size of $64 \times 64$ (built out of single and double excitations arising from 3 occupied and 7 virtual orbitals) in the STO-6G basis, with $\mathcal{\kappa}=64.07$. 
\item KH: 26-qubit circuit with an $A$ matrix size of $256 \times 256$ (built out of single and double excitations arising from 6 occupied and 6 virtual orbitals) in the Sapporo double zeta basis, with $\mathcal{\kappa}=365.40$. 
\end{itemize} 

We pick these two systems, since in one case, $\mathcal{\kappa}$ is less than 100, and in the other much larger at over 300. Repeating the calculations for RbH and CsH may not give any new insights, and hence we restrict the results only to Li$_{\mathrm{2}}$ and KH. For Li$_{\mathrm{2}}$, the condition number is very small, and hence we only study the performance of HHL and Psi-HHL between 500 and 10000 shots. The data clearly shows that for Li$_{\mathrm{2}}$, the Psi-HHL approach always has a smaller spread in PFDs when compared to HHL. For KH, we observe that at 1000 shots, HHL fails, while Psi-HHL does not, but comes with large error bars. As we increase the number of shots, the error bars for both HHL and Psi-HHL are seen to become much smaller, but Psi-HHL consistently has smaller error bars than HHL. \\ 

We now move to our next set of results, namely the PEC for the KH molecule. Since it is computationally not feasible to pick the larger system sizes from Figure \ref{fig:kappa}, we choose the 6 occupied and 6 virtual orbitals case. This choice corresponds to a 26-qubit calculation with an $A$ matrix size of $256 \times 256$. We consider geometries in the neighbourhood of the equilibrium bond length (3 geometries in steps of 0.47 Bohr on either side of the equilibrium bond length). With such a choice, the correlation energies required to be captured ranges from about 0.3 mHa (milli-Hartree) for 5.65 Bohr geometry to 3 mHa for the 2.82 Bohr geometry (the two extreme points in the 7 geometries considered), thus allowing us to check the performance of HHL and Psi-HHL across a range of correlation energy values. Calculating energies across a range of bond lengths further allows us to check across a (limited) range of condition numbers (the smallest being 365.40 and the largest 393.21) the behaviour of HHL and Psi-HHL. Each calculation involves 50000 shots and our reported results are averaged over 30 repetitions. For our Psi-HHL computations, we pick $\alpha$ of 60 degrees. Figure \ref{fig:PEC}(a) (and the accompanying Table \ref{tab:A500} of the Appendix) shows the results for $\Delta E$, which is the difference between HHL (Psi-HHL) total energy and and the classical LCC value for the considered geometries. We find that Psi-HHL performs better than HHL in predicting correlation energies, especially on the extreme ends of our PEC when $\mathcal{\kappa}$ values are relatively larger than for those points around the equilibrium geometry. When we examine Figure \ref{fig:PEC}(b), we see that Psi-HHL performs consistently better than HHL. Furthermore, the spread in results across 30 repetitions follows a trend globally: it increases from 2.82 through 5.65 Bohr calculations. We recall that the correlation energies themselves decrease monotonically and by as much as an order of magnitude as we move from 2.82 Bohr to 5.65 Bohr geometries. Thus, although the condition numbers are comparable (for example, 374.93 and 371.38 for the 3.29 and the 4.71 Bohr cases respectively), since the quantity to be captured (the correlation energy) itself is much smaller for the latter, we need to supply more shots to restrict the spread in PFD across repeats. However, since we keep the number of shots fixed throughout the PEC, we observe a larger spread in PFDs. \\ 

\begin{figure}[t]
    \centering
    \begin{tabular}{c}     \includegraphics[scale=0.8]{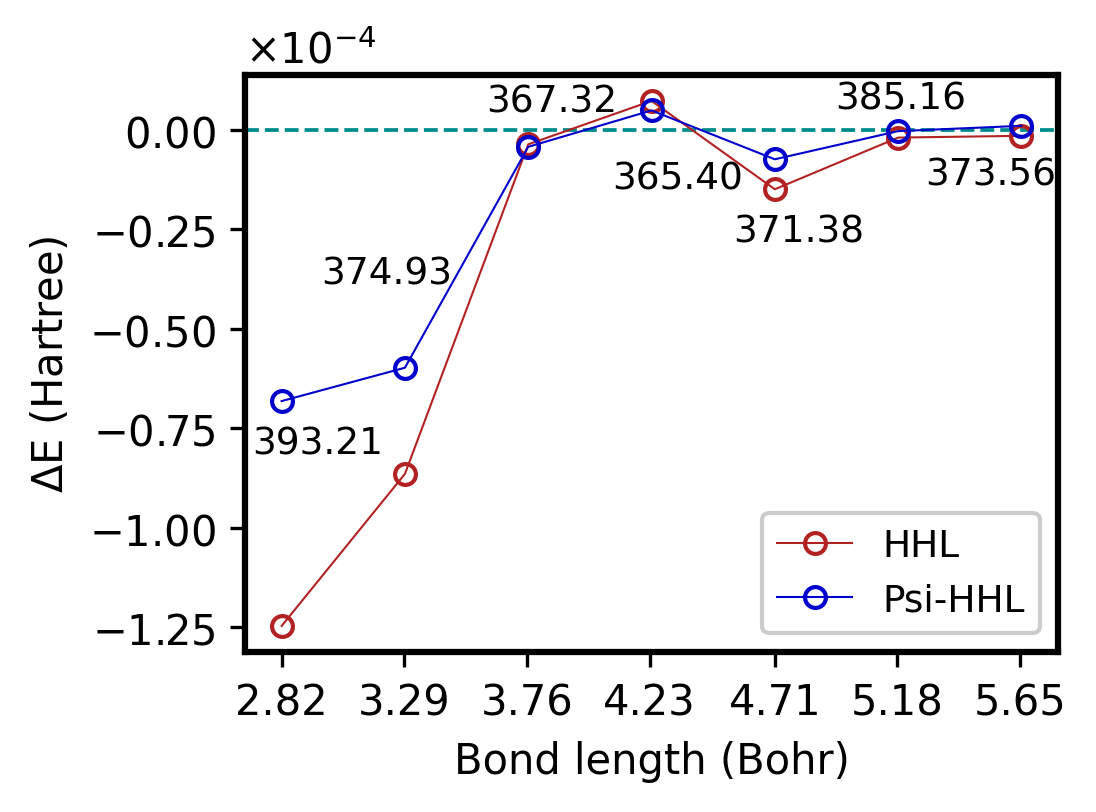} \\
    (a) \\
\hspace{0.4cm}\includegraphics[scale=0.8]{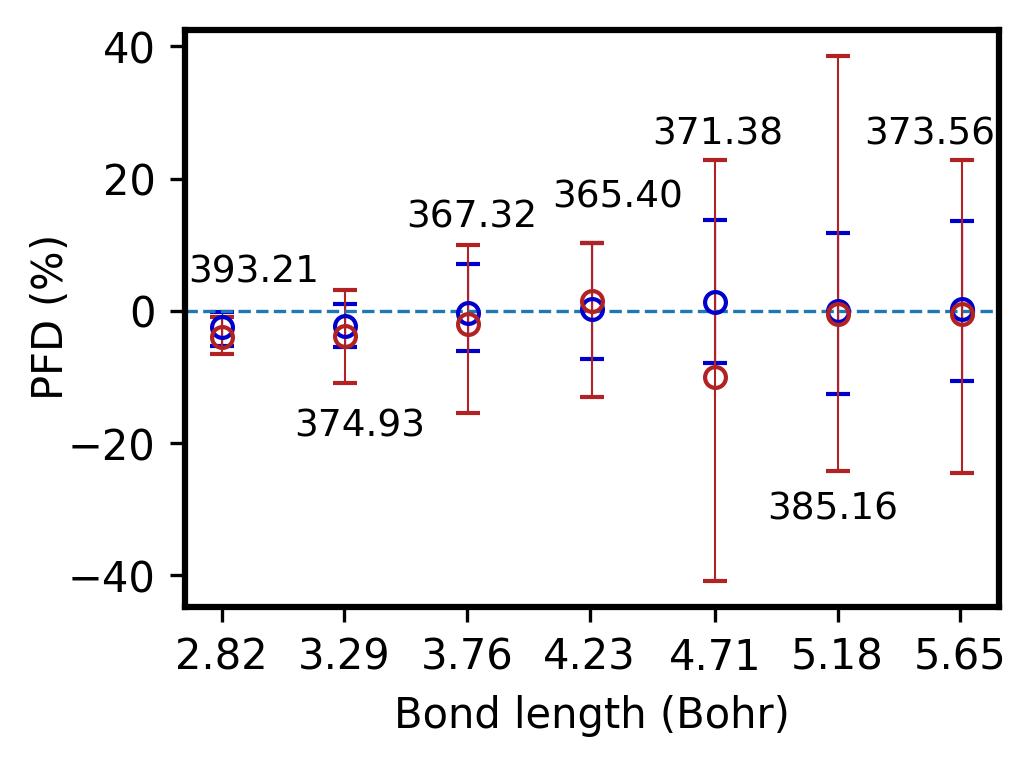} \\
(b)
    \end{tabular}
\caption{Figure illustrating the energy difference ($\Delta E$) between the total HHL/Psi-HHL energy and the classical value of the LCCSD total energy (sub-figure (a)) and PFDs versus bond lengths (sub-figure (b)) of the KH molecule obtained from HHL and Psi-HHL methods (26-qubit calculations involving $A$ matrix size of $256 \times 256$; red and blue circles indicate HHL and Psi-HHL calculations respectively, averaged over 30 repetitions and with 50000 shots). The associated value of $\mathcal{\kappa}$ is shown above or below each data point. } 
\label{fig:PEC}
\end{figure}

\subsection{Scope of Psi-HHL}\label{scope}
    
We begin by recalling the salient features of Psi-HHL: 
\begin{itemize}
\item Superior performance in the large $\mathcal{\kappa}$ regime, as discussed in sub-section \ref{Subsec:Complexity}, and its ability to yield accurate predictions, as indicated by our results from numerical simulations on toy matrices whose sizes range from $4 \times 4$ through $64 \times 64$. 
\item The approach works for extracting overlaps and expectation values (Section \ref{Subsec:Transition}) involving $\ket{x}$. Psi-HHL also works for extracting features of a state of which $\ket{x}$ is a part, as expounded in sub-section \ref{Subsec:Primitive} with a representative example. 
\end{itemize} 

A feature of $\ket{x}$ relies on the properties of $A$ and $\ket{b}$, since $\ket{x}$ is $A^{-1}\ket{b}$. This translates to the feature being affected by the norm of $|| \ket{x}_{\rm un}||$ (which in turn is impacted by the measurement probabilities on the HHL ancillary qubit) and the condition number of $A$. Thus, if the condition number of $A$ is large, Psi-HHL fares substantially better in extracting this quantity (in fact, HHL in its traditional form does not even allow extracting a feature of $\ket{x}$ when $\mathcal{\kappa}$ is large, since obtaining $\ket{x}$ itself is problematic), whereas in low condition number regimes, HHL is superior. \\ 

We now comment on the scope of Psi-HHL in four parts: Requirements on $A$, the role of $\vec{b}$ in HHL and Psi-HHL results, the limitation versus usefulness of Psi-HHL in the low and large condition number regimes respectively, and the role of the magnitude of the overlap itself in the feature to be extracted. 

\subsubsection{Requirements on $A$}\label{Requirements} 

We first address the requirements on $A$ from Psi-HHL in particular. Then, we mention
those requirements on $A$ that Psi-HHL and HHL need in common. \\ 

We begin with a requirement specific to Psi-HHL: Psi-HHL proposes to efficiently
capture a feature of the solution vector when the maximum and minimum eigenvalues of $A$ are well-separated. This is because the approach is designed to leverage the fact that a low probability of obtaining outcome $1$ on the ancilla qubit means that the probability of obtaining outcome $0$ is large. Thus, an additional requirement on $A$ from Psi-HHL is that its condition number be large for the approach to be useful. In such cases, Psi-HHL can efficiently (in the number of shots supplied for an experiment) extract a feature of $\vec{x}$. \\ 

We now move to requirements common to Psi-HHL and HHL: The Psi-HHL approach is built on the subtle notion of carrying out two HHL calculations and subtracting the resultant output signals to extract the correct signal, and thus requirements on $A$ such as $\mathcal{\kappa}$ and $s$ of the matrix having to scale polylogarithmically in system size (in order for an application to be useful via an HHL calculation) carries over from HHL to Psi-HHL. That is, although the HHL and the Psi-HHL approaches accommodate a wide range of applications in principle, the aforementioned constraints on $A$ limit the range of problems to apply the HHL and Psi-HHL algorithms to. \\ 

\subsubsection{The role of $\vec{b}$ on HHL and Psi-HHL results}

While the numerical examples from earlier sections showed the efficacy of Psi-HHL, it is worth noting that the choice of $\vec{b}$ affects the quality of results via P(0) in Eq.~\ref{eqn:p0}. To illustrate this point, we pick an example of a $2 \times 2$ matrix of the form $A = \begin{pmatrix}
    2^{-10}  & 0 \\ 0 & 1 \end{pmatrix}$ with $\mathcal{\kappa}$ thus set to 1024, and $\vec{b}$ being an extreme case at $\begin{pmatrix}
    0\\ 1
\end{pmatrix}$. Our results are presented in Figure \ref{fig:lowkappa}(a). We find that Psi-HHL does better than HHL always in terms of the spread in the PFD across 10 repetitions. However, when we consider the other extreme case, that is, $\vec{b} = \begin{pmatrix}
    1\\ 0
\end{pmatrix}$, $P(1)=1$ always, and thus HHL gives excellent results while Psi-HHL performs marginally worse (we note that the y-axis is between -0.6 and 0.4 percent for Figure \ref{fig:lowkappa}(b) as opposed to -200 to 100 percent for Figure \ref{fig:lowkappa}(a)), as Figure \ref{fig:lowkappa}(b) shows. \\ 

\subsubsection{Analysis of limitation of Psi-HHL in low $\mathcal{\kappa}$ regime versus its usefulness in large $\mathcal{\kappa}$ regime} 

\begin{figure*}[t]
    \centering
    \includegraphics[scale=0.65]{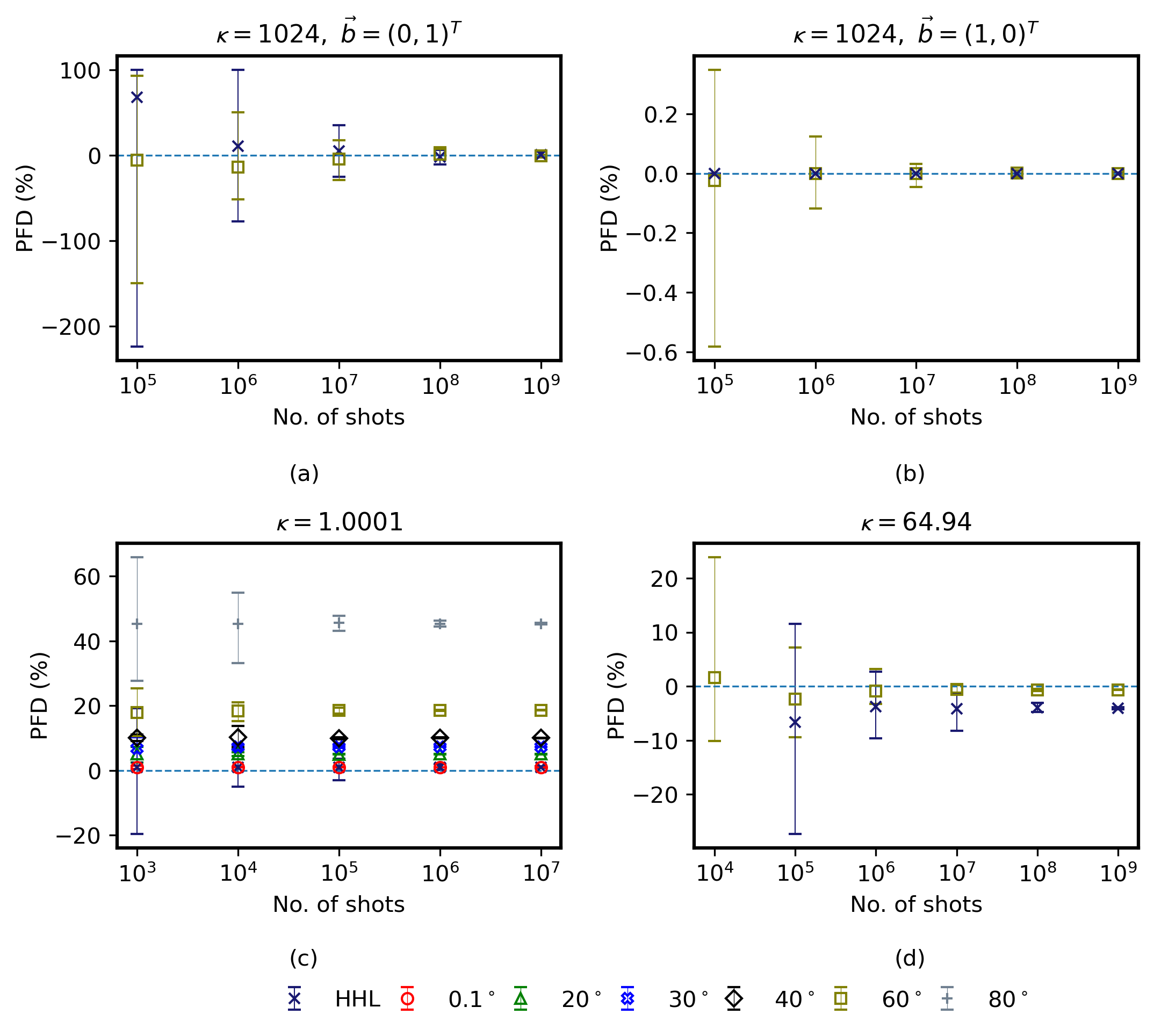}
\caption{Figures presenting our results for (a) a $2 \times 2$ matrix with $\vec{b}=\begin{pmatrix}
0\\1
\end{pmatrix}$, while sub-figure (b) considers the other extreme case of $\vec{b}=\begin{pmatrix}
1\\0
\end{pmatrix}$. Sub-figure (c) presents HHL and Psi-HHL results for a $2 \times 2$ matrix whose $\mathcal{\kappa}$=1.0001, whereas sub-figure (d) gives results for Psi-HHL calculations on a $16\times16$ matrix whose $\mathcal{\kappa}$ is 64. Sub-figures (a) and (b) illustrate the role of $\vec{b}$ in influencing a result, whereas the third indicates the limitation in low $\mathcal{\kappa}$ regime. The last subfigure is an example of the opposite of the third, where HHL yields incorrect PFDs with low spread. In all the sub-figures, each data point is an average over 10 repetitions. } 
\label{fig:lowkappa}
\end{figure*} 

We now comment on the limitation of Psi-HHL when $\mathcal{\kappa}$ is low, and contrast it with its usefulness when $\mathcal{\kappa}$ is large. For this purpose, we pick two representative examples: a simple toy matrix of size $2 \times 2$, where $ A =\begin{pmatrix} 0.25&0.00\\0.00&0.250025\end{pmatrix}$ (thus, $\mathcal{\kappa}$ is 1.0001) and $\vec{b}$ is $\begin{pmatrix} 0.10\\0.20\end{pmatrix}$ for the low $\mathcal{\kappa}$ example, and our $16 \times 16$ toy matrix from earlier sections with a condition number of 64.94 for the large $\mathcal{\kappa}$ scenario. We set the number of required clock register qubits to 11 and 7 for the $\mathcal{\kappa}$=1.0001 and 64.94 cases respectively. In the following paragraphs, we will discuss the results from our calculations, and then provide the rationale for the observed trends. \\ 

Figures \ref{fig:lowkappa}(c) and (d) presents the results for $\mathcal{\kappa}$=1.0001 and 64.94 cases respectively. The minimum number of shots is set to $10^3$ for Figure \ref{fig:lowkappa}(c) and $10^4$ for Figure \ref{fig:lowkappa}(d), since below that value, we sometimes run into the `nan' problem. It is immediately clear from the data points that as the number of shots increase, the spread in the average PFD across 10 repetitions decreases. This is unsurprising, and is consistent with results from earlier simulations. However, we observe that while the choice of $\alpha$ does not influence the average PFDs themselves for large $\mathcal{\kappa}$ case (Figure \ref{fig:lowkappa}(d)), it has a significant impact when $\mathcal{\kappa}$=1.0001 (Figure \ref{fig:lowkappa}(c)). We verify that the observation is a result of insufficient number of shots by carrying out calculations while assuming that we are in the infinite shot limit. We find that in such a case, the PFDs are consistent with 0 for all $0\degree<\alpha<90\degree$. This leads us to expect that in the $\mathcal{\kappa} \rightarrow 1$ limit, both the PFD and the spread can depend on the mixing angle, $\alpha$. \\ 

We now explain the reasoning behind the observations made in Figures \ref{fig:lowkappa}(c) and (d). We specifically focus on two aspects by scrutinizing the observed and theoretical probabilities: the PFD values themselves, which involve data that we average over 10 repetitions, and the spread in PFDs, for which we examine the range in the data from 10 repetitions. Examining the probabilities matter for analyzing the quality of results because for HHL, the relevant probabilities impact the results via $|| \ket{x}_{\rm un}||$ (which is $\sqrt{P(1)}$) when we extract the feature, which we recall is given by $- \|\ket{b}_{\rm un}\|^2 \ \|\ket{x}_{\rm un}\|\ \ |\langle b | x \rangle|$. On the other hand, for Psi-HHL, the quantity that decides the result is $\| \ket{x'_m}\|$ (which is $\sqrt{P'(1)}$).  \\ 

We begin with the PFD values obtained from HHL and Psi-HHL calculations in Figure \ref{fig:lowkappa}(d), that is, the $\mathcal{\kappa}$=64.94 case. The incorrect PFD values predicted by HHL even at 1 billion shots can be backtracked to the discrepancy in the obtained $P(1)$ value of 0.007 upon averaging over 10 repetitions, as opposed to the theoretical value of 0.01 (obtained by using the relevant expression from Section \ref{hhl-algo}). This means that one needs to supply more shots to be able to recover the correct $P(1)$. This is in stark contrast to the obtained $P'(1)$ of 0.7668 from HHL$\mathrm{_2}$ (the HHL$_{\mathrm{1}}$ module always has adequate shots in the large $\mathcal{\kappa}$ regime, and thus we only need to focus on HHL$\mathrm{_2}$) that is comparable to the theoretical value of 0.7674. In fact, the observed (calculated using simulation) and the theoretical $P'(1)$ values are the very same at 0.7668 and 0.7674 respectively even with $10^7$ shots. \\ 

Next, we discuss the spread in the PFD values across 10 repetitions observed in Figure \ref{fig:lowkappa}(d). For this case, we see that the spread in $P(1)$ is very low, and is between 0.0070511 and 0.0070574, and that for $P'(1)$ is between 0.76676 and 0.76678. Thus, the PFDs also show minimal spread in the plots. \\ 

We now move to the PFD values obtained from HHL and Psi-HHL computations for the $\mathcal{\kappa}$=1.0001 case, with the results presented in Figure \ref{fig:lowkappa}(c). Here, the theoretical value of $P(0)$ is  0.00015 whereas $P(0)$ from our simulation is 0.0171 even at $10^7$ shots. On the other hand, $P(1)$ is 0.999 and 0.982 for theoretical and simulated values respectively. \\ 

We check the spread in the PFD values across 10 repetitions for the data pertaining to Figure \ref{fig:lowkappa}(c), for $10^7$ shots case. Here too, the $P(0)$ values lie in a narrow range (between 0.001137 and 0.001166), as do the $P'(1)$ values (between 0.99883 and 0.99886), hence giving the low spread in PFDs. \\  

Although the considerations above place a limitation on Psi-HHL in the low condition number regimes, this is unlikely to be a limiting factor to the scope of problems that one can apply the algorithm to in practice. This is because for sufficiently large system sizes where classical computers struggle and where Psi-HHL is expected to become useful, it is very unlikely that $\mathcal{\kappa}$ would be small. Furthermore, one could always begin with HHL and check the output histograms for both 0 and 1, and opt for HHL$_2$ if the former is sufficiently small (with `sufficiently' being decided by the application of interest) or go ahead with the HHL computation if the former is sufficiently large. 

For completeness, we discuss the $\mathcal{\kappa}$=1 extreme case, since $C/\lambda_i$=1 $\forall i$, it is easy to verify that the wrong signal described in Eq. \ref{wrongg} as well as the wrong part of the mixed signal from Eq. \ref{mixedd} are zero, thus leaving behind only the correct signal. The additional cosine term in the correct part of the mixed signal is cancelled out due to the sine and the cotangent present in Eq. \ref{eqn:EcorrpsiHHL}. \\ 

\subsubsection{The role of the magnitude of the overlap to be extracted}

\begin{figure}
\centering
    \includegraphics[scale=0.7]{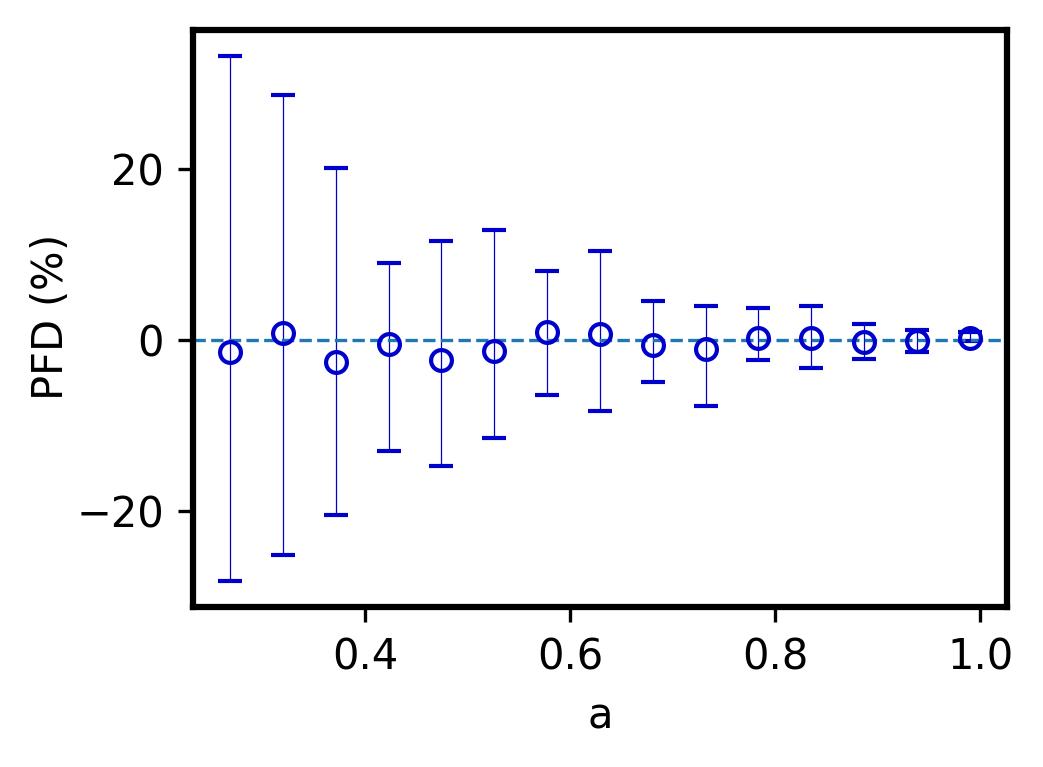}
\caption{Figure showing average PFDs and range of PFDs (over 20 repetitions) for different values of $a$ for the overlap, $|\langle 0|(a|0\rangle+b|1\rangle)|$. We fix the number of shots to 1000 for generating all of the data.} 
\label{overlapcheck}
\end{figure}

We discussed the roles of $\vec{b}$ and the role of the normalization factor associated with the solution vector on the quality of our results when we extract a feature. We finally move to the case of the overlap value itself being small. This analysis is common for HHL and Psi-HHL. For this purpose, we consider a simple toy calculation, where we evaluate the overlap $|\langle 0|(a|0\rangle+b|1\rangle)|$ using the HOM circuit, with 1000 shots and average each of our results over 20 repetitions. $a$ is varied from 0.26 to 0.99 (below 0.26, we run into the `nan' issue discussed earlier). This exercise also verifies the trend in Figure \ref{fig:PEC}(b) (see Figure \ref{overlapcheck}). We conclude from the analysis that if the value of the overlap itself is small, a user needs to supply more shots to extract the feature even if the other aforementioned aspects such as $\vec{b}$ and $\mathcal{\kappa}$ are conducive. \\ 

\section{Conclusion}\label{Sec:Conclusion} 

We propose the Psi-HHL algorithm to efficiently handle problems involving $A$ matrices with large condition numbers ($\mathcal{\kappa}$). The approach involves post-selecting $0$ in HHL (the `wrong signal'), followed by post-selecting $1$ from an HHL-like circuit that contains an additional mixing $2 \times 2$ unitary (the `mixed signal'), and taking the difference between the wrong and the mixed signals to extract the correct signal. \\  

We present arguments to show that the complexity of Psi-HHL is optimal in condition number for large $\mathcal{\kappa}$ regimes; the complexity of the algorithm scales as $\mathcal{O}(\mathcal{\kappa})$. This is an improvement over earlier approaches in the literature, such as HHL with amplitude amplification (scales quadratically, and incurs several additional gates due to its iterative nature; Psi-HHL practically involves executing two circuits of almost the same depth as HHL) and HHL with variable time amplitude amplification (scales as $\mathcal{\kappa}\mathrm{log}^3\mathcal{\kappa}$). It is also important to note that while the optimal $\mathcal{\kappa}$ behaviour can be recovered using a discrete quantum adiabatic approach based quantum linear solver algorithm \cite{Costa}, we achieve the same result using Psi-HHL, which is a simple modification to the HHL algorithm. We also discuss the case of applying Psi-HHL to singular matrices. \\ 

We apply our method to different toy matrices, where we systematically vary the condition number all the way to about 1 million and keeping the number of shots fixed to 1 million for each of the computations, and show that while HHL fails beyond a certain value of $\mathcal{\kappa}$, Psi-HHL continues to predict good results that we quantify via percentage fraction difference (PFD) relative to the classical value as well as via spread in the PFD across many repetitions. We also demonstrate the performance of our proposed Psi-HHL approach for quantum chemistry examples with up to 26-qubit computations, and we once again find that Psi-HHL yields better quality results in much fewer number of shots. En route, we also carry out a qualitative analysis to check suitability of treating LCC equations using the HHL algorithm. \\ 

We anticipate that our work will pave the way for development of efficient quantum algorithms with Psi-HHL integrated into it as an algorithmic primitive. \\ 

\begin{acknowledgements}
We acknowledge National Supercomputing Mission (NSM) for providing computing resources of `PARAM Siddhi-AI', under National PARAM Supercomputing Facility (NPSF), C-DAC, Pune and supported by the Ministry of Electronics and Information Technology (MeitY) and Department of Science and Technology (DST), Government of India. VSP thanks Prof. Aram Harrow for a very fruitful discussion on HHL as well as HHL-LCC. VSP and AJ thank Dr. Shibdas Roy for nice discussions on complexity. PBT thank Prof. Debashis Mukherjee for useful discussions on the quantum chemical sections of the work. VSP and PBT thank their group members for carefully reading the manuscript. KS acknowledges support from Quantum Leap Flagship Program (Grant No. JPMXS0120319794) from Ministry of Education, Culture, Sports, Science and Technology (MEXT), Japan; Center of Innovations for Sustainable Quantum AI (JPMJPF2221) from Japan Science and Technology Agency (JST), Japan; and Grants-in-Aid for Scientific Research C (21K03407) and for Transformative Research Area B (23H03819) from Japan Society for the Promotion of Sciences (JSPS), Japan. 
\end{acknowledgements}

\bibliography{references}

\newpage

\begin{appendices}

\section*{Appendix}

\renewcommand{\thesection}{A\arabic{section}}
\renewcommand{\theequation}{A\arabic{equation}}
\renewcommand{\thefigure}{A\arabic{figure}}
\renewcommand{\thetable}{A\arabic{table}}

\setcounter{section}{0} 
\setcounter{figure}{0} 
\setcounter{table}{0} 

\section{Psi-HHL for overlap calculation with sign information} \label{SM:ocal}

\subsection{Extracting the wrong signal} 

We briefly outline our protocol for the circuit given in Figures \ref{fig:ohhl}(a) and \ref{fig:ohhl}(b) below. We assume here and in the subsequent sub-sections with no loss of generality that the eigenvalues $\lambda_i$ are precisely captured with adequate number of QPE clock register qubits, $n_r$. We start with Figure \ref{fig:ohhl}(a) which helps to calculate $Re(\langle  b \ket{x_{\rm w}})$ and $Im(\langle  b \ket{x_{\rm w}})$.
\begin{enumerate}
\item We begin with $\ket{0} \otimes \ket{0} \otimes \ket{0^{n_r}}\otimes \ket{b}$, and upon applying $H \otimes I^{\otimes(1+n_r + n_b)}$, we get $ \frac{1}{\sqrt{2}}(\ket{0} + \ket{1}) \otimes \ket{0} \otimes\ket{0^{n_r}}\otimes \ket{b}$. 
\item This is followed by the operation $|0\rangle\langle 0 |\otimes I + |1\rangle\langle 1|\otimes\text{HHL}_1\text{(No measurement and post-selection)}$, which results in 
\begin{eqnarray}
    && \frac{1}{\sqrt{2}}\ket{0}  \otimes \ket{0} \otimes \ket{0^{n_r}}\otimes \ket{b} \nonumber \\ 
&+& \frac{1}{\sqrt{2}}\ket{1} \otimes\left(\sum_i b_i \sqrt{1 - \frac{C^2}{\lambda_i^2}}\ket{0}  + b_i\frac{C}{\lambda_i}\ket{1}\right)\otimes\ket{0^{n_r}}\otimes \ket{\nu_i}. \nonumber
\end{eqnarray} 
\item Measuring the second qubit from the top and post-selecting the outcome $0$, we get the state-vector, 
\begin{eqnarray}
    &&\frac{1}{\sqrt{2 \mathfrak{\Gamma}}}\ket{0}\otimes\ket{0}\otimes\ket{0^{n_r}}\otimes\ket{b} \nonumber \\ &+&  \frac{1}{\sqrt{2\mathfrak{\Gamma}}}\ket{1}\otimes\ket{0}\otimes\ket{0^{n_r}}\otimes\ket{x_{\rm w}}. \nonumber 
\end{eqnarray}
where $1/\sqrt{\mathfrak{\Gamma}}$ is the normalisation factor after measurement and $\mathfrak{\Gamma} = \frac{1}{2}(\|\ket{b}\|^2 + \|\ket{x_{\rm w }}\|^2) = \frac{1}{2}(1 + \|\ket{x_{\rm w }}\|^2).$
\item Acting the resulting state upon by $H\otimes I^{\otimes(1+n_r+n_b)}$, we obtain 
    \begin{eqnarray}
    &&\frac{1}{2\sqrt{\mathfrak{\Gamma}}}\left(\ket{0}\otimes\ket{0}\otimes \ket{0^{n_r}}\otimes(\ket{b}+\ket{x_{\rm w}}) \right. \nonumber \\  &+& \left.\ket{1}\otimes\ket{0}\otimes\ket{0^{n_r}}\otimes(\ket{b} -\ket{x_{\rm w}})\right). \nonumber 
\end{eqnarray}
\item The probability of measuring $0$ on the first qubit from the top is \begin{eqnarray}
\mathcal{P}_R(0) &=& \left \| \frac{1}{2\sqrt{\mathfrak{\Gamma}}}(\ket{b} + \ket{x_{\rm w}}) \right \|^2 \nonumber \\
&=&  \frac{1}{2}\left( 1+\frac{2Re(\langle b  \ket{x_{\rm w}}}{1+\|\ket{x_{\rm w }}\|^2}\right). \nonumber 
\end{eqnarray}
Similarly, the introduction of $S^{\dagger}$ gate in the circuit in Figure \ref{fig:ohhl}(a) helps us find 
\begin{eqnarray}
\mathcal{P}_I(0) &=&  \left\| \frac{1}{2\sqrt{\mathfrak{\Gamma}}} (\ket{b} - i \ket{x_{\rm w}}) \right\|^2 \nonumber \\
&=&  \frac{1}{2}\left( 1+\frac{2Im(\langle b  \ket{x_{\rm w}})}{1 + \|\ket{x_{\rm w }}\|^2}\right). \nonumber 
\end{eqnarray}

Through the last these expressions, we find that 
\begin{equation}
    \langle b | x_{\rm w} \rangle = ((2\mathcal{P}_R(0) - 1) +i(2\mathcal{P}_I(0) - 1))\mathfrak{\Gamma}.
\end{equation}
\end{enumerate}

\subsection{Extracting the mixed signal}

Now we give the protocol which helps us to calculate $Re(\langle b \ket{x'_{\rm m}})$ and $Im(\langle b \ket{x'_{\rm m}})$ from Figure \ref{fig:ohhl}(b).
\begin{enumerate}
\item We begin with $\ket{0} \otimes \ket{0} \otimes \ket{0^{n_r}}\otimes \ket{b}$, and upon applying $H \otimes I^{\otimes (1+n_r + n_b)}$, we get $ \frac{1}{\sqrt{2}}(\ket{0} + \ket{1}) \otimes \ket{0} \otimes\ket{0^{n_r}}\otimes \ket{b}$. 
\item This is followed by the operation $|0\rangle\langle 0 |\otimes I + |1\rangle\langle 1|\otimes\text{HHL}_2 \text{(No measurement and post-selection)}$, which results in 
\begin{eqnarray}
    && \frac{1}{\sqrt{2}}\ket{0}  \otimes \ket{0} \otimes \ket{0^{n_r}} \otimes \ket{b} \nonumber \\ 
&+& \frac{1}{\sqrt{2}}\ket{1} \otimes \left[\left(\sum_i b_i \sqrt{1 - \frac{C^2}{\lambda_i^2}}\cos(\alpha)  - b_i\frac{C}{\lambda_i}\sin(\alpha)\right)\ket{0} \right.\nonumber \\
&+& \left(\left.\sum_i b_i \sqrt{1 - \frac{C^2}{\lambda_i^2}}\sin(\alpha)  + b_i\frac{C}{\lambda_i}\cos(\alpha)\right)\ket{1}
\right] \otimes \ket{0^{n_r}} \otimes \ket{\nu_i}. \nonumber
\end{eqnarray} 
\item Then, we apply the operation $(|0\rangle\langle 0 |\otimes X + |1\rangle\langle 1 | \otimes I)\otimes I^{\otimes (n_r+n_b)} $ on the first and second qubits from the top and get 
\begin{eqnarray}
    && \frac{1}{\sqrt{2}}\ket{0} \otimes \ket{1}\otimes \ket{0^{n_r}}\otimes \ket{b} \nonumber \\ 
&+& \frac{1}{\sqrt{2}}\ket{1} \otimes \left[\left(\sum_i b_i \sqrt{1 - \frac{C^2}{\lambda_i^2}}\cos(\alpha)  - b_i\frac{C}{\lambda_i}\sin(\alpha)\right)\ket{0} \right.\nonumber \\
&+& \left(\left.\sum_i b_i \sqrt{1 - \frac{C^2}{\lambda_i^2}}\sin(\alpha)  + b_i\frac{C}{\lambda_i}\cos(\alpha)\right)\ket{1}
\right]\otimes \ket{0^{n_r}} \otimes \ket{\nu_i}. \nonumber 
\end{eqnarray} 
\item Measuring the second qubit from the top and post-selecting the outcome 1, we get the state-vector, 
\begin{eqnarray}
    &&\frac{1}{\sqrt{2\beta}}\left(\ket{0}\otimes \ket{1}\otimes \ket{0^{n_r}}\otimes\ket{b} \right.\nonumber \\ &+& \left. \ket{1} \otimes \ket{1} \otimes \ket{0^{n_r}} \otimes \ket{x'_{\rm m}}\right), \nonumber  
\end{eqnarray}
where $1/\sqrt{\beta}$ is the normalisation factor after measurement and $\beta = \frac{1}{2}(\|\ket{b}\|^2 + \|\ket{x'_{\rm m }}\|^2) = \frac{1}{2}(1 + \|\ket{x'_{\rm m }}\|^2).$
\item Acting the resulting state upon by $H\otimes I^{\otimes(1+n_r+n_b)}$, we obtain 
    \begin{eqnarray}
    &&\frac{1}{2\sqrt{\beta}}\left(\ket{0}\otimes\ket{1} \otimes \ket{0^{n_r}}\otimes (\ket{b}+\ket{x'_{\rm m}}) \right. \nonumber \\  &+& \left.\ket{1}\otimes \ket{1}\otimes \ket{0^{n_r}}\otimes (\ket{b} -\ket{x'_{\rm m}})\right). \nonumber 
\end{eqnarray}
\item The probability of measuring 0 on the first qubit from the top is \begin{eqnarray}
\mathcal{P}'_R(0) &=& \left \| \frac{1}{2\sqrt{\beta}} (\ket{b} + \ket{x'_{\rm m}}) \right \|^2 \nonumber \\
&=&  \frac{1}{2}\left( 1+\frac{2Re(\langle b  \ket{x'_{\rm m}})}{1 + \|\ket{x'_{\rm m }}\|^2}\right)\nonumber 
\end{eqnarray}
Similarly, the introduction of $S^{\dagger}$ gate in the circuit in Figure \ref{fig:ohhl}(b) helps us find 
\begin{eqnarray}
\mathcal{P}'_I(0) &=& \left \| \frac{1}{2\sqrt{\beta}} (\ket{b} - i  \ket{x'_{\rm m}}) \right \|^2 \nonumber \\
&=&  \frac{1}{2}\left( 1+\frac{2Im(\langle b  \ket{x'_{\rm m}})}{1 + \|\ket{x'_{\rm m }}\|^2}\right)\nonumber 
\end{eqnarray}

Through the last two expressions, we can find that 
\begin{equation}
    \langle b | x'_{\rm m} \rangle = ((2\mathcal{P}'_R(0) - 1) +i(2\mathcal{P}'_I(0) - 1))\beta.
\end{equation}
\end{enumerate} 

\section{Psi-HHL for expectation value calculation: details} \label{SM:ev}

\subsection{Extraction of wrong signal} 

We discuss below the protocol to extract the wrong signal as shown in Figure \ref{fig:ev}(a).

\begin{enumerate}
\item We begin with $\ket{0} \otimes \ket{0} \otimes \ket{0^{n_r}}\otimes \ket{b}$, and upon applying $H \otimes \text{HHL}_1$, we get $ \frac{1}{\sqrt{2}\|\ket{x_{\rm w}}\|}(\ket{0} + \ket{1}) \otimes \ket{0} \otimes \ket{0^{n_r}}\otimes \ket{x_{\rm w}}$, where $1/\|\ket{x_{\rm w}}\|$ is the normalisation factor after measurement in HHL$_1$. 
\item This is followed by the operation $|0\rangle\langle 0 |\otimes I^{\otimes(1+n_r+n_b)} + |1\rangle\langle 1| \otimes I^{\otimes(1+n_r)} \otimes \mathcal{U}$, which results in 
\begin{equation}
    \frac{1}{\sqrt{2}\|\ket{x_{\rm w}}\|}(\ket{0} \otimes \ket{0} \otimes \ket{0^{n_r}}\otimes \ket{x_{\rm w}} + \ket{1} \otimes \ket{0} \otimes \ket{0^{n_r}}\otimes (\mathcal{U}\ket{x_{\rm w}})). \nonumber
\end{equation}
\item Acting the resulting state upon by $H\otimes I^{\otimes(1+n_r+n_b)}$, we obtain 
    \begin{eqnarray}
    &&\frac{1}{2\|\ket{x_{\rm w}}\|}\left(\ket{0}\otimes\ket{0}\otimes\ket{0^{ n_r}}\otimes(\ket{x_{\rm w}}+\mathcal{U}\ket{x_{\rm w}}) \right. \nonumber \\  &+& \left.\ket{1}\otimes\ket{0}\otimes\ket{0^{ n_r}}\otimes(\ket{x_{\rm w}} -\mathcal{U}\ket{x_{\rm w}})\right). \nonumber 
\end{eqnarray}

\item The probability of measuring 0 on the first qubit from the top is \begin{eqnarray}
\mathcal{P}_R(0) &=& \left \| \frac{1}{2\|\ket{x_{\rm w}}\|} (\ket{x_{\rm w}} + \mathcal{U} \ket{x_{\rm w}}) \right \|^2 \nonumber \\
&=&  \frac{1}{2}\left( 1+\frac{Re(\bra{x_{\rm w}} \mathcal{U} \ket{x_{\rm w}})}{\|\ket{x_{\rm w}}\|^2 }\right). \nonumber 
\end{eqnarray}

Similarly, the introduction of $S^{\dagger}$ gate in the circuit in Figure \ref{fig:ev}(a) helps us find 
\begin{eqnarray}
\mathcal{P}_I(0) &=& \left \| \frac{1}{2\|\ket{x_{\rm w}}\|} (\ket{x_{\rm w}} -i \mathcal{U} \ket{x_{\rm w}}) \right \|^2 \nonumber \\
&=&  \frac{1}{2}\left( 1+\frac{Im(\bra{x_{\rm w}} \mathcal{U} \ket{x_{\rm w}})}{\|\ket{x_{\rm w}}\|^2 }\right). \nonumber 
\end{eqnarray}

Through the last two expressions, we can find that 
\begin{equation}
    \bra{x_{\rm w}}  \mathcal{U} \ket{x_{\rm w}} = ((2\mathcal{P}_R(0) - 1) + i(2\mathcal{P}_I(0) - 1))\|\ket{x_{\rm w}}\|^2\nonumber.
\end{equation}
\end{enumerate}

\subsection{Extraction of mixed signal}

We discuss below the protocol for extracting the mixed signal as shown in Figure \ref{fig:ev}(b). We note that the mixed signal has in it a cross-term. 

\begin{enumerate}
\item We begin with $\ket{0} \otimes \ket{0} \otimes \ket{0^{n_r}}\otimes \ket{b}$, and upon applying $H \otimes \text{HHL}_2$, we get $ \frac{1}{\sqrt{2}\|\ket{x'_{\rm m}}\|}(\ket{0} + \ket{1}) \otimes \ket{1} \otimes \ket{0^{n_r}}\otimes \ket{x'_{\rm m}}$, where $1/\|\ket{x'_{\rm m}}\|$ is the normalisation factor after measurement in HHL$_2$. 
\item This is followed by the operation $|0\rangle\langle 0 |\otimes I^{\otimes(1+n_r+n_b)} + |1\rangle\langle 1| \otimes I^{\otimes(1+n_r)} \otimes \mathcal{U}$, which results in 
\begin{equation}
    \frac{1}{\sqrt{2}\|\ket{x'_{\rm m}}\|}(\ket{0} \otimes \ket{1} \otimes \ket{0^{n_r}}\otimes \ket{x'_{\rm m}} + \ket{1} \otimes \ket{1} \otimes \ket{0^{n_r}}\otimes (\mathcal{U}\ket{x'_{\rm m}})). \nonumber
\end{equation}
\item Acting the resulting state upon by $H\otimes I^{\otimes(1+n_r+n_b)}$, we obtain 
    \begin{eqnarray}
    &&\frac{1}{2\|\ket{x'_{\rm m}}\|}\left(\ket{0}\otimes\ket{1}\otimes\ket{0^{ n_r}}\otimes(\ket{x'_{\rm m}}+\mathcal{U}\ket{x'_{\rm m}}) \right. \nonumber \\  &+& \left.\ket{1}\otimes\ket{1}\otimes\ket{0^{ n_r}}\otimes(\ket{x'_{\rm m}} -\mathcal{U}\ket{x'_{\rm m}})\right). \nonumber 
\end{eqnarray}

\item The probability of measuring 0 on the first qubit from the top is \begin{eqnarray}
\mathcal{P}'_R(0) &=& \left \| \frac{1}{2\|\ket{x'_{\rm m}}\|} (\ket{x'_{\rm m}} + \mathcal{U} \ket{x'_{\rm m}}) \right \|^2 \nonumber \\
&=&  \frac{1}{2}\left( 1+\frac{Re(\bra{x'_{\rm m}} \mathcal{U} \ket{x'_{\rm m}})}{\|\ket{x'_{\rm m}}\|^2 }\right).\nonumber 
\end{eqnarray}
Similarly, the introduction of $S^{\dagger}$ gate in the circuit in Figure \ref{fig:ev}(a) helps us find 
\begin{eqnarray}
\mathcal{P}'_I(0) &=& \left \| \frac{1}{2\|\ket{x'_{\rm m}}\|} (\ket{x'_{\rm m}} - i \mathcal{U} \ket{x'_{\rm m}}) \right \|^2 \nonumber \\
&=&  \frac{1}{2}\left( 1+\frac{Im(\bra{x'_{\rm m}} \mathcal{U} \ket{x'_{\rm m}})}{\|\ket{x'_{\rm m}}\|^2 }\right).\nonumber 
\end{eqnarray}
\begin{equation}
\bra{x'_{\rm m}}  \mathcal{U} \ket{x'_{\rm m}} = ((2\mathcal{P}'_R(0) - 1) + i(2\mathcal{P}'_I(0) - 1))\|\ket{x'_{\rm m}}\|^2.\nonumber
\end{equation}
\end{enumerate}

\subsection{Extracting the cross-term} 

We present below our protocol for the circuit given in Figure \ref{fig:ev}(c), using which we obtain the cross-term. 

\begin{enumerate}
\item We begin with $\ket{0} \otimes \ket{0} \otimes \ket{0^{n_r}}\otimes \ket{b}$, and upon applying $H \otimes \text{HHL\ (without post-selection)}$, we get $ \frac{1}{\sqrt{2}}(\ket{0} + \ket{1}) \otimes \left(\sum_i b_i \sqrt{1 - \frac{C^2}{\lambda_i^2}} \ket{0} + \frac{C}{\lambda_i} \ket{1}\right)\otimes \ket{0^{n_r}}\otimes \ket{\nu_i}$. 
\item This is followed by the operation $(|0\rangle\langle 0 |\otimes I + |1\rangle\langle 1 |\otimes R_Y(2\alpha))\otimes I^{\otimes (n_r+n_b)}$, which results in 
\begin{eqnarray}
    && \frac{1}{\sqrt{2}}\ket{0} \otimes \left(\sum_i b_i \sqrt{1 - \frac{C^2}{\lambda_i^2}} \ket{0} + b_i\frac{C}{\lambda_i} \ket{1}\right)\otimes\ket{0^{n_r}}\ket{\nu_i} \nonumber \\ 
&+& \frac{1}{\sqrt{2}}\ket{1} \otimes \left[\left(\sum_i b_i \sqrt{1 - \frac{C^2}{\lambda_i^2}}\cos(\alpha)  - b_i\frac{C}{\lambda_i}\sin(\alpha)\right)\ket{0} \right.\nonumber \\
&+& \left(\left.\sum_i b_i \sqrt{1 - \frac{C^2}{\lambda_i^2}}\sin(\alpha)  + b_i\frac{C}{\lambda_i}\cos(\alpha)\right)\ket{1}
\right]\otimes\ket{0^{n_r}}\otimes \ket{\nu_i}. \nonumber
\end{eqnarray} 
\item Then, we apply the operation $(|0\rangle\langle 0 |\otimes X + |1\rangle\langle 1 | \otimes I)\otimes I^{\otimes (n_r+n_b)} $ on the first and second qubits from the top and get 
\begin{eqnarray}
    && \frac{1}{\sqrt{2}}\ket{0} \otimes \left(\sum_i b_i \sqrt{1 - \frac{C^2}{\lambda_i^2}} \ket{1} + b_i\frac{C}{\lambda_i} \ket{0}\right)\otimes \ket{0^{n_r}}\otimes\ket{\nu_i} \nonumber \\ 
&+& \frac{1}{\sqrt{2}}\ket{1} \otimes \left[\left(\sum_i b_i \sqrt{1 - \frac{C^2}{\lambda_i^2}}\cos(\alpha)  - b_i\frac{C}{\lambda_i}\sin(\alpha)\right)\ket{0} \right.\nonumber \\
&+& \left(\left.\sum_i b_i \sqrt{1 - \frac{C^2}{\lambda_i^2}}\sin(\alpha)  + b_i\frac{C}{\lambda_i}\cos(\alpha)\right)\ket{1}
\right]\otimes \ket{0^{n_r}} \otimes \ket{\nu_i}. \nonumber 
\end{eqnarray} 
\item Measuring the second qubit from the top and post-selecting the outcome 1, we get the state-vector, 
\begin{eqnarray}
    &&\frac{1}{\sqrt{\delta}}\left(\frac{1}{\sqrt{2}}\ket{0}\otimes\ket{1}\otimes\ket{0^{ n_r}}\otimes\ket{x_{\rm w}} \right.\nonumber \\ &+& \left. \frac{1}{\sqrt{2}}\ket{1}\otimes\ket{1}\otimes\ket{0^{ n_r}}\otimes\ket{x'_{\rm m}}\right), \nonumber  
\end{eqnarray}
where $1/\sqrt{\delta}$ is the normalisation factor after measurement and $\delta = \frac{1}{2}(\|\ket{x_{\rm w}}\|^2 + \|\ket{x'_{\rm m }}\|^2).$
\item Applying $|0\rangle\langle 0 |\otimes I^{\otimes(1+n_r+n_b)} + |1\rangle\langle 1| \otimes I^{\otimes(1+n_r)} \otimes \mathcal{U}$, we obtain 
\begin{eqnarray}
    &&\frac{1}{\sqrt{\delta}}\left(\frac{1}{\sqrt{2}}\ket{0}\otimes\ket{1}\otimes\ket{0^{ n_r}}\otimes\ket{x_{\rm w}} \right.\nonumber \\ &+&  \left.\frac{1}{\sqrt{2}}\ket{1}\otimes\ket{1}\otimes\ket{0^{ n_r}}\otimes(\mathcal{U}\ket{x'_{\rm m}})\right). \nonumber 
\end{eqnarray} 
\item Acting the resulting state upon by $H\otimes I^{\otimes(1+n_r+n_b)}$, we obtain 
    \begin{eqnarray}
    &&\frac{1}{\sqrt{\delta}}\left(\frac{1}{2}\ket{0}\otimes\ket{1}\otimes\ket{0^{ n_r}}\otimes(\ket{x_{\rm w}}+\mathcal{U}\ket{x'_{\rm m}}) \right. \nonumber \\  &+& \left.\frac{1}{2}\ket{1}\otimes\ket{1}\otimes\ket{0^{ n_r}}\otimes(\ket{x_{\rm w}} -\mathcal{U}\ket{x'_{\rm m}})\right). \nonumber 
\end{eqnarray}
\item The probability of measuring 0 on the first qubit from the top is \begin{eqnarray}
\bar{P}(0) &=& \left \| \frac{1}{2\sqrt{\delta}} (\ket{x_{\rm w}} + \mathcal{U} \ket{x'_{\rm m}}) \right \|^2 \nonumber \\
&=&  \frac{1}{2}\left( 1+\frac{2Re(\bra{x_{\rm w}} \mathcal{U} \ket{x'_{\rm m}})}{\|\ket{x_{\rm w}}\|^2 + \|\ket{x'_{\rm m }}\|^2}\right)\nonumber 
\end{eqnarray}
Through the last equation and the expression of $\bra{x_{\rm w}}  \mathcal{U} \ket{x_{\rm w}}$ computed earlier, we can find that
\begin{equation}
Re(\bra{x_{\rm w}}  \mathcal{U} \ket{x_{\rm r}}) = \frac{\delta}{\cos(\alpha)} (2\bar{P}(0) - 1) - \frac{\sin(\alpha)}{\cos(\alpha)}(2\mathcal{P}_R(0)-1) \|\ket{x_{\rm w}}\|^2.\nonumber
\end{equation}
\end{enumerate}

\section{Details on calculations involving toy matrices from size $4 \times 4$ through $64 \times 64$} \label{SM: toy-matrices} 

\begin{figure*}
\centering
\begin{tabular}{cc}
\includegraphics[scale=0.30]{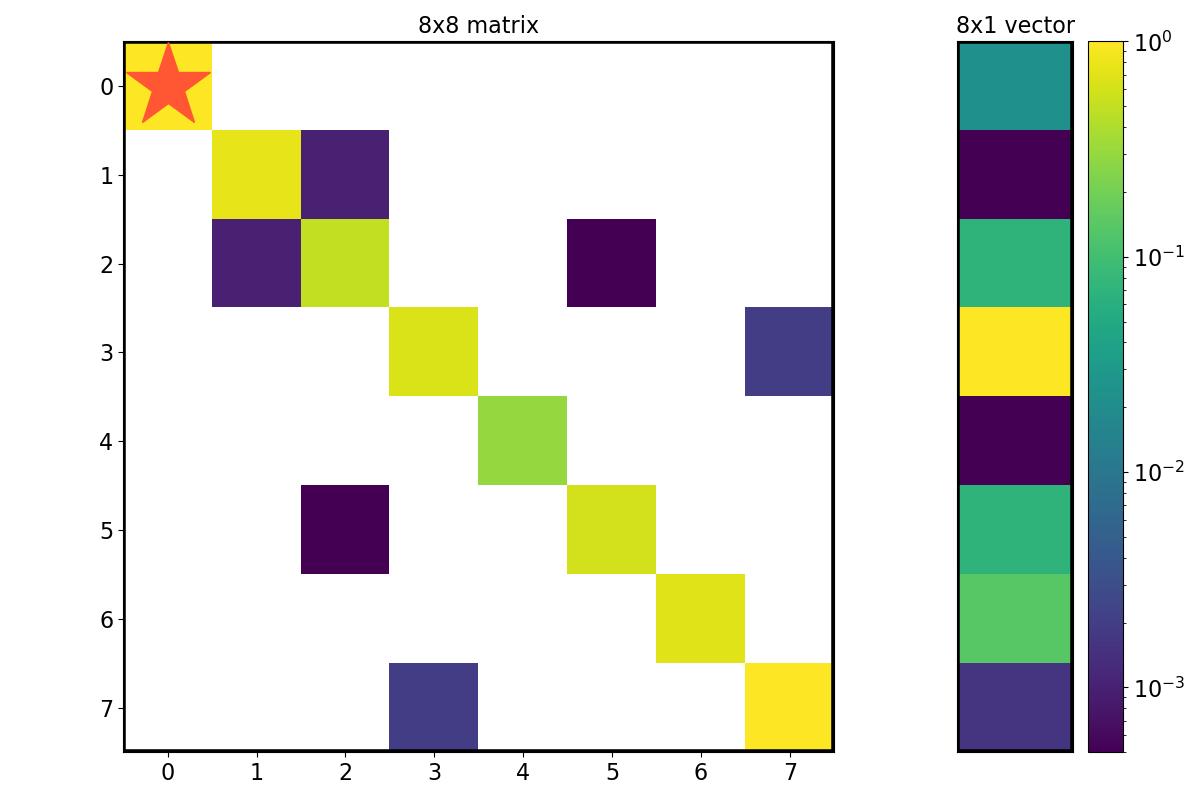}&
\includegraphics[scale=0.30]{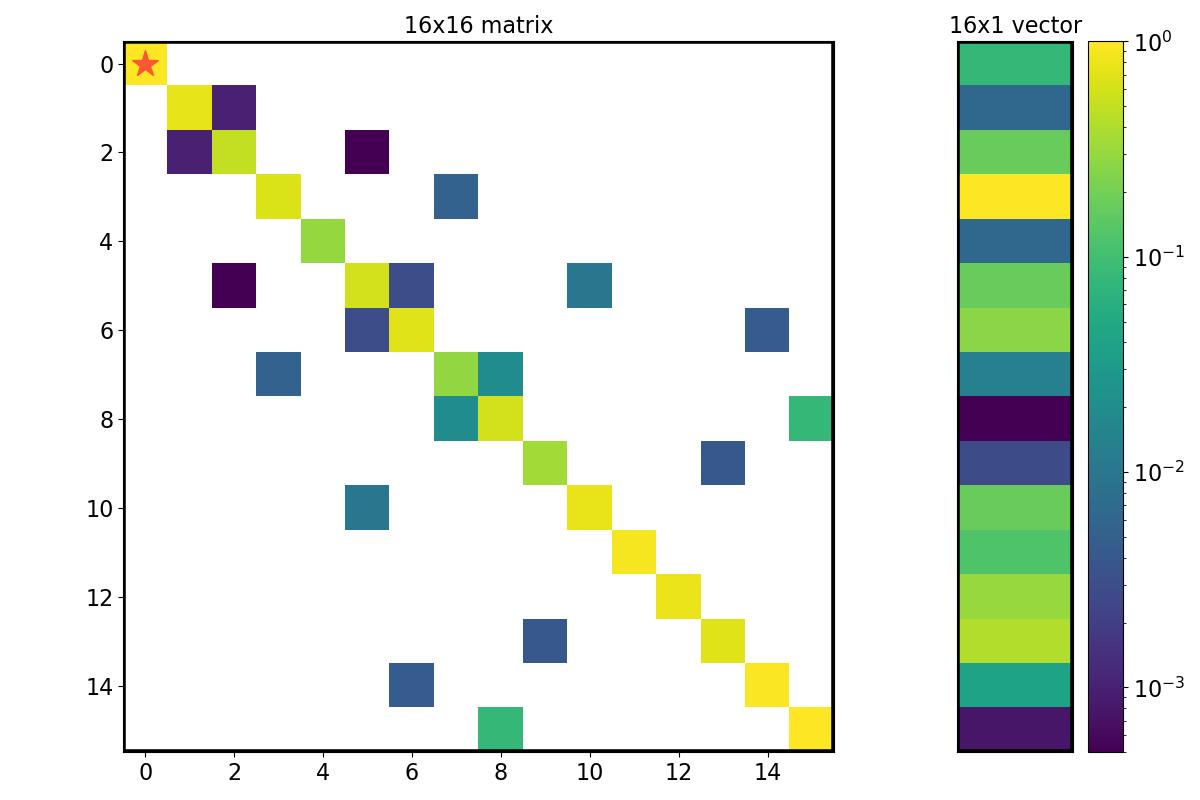} \\ 
(a)&(b)\\ 
\includegraphics[scale=0.30]{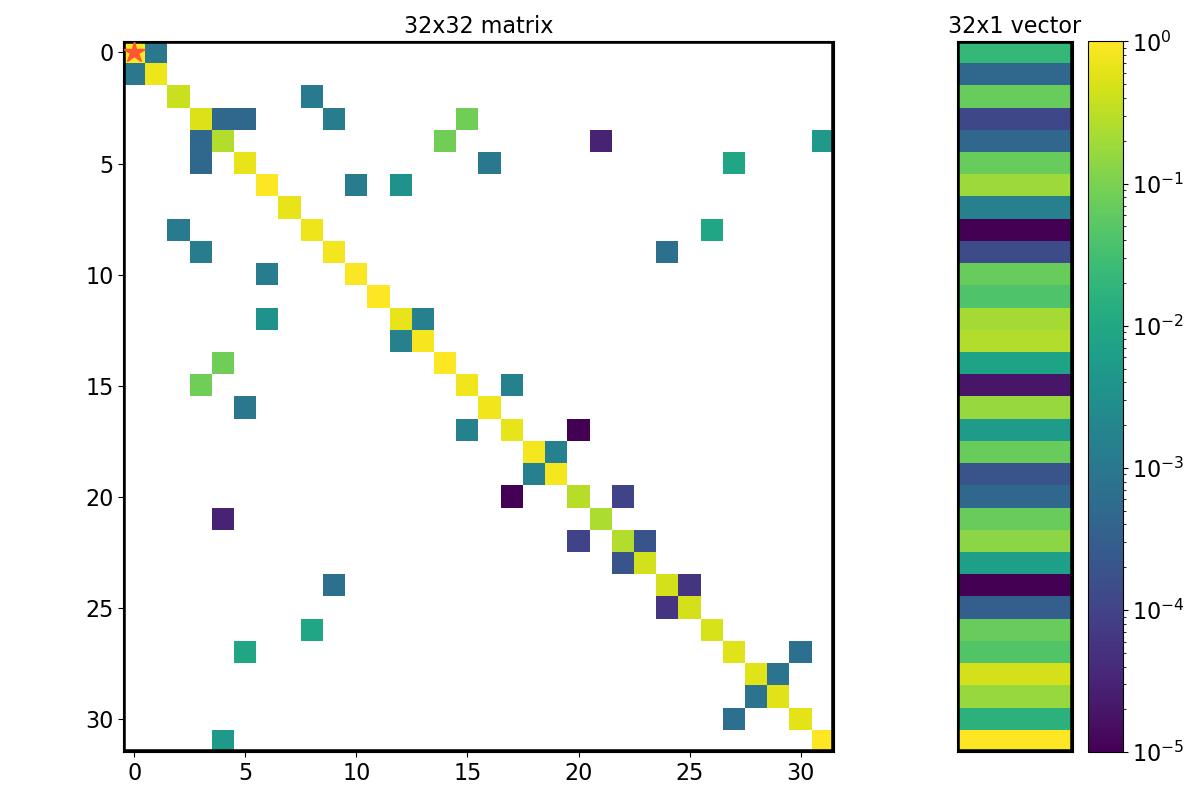}&\includegraphics[scale=0.30]{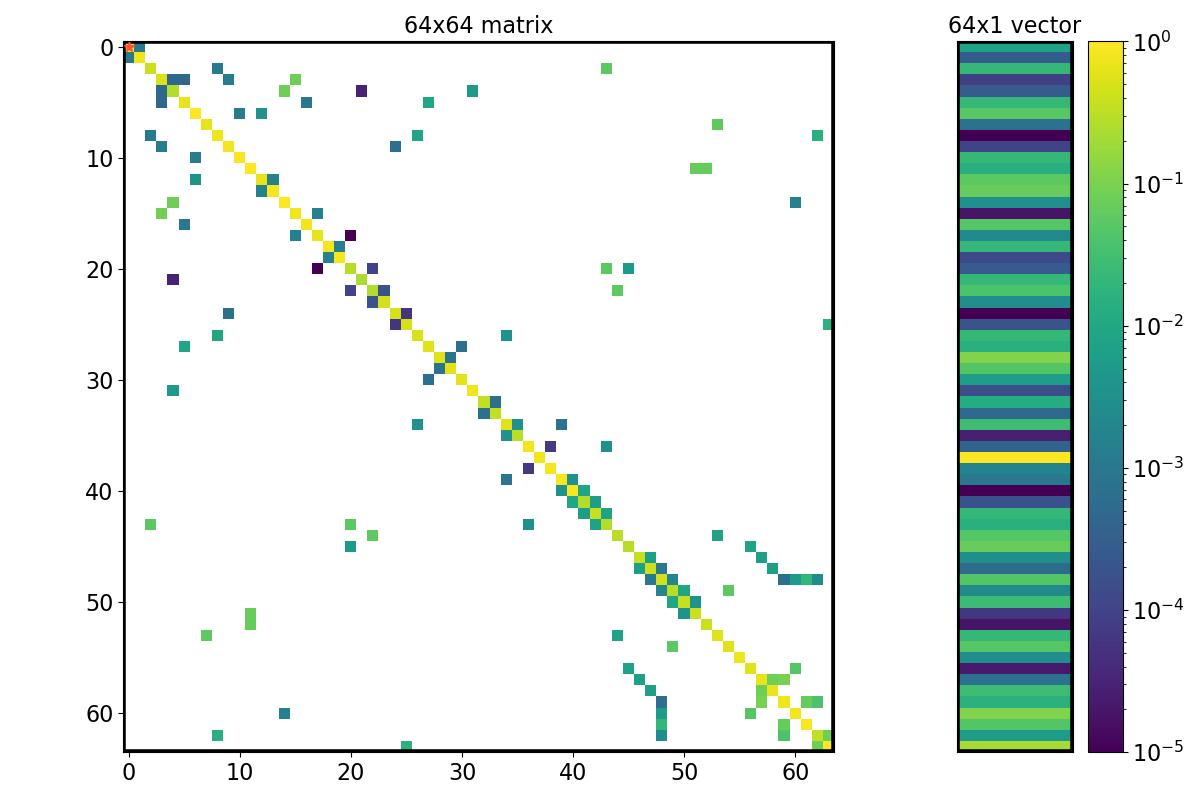}\\
(c)&(d)\\   
\end{tabular}
\caption{Our choices for (a) $8 \times 8$, (b) $16 \times 16$, (c) $32 \times 32$, and (d) $64 \times 64$ toy $A$ matrices and un-normalized $\vec{b}$ for comparing the performance of HHL and the Psi-HHL algorithms. The cell with a star symbol denotes the matrix element with variable $n_r$ (the number of clock register qubits from the HHL circuit), which sets $\mathcal{\kappa}$. A white cell denotes a zero matrix element. }
\label{fig:mat-sizes}
\end{figure*}

\begin{figure*}[t]
\centering
    \begin{tabular}{ccc}
\includegraphics[scale=0.65]{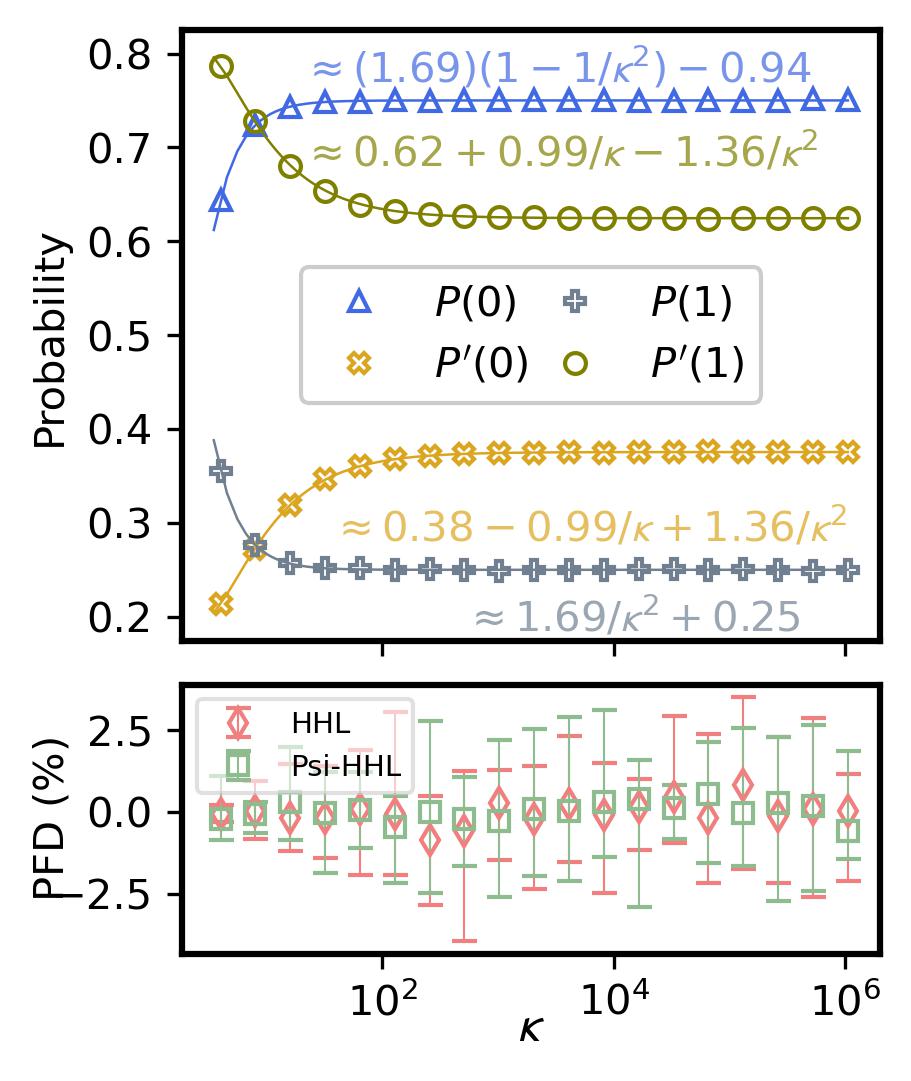}&
\includegraphics[scale=0.65]{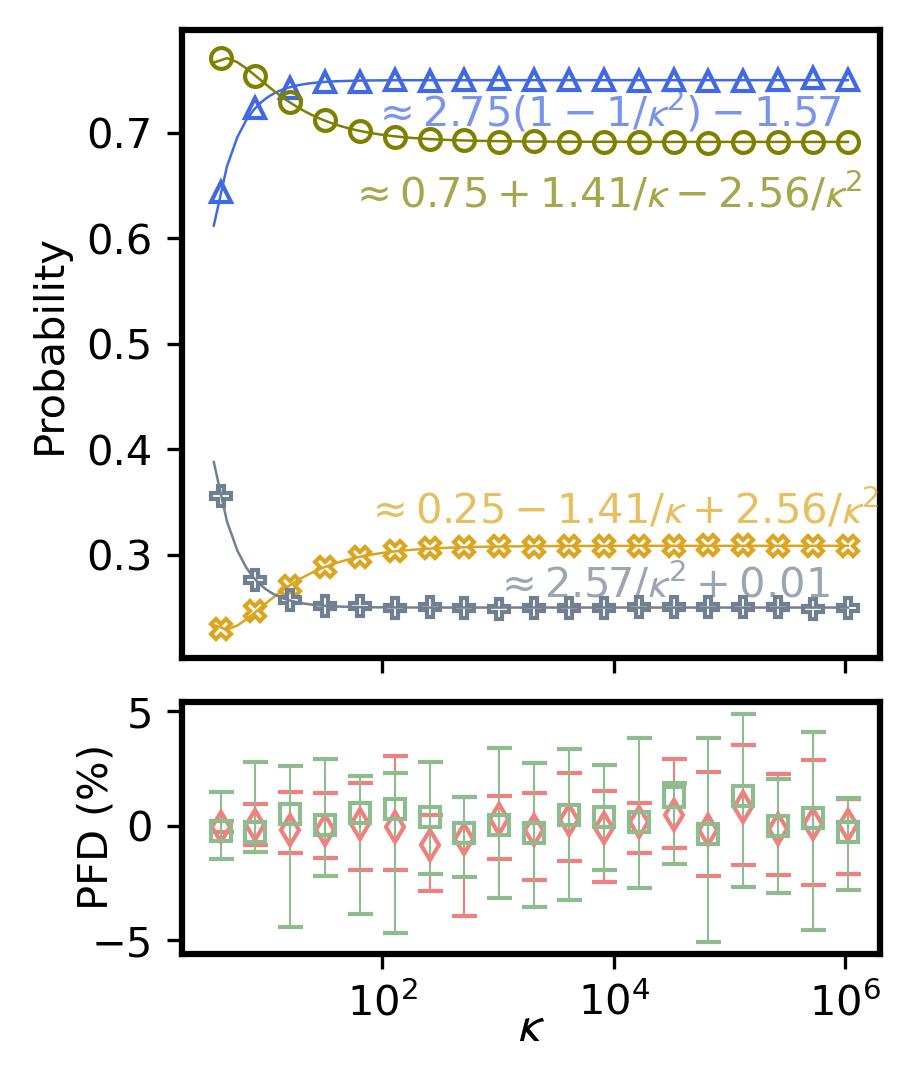}&\includegraphics[scale=0.65]{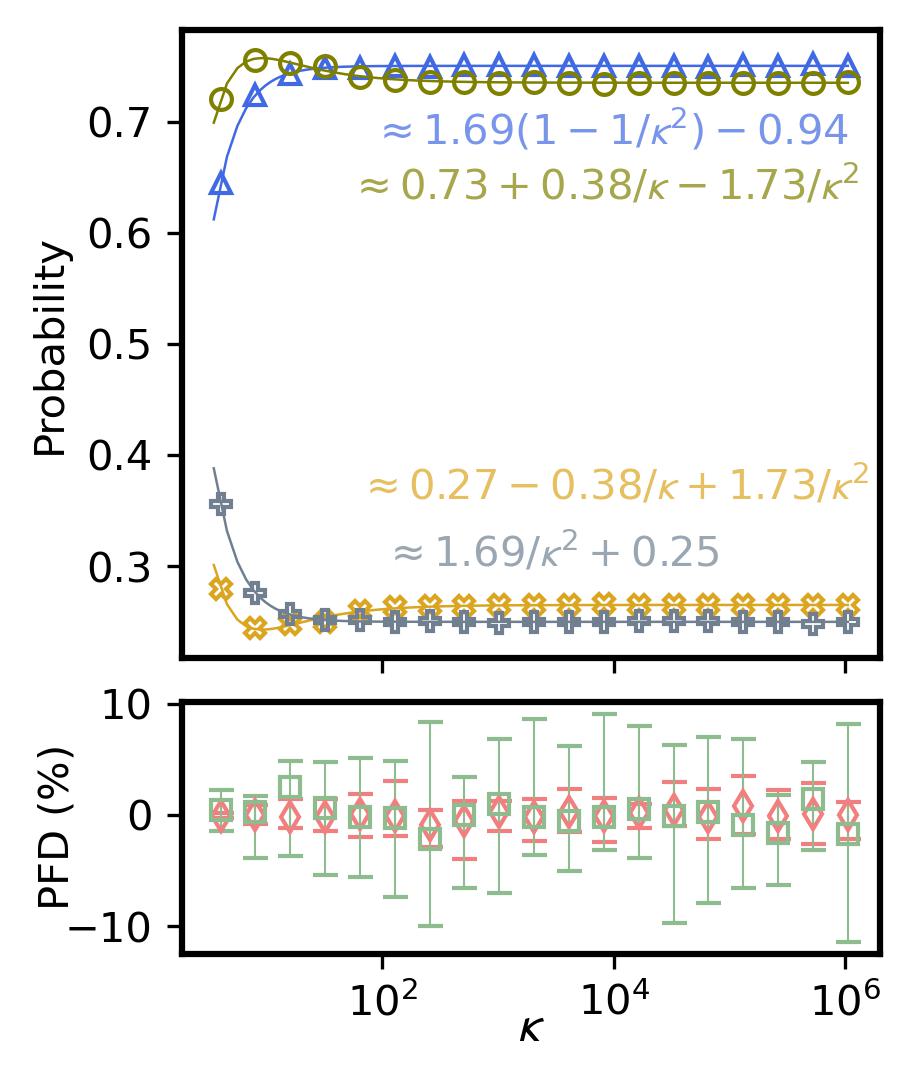}\\(a) $\alpha=60^\circ$ &(b) $\alpha=70^\circ$ &(c) 
 $\alpha=80^\circ$ 
\end{tabular}
\begin{tabular}{ccc}
\includegraphics[scale=0.65]{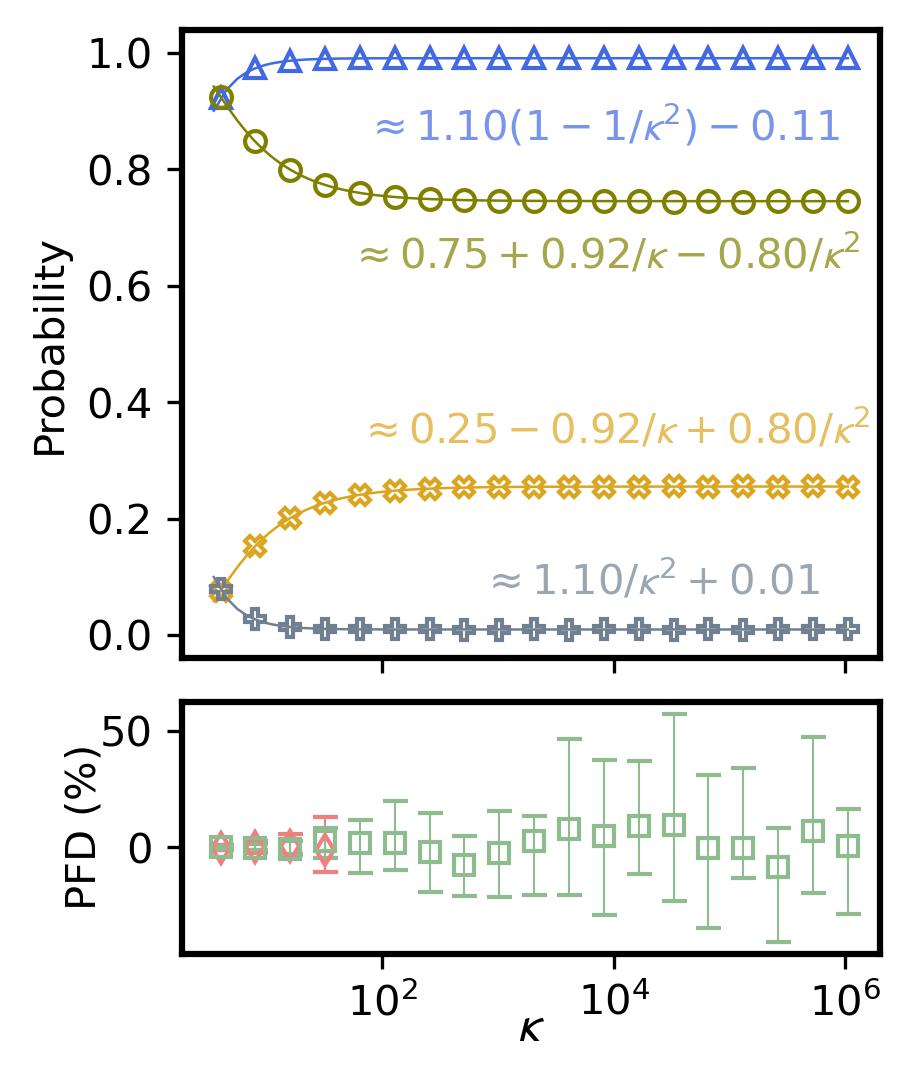}&
\includegraphics[scale=0.65]{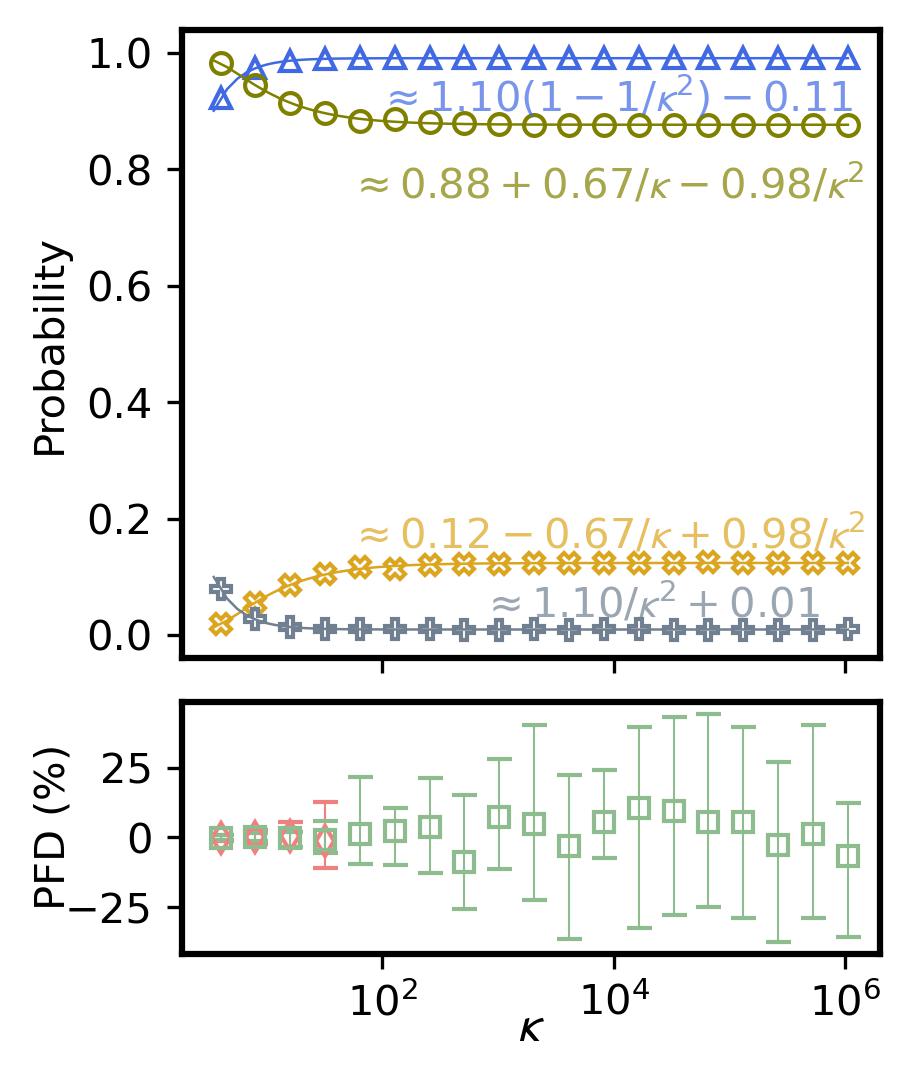}&\includegraphics[scale=0.65]{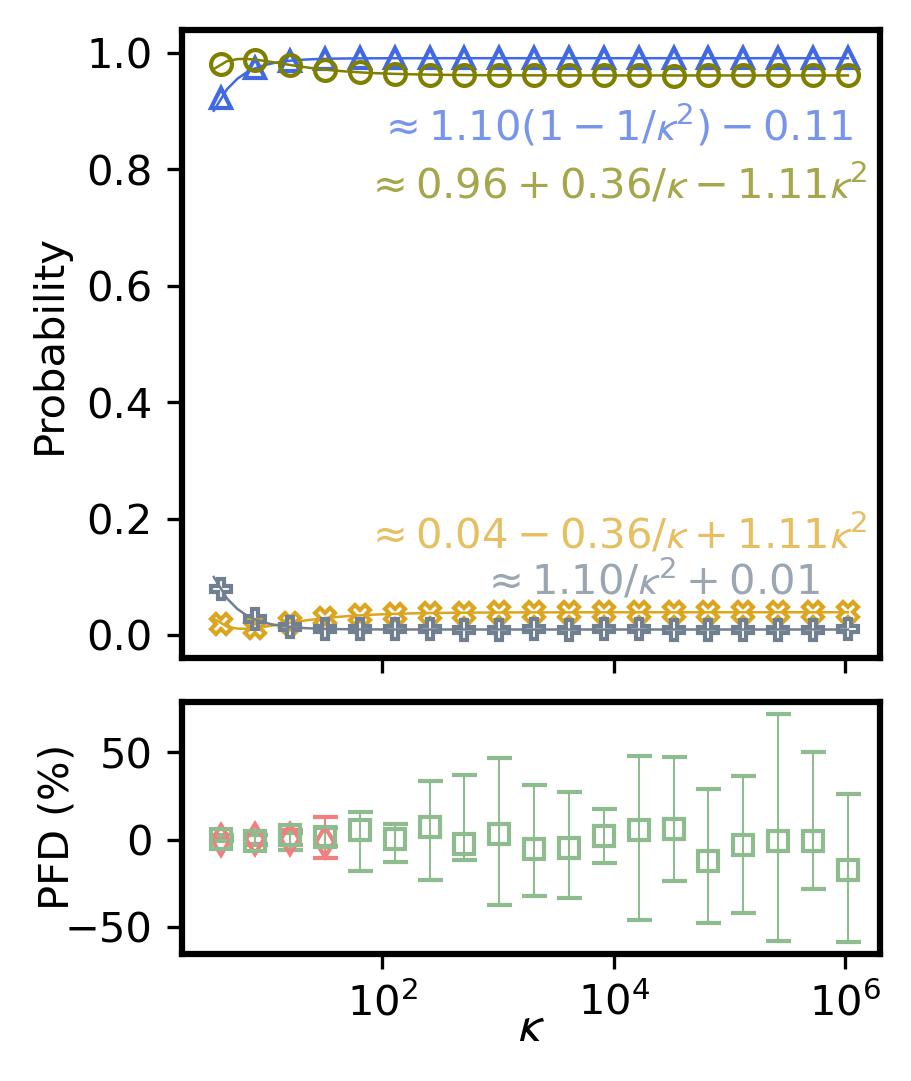}\\(d) $\alpha=60^\circ$ &(e) $\alpha=70^\circ$ &(f) 
 $\alpha=80^\circ$ 
\end{tabular}
\begin{tabular}{ccc}
\includegraphics[scale=0.65]{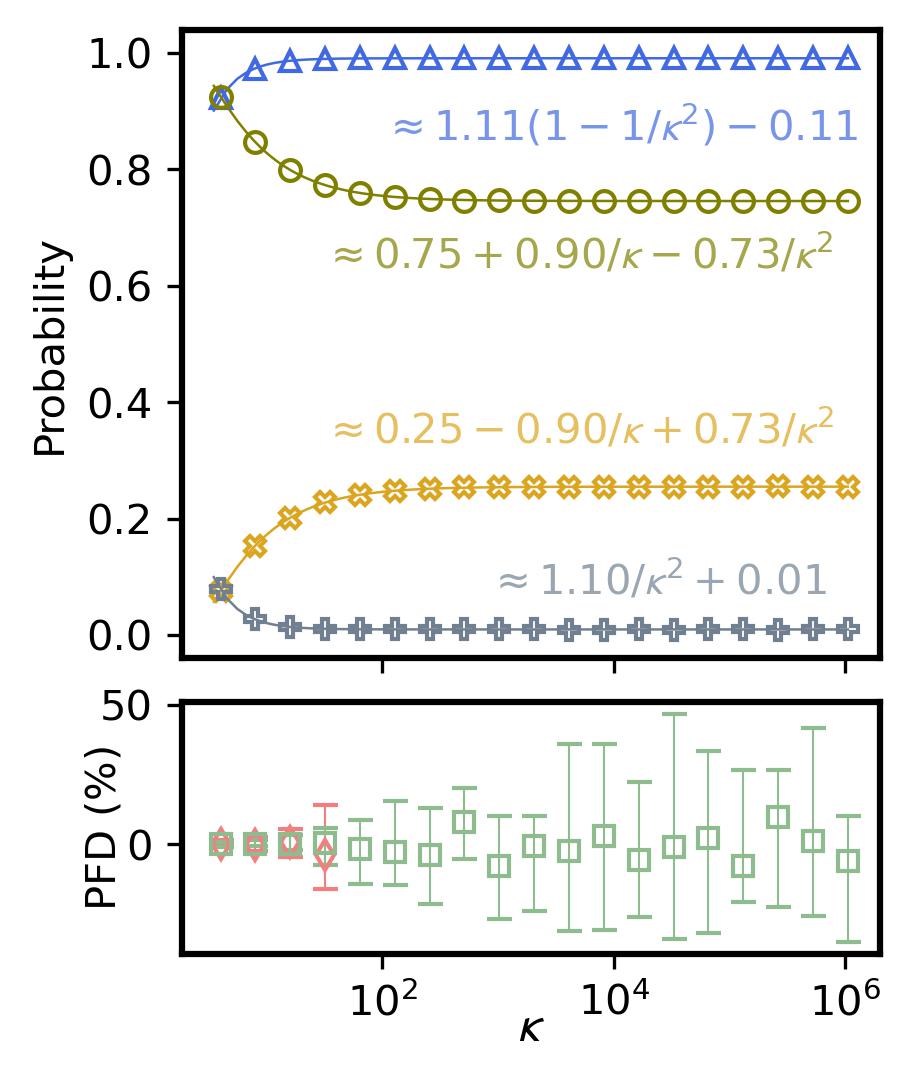}&
\includegraphics[scale=0.65]{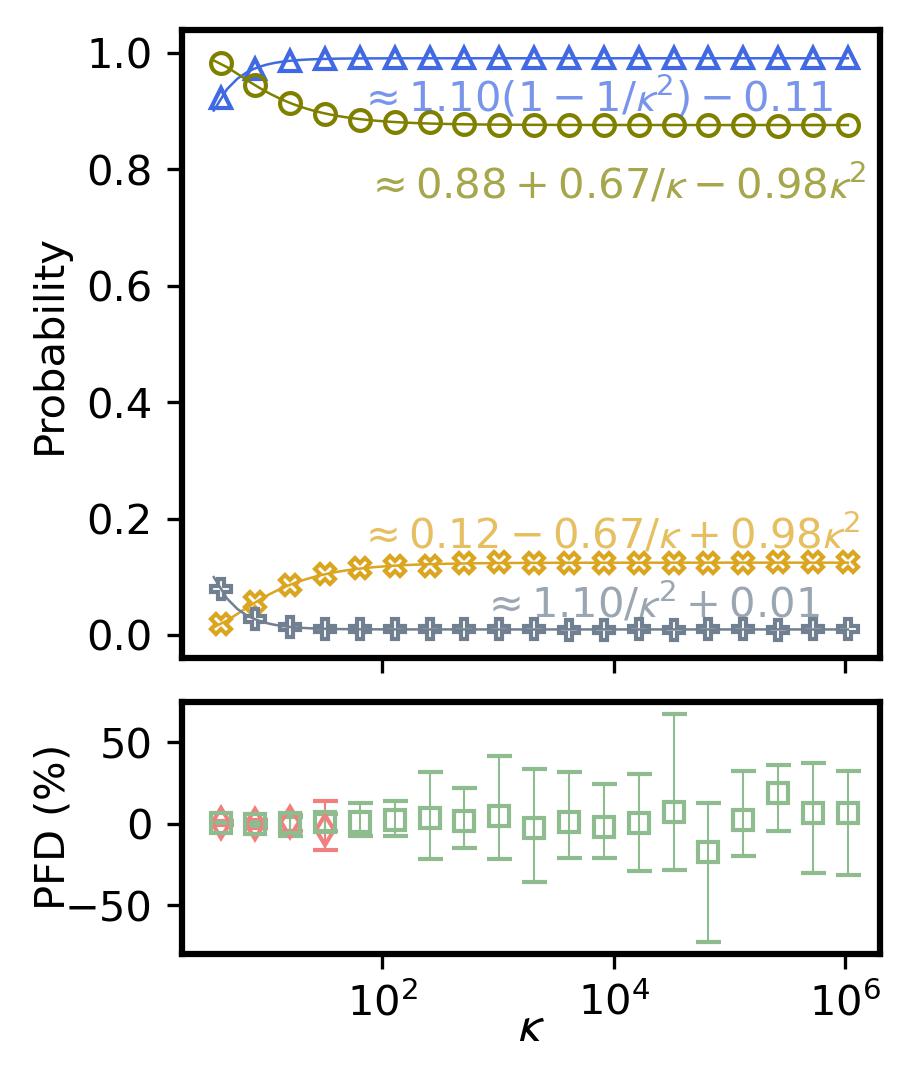}&\includegraphics[scale=0.65]{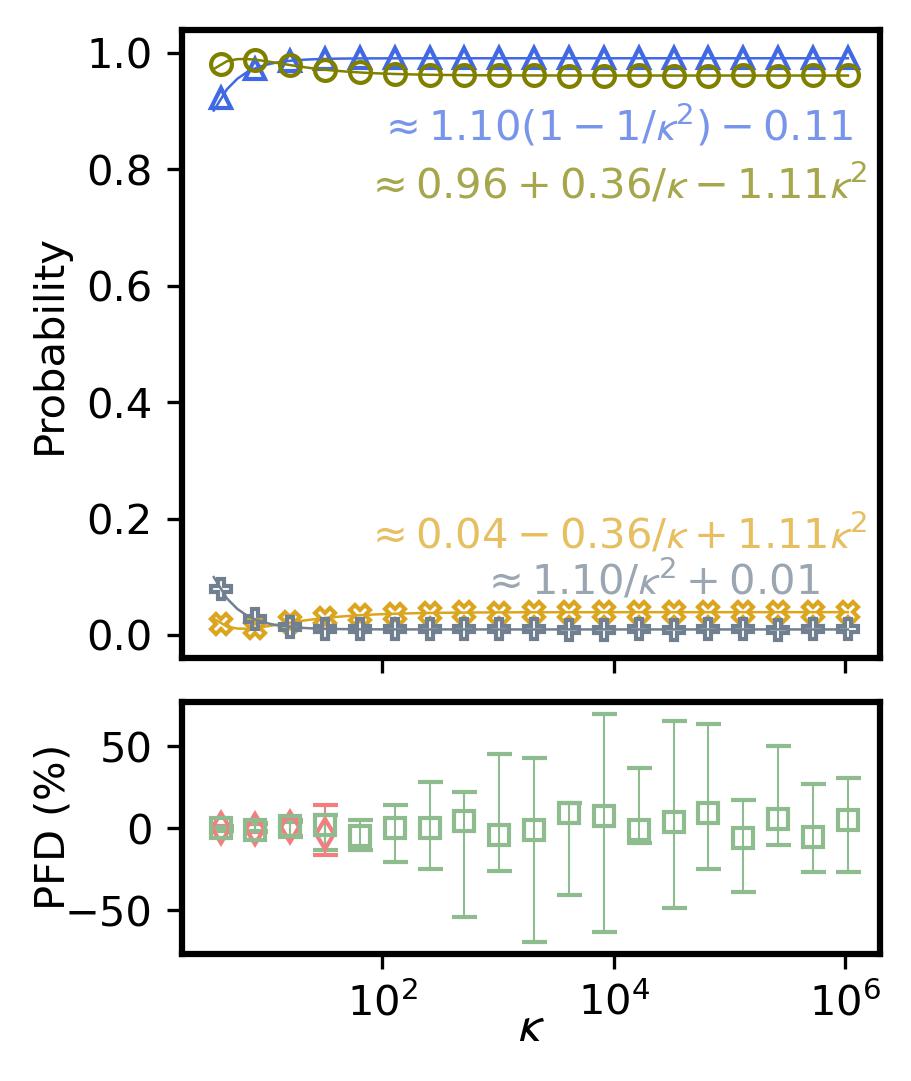}\\(g) $\alpha=60^\circ$ &(h) $\alpha=70^\circ$ &(i) 
 $\alpha=80^\circ$ 
\end{tabular}
\caption{Sub-figures presenting the average probabilities of HHL and Psi-HHL with varying condition number $\mathcal{\kappa}$ of $A$, for $4 \times 4$ matrices and using $10^5$ shots throughout. Sub-figures (a) through (c) consider the case where $A$ is diagonal and $\vec{b}$ is in equal superposition. Sub-figures (d) through (f) consider the case where $A$ is diagonal and $\vec{b}$ is in unequal superposition, and Sub-figures (g) through (i) consider the case where $A$ is not diagonal, and $\vec{b}$ is in unequal superposition. Each data point is generated by taking the average over 10 repetitions.}
\label{fig:4by4-1e5shots10reps}
\end{figure*}

\begin{figure*}[t]
\centering
    \begin{tabular}{ccc}
\includegraphics[scale=0.65]{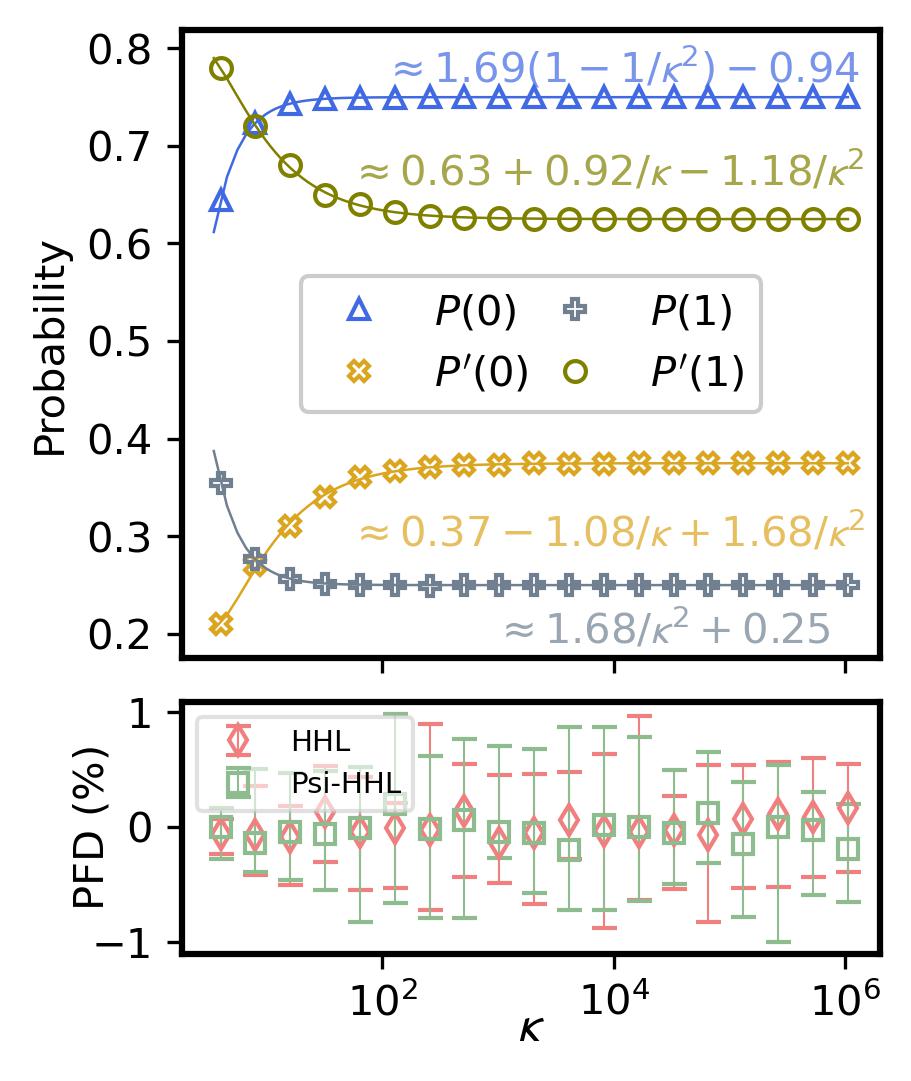}&
\includegraphics[scale=0.65]{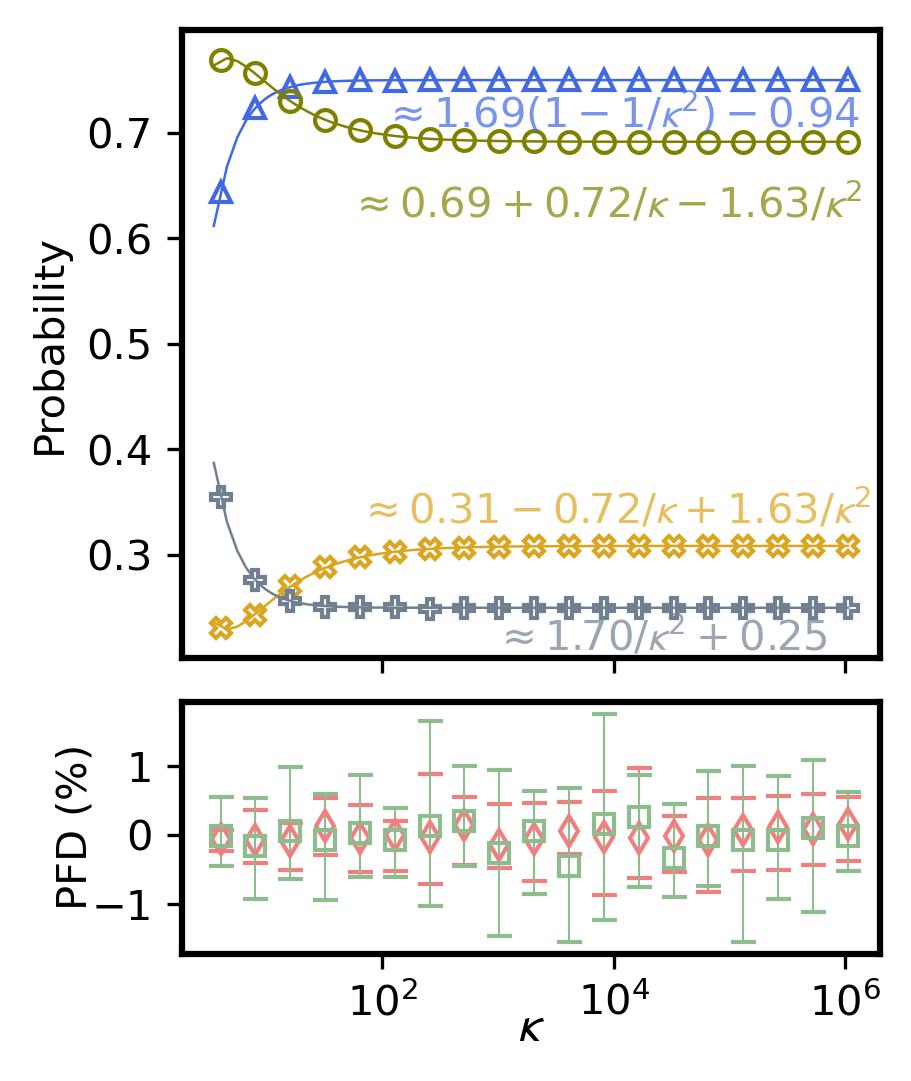}&\includegraphics[scale=0.65]{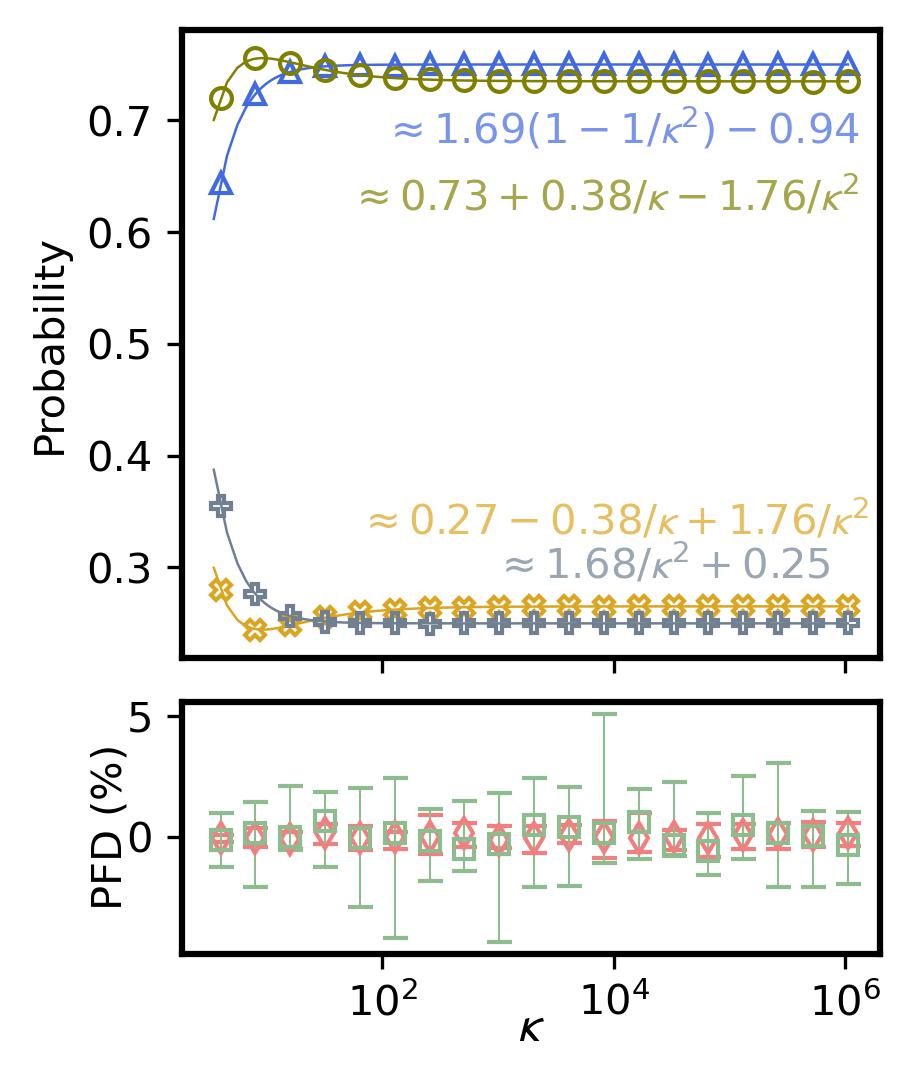}\\(a) $\alpha=60^\circ$ &(b) $\alpha=70^\circ$ &(c) 
 $\alpha=80^\circ$ 
\end{tabular}
\begin{tabular}{ccc}
\includegraphics[scale=0.65]{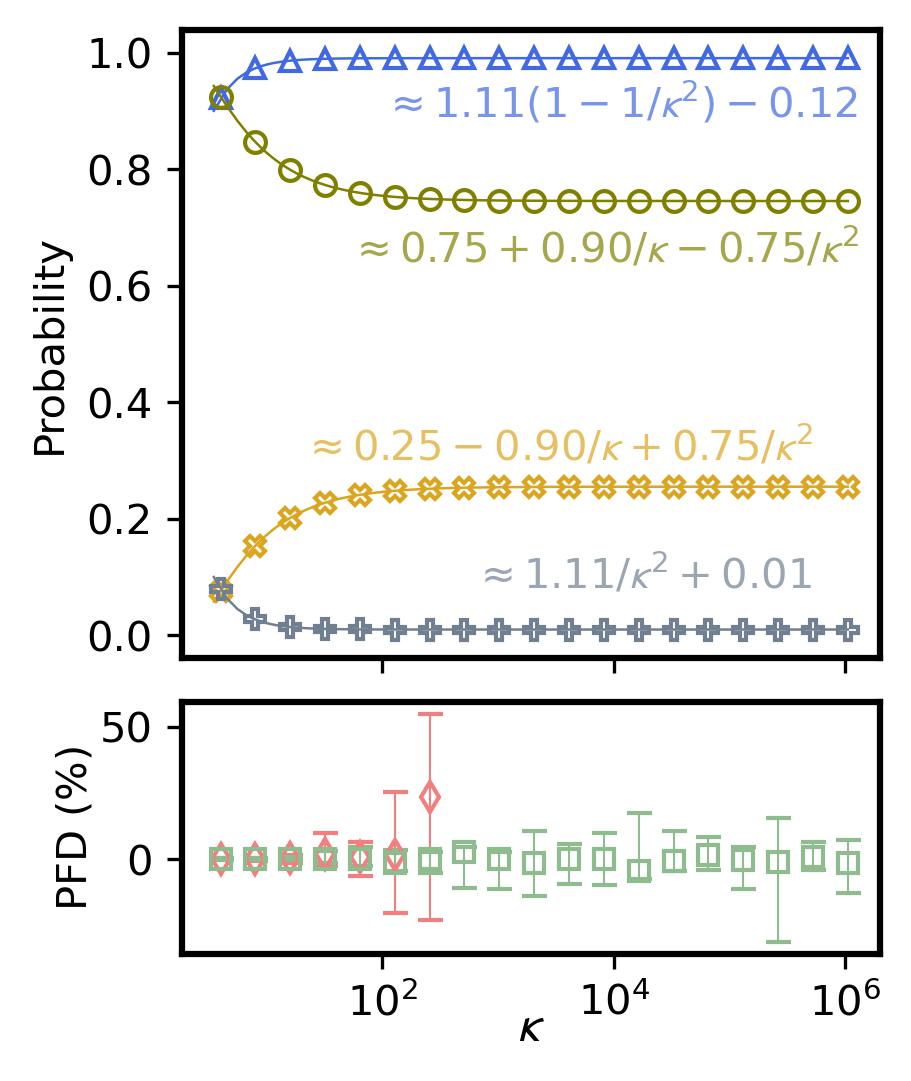}&
\includegraphics[scale=0.65]{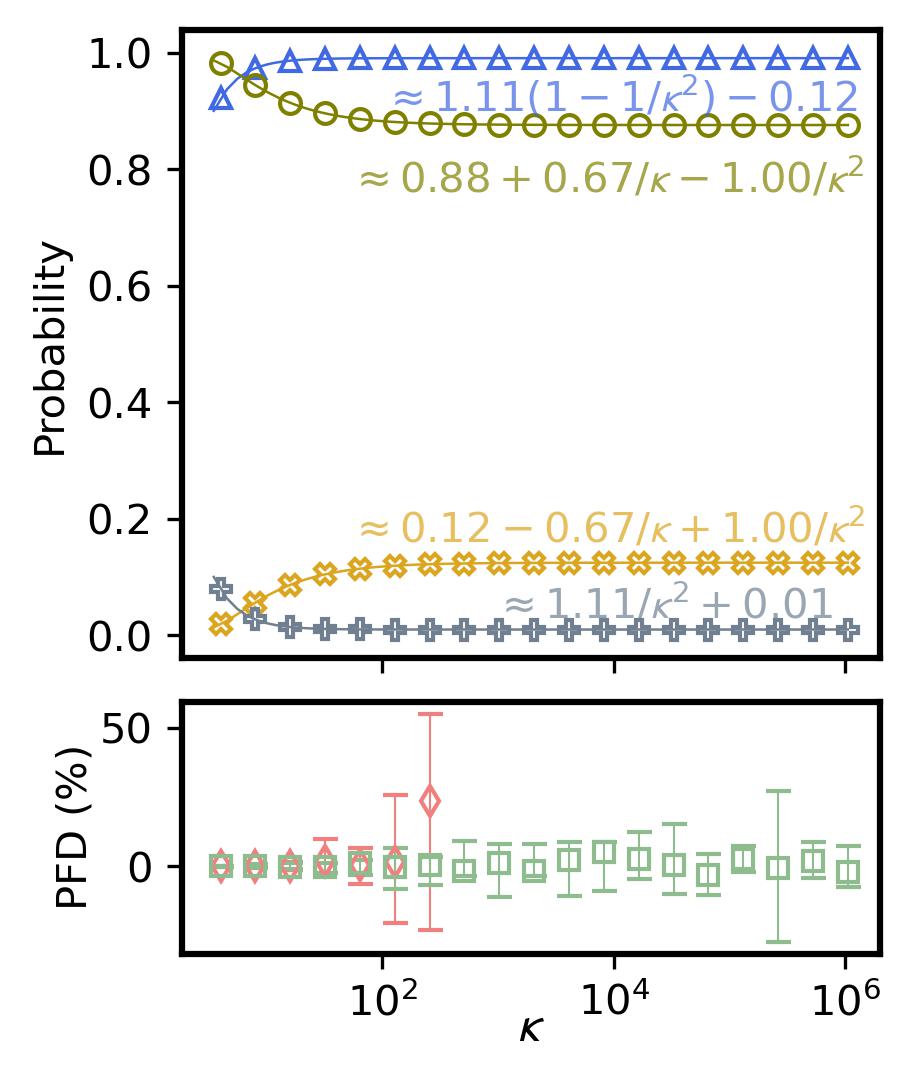}&\includegraphics[scale=0.65]{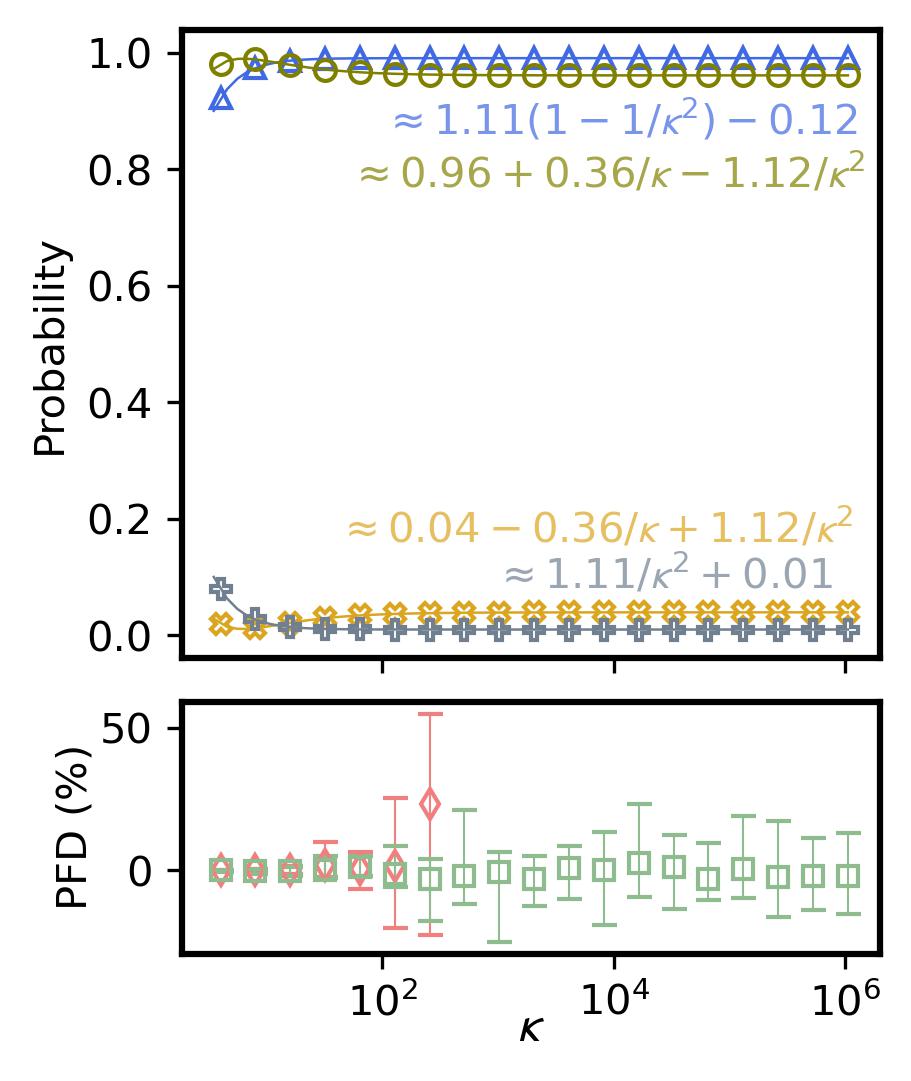}\\(d) $\alpha=60^\circ$ &(e) $\alpha=70^\circ$ &(f) 
 $\alpha=80^\circ$ 
\end{tabular}
\begin{tabular}{ccc}
\includegraphics[scale=0.65]{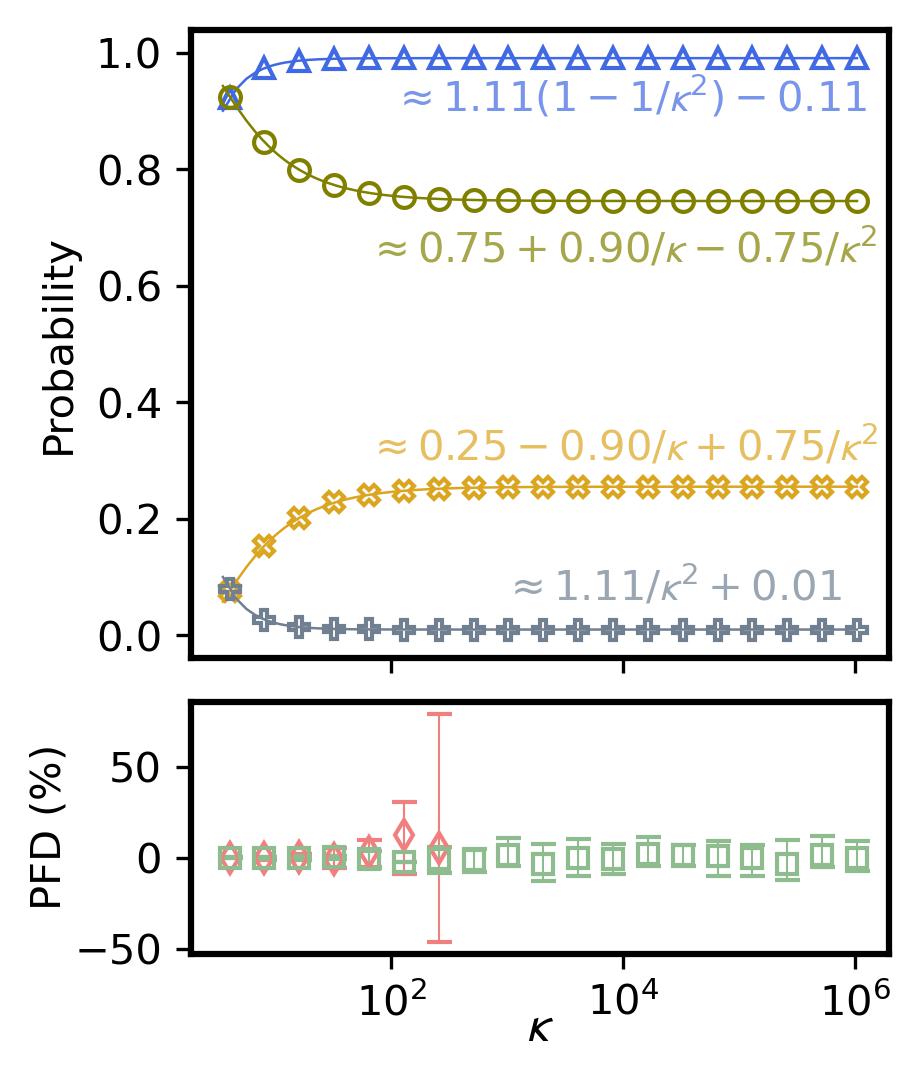}&
\includegraphics[scale=0.65]{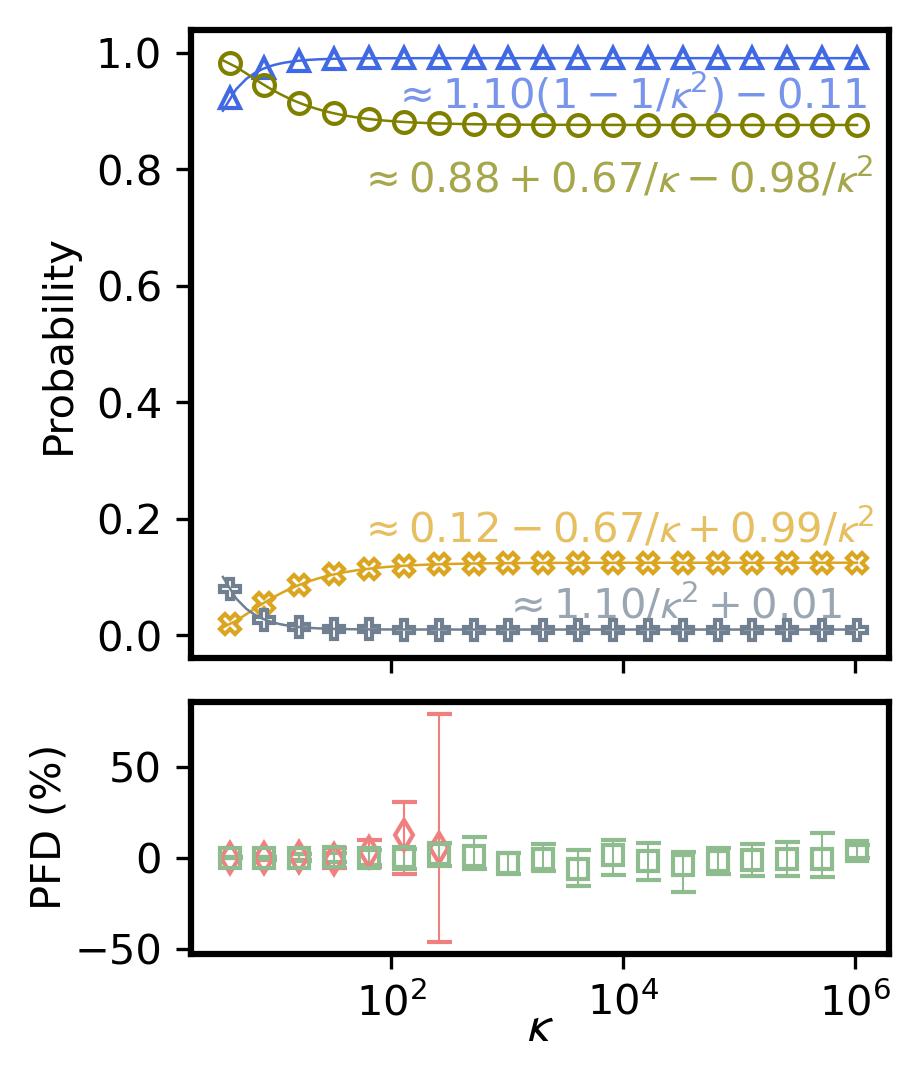}&\includegraphics[scale=0.65]{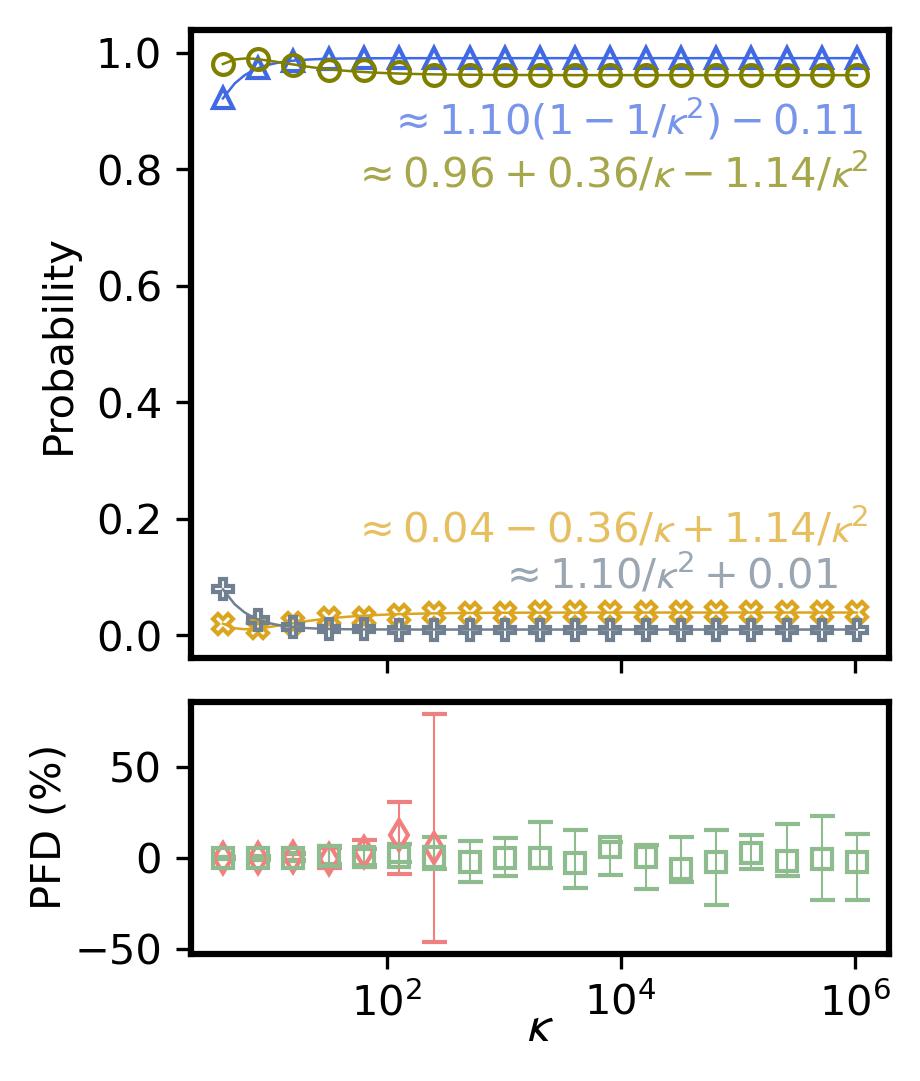}\\(g) $\alpha=60^\circ$ &(h) $\alpha=70^\circ$ &(i) 
 $\alpha=80^\circ$ 
\end{tabular}
\caption{Sub-figures presenting the average probabilities of HHL and Psi-HHL with varying condition number $\mathcal{\kappa}$ of $A$, for $4 \times 4$ matrices and using $10^6$ shots throughout. Sub-figures (a) through (c) consider the case where $A$ is diagonal and $\vec{b}$ is in equal superposition. Sub-figures (d) through (f) consider the case where $A$ is diagonal and $\vec{b}$ is in unequal superposition, and Sub-figures (g) through (i) consider the case where $A$ is not diagonal, and $\vec{b}$ is in unequal superposition. Each data point is generated by taking the average over 10 repetitions. }
\label{fig:4by4-1e6shots10reps}
\end{figure*}

\begin{table*}[]
\centering
\caption{Table presenting the data for HHL and Psi-HHL algorithms (with $\alpha=60^\circ$) for $4 \times 4$ matrices, and in particular, for three cases: $A$ diagonal and $\vec{b}$ in equal superposition (denoted in the table as `$A$ diag, $\vec{b}$ equal'), $A$ diagonal and $\vec{b}$ not in equal superposition (denoted in the table as `$A$ diag, $\vec{b}$ unequal'), and $A$ not diagonal and $\vec{b}$ in an unequal superposition (denoted in the table as `$A$ not diag, $\vec{b}$ unequal'). The string `nan’ refers to situations where we get the square root of a negative quantity for energy, which is unphysical, and hence is not a valid result. Each data point is generated by taking the average over 10 repetitions. '\# shots' refers to the number of shots, while `\# qubits' denotes the number of qubits. }
\resizebox{\textwidth}{!}{
\begin{tabular}{c|cccccccc}
\hline \hline 
 &  & \multicolumn{1}{c}{} & \multicolumn{2}{c}{$A$ diag, $\vec{b}$ equal} & \multicolumn{2}{c}{$A$ diag, $\vec{b}$ unequal} & \multicolumn{2}{c}{$A$ not diag, $\vec{b}$ unequal} \\ 
\multirow{-2}{*}{\# shots} & \multirow{-2}{*}{$\mathcal{\approx \mathcal{\kappa}}$} & \multicolumn{1}{c}{\multirow{-2}{*}{\# qubits}} & \multicolumn{1}{c}{PFD ($P(0), P(1)$)} & PFD ($P'(0), P'(1)$) & \multicolumn{1}{c}{PFD ($P(0), P(1))$} & PFD ($P'(0), P'(1)$) & \multicolumn{1}{c}{PFD ($P(0), P(1)$)} & PFD ($P'(0), P'(1)$) \\ \hline
 & $2^2$ & 8 & \multicolumn{1}{c}{-0.04 (0.65, 0.35)} & \multicolumn{1}{c}{-0.22 (0.21, 0.79)} & \multicolumn{1}{c}{-0.25 (0.92, 0.08)} & \multicolumn{1}{c}{-0.10 (0.06, 0.94)} & \multicolumn{1}{c}{ 0.27 (0.92,  0.08)} & \multicolumn{1}{c}{ -0.04 (0.08, 0.92)} \\ 
 &  $2^3$ & 9 & \multicolumn{1}{c}{0.02 (0.72, 0.28)} & \multicolumn{1}{c}{-0.07 (0.27, 0.73)} & \multicolumn{1}{c}{ 0.27 (0.97, 0.03)} & \multicolumn{1}{c}{-0.32 (0.15, 0.85)} & \multicolumn{1}{c}{-0.32 (0.97, 0.03)} & \multicolumn{1}{c}{ 0.08 (0.15, 0.85)} \\ 
 &  $2^4$ & 10 & \multicolumn{1}{c}{-0.19 (0.74, 0.26)} & \multicolumn{1}{c}{ 0.32 (0.32, 0.68)} & \multicolumn{1}{c}{ 0.37 (0.99, 0.01)} & \multicolumn{1}{c}{ -0.69 (0.20, 0.80)} & \multicolumn{1}{c}{ 0.96 (0.99, 0.01)} & \multicolumn{1}{c}{0.26 (0.20, 0.80)} \\ 
 &  $2^5$ & 11 & \multicolumn{1}{c}{-0.17 (0.75, 0.25)} & \multicolumn{1}{c}{-0.02 (0.35, 0.65)} & \multicolumn{1}{c}{  -1.20 (0.99, 0.01)} & \multicolumn{1}{c}{2.85 (0.23, 0.77)} & \multicolumn{1}{c}{-3.65 (0.99, 0.01)} & \multicolumn{1}{c}{ 0.32 (0.22, 0.78)} \\ 
 &  $2^6$ & 12 & \multicolumn{1}{c}{ 0.08 (0.75, 0.25)} & \multicolumn{1}{c}{0.05 (0.37, 0.63)} & \multicolumn{1}{c}{ nan (0.99, 0.01)} & \multicolumn{1}{c}{1.66 (0.24, 0.76)} & \multicolumn{1}{c}{nan (0.99, 0.01)} & \multicolumn{1}{c}{ -1.56 (0.24, 0.76)} \\
\hline \hline 
\end{tabular}
}
\label{tab:A210reps}
\end{table*}

\begin{table*}[]
\centering
\caption{Table providing the data for HHL and Psi-HHL algorithms (with $\alpha=70^\circ$) for $4 \times 4$ matrices. The notations are the same as those used in Table \ref{tab:A210reps}. Each data point is generated by taking the average over 10 repetitions. }
\resizebox{\textwidth}{!}{
\begin{tabular}{c|cccccccc}
\hline \hline 
 &
   &
   &
  \multicolumn{2}{c}{$A$ diag, $\vec{b}$ equal} &
  \multicolumn{2}{c}{$A$ diag, $\vec{b}$ unequal} &
  \multicolumn{2}{c}{$A$ not diag, $\vec{b}$ unequal} \\  
\multirow{-2}{*}{\# shots} &
  \multirow{-2}{*}{$\mathcal{\approx \mathcal{\kappa}}$} &
  \multirow{-2}{*}{\# qubits} &
  \multicolumn{1}{c}{PFD ($P(0), P(1)$)} &
  PFD ($P'(0), P'(1)$) &
  \multicolumn{1}{c}{PFD ($P(0), P(1)$)} &
  \multicolumn{1}{c}{PFD ($P'(0), P'(1)$)} &
  \multicolumn{1}{c}{PFD ($P(0), P(1)$)} &
  PFD ($P'(0), P'(1)$) \\ \hline
 &
  $2^2$ &
  8 &
  \multicolumn{1}{c}{-0.04 (0.65, 0.35)} &
  -0.24 (0.23, 0.77) &
  \multicolumn{1}{c}{-0.25 (0.92, 0.08)} &
 -0.03 (0.02, 0.98) &
  \multicolumn{1}{c}{0.27 (0.92,  0.08)} &
0.58 (0.02, 0.98) \\ 
 &
$2^3$ &
  9 &
  \multicolumn{1}{c}{0.02 (0.72, 0.28)} &
  -0.28 (0.25, 0.75) &
  \multicolumn{1}{c}{0.27 (0.97, 0.03)} &
 0.14 (0.06, 0.94) &
  \multicolumn{1}{c}{-0.32 (0.97, 0.03)} &
 -0.41 (0.06, 0.94) \\
 &
  $2^4$ &
  10 &
  \multicolumn{1}{c}{-0.19 (0.74, 0.26)} &
  0.51 (0.27, 0.73) &
  \multicolumn{1}{c}{0.37 (0.99, 0.01)} &
  -0.35 (0.12, 0.88) &
  \multicolumn{1}{c}{0.96 (0.99, 0.01)} &
   0.44 (0.09, 0.91) \\ 
 &
 $2^5$ &
  11 &
  \multicolumn{1}{c}{-0.17 (0.75, 0.25)} &
  0.05 (0.29, 0.71) &
  \multicolumn{1}{c}{-1.20 (0.99, 0.01)} &
 -0.99 (0.10, 0.90) &
  \multicolumn{1}{c}{-3.65 (0.99, 0.01)} &
   1.17 (0.10, 0.90) \\  
 &
  $2^6$ &
  12 &
  \multicolumn{1}{c}{0.08 (0.75, 0.25)} &
  0.54 (0.30, 0.70) &
  \multicolumn{1}{c}{nan (0.99, 0.01)} &
  1.40 (0.12, 0.88) &
  \multicolumn{1}{c}{nan (0.99, 0.01)} &
  0.99 (0.11, 0.89) \\  
 &
  $2^7$ &
  13 &
  \multicolumn{1}{c}{-0.05 (0.75, 0.25)} &
 0.74 (0.30, 0.70) &
  \multicolumn{1}{c}{nan (0.99, 0.01)} &
  2.34 (0.12, 0.88) &
  \multicolumn{1}{c}{nan (0.99, 0.01)} &
  2.20 (0.02, 0.88) \\
\hline \hline 
\end{tabular}
}
\label{tab:A310reps}
\end{table*}

\begin{table*}[]
\centering
\caption{Table providing the data for HHL and Psi-HHL algorithms (with $\alpha=80^\circ$) for $4 \times 4$ matrices. The notations are the same as those used in Tables \ref{tab:A210reps} and \ref{tab:A310reps}. Each data point is generated by taking the average over 10 repetitions. }
\resizebox{\textwidth}{!}{
\begin{tabular}{c|cccccccc}
\hline \hline 
 &
   &
   &
  \multicolumn{2}{c}{$A$ diag, $\vec{b}$ equal} &
  \multicolumn{2}{c}{$A$ diag, $\vec{b}$ unequal} &
  \multicolumn{2}{c}{$A$ not diag, $\vec{b}$ unequal} \\
\multirow{-2}{*}{\# shots} &
  \multirow{-2}{*}{$\mathcal{\approx \mathcal{\kappa}}$} &
  \multirow{-2}{*}{\# qubits} &
  \multicolumn{1}{c}{PFD ($P(0), P(1)$)} &
  PFD ($P'(0), P'(1)$) &
  \multicolumn{1}{c}{PFD ($P(0), P(1)$)} &
  PFD ($P'(0), P'(1)$) &
  \multicolumn{1}{c}{PFD ($P(0), P(1)$)} &
  PFD ($P'(0), P'(1)$) \\ \hline
\multicolumn{1}{c|}{} &
  $2^2$ &
  8 &
  \multicolumn{1}{c}{-0.04 (0.65, 0.35)} &
  0.43 (0.21, 0.79) &
  \multicolumn{1}{c}{-0.25 (0.92, 0.08)} &
  0.06 (0.01, 0.99) &
  \multicolumn{1}{c}{0.27 (0.92,  0.08)} &
  0.31 (0.02, 0.98) \\ 
\multicolumn{1}{c|}{} &
  $2^3$ &
  9 &
  \multicolumn{1}{c}{0.02 (0.72, 0.28)} &
  0.29 (0.27, 0.73) &
  \multicolumn{1}{c}{ 0.27 (0.97, 0.03)} &
  -0.59 (0.02, 0.98) &
  \multicolumn{1}{c}{-0.32 (0.97, 0.03)} &
  -0.92 (0.02, 0.98) \\  
\multicolumn{1}{c|}{} &
  $2^4$ &
  10 &
  \multicolumn{1}{c}{-0.19 (0.74, 0.26)} &
  2.47 (0.32, 0.68) &
  \multicolumn{1}{c}{0.37 (0.99, 0.01)} &
 2.59 (0.02, 0.98) &
  \multicolumn{1}{c}{ 0.96 (0.99, 0.01)} &
  1.55 (0.02, 0.98) \\ 
\multicolumn{1}{c|}{} &
  $2^5$ &
  11 &
  \multicolumn{1}{c}{-0.17 (0.75, 0.25)} &
  0.67 (0.35, 0.65) &
  \multicolumn{1}{c}{-1.20 (0.99, 0.01)} &
  1.63 (0.03, 0.97) &
  \multicolumn{1}{c}{-3.65 (0.99, 0.01)} &
  1.89 (0.03, 0.97) \\  
\multicolumn{1}{c|}{} &
 $2^6$ &
  12 &
  \multicolumn{1}{c}{0.08 (0.75, 0.25)} &
 -0.17 (0.36, 0.64) &
  \multicolumn{1}{c}{nan (0.99, 0.01)} &
  5.34 (0.04, 0.96) &
  \multicolumn{1}{c}{nan (0.99, 0.01)} &
  -4.59 (0.03, 0.97) \\ 
\multicolumn{1}{c|}{} &
  $2^7$ &
  13 &
  \multicolumn{1}{c}{-0.05 (0.75, 0.25)} &
  -0.30 (0.37, 0.63) &
  \multicolumn{1}{c}{nan (0.99, 0.01)} &
  0.55 (0.04, 0.96) &
  \multicolumn{1}{c}{nan (0.99, 0.01)} &
  0.37 (0.04, 0.96) \\
\hline \hline 
\end{tabular}
}
\label{tab:A410reps}
\end{table*}

In our Results section in the main text, we discussed the case of $A$ not diagonal and the amplitude encoded state corresponding to $\vec{b}$ in an unequal superposition. We also provide our $4 \times 4$ through $64 \times 64$ matrices that we considered in the main text in Figure \ref{fig:mat-sizes}. 

We now provide additional data on two more choices for our $4 \times 4$ $A$ matrices: 

\begin{enumerate}
\item $A$ is diagonal and the amplitude encoded state corresponding to $\vec{b}$ is in an equal superposition. In particular, 	$A=%
$, and the un-normalized $\vec{b}=%
$. The sparsity, $s$, of $A$ is  1. 
\item $A$ is diagonal and the amplitude encoded state associated with $\vec{b}$ is in an unequal superposition. In particular, $A$ is the same as in case 1 given above, and $\vec{b}=%
$. 
\end{enumerate} 

\begin{figure*}[t]
\centering
    \begin{tabular}{ccc}
\includegraphics[scale=0.65]{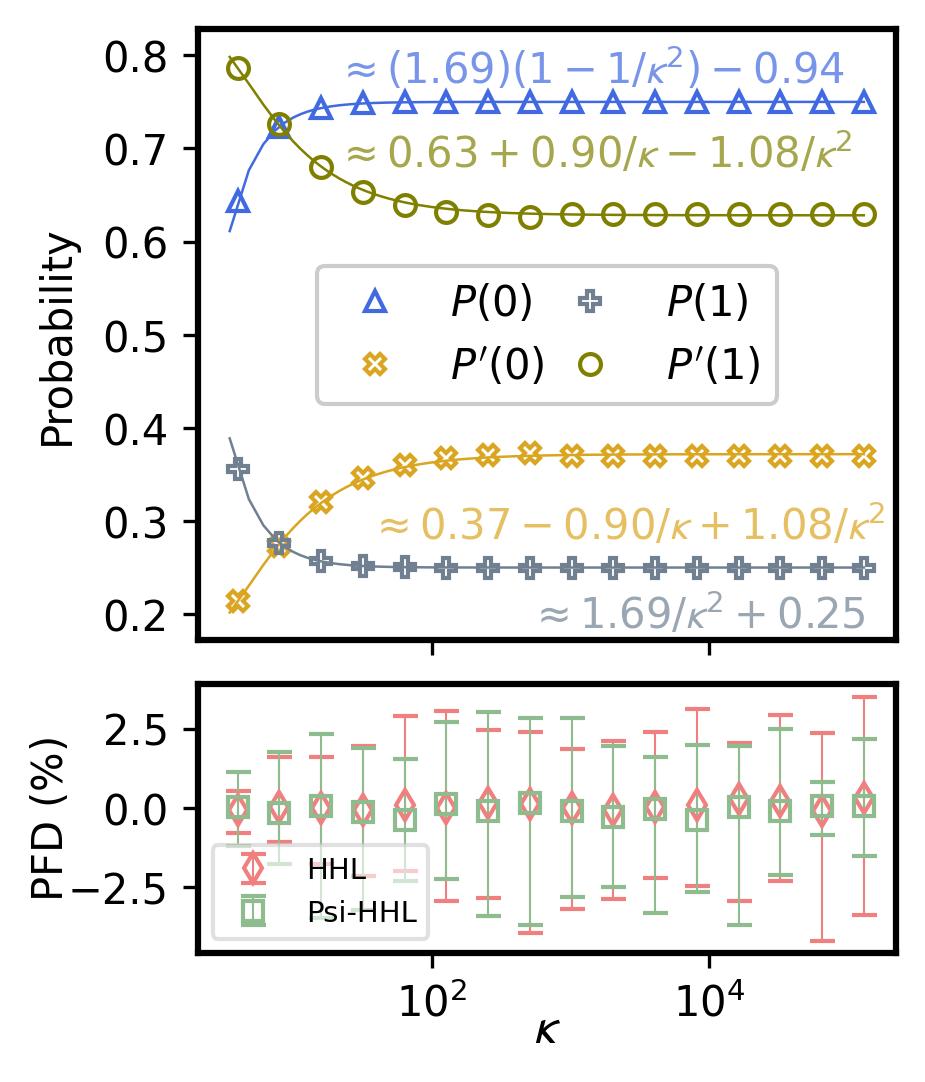}&
\includegraphics[scale=0.65]{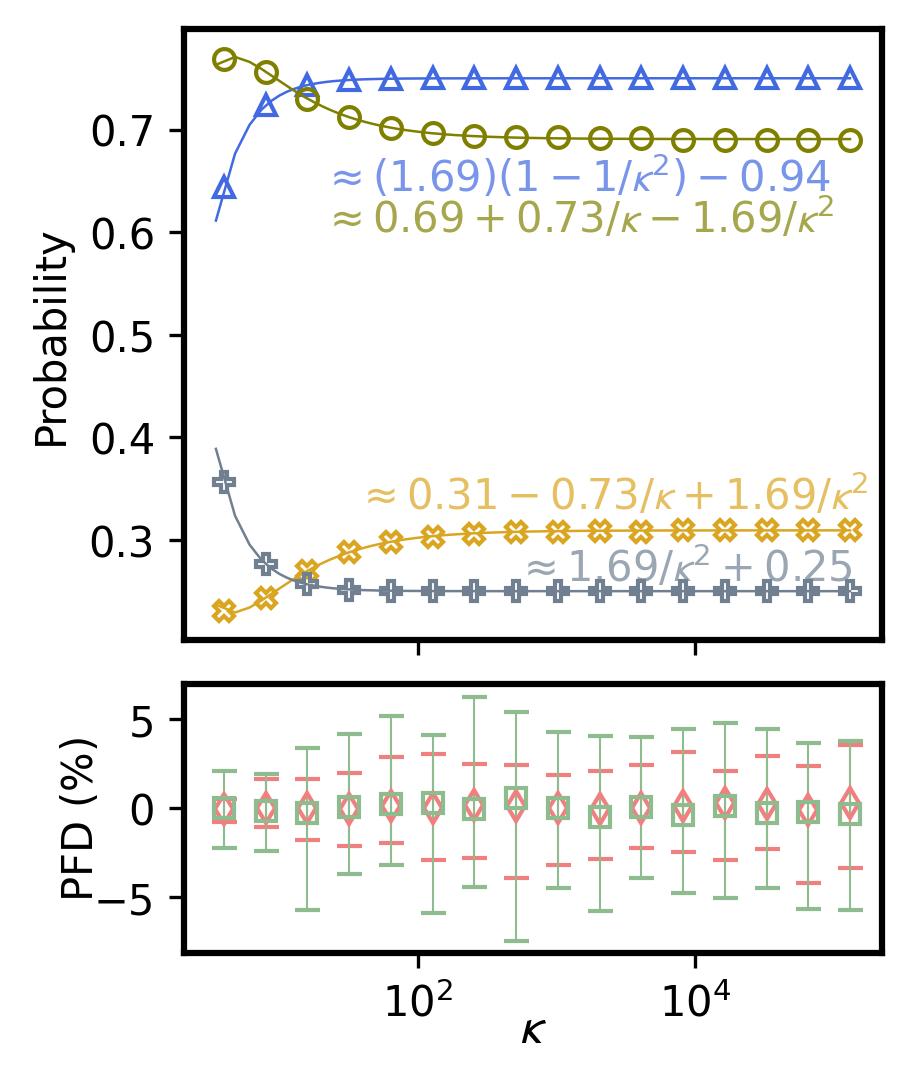}&\includegraphics[scale=0.65]{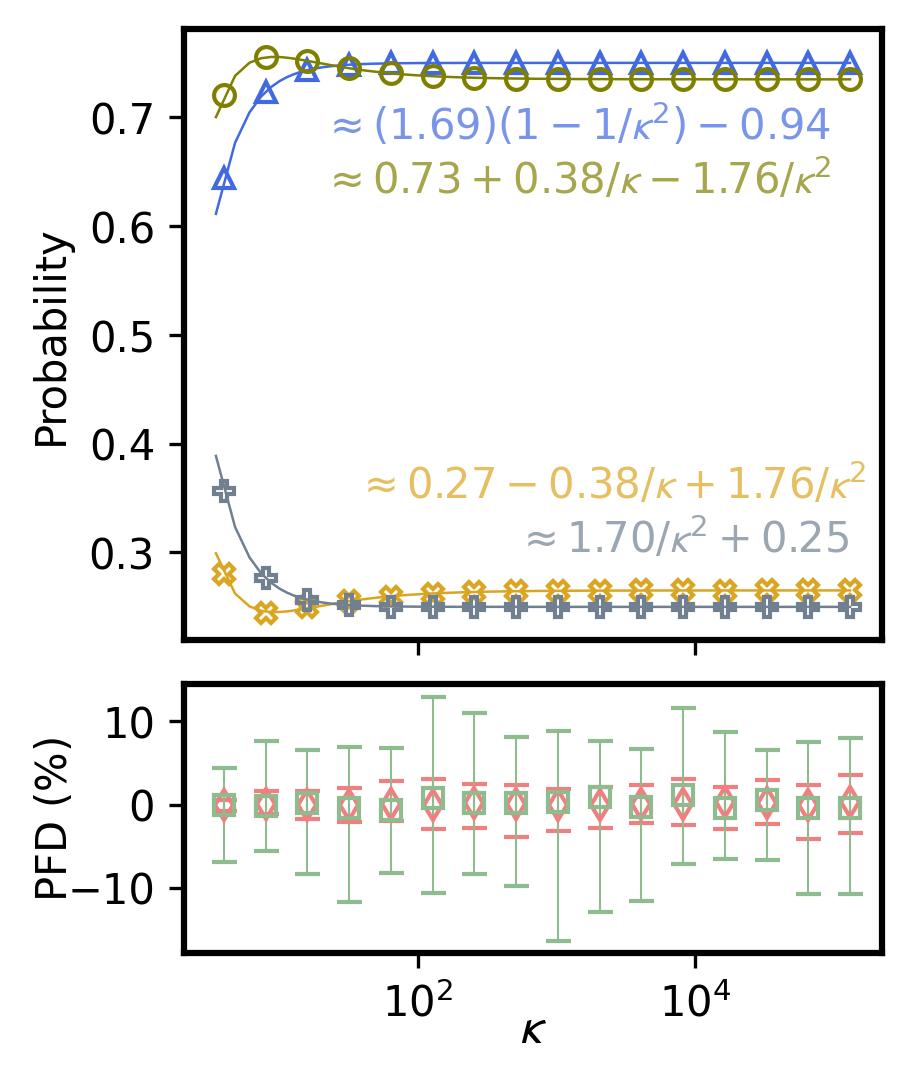}\\(a) $\alpha=60^\circ$ &(b) $\alpha=70^\circ$ &(c) 
 $\alpha=80^\circ$ 
\end{tabular}
\begin{tabular}{ccc}
\includegraphics[scale=0.65]{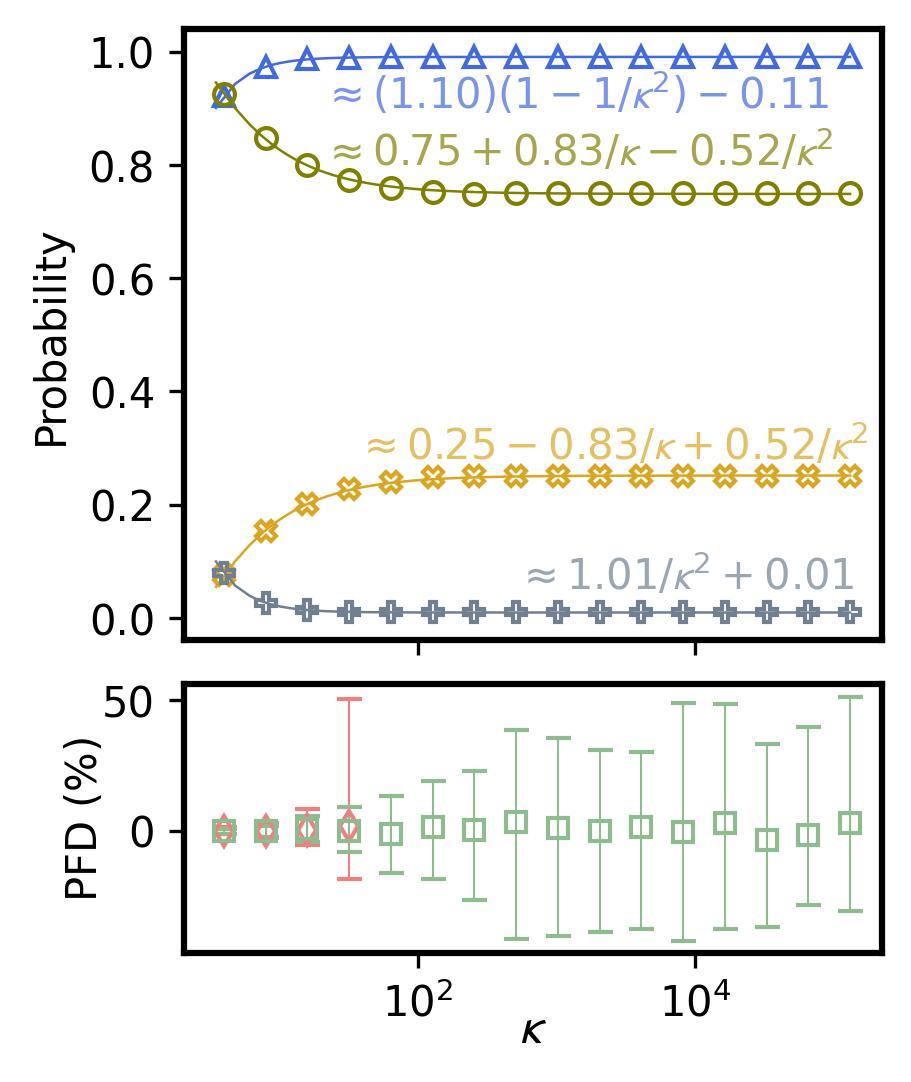}&
\includegraphics[scale=0.65]{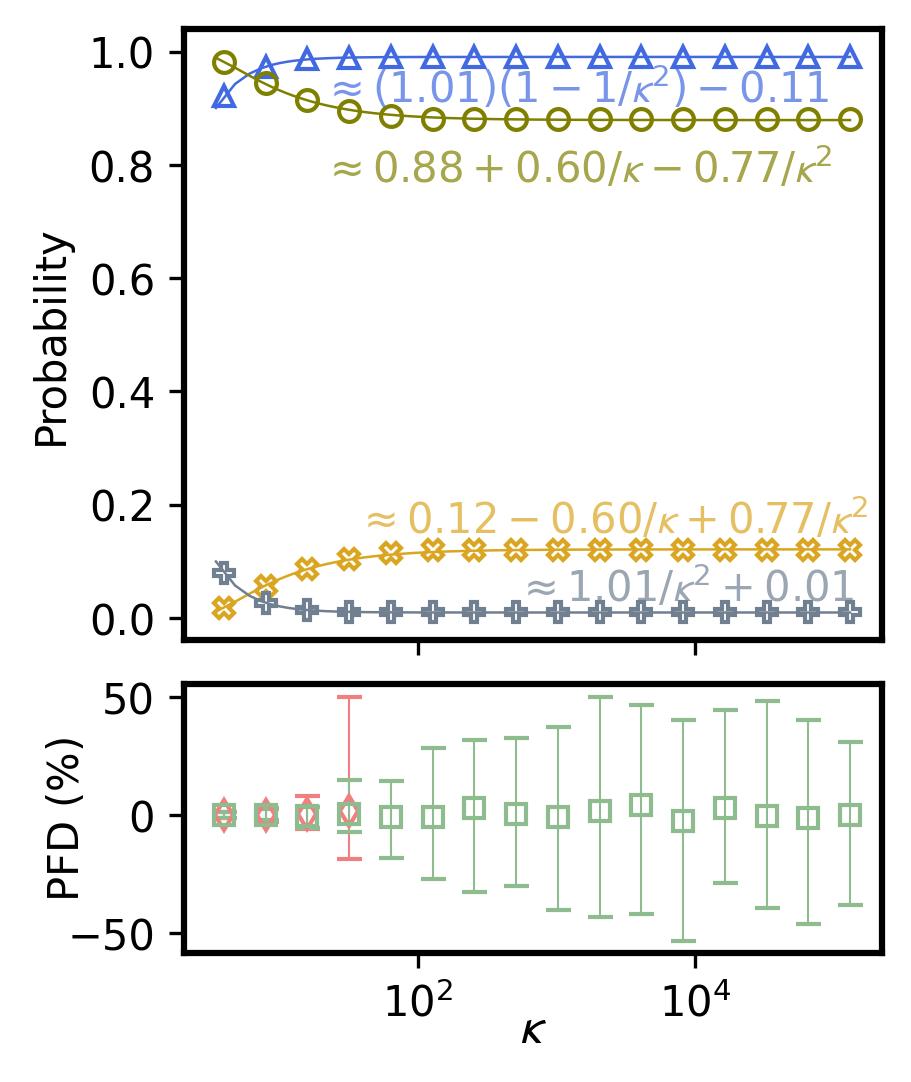}&\includegraphics[scale=0.65]{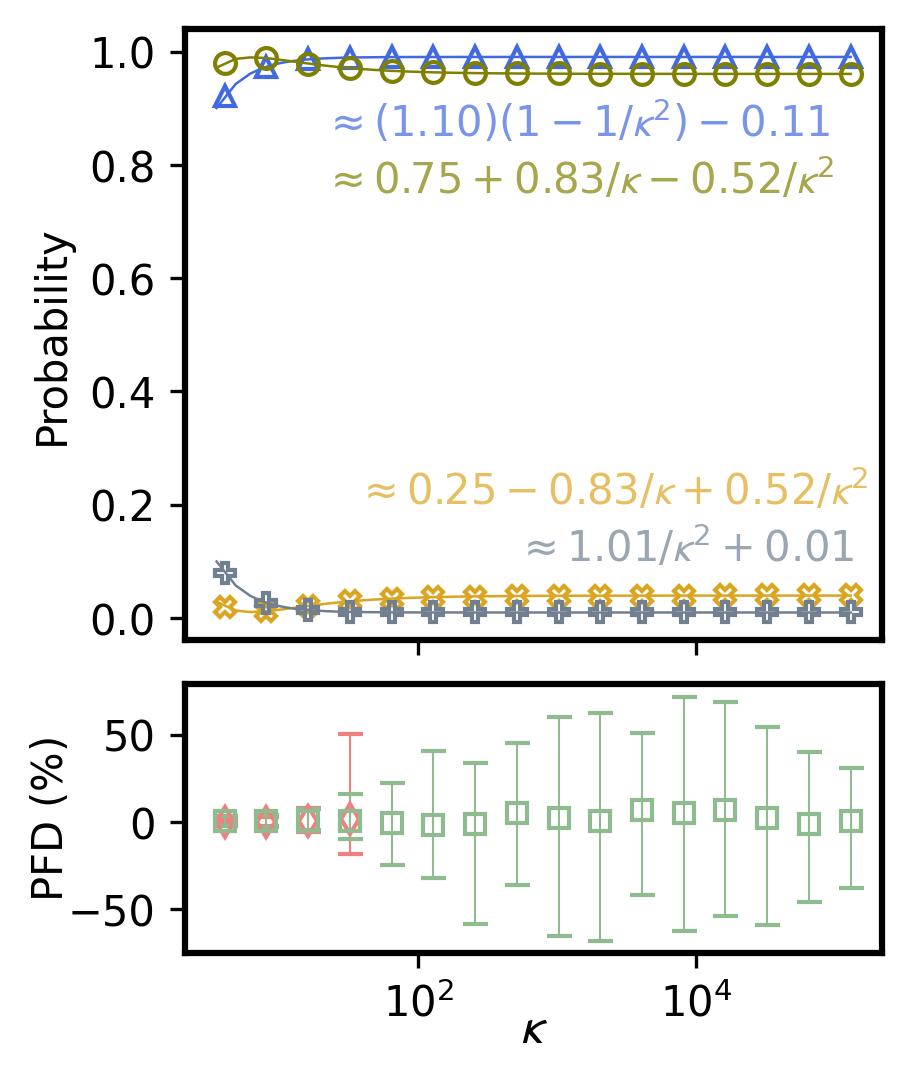}\\(d) $\alpha=60^\circ$ &(e) $\alpha=70^\circ$ &(f) 
 $\alpha=80^\circ$ 
\end{tabular}
\begin{tabular}{ccc}
\includegraphics[scale=0.65]{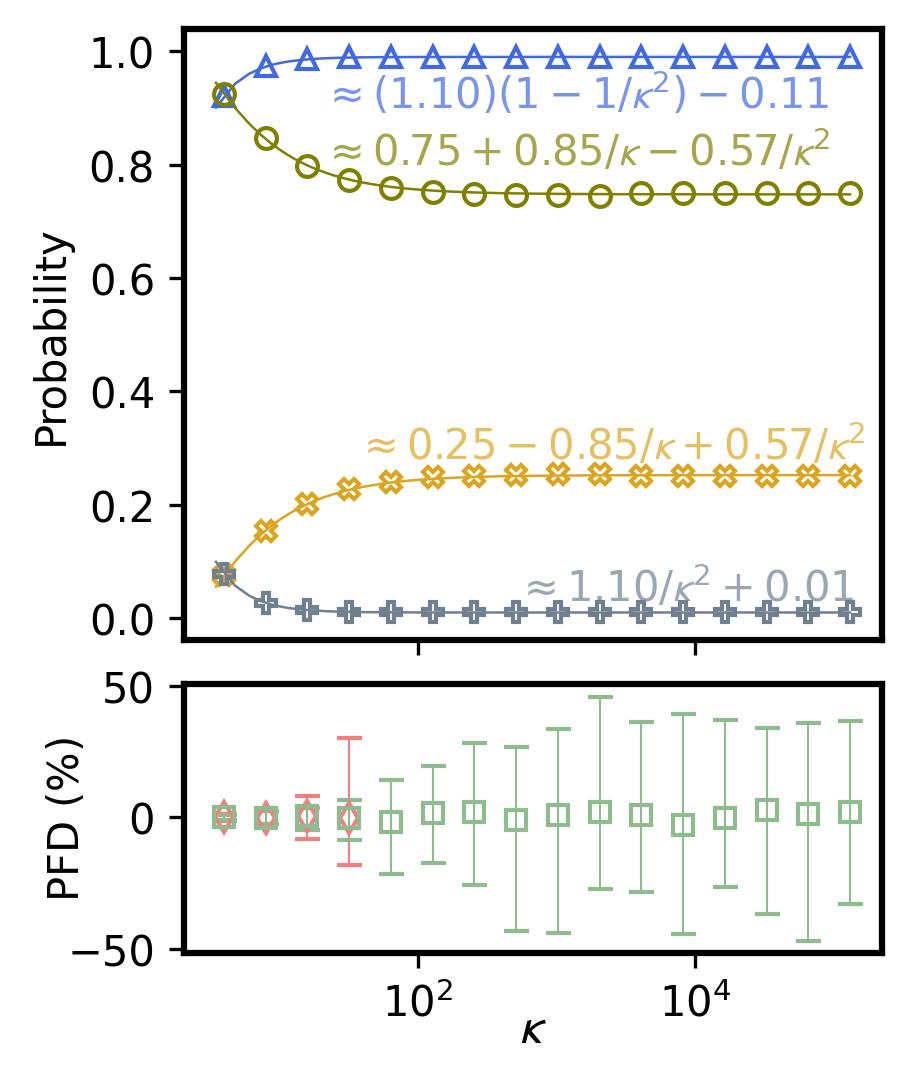}&
\includegraphics[scale=0.65]{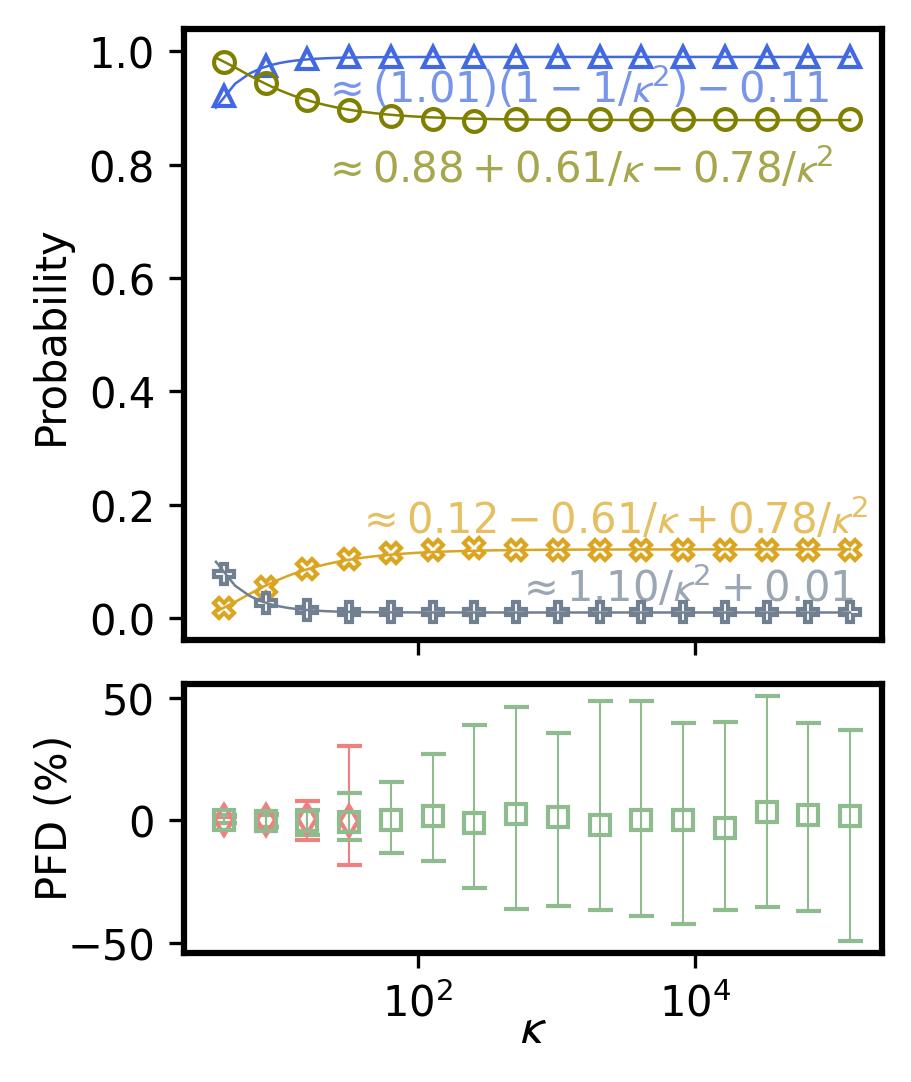}&\includegraphics[scale=0.65]{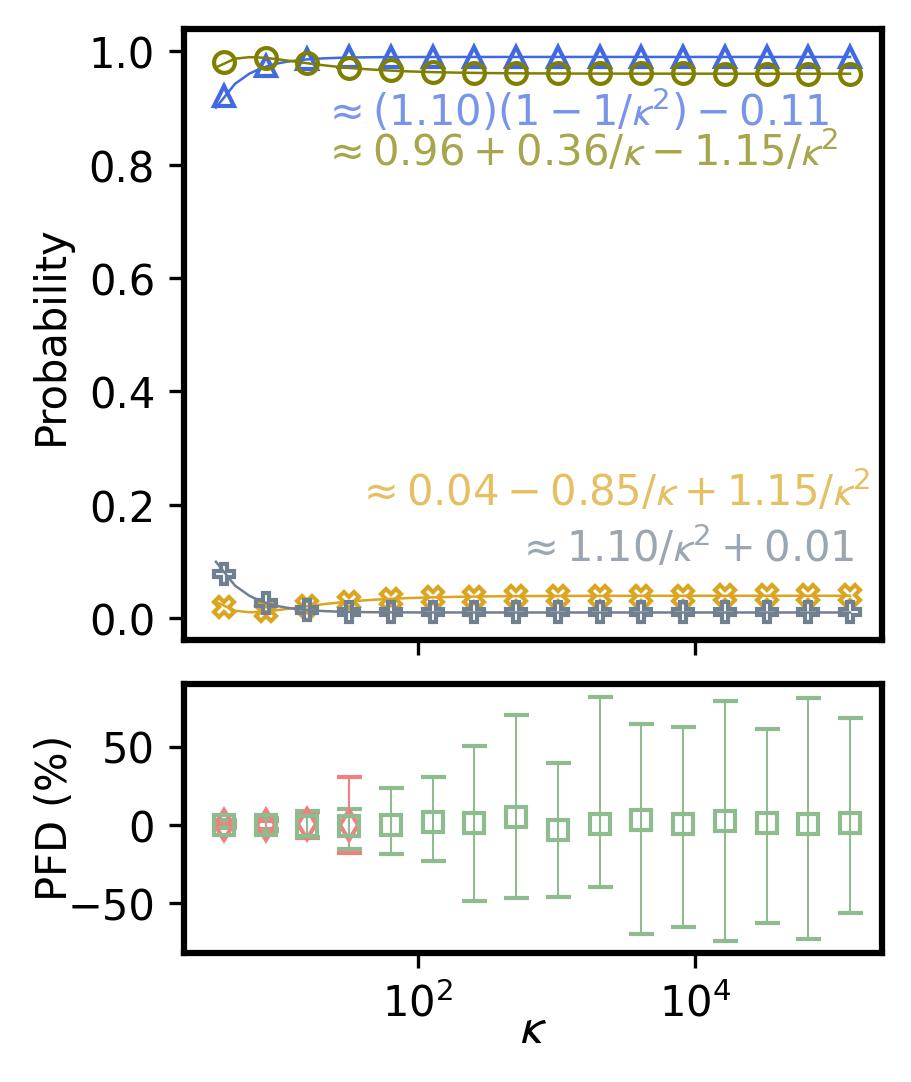}\\(g) $\alpha=60^\circ$ &(h) $\alpha=70^\circ$ &(i) 
 $\alpha=80^\circ$ 
\end{tabular}
\caption{Sub-figures presenting the average probabilities of HHL and Psi-HHL with varying condition number $\mathcal{\kappa}$ of $A$, for $4 \times 4$ matrices and using $10^5$ shots throughout. Sub-figures (a) through (c) consider the case where $A$ is diagonal and $\vec{b}$ is in equal superposition. Sub-figures (d) through (f) consider the case where $A$ is diagonal and $\vec{b}$ is in unequal superposition, and Sub-figures (g) through (i) consider the case where $A$ is not diagonal and $\vec{b}$ is in unequal superposition. Each data point is generated by taking the average over 50 repetitions. }
\label{fig:4by4-1e5shots}
\end{figure*}

\begin{figure*}[t]
\centering
    \begin{tabular}{ccc}
\includegraphics[scale=0.65]{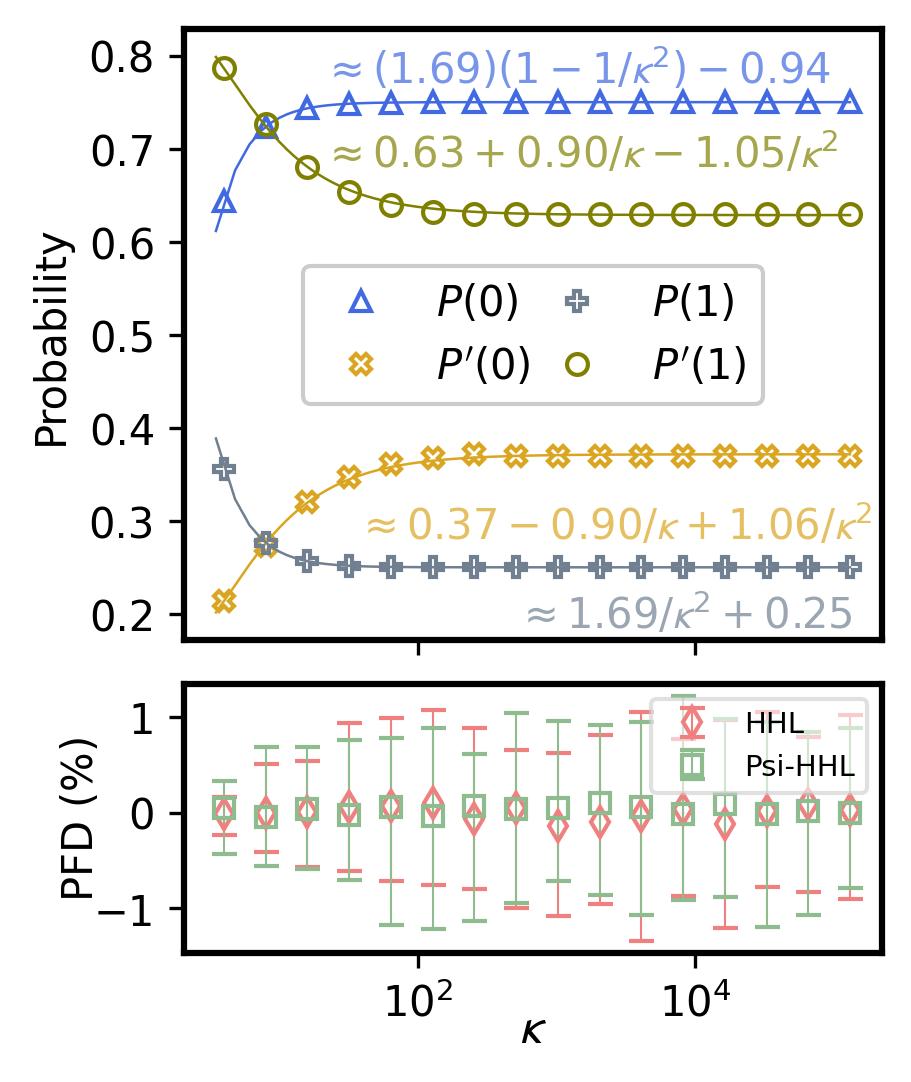}&
\includegraphics[scale=0.65]{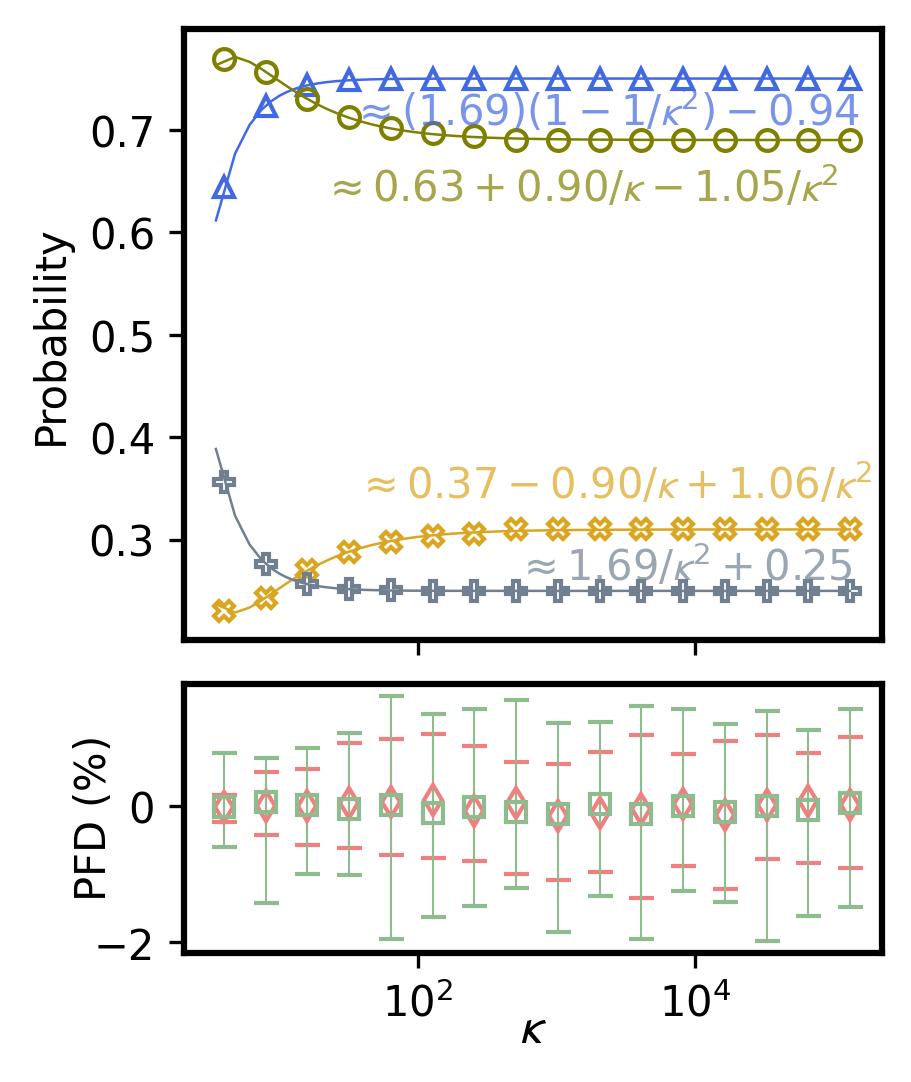}&\includegraphics[scale=0.65]{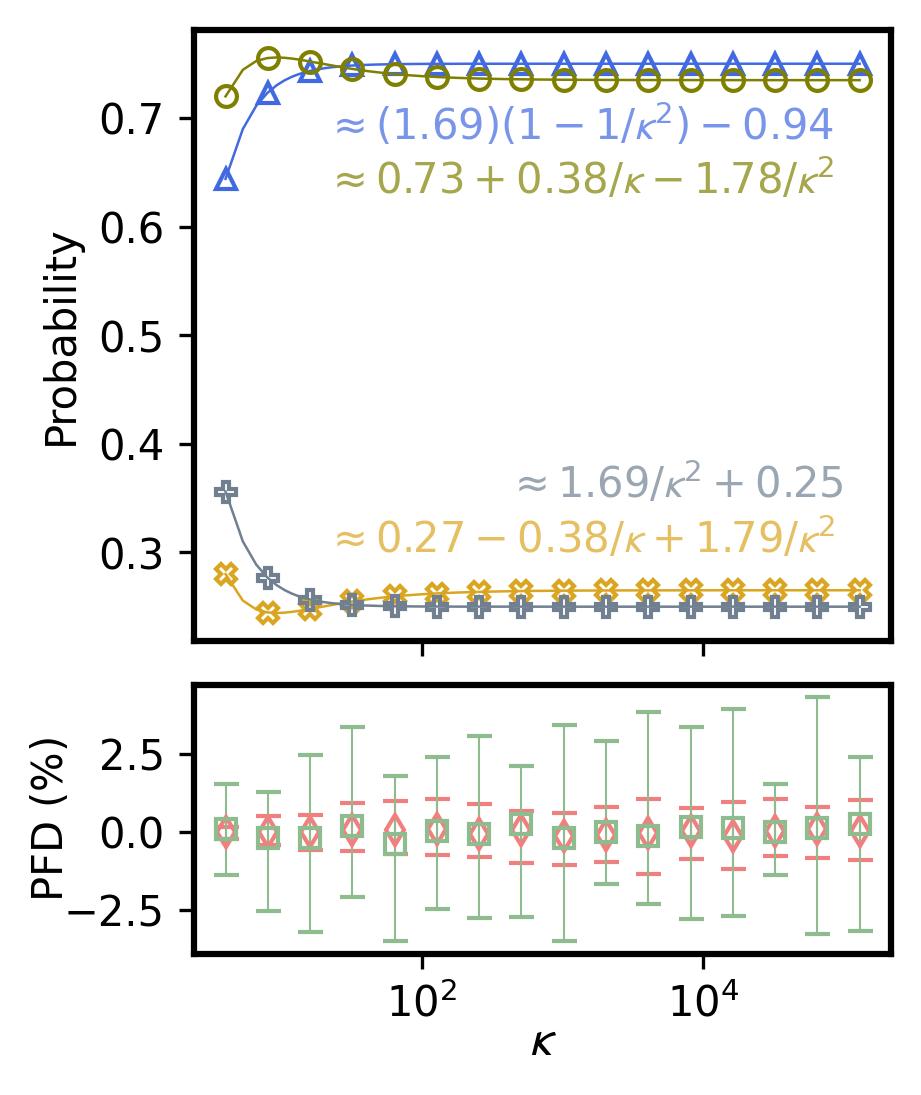}\\(a) $\alpha=60^\circ$ &(b) $\alpha=70^\circ$ &(c) 
 $\alpha=80^\circ$ 
\end{tabular}
\begin{tabular}{ccc}
\includegraphics[scale=0.65]{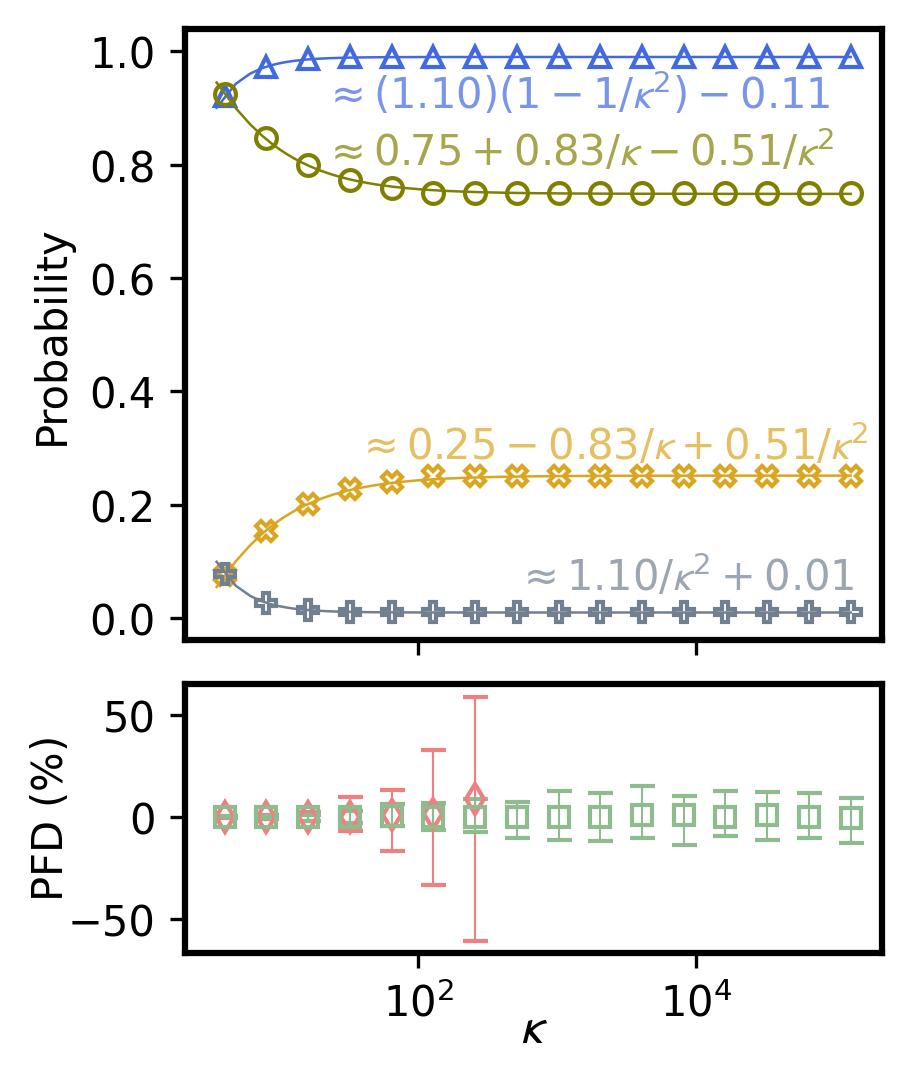}&
\includegraphics[scale=0.65]{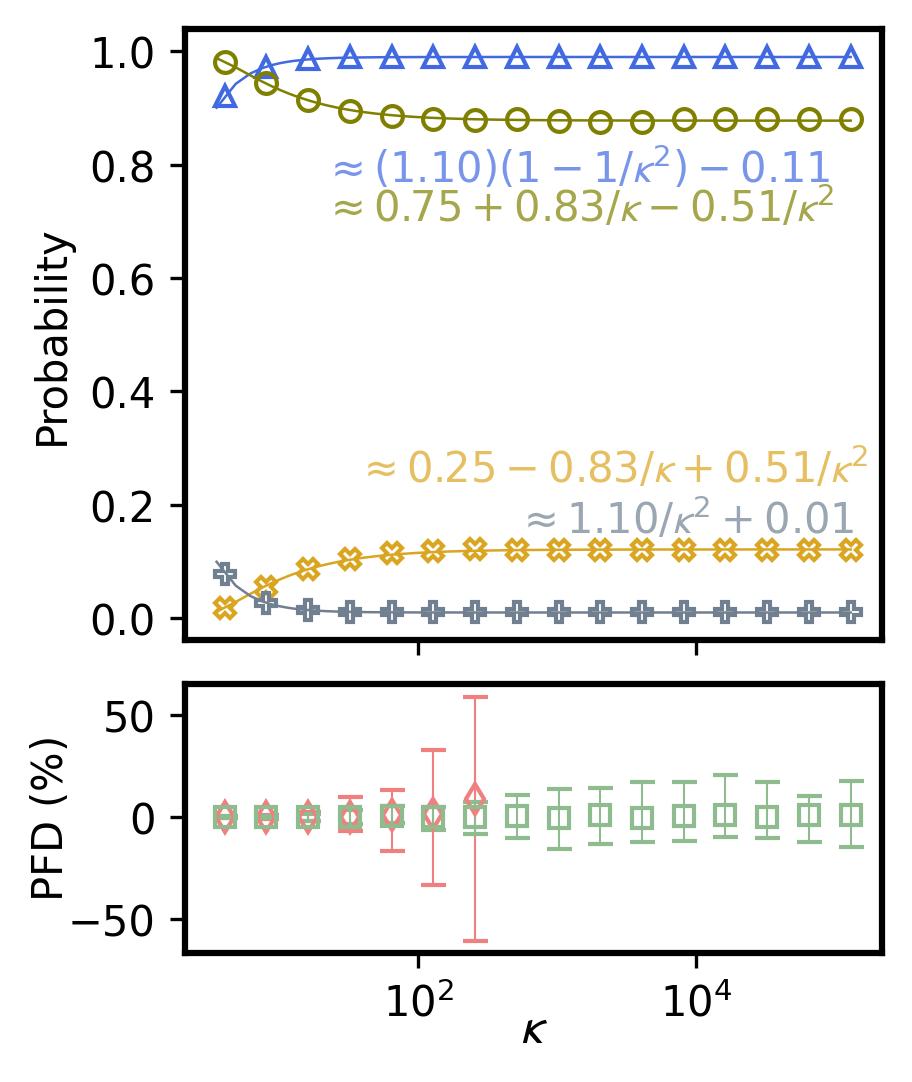}&\includegraphics[scale=0.65]{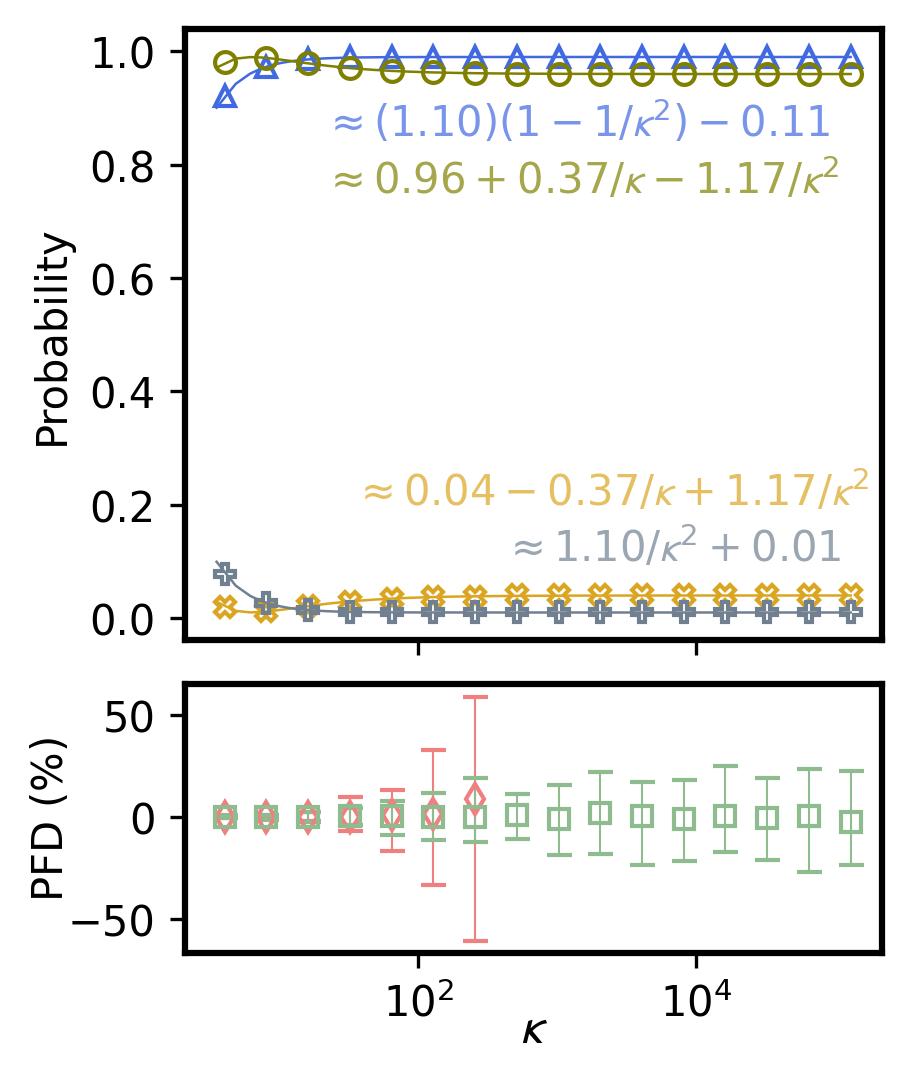}\\(d) $\alpha=60^\circ$ &(e) $\alpha=70^\circ$ &(f) 
 $\alpha=80^\circ$ 
\end{tabular}
\begin{tabular}{ccc}
\includegraphics[scale=0.65]{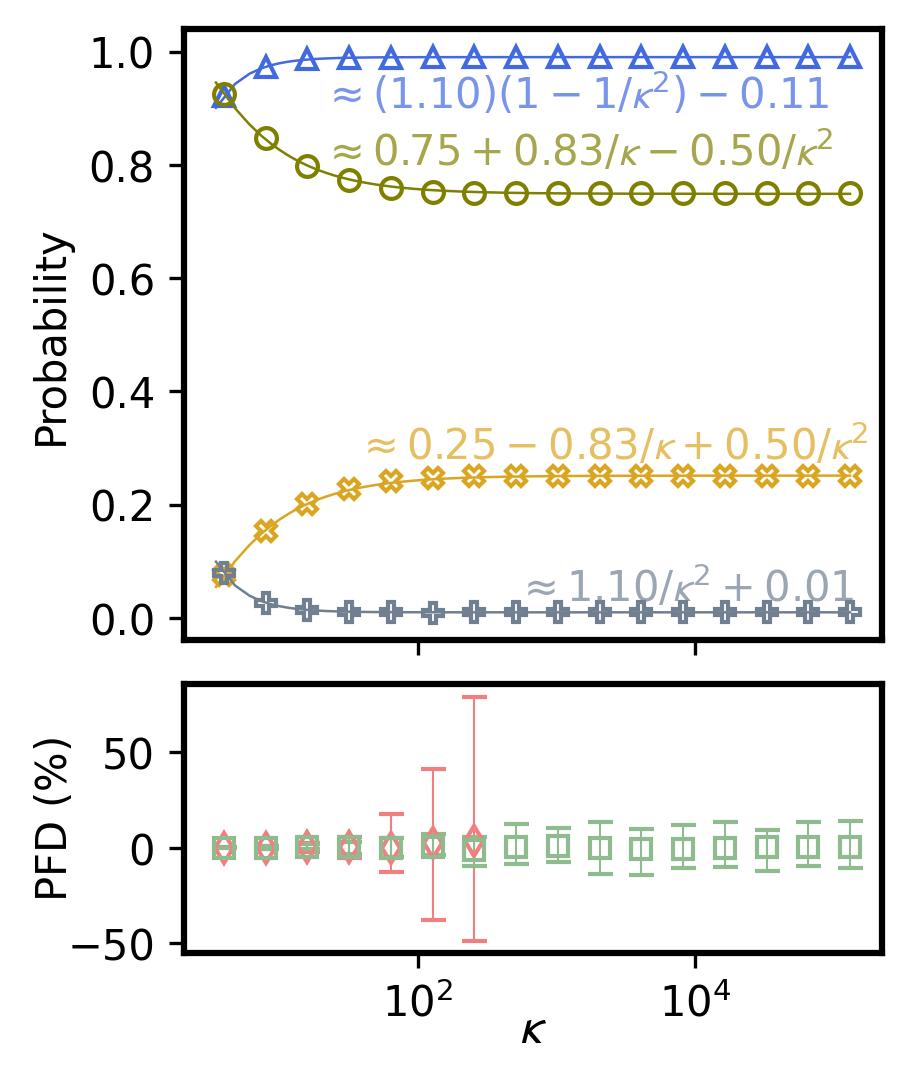}&
\includegraphics[scale=0.65]{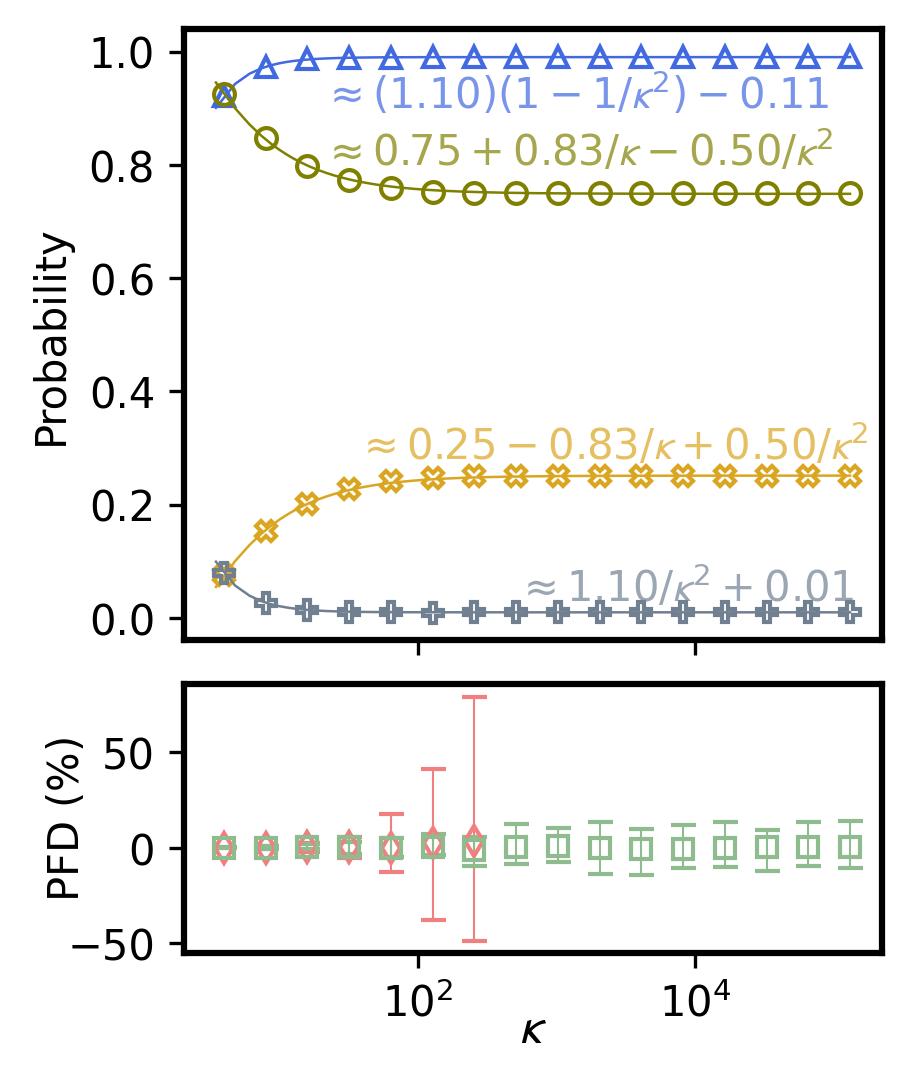}&\includegraphics[scale=0.65]{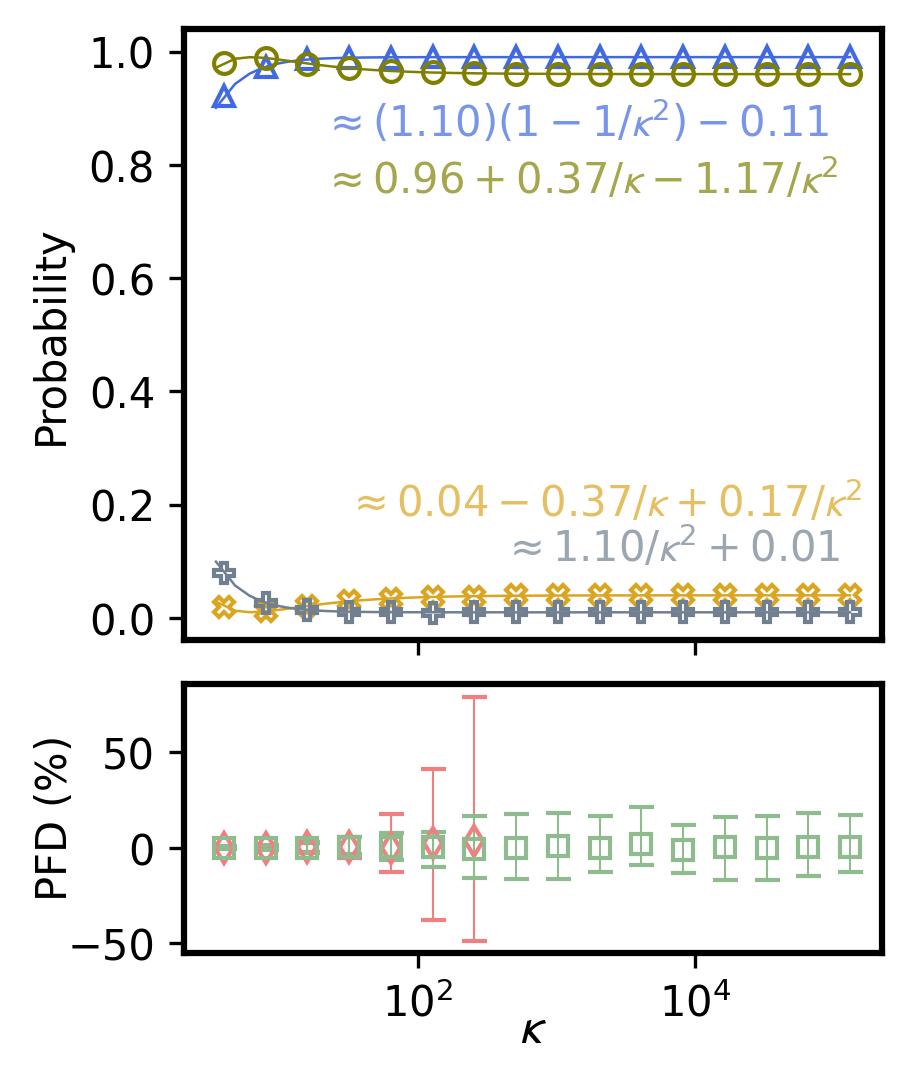}\\(g) $\alpha=60^\circ$ &(h) $\alpha=70^\circ$ &(i) 
 $\alpha=80^\circ$ 
\end{tabular}
\caption{Sub-figures presenting the average probabilities of HHL and Psi-HHL with varying condition number $\mathcal{\kappa}$ of $A$, for  $4 \times 4$  matrices and using $10^6$ throughout.  Sub-figures (a) through (c) consider the case where $A$ is diagonal, and $\vec{b}$ is in equal superposition. Sub-figures (d) through (f) consider the case where $A$ is diagonal and $\vec{b}$ is in unequal superposition, and Sub-figures (g) through (i) consider the case where $A$ is not diagonal, and $\vec{b}$ is in unequal superposition. Each data point is generated by taking the average over 50 repetitions. }
\label{fig:4by4-1e6shots}
\end{figure*}

We also present these results for $\alpha$ set to 60, 70, and 80 degrees (the main text only discusses the first case). Furthermore, while the main text considers 1 million shots, we also perform the same analyses for $10^5$ shots. All of these results are presented in Figures \ref{fig:4by4-1e5shots} and \ref{fig:4by4-1e6shots} for the cases considering $10^5$ and $10^6$ shots respectively. The accompanying data is provided in Tables \ref{tab:A2}, \ref{tab:A3}, and \ref{tab:A4}, for 60, 70, and 80 degrees examples respectively. 

\begin{table*}[]
\centering
\caption{Table presenting the data for HHL and Psi-HHL algorithms (with $\alpha=60^\circ$) for $4 \times 4$ matrices, and in particular, for three cases: $A$ diagonal and $\vec{b}$ in equal superposition (denoted in the table as `$A$ diag, $\vec{b}$ equal'), $A$ diagonal and $\vec{b}$ not in equal superposition (denoted in the table as `$A$ diag, $\vec{b}$ unequal'), and $A$ not diagonal and $\vec{b}$ in an unequal superposition (denoted in the table as `$A$ not diag, $\vec{b}$ unequal'). The string `nan’ refers to situations where we get the square root of a negative quantity for energy, which is unphysical, and hence is not a valid result. Each data point is generated by taking the average over 50 repetitions. }
\resizebox{\textwidth}{!}{
\begin{tabular}{
c| cc c c c c c c c }
\hline \hline
\multicolumn{1}{c|}{} &
  &
  \multicolumn{1}{c}{} &
  \multicolumn{2}{c}{$A$ diag $\vec{b}$ equal} &
  \multicolumn{2}{c}{$A$ diag $\vec{b}$ unequal} &
  \multicolumn{2}{c}{$A$ not diag $\vec{b}$ unequal} &
   \\
\multicolumn{1}{l|}{\multirow{-2}{*}{\# shots}} &
  \multirow{-2}{*}{$\approx \kappa$} &
  \multicolumn{1}{l}{\multirow{-2}{*}{\# qubits}} &
  PFD ($P(0), P(1)$) &
  PFD ($P'(0), P'(1)$) &
  PFD ($P(0), P(1)$) &
  PFD ($P'(0), P'(1)$) &
  PFD ($P(0), P(1)$) &
  PFD ($P'(0), P'(1)$) &
   \\\hline
 &
  $2^2$ &
  8 &
  -0.08 (0.64, 0.36) &
  0.01 (0.21, 0.79) &
  0.03 (0.92, 0.08) &
  -0.11 (0.08, 0.92) &
  0.07 (0.92,  0.08) &
  0.01 (0.08, 0.92) &
   \\
 &
  $2^3$ &
  9 &
  0.01 (0.72, 0.28) &
  -0.17 (0.27, 0.73) &
  -0.05 (0.97, 0.03) &
  -0.09 (0.15, 0.85) &
  -0.02 (0.97, 0.03) &
  -0.05 (0.15, 0.85) &
   \\
 &
  $2^4$ &
  10 &
  -0.01 (0.74, 0.26) &
  0.06 (0.32, 0.68) &
  0.31 (0.99, 0.01) &
  0.47 (0.20, 0.80) &
  0.48 (0.99, 0.01) &
  0.39 (0.20, 0.80) &
   \\
 &
  $2^5$ &
  11 &
  -0.06 (0.75, 0.25) &
  -0.14 (0.35, 0.65) &
  1.85 (0.99, 0.01) &
  -0.04 (0.23, 0.77) &
  -0.10 (0.99, 0.01) &
  -0.15 (0.22, 0.78) &
   \\
 &
  $2^6$ &
  12 &
  0.08 (0.75, 0.25) &
  -0.38 (0.37, 0.63) &
  nan (0.99, 0.01) &
  -1.41 (0.24, 0.76) &
  nan (0.99, 0.01) &
  -1.75 (0.24, 0.76) &
   \\
 &
  $2^7$ &
  13 &
  0.00 (0.75, 0.25) &
  0.10 (0.37, 0.63) &
  nan (0.99, 0.01) &
  1.39 (0.25, 0.75) &
  nan (0.99, 0.01) &
  1.85 (0.25, 0.75) &
   \\
\hline\hline
\end{tabular}
}\label{tab:A2}
\end{table*}

\begin{table*}[]
\centering
\caption{Table providing the data for HHL and Psi-HHL algorithms (with $\alpha=70^\circ$) for $4 \times 4$ matrices. The notations are the same as those used in Table \ref{tab:A2}.  Each data point is generated by taking the average over 50 repetitions.}
\resizebox{\textwidth}{!}{
\begin{tabular}{c |c c c c c c c c c}
\hline\hline
\multicolumn{1}{l|}{} &
   &
   &
  \multicolumn{2}{c}{$A$ diag $\vec{b}$ equal} &
  \multicolumn{2}{c}{$A$ diag $\vec{b}$ unequal} &
  \multicolumn{2}{c}{$A$ not diag $\vec{b}$ unequal} &
   \\
\multicolumn{1}{l|}{\multirow{-2}{*}{\# shots}} &
  \multirow{-2}{*}{$\approx \kappa$} &
  \multirow{-2}{*}{\# qubits} &
  PFD ($P(0), P(1)$) &
  PFD ($P'(0), P'(1)$) &
  PFD ($P(0), P(1)$) &
  PFD ($P'(0), P'(1)$) &
  PFD ($P(0), P(1)$) &
  PFD ($P'(0), P'(1)$) &
   \\\hline
 &
  $2^2$ &
  8 &
  -0.08 (0.64, 0.36) &
  -0.03 (0.23, 0.77) &
  0.03 (0.92, 0.08) &
  -0.09 (0.02, 0.98) &
  0.07 (0.92,  0.08) &
  -0.05 (0.02, 0.98) &
   \\
&
  $2^3$ &
  9 &
  0.01 (0.72, 0.28) &
  -0.17 (0.24, 0.76) &
  -0.05 (0.97, 0.03) &
  0.12 (0.06, 0.94) &
  -0.02 (0.97, 0.03) &
  -0.22 (0.06, 0.94) &
   \\
 &
  $2^4$ &
  10 &
  -0.02 (0.74, 0.26) &
  -0.26 (0.27, 0.73) &
  0.31 (0.99, 0.01) &
  -0.48 (0.12, 0.88) &
  0.48 (0.99, 0.01) &
  -0.17 (0.09, 0.91) &
   \\
 &
  $2^5$ &
  11 &
  -0.06 (0.75, 0.25) &
  0.07 (0.29, 0.71) &
  1.85 (0.99, 0.01) &
  0.70 (0.10, 0.90) &
  -0.10 (0.99, 0.01) &
  -0.68 (0.10, 0.90) &
   \\
 &
  $2^6$ &
  12 &
  0.09 (0.75, 0.25) &
  0.23 (0.30, 0.70) &
  nan (0.99, 0.01) &
  -0.55 (0.12, 0.88) &
  nan (0.99, 0.01) &
  0.18 (0.11, 0.89) &
   \\
 &
  $2^7$ &
  13 &
  -0.01 (0.75, 0.25) &
  0.28 (0.30, 0.70) &
  nan (0.99, 0.01) &
  0.74 (0.12, 0.88) &
  nan (0.99, 0.01) &
  1.88 (0.11, 0.88) &
   \\
\hline\hline
\end{tabular}
}
\label{tab:A3}
\end{table*}

\begin{table*}[]
\centering
\caption{Table providing the data for HHL and Psi-HHL algorithms (with $\alpha=80^\circ$) for $4 \times 4$ matrices. The notations are the same as those used in Tables \ref{tab:A2} and \ref{tab:A3}.  Each data point is generated by taking the average over 50 repetitions. }
\resizebox{\textwidth}{!}{
\begin{tabular}{
c |
c 
c 
c 
c 
c 
c 
c 
c 
c}
\hline\hline 
&
   &
   &
  \multicolumn{2}{c}{$A$ diag $\vec{b}$ equal} &
  \multicolumn{2}{c}{$A$ diag $\vec{b}$ unequal} &
  \multicolumn{2}{c}{$A$ not diag $\vec{b}$ unequal} &
   \\
\multirow{-2}{*}{\# shots} &
  \multirow{-2}{*}{$\approx \kappa$} &
  \multirow{-2}{*}{\# qubits} &
  \multicolumn{1}{l}{PFD ($P(0), P(1))$} &
  PFD ($P'(0), P'(1)$) &
  PFD ($P(0), P(1)$) &
  PFD ($P'(0), P'(1)$) &
  PFD ($P(0), P(1)$) &
  PFD ($P'(0), P'(1)$) &
   \\\hline
 &
  $2^2$ &
  8 &
  \multicolumn{1}{l}{-0.08 (0.64, 0.36)} &
  -0.08 (0.28, 0.72) &
  0.04 (0.92, 0.08) &
  0.27 (0.02, 0.98) &
  0.07 (0.92,  0.08) &
  0.26 (0.02, 0.98) &
   \\
 &
  $2^3$ &
  9 &
  0.01 (0.72, 0.28) &
  -0.14 (0.24, 0.76) &
  -0.06 (0.97, 0.03) &
  0.14 (0.01, 0.99) &
  -0.02 (0.97, 0.03) &
  0.05 (0.02, 0.98) &
   \\
 &
  $2^4$ &
  10 &
  -0.01 (0.74, 0.26) &
  0.13 (0.24, 0.76) &
  0.31 (0.99, 0.01) &
  0.14 (0.02, 0.98) &
  0.48 (0.99, 0.01) &
  -0.37 (0.02, 0.98) &
   \\
&
  $2^5$ &
  11 &
  -0.06 (0.75, 0.25) &
  -0.41 (0.26, 0.74) &
  1.85 (0.99, 0.01) &
  0.27 (0.03, 0.97) &
  -0.10 (0.99, 0.01) &
  -0.33 (0.03, 0.97) &
   \\
 &
  $2^6$ &
  12 &
  0.08 (0.75, 0.25) &
  -0.62 (0.26, 0.74) &
  nan (0.99, 0.01) &
  -0.48 (0.04, 0.96) &
  nan (0.99, 0.01) &
  -0.28 (0.03, 0.97) &
   \\
 &
  $2^7$ &
  13 &
  -0.01 (0.75, 0.25) &
  0.76 (0.26, 0.74) &
  nan (0.99, 0.01) &
  -1.63 (0.04, 0.96) &
  nan (0.99, 0.01) &
  1.86 (0.04, 0.96) &
   \\
\hline\hline  
\end{tabular}
}
\label{tab:A4}
\end{table*} 

\begin{table*}[t]
\centering
\caption{Summary of the fitted probabilities obtained from HHL and Psi-HHL for the 4x4 case of $A$ not diagonal and $\vec{b}$ in an unequal superposition, for $\alpha=60^\circ, 70^\circ, 80^\circ$. The number of shots is fixed to $10^6$. }
\begin{tabular}{cccccccc}
\hline \hline 
        $\alpha$ & $\approx \mathcal{\kappa}$ & $P_{\text{Fit}}(0)$& $P_{\text{Fit}}(1)$& $P_{\text{Fit}}(0)+P_{\text{Fit}}(1)$ & $P'_\text{Fit}(0)$ & $P'_{\text{Fit}}(1)$ & $P'_{\text{Fit}}(0)+P'_{\text{Fit}}(1)$ \\ \hline  
        \multirow{5}{*}{$60^\circ$} & 4 & 0.92 & 0.08 & 1.00 & 0.08 & 0.92 & 1.00 \\
        ~ & 32 & 0.99 & 0.01 & 1.00 & 0.22 & 0.78 & 1.00 \\ 
        ~ & 64 & 0.99 & 0.01 & 1.00 & 0.24 & 0.76 & 1.00 \\ 
        ~ & 1024 & 0.99 & 0.01 & 1.00 & 0.25 & 0.75 & 1.00 \\ 
        ~ & 2048 & 0.99 & 0.01 & 1.00 & 0.25 & 0.75 & 1.00 \\ \cline{1-8}
        \multirow{5}{*}{$70^\circ$} & 4 & 0.92 & 0.08 & 1.00 & 0.02 & 0.98 & 1.00 \\ 
        ~ & 32 & 0.99 & 0.01 & 1.00 & 0.10 & 0.90 & 1.00 \\ 
        ~ & 128 & 0.99 & 0.01 & 1.00 & 0.11 & 0.89 & 1.00 \\ 
        ~ & 1024 & 0.99 & 0.01 & 1.00 & 0.12 & 0.88 & 1.00 \\ 
        ~ & 2048 & 0.99 & 0.01 & 1.00 & 0.12 & 0.88 & 1.00 \\\cline{1-8}
        \multirow{5}{*}{$80^\circ$} & 4 & 0.92 & 0.08 & 1.00 & 0.02 & 0.98 & 1.00 \\ 
        ~ & 32 & 0.99 & 0.01 & 1.00 & 0.03 & 0.97 & 1.00 \\ 
        ~ & 128 & 0.99 & 0.01 & 1.00 & 0.03 & 0.97 & 1.00 \\ 
        ~ & 1024 & 0.99 & 0.01 & 1.00 & 0.04 & 0.96 & 1.00 \\ 
        ~ & 2048 & 0.99 & 0.01 & 1.00 & 0.04 & 0.96 & 1.00 \\ \hline \hline 
\end{tabular}
\label{tab:A5}
\end{table*}

\begin{figure*}[t]
\centering
    \begin{tabular}{ccc}
\includegraphics[scale=0.65]{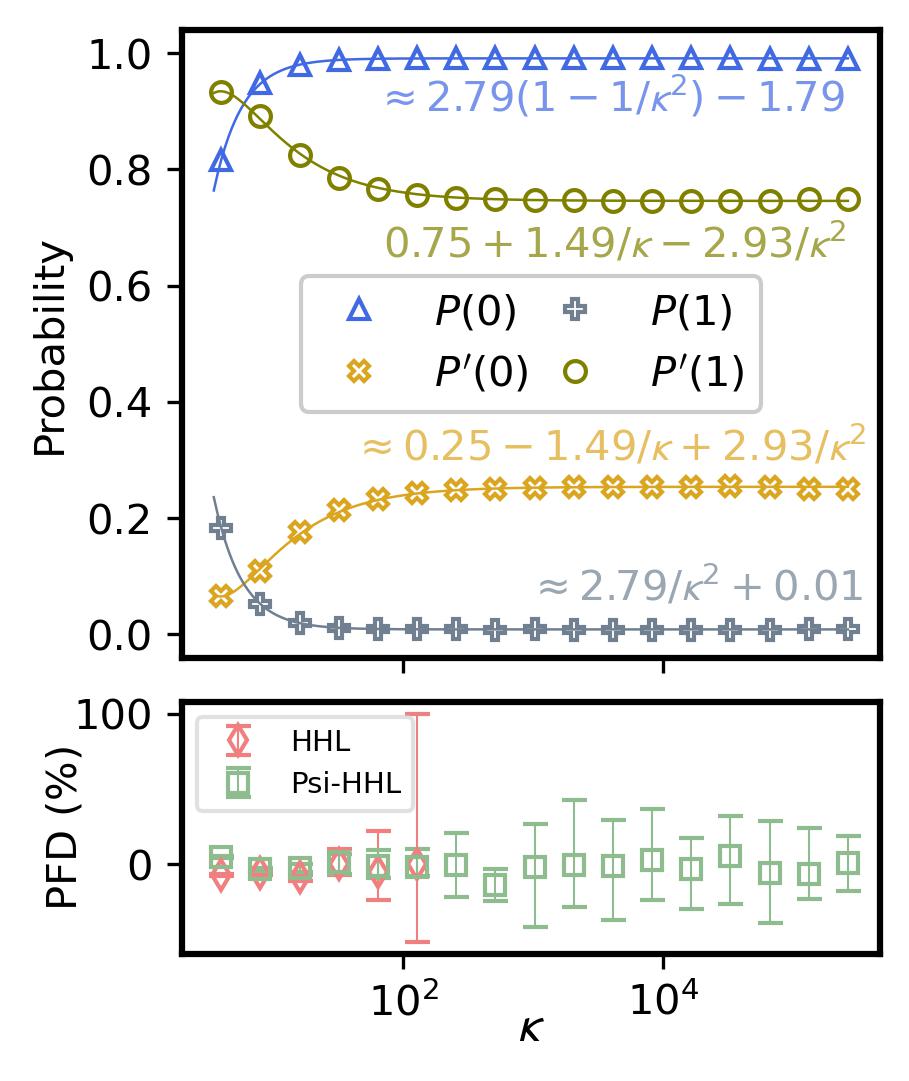}&
\includegraphics[scale=0.65]{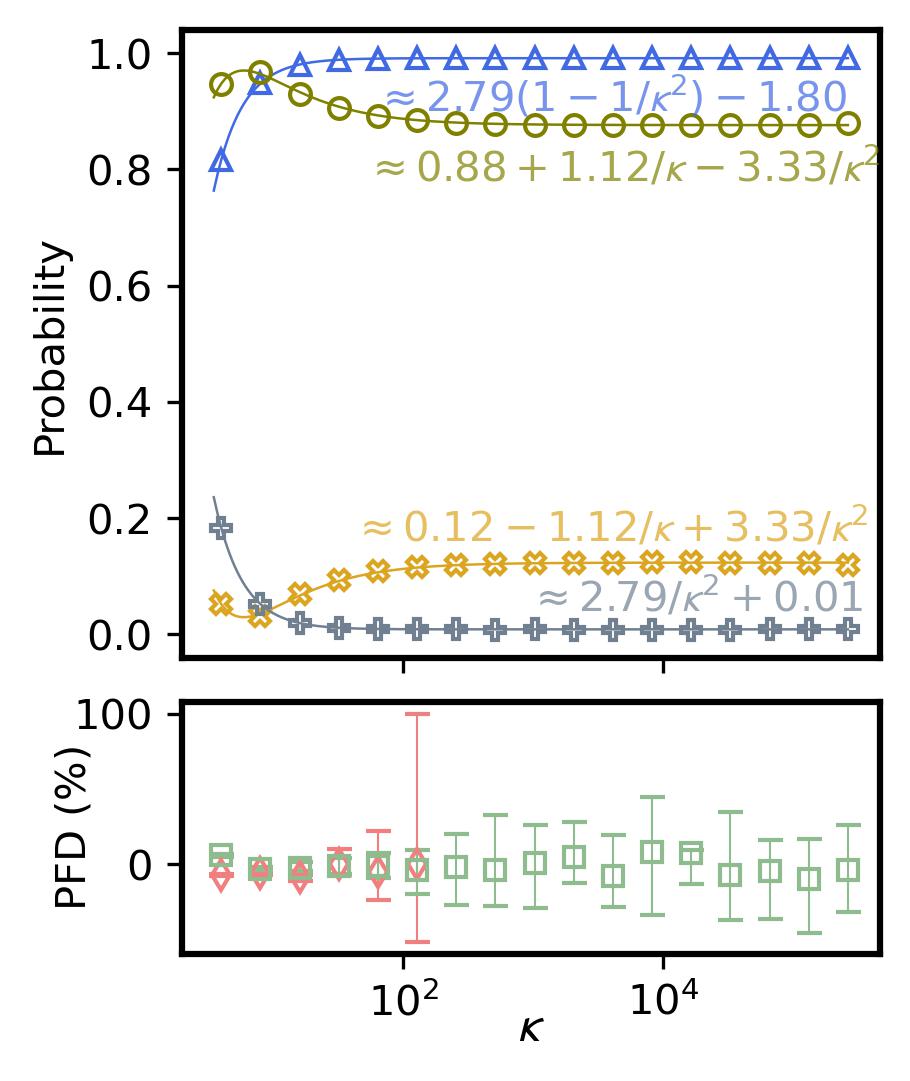}&\includegraphics[scale=0.65]{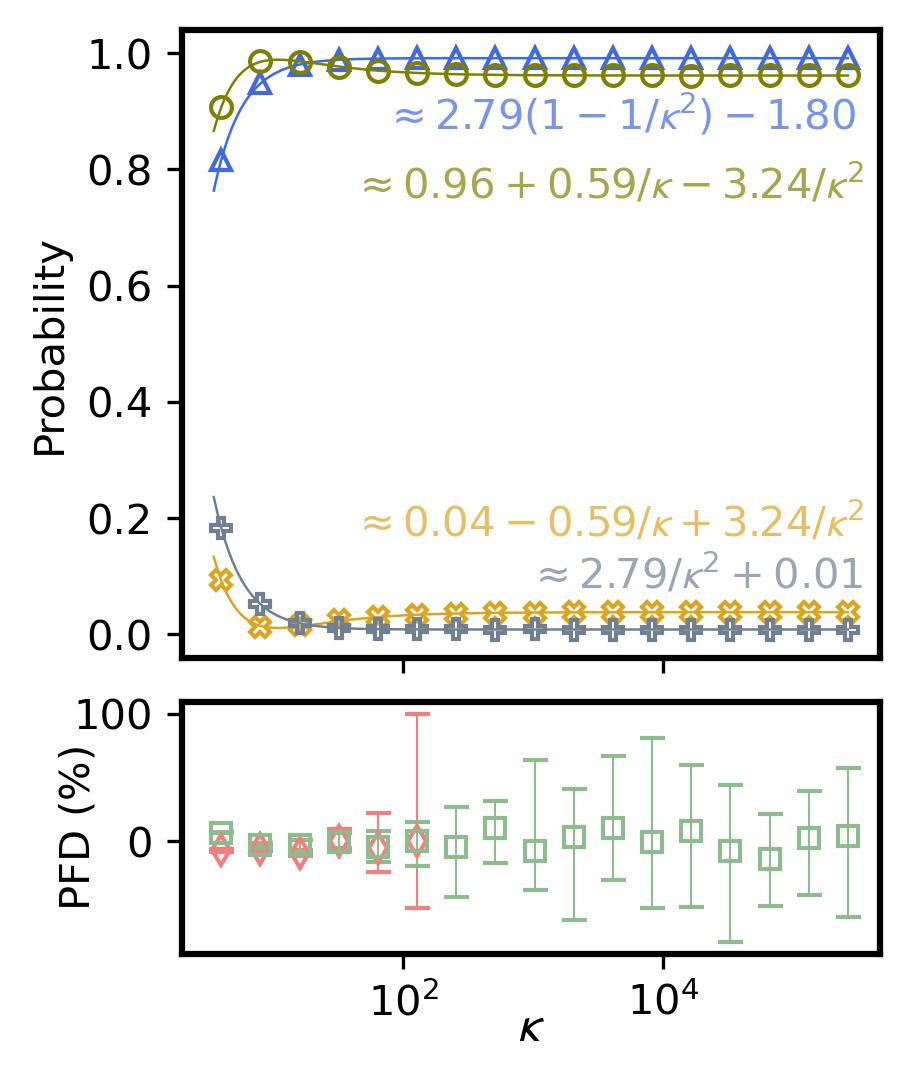}\\(a) $\alpha=60^\circ$ &(b) $\alpha=70^\circ$ &(c) $\alpha=80^\circ$ 
\end{tabular}
    \begin{tabular}{ccc}
\includegraphics[scale=0.65]{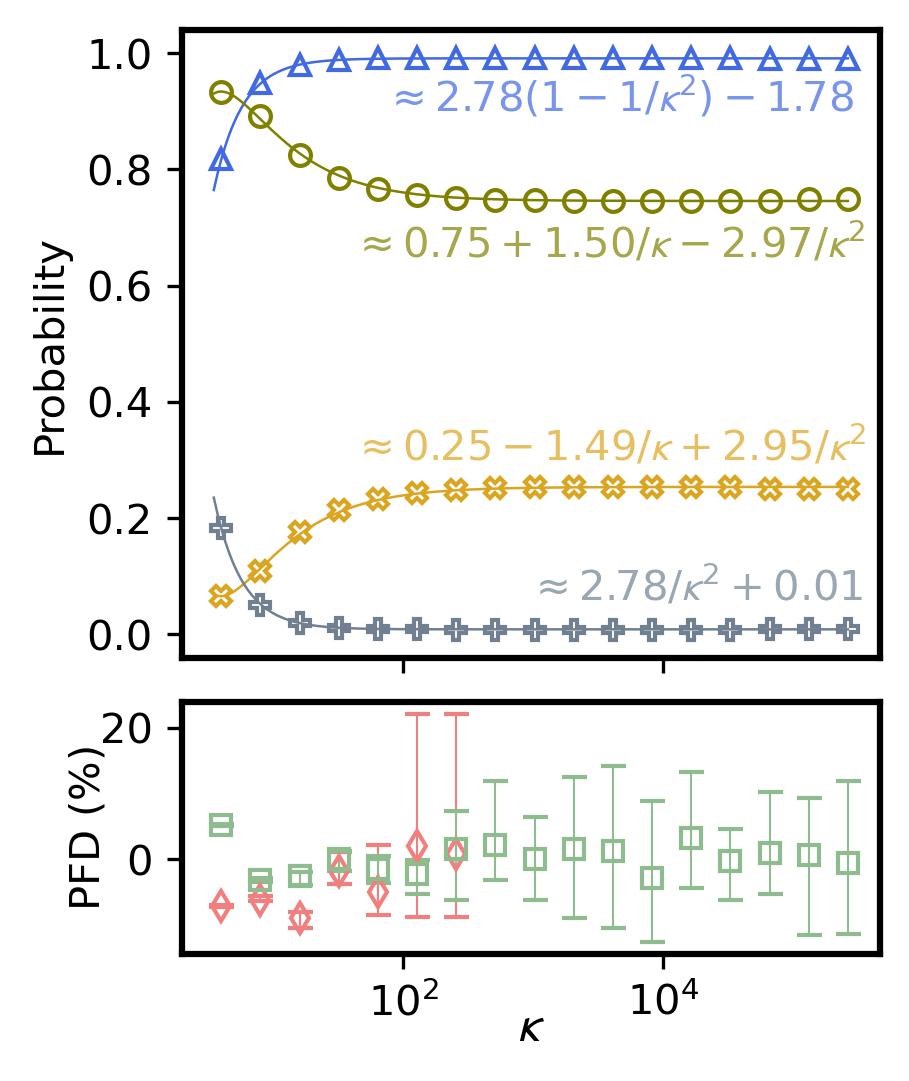}&
\includegraphics[scale=0.65]{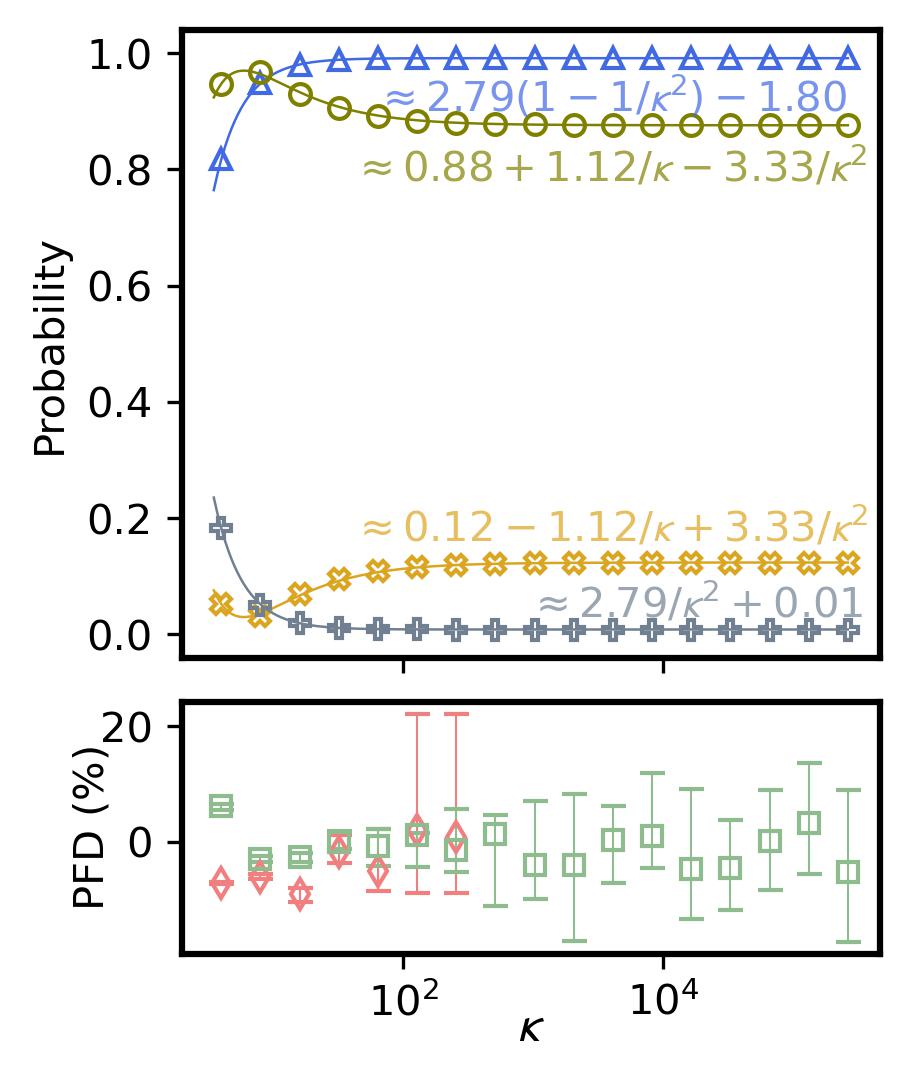}&\includegraphics[scale=0.65]{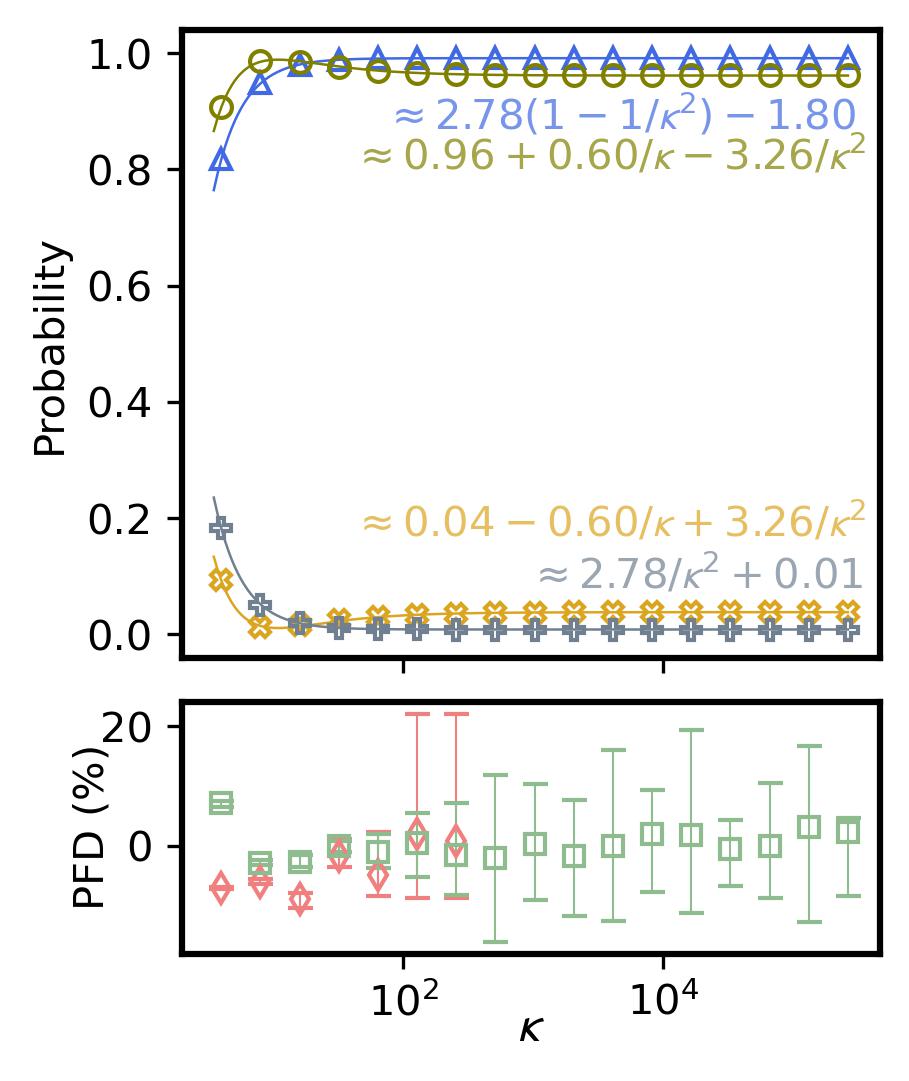}\\(d) $\alpha=60^\circ$ &(e) $\alpha=70^\circ$ &(f) $\alpha=80^\circ$ 
\end{tabular}
\caption{Sub-figures presenting the average probabilities, PFDs and range of PFDs of HHL and Psi-HHL with varying condition number $\mathcal{\kappa}$ of $A$, for  $8 \times 8$  matrices using $10^5$ shots ((a) through (c)) and using $10^6$ ((d) through (f)). In all of the examples, $A$ is chosen to be a non-diagonal matrix and $\vec{b}$ is set in an unequal superposition. Each data point is an average over 10 repetitions. }
\label{fig:8by8-1e6shots}
\end{figure*} 

\begin{figure*}[t]
\centering
    \begin{tabular}{ccc}
\includegraphics[scale=0.65]{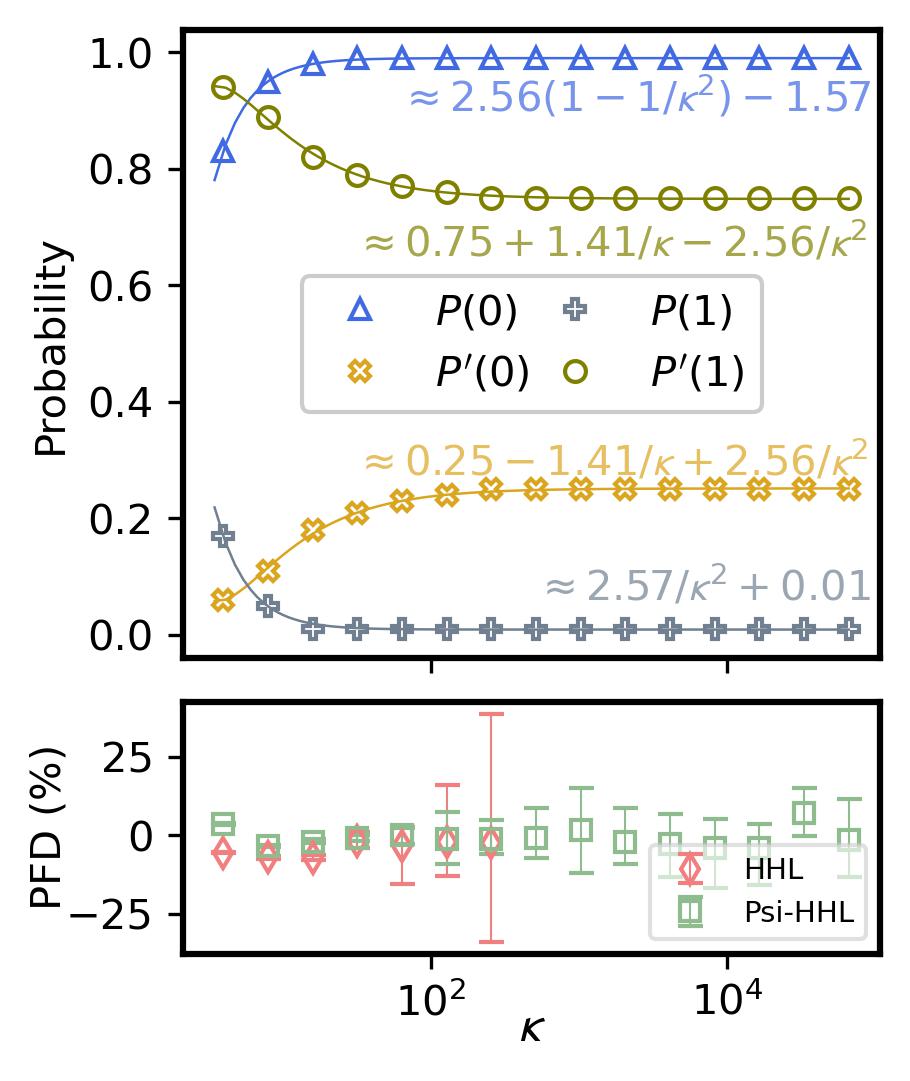}&
\includegraphics[scale=0.65]{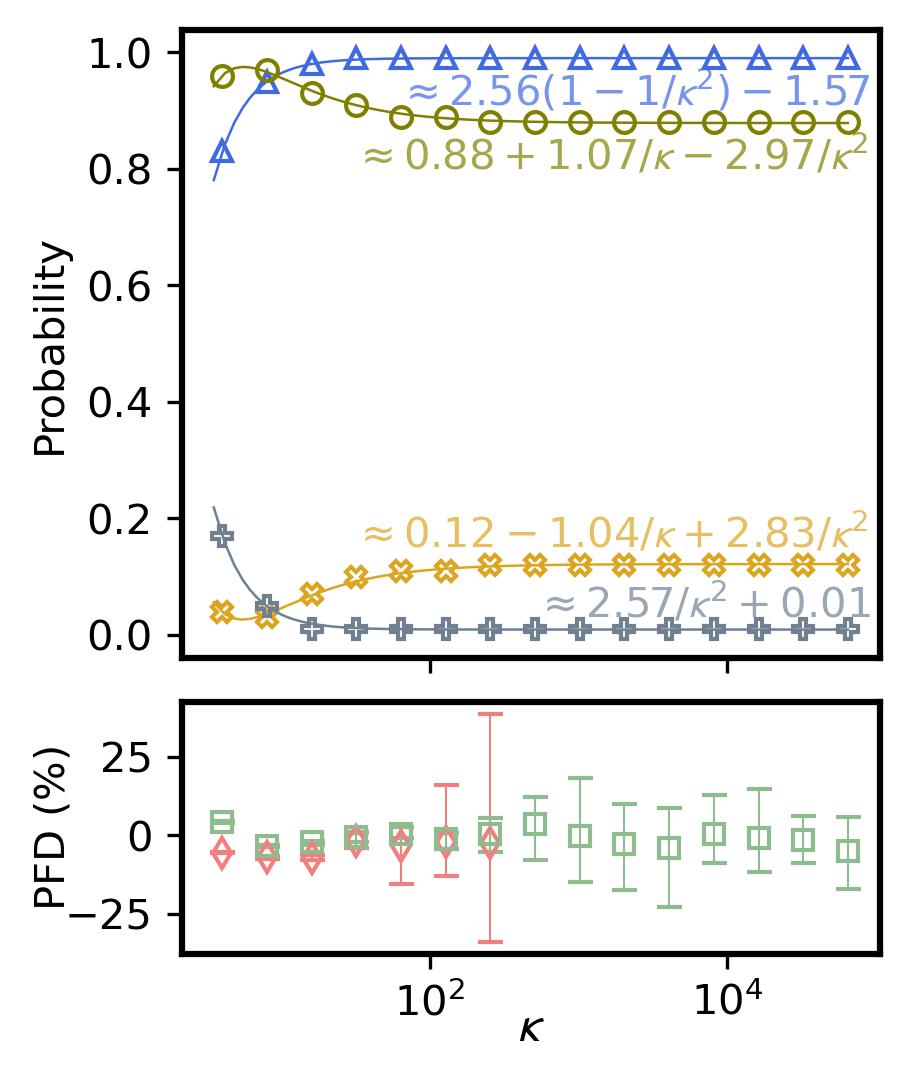}&\includegraphics[scale=0.65]{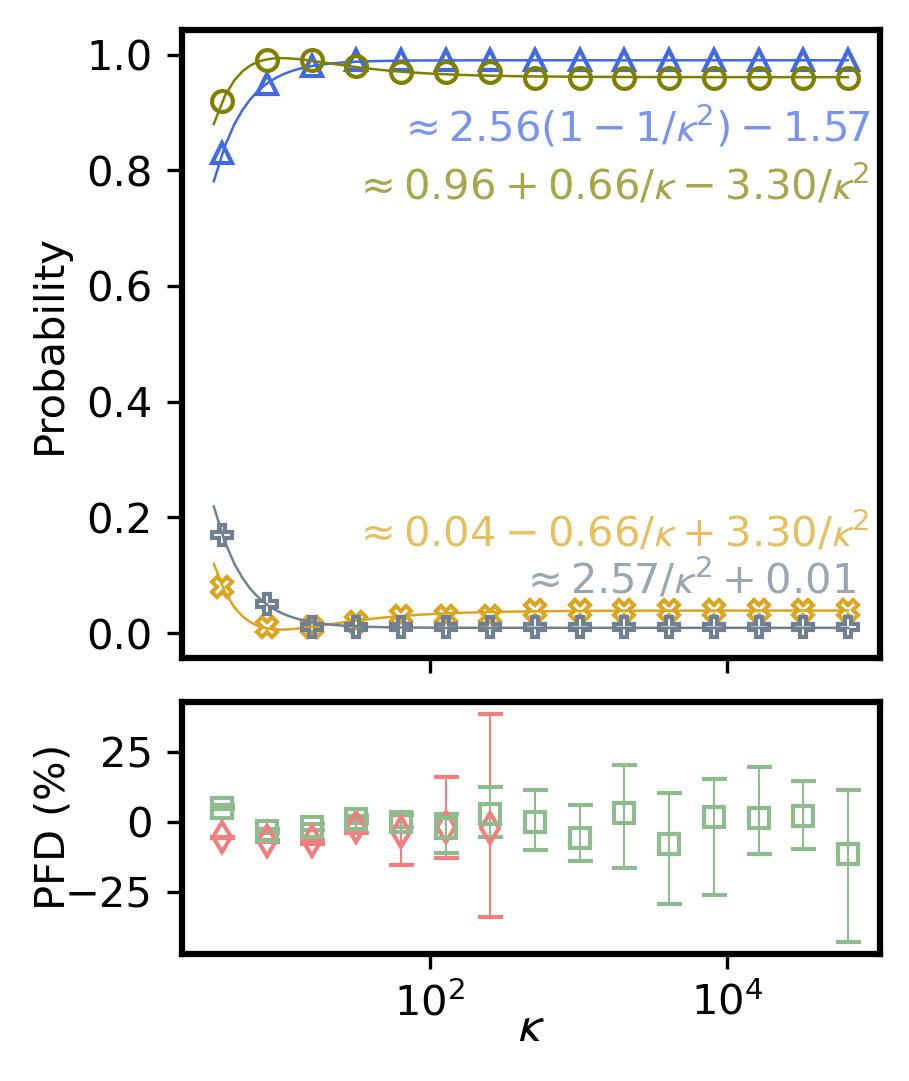}\\(a) $\alpha=60^\circ$ &(b) $\alpha=70^\circ$ &(c) $\alpha=80^\circ$ 
\end{tabular}
\caption{Sub-figures presenting the average probabilities, PFDs and range of PFDs of HHL and Psi-HHL with varying condition number $\mathcal{\kappa}$ of $A$, for  $16 \times 16$  matrices and using $10^6$ shots throughout. In all of the examples, $A$ is chosen to be a non-diagonal matrix and $\vec{b}$ is set in an unequal superposition. Each data point is an average over 10 repetitions. } 
\label{fig:16by16-1e6shots}
\end{figure*} 

\begin{figure*}[t]
\centering
    \begin{tabular}{ccc}
\includegraphics[scale=0.65]{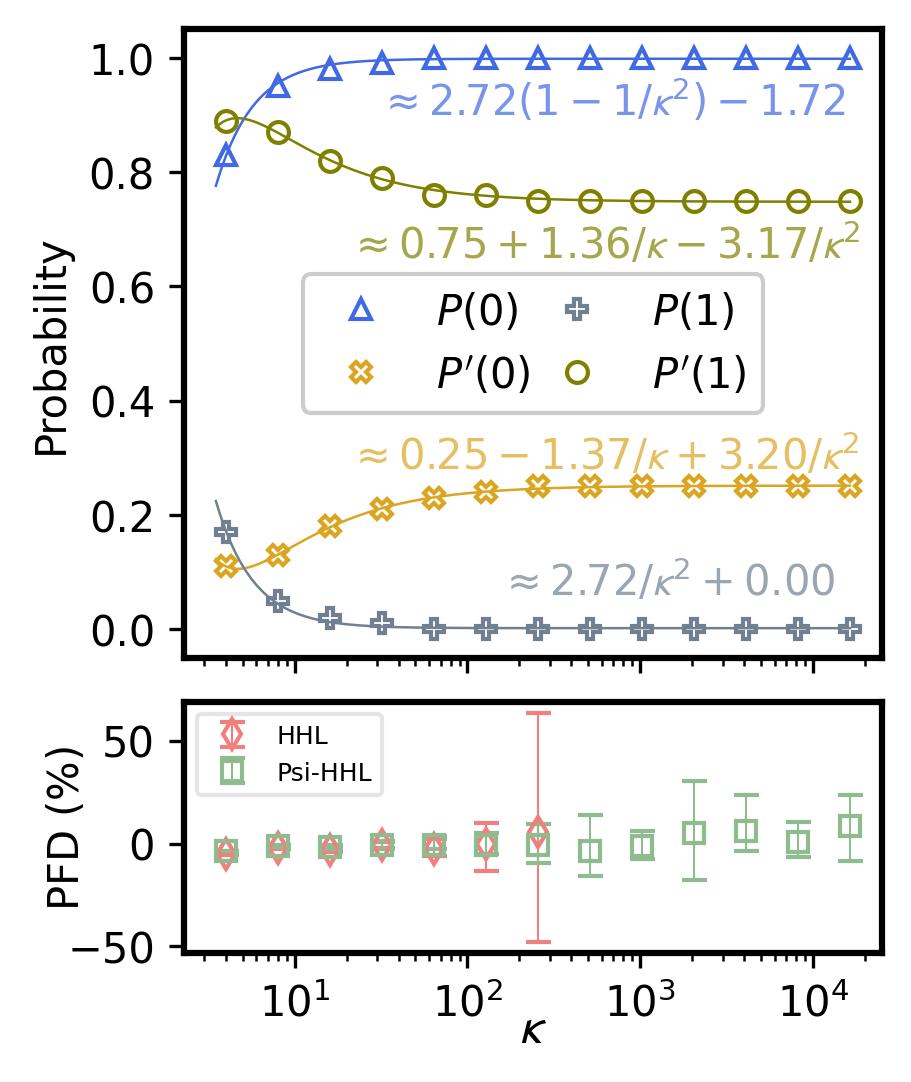}&
\includegraphics[scale=0.65]{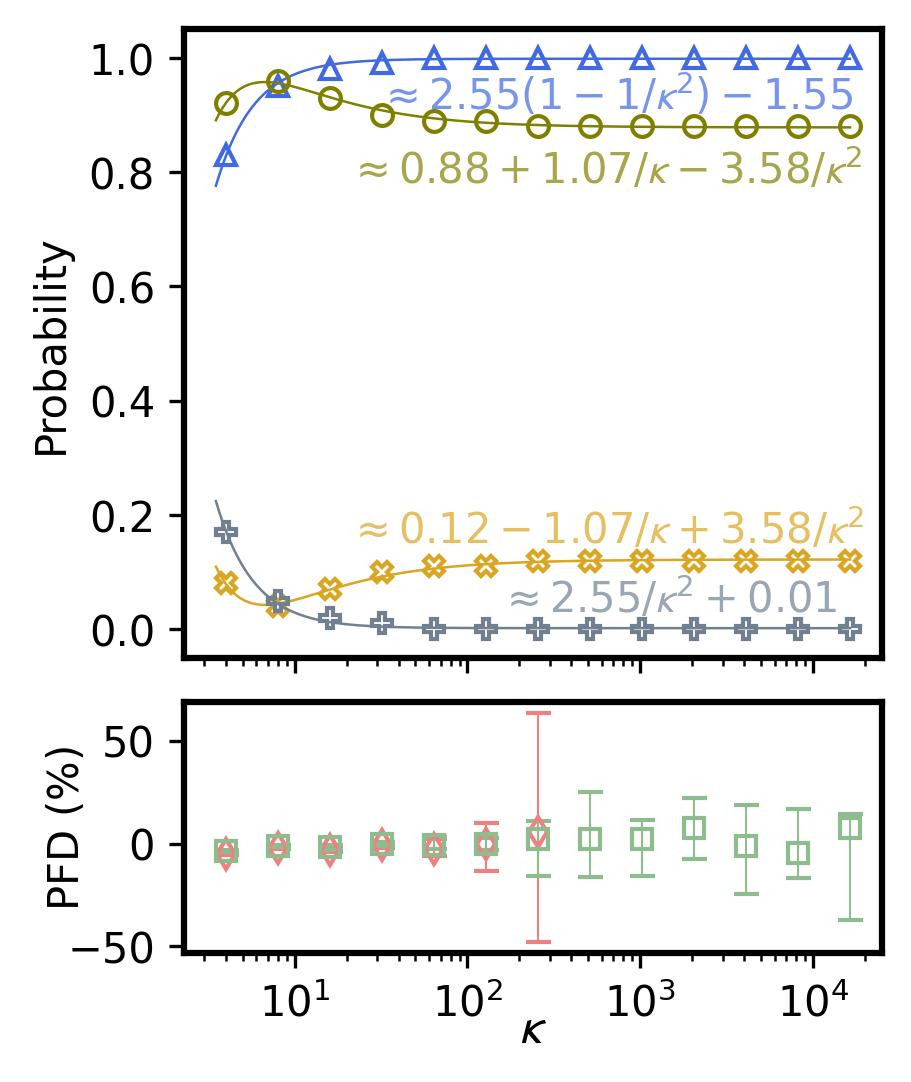}&\includegraphics[scale=0.65]{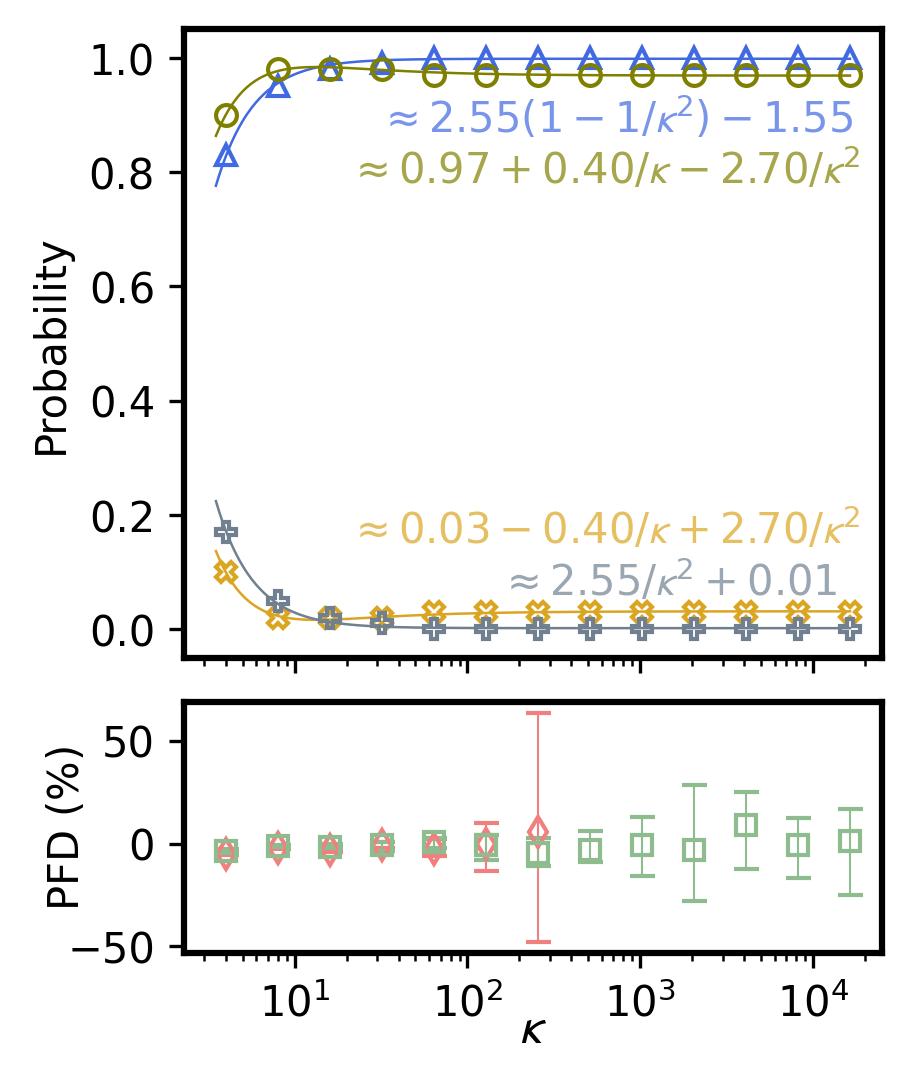}\\(a) $\alpha=60^\circ$ &(b) $\alpha=70^\circ$ &(c) $\alpha=80^\circ$ 
\end{tabular}
\caption{Data for $32 \times 32$ matrices. Note that we have again reproduced here the sub-figure (a) with 1 million shots (which are already present in the main text). In all of the examples, $A$ is chosen to be a non-diagonal matrix and $\vec{b}$ is set in an unequal superposition. Each data point is an average over 10 repetitions. }
\label{fig:32by32-1e6shots}
\end{figure*} 

\begin{figure*}[t]
\centering
    \begin{tabular}{ccc}
\includegraphics[scale=0.65]{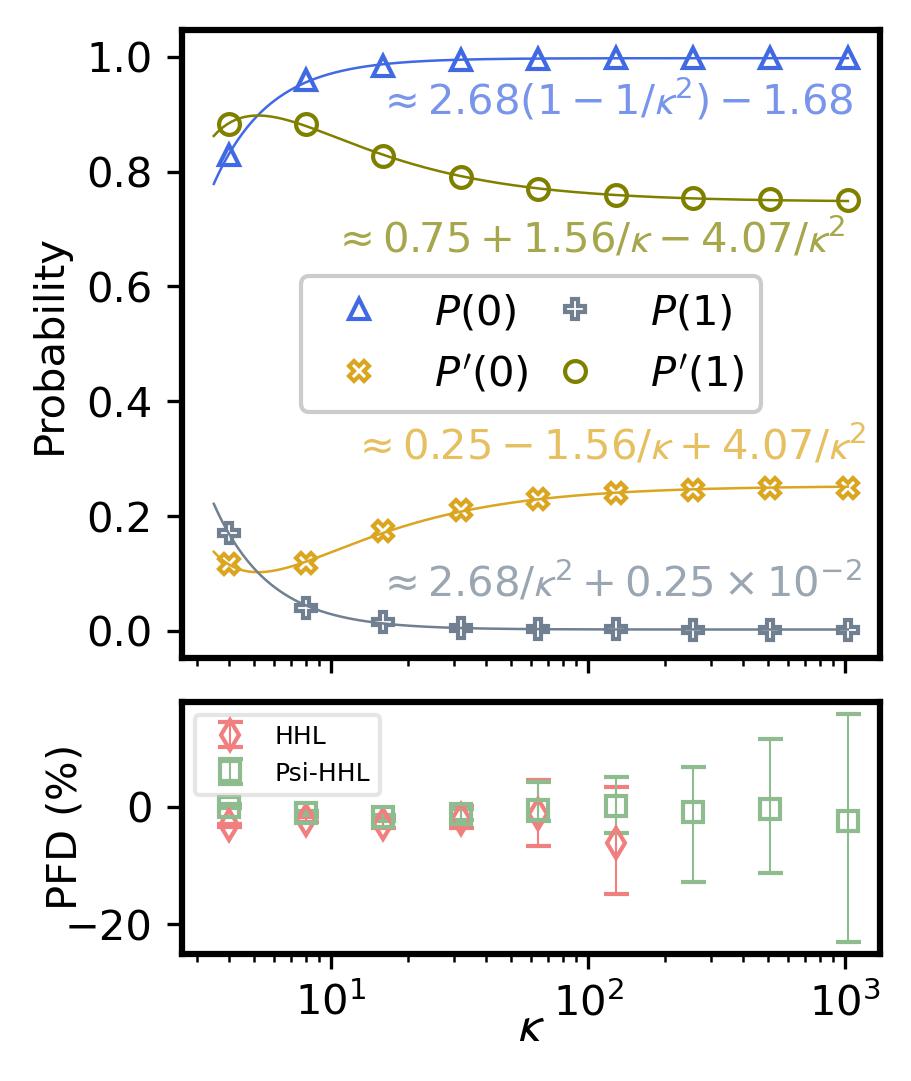}&
\includegraphics[scale=0.65]{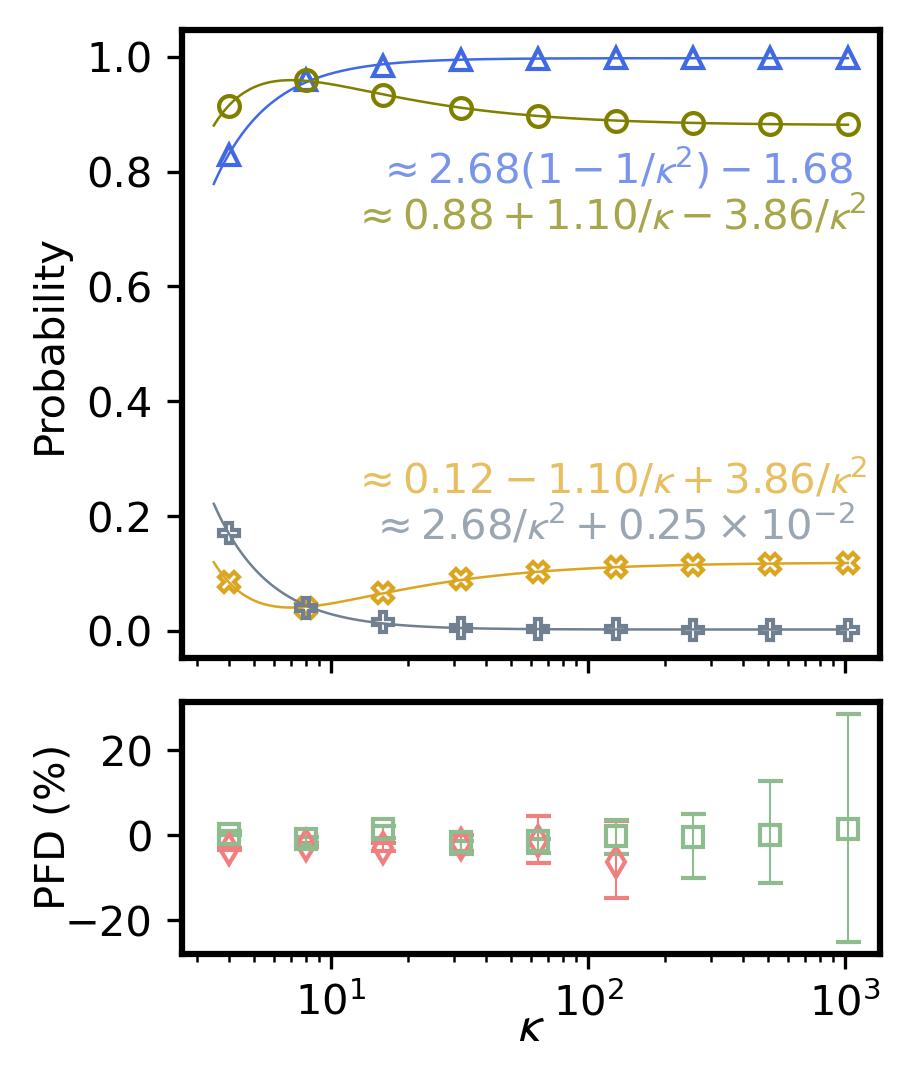}&\includegraphics[scale=0.65]{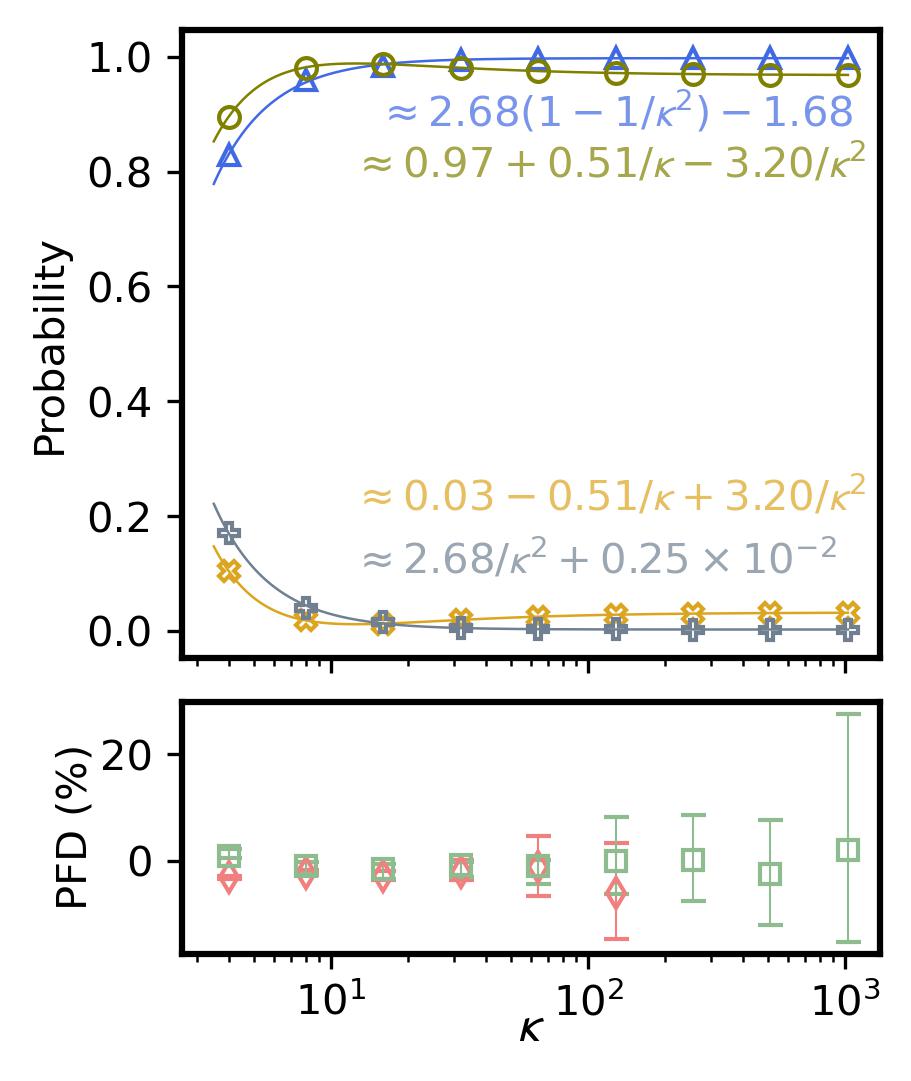}\\(a) $\alpha=60^\circ$ &(b) $\alpha=70^\circ$ &(c) $\alpha=80^\circ$ 
\end{tabular}
\caption{Data for $64 \times 64$ matrices. Note that we have again reproduced here the sub-figure (a) with 1 million shots (which are already present in the main text). In all of the examples, $A$ is chosen to be a non-diagonal matrix and $\vec{b}$ is set in an unequal superposition. Each data point is an average over 10 repetitions. }
\label{fig:64by64-1e6shots}
\end{figure*} 

\begin{figure}[t]
\centering    \includegraphics[scale=0.2]{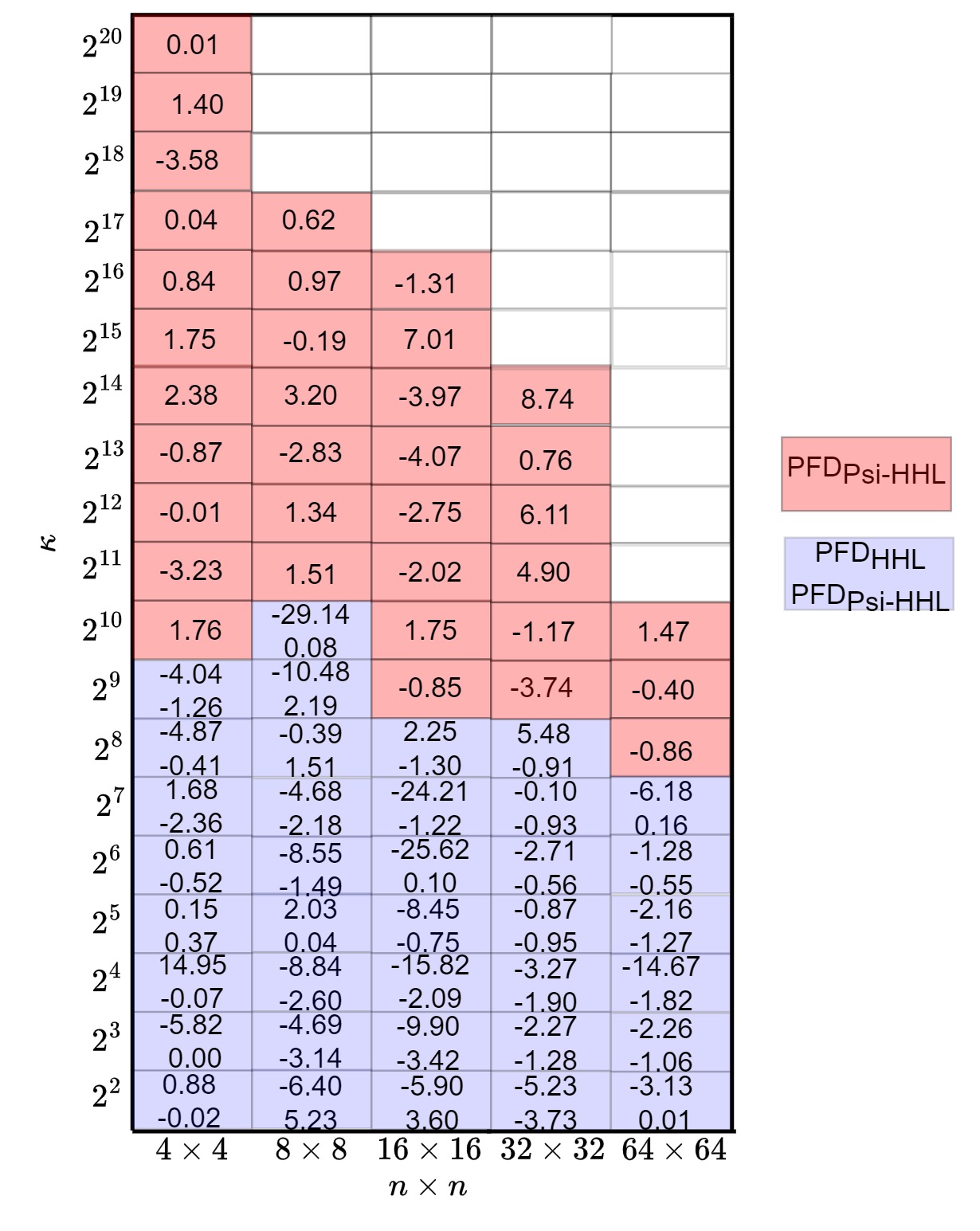}
\caption{Figure showing average (over 10 repetitions) values of percentage fraction difference (PFD) obtained from HHL and Psi-HHL, as the condition number, $\mathcal{\kappa}$, grows from $2^2$ to $2^{20}$, for fixed $A$ matrix sizes going from $4 \times 4$ to $64 \times 64$. Only the case where $\alpha=60^\circ$, $A$ not diagonal, $\vec{b}$ in an unequal superposition, and number of shots = $10^6$ is considered. The blue cells indicate cases where HHL gives outputs, and red denotes cases where HHL fails. The white cells denote those cases for which we did not carry out a calculation, due to computational cost. } 
\label{fig:hhlvspsihhl1}
\end{figure} 

\begin{table*}[]
\caption{Table providing the data for HHL and Psi-HHL algorithms (with $\alpha=60^\circ$, $70^\circ$, $80^\circ$) for $8 \times 8$ matrices. $A$ is a non-diagonal matrix and $\vec{b}$ is an unequal superposition. Each data point is an average over 10 repetitions. } 
\begin{tabular}{c|cccccccc}
\hline\hline
 & \multicolumn{1}{c}{} &  & HHL & \multicolumn{3}{c}{Psi-HHL} &  &  \\
 & \multicolumn{1}{c}{} &  & \multicolumn{1}{c}{} & $\alpha=60^\circ$ & $\alpha=70^\circ$ & $\alpha=80^\circ$ &  &  \\
\multirow{-3}{*}{\# shots} & \multicolumn{1}{c}{\multirow{-3}{*}{$\approx \mathcal{\kappa}$}} & \multirow{-3}{*}{\# qubits} & PFD ($P(0), P(1)$) &PFD ($P'(0), P'(1)$) & PFD ($P'(0), P'(1)$) & PFD ($P'(0), P'(1)$) &  &  \\\hline
 & $2^2$ & 10 & \multicolumn{1}{c}{-7.05 (0.82, 0.18)} & 4.90 (0.07, 0.93) & \multicolumn{1}{c}{5.99 (0.05, 0.95)} & 6.97 (0.10, 0.90) &  &  \\
 &$2^3$ & 11 & -6.18 (0.95, 0.05) & -3.11 (0.11, 0.89) & -2.93 (0.03, 0.97)&-2.93 (0.01, 0.99) &  &  \\
 & $2^4$ & 12 & \multicolumn{1}{c}{-8.78 (0.98, 0.02)} & -2.75 (0.18, 0.82) & \multicolumn{1}{c}{-2.48 (0.07, 0.93)} & -2.54 (0.02, 0.98) &  &  \\
 &$2^5$ & 13 & \multicolumn{1}{c}{0.42 (0.99, 0.01)} &  1.60 (0.21, 0.79) & \multicolumn{1}{c}{-0.90 (0.10, 0.90)} & -0.09 (0.02, 0.98) &  &  \\
 & $2^6$ & 14 & 0.12 (0.99, 0.01) & -5.01 (0.23, 0.77) & \multicolumn{1}{c}{-1.21 (0.11, 0.89)} & -4.35 (0.03, 0.97) &  &  \\
 & $2^7$ & 15 & 0.09 (0.99, 0.01) & 0.09 (0.24, 0.76) & \multicolumn{1}{c}{-3.83 (0.12, 0.88)} & 0.40 (0.03, 0.97) &  &  \\
 & $2^8$ & 16 & nan (0.99, 0.01) & -0.33 (0.25, 0.75) & -1.74 (0.12, 0.88) &-4.53 (0.03, 0.97) &  &  \\
 & $2^9$ & 17 & nan (0.99, 0.01) & -13.83 (0.25, 0.75) & -3.82 (0.12, 0.88) & 10.35 (0.04, 0.96) &  &  \\
 & $2^{10}$ & 18 & nan (0.99, 0.01) & -1.66 (0.25, 0.75) & 0.94 (0.12, 0.88) & -7.44(0.04, 0.96) &  &  \\
 & $2^{11}$ & 19 & nan (0.99, 0.01) & -0.82 (0.25, 0.75) & 4.90 (0.12, 0.88) & 3.13 (0.04, 0.96) &  &  \\
 & $2^{12}$ & 20 & nan (0.99, 0.01) & -0.91 (0.25, 0.75) & -7.52 (0.12, 0.88) & 10.89 (0.04, 0.96) &  &  \\
 & $2^{13}$ & 21 & nan (0.99, 0.01) & 2.62 (0.25, 0.75) & 8.41 (0.12, 0.88) & -0.26 (0.04, 0.96) &  &  \\
 & $2^{14}$ & 22 & nan (0.99, 0.01) & -3.20 (0.25, 0.75) & 7.12 (0.12, 0.88) & 8.53 (0.04, 0.96) &  &  \\
 & $2^{15}$ & 23 & nan (0.99, 0.01) & 5.74 (0.25, 0.75) & -7.34 (0.12, 0.88) & -7.60 (0.04, 0.96) &  &  \\
 & $2^{16}$ & 24 & nan (0.99, 0.01) & -5.89 (0.25, 0.75) & -4.65 (0.12, 0.88) & -13.89 (0.04, 0.96) &  &  \\
 \multirow{-17}{*}{$10^5$} & $2^{17}$ & 25 & nan (0.99, 0.01) & -6.58 (0.25, 0.75) & \multicolumn{1}{c}{-10.10 (0.12, 0.88)} & 2.78 (0.04, 0.96) &  &  \\\hline
 & $2^2$ & 10 &  -7.10 (0.82, 0.18)& 5.23 (0.10, 0.99) & 6.13 (0.05, 0.95) & 7.15 (0.10, 0.90) &  &  \\
 & $2^3$ & 11 & -6.10 (0.95, 0.05) & -3.14 (0.10, 0.90) & -3.02 (0.03, 0.97) & -2.83 (0.01, 0.99) &  &  \\
 & $2^4$ & 12 & -8.97 (0.98, 0.02) & -2.60 (0.18, 0.82) & -2.58 (0.07, 0.93) & -2.71 (0.02, 0.98) &  &  \\
 & $2^5$ & 13 & -1.76 (0.99, 0.01) & 0.04 (0.21, 0.79) & 0.03 (0.09, 0.91) & -0.04 (0.02, 0.98) &  &  \\
 & $2^6$ & 14 & -5.00 (0.99, 0.01) & -1.49 (0.23, 0.77) & -0.73 (0.11, 0.89) & -1.09 (0.03, 0.97) &  &  \\
 & $2^7$ & 15 &  1.96 (0.99, 0.01) & -2.18 (0.24, 0.76) & 1.16 (0.12, 0.88) & 0.36 (0.03, 0.97) &  &  \\
 &$2^8$ & 16 & 0.78 (0.99, 0.01) & 1.51 (0.25, 0.75) & -1.35 (0.12, 0.88) &-1.61 (0.04, 0.96) &  &  \\
 &$2^9$ & 17 & nan (0.99, 0.01) & 2.19 (0.25, 0.75) & 1.34 (0.12, 0.88) & -1.99 (0.04, 0.96) &  &  \\
 &$2^{10}$ & 18 & nan (0.99, 0.01) &  0.08 (0.25, 0.75) & -3.98 (0.12, 0.88) & 0.33 (0.04, 0.96) &  &  \\
 & $2^{11}$ & 19 & nan (0.99, 0.01) & 1.51 (0.25, 0.75) & -4.00 (0.12, 0.88) & -1.72 (0.04, 0.96) &  &  \\
 & $2^{12}$ & 20 & nan (0.99, 0.01) & 1.34 (0.25, 0.75) & 0.28 (0.12, 0.88) & 0.00 (0.04, 0.96) &  &  \\
 &$2^{13}$ & 21 & nan (0.99, 0.01) & -2.83 (0.25, 0.75) & 1.03 (0.12, 0.88) & 1.99 (0.04, 0.96) &  &  \\
 & $2^{14}$ & 22 & nan (0.99, 0.01) & 3.20 (0.25, 0.75) & -4.70 (0.12, 0.88) & 1.75 (0.04, 0.96) &  &  \\
 &$2^{15}$ & 23 & nan (0.99, 0.01) & -0.19 (0.25, 0.75) & -4.56 (0.12, 0.88) & -0.58 (0.04, 0.96) &  &  \\
 & $2^{16}$ & 24 & nan (0.99, 0.01) & 0.97 (0.25, 0.75) & 0.14 (0.12, 0.88) & -0.11 (0.04, 0.96) &  &  \\
 \multirow{-16}{*}{$10^6$}  & $2^{17}$ & 25 & nan (0.99, 0.01) & 0.62 (0.25, 0.75) & \multicolumn{1}{c}{3.21 (0.12, 0.88)} & 3.19 (0.04, 0.96) &  &  \\\hline\hline
\end{tabular}
\label{tab:a8x8}
\end{table*}

\begin{table*}[]
\caption{Table providing the data for HHL and Psi-HHL algorithms (with $\alpha=60^\circ$, $70^\circ$, $80^\circ$) for $16 \times 16$ matrices.  $A$ is a non-diagonal matrix and $\vec{b}$ is an unequal superposition. Each data point is an average over 10 repetitions. } 
\begin{tabular}{c|cccccccc}
\hline\hline
\multicolumn{1}{c|}{} & \multicolumn{1}{c}{} & \multicolumn{1}{c}{} &\multicolumn{1}{c}{HHL} & \multicolumn{3}{c}{Psi-HHL} &  &  \\
\multicolumn{1}{c|}{} & \multicolumn{1}{c}{} & \multicolumn{1}{c}{} &  & \multicolumn{1}{c}{$60^\circ$} & \multicolumn{1}{c}{$70^\circ$} & \multicolumn{1}{c}{$80^\circ$} &  &  \\
\multicolumn{1}{c|}{\multirow{-3}{*}{\# shots}} & \multicolumn{1}{c}{\multirow{-3}{*}{$ \approx \kappa$}} & \multicolumn{1}{c}{\multirow{-3}{*}{\# qubits}} & \multicolumn{1}{c}{PFD ($P(0), P(1)$)} & \multicolumn{1}{c}{PFD ($P'(0), P'(1)$)} & \multicolumn{1}{c}{PFD ($P'(0), P'(1)$)} & \multicolumn{1}{c}{PFD ($P'(0), P'(1)$)} &  &  \\\hline
\multicolumn{1}{c|}{} & \multicolumn{1}{c}{$2^2$} & \multicolumn{1}{c}{12} & \multicolumn{1}{c}{-5.50 (0.83, 0.17)} & \multicolumn{1}{c}{3.60 (0.06, 0.94)} & \multicolumn{1}{c}{4.29 (0.04, 0.96)} & \multicolumn{1}{c}{5.11 (0.08, 0.92)} &  &  \\
\multicolumn{1}{c|}{} & \multicolumn{1}{c}{$2^3$} & \multicolumn{1}{c}{13} & \multicolumn{1}{c}{-7.02 (0.95, 0.05)} & \multicolumn{1}{c}{-3.42 (0.11, 0.89)} & \multicolumn{1}{c}{-3.26 (0.03, 0.97)} & \multicolumn{1}{c}{-3.19 (0.01, 0.99)} &  &  \\
\multicolumn{1}{c|}{} & \multicolumn{1}{c}{$2^4$} & \multicolumn{1}{c}{14} & \multicolumn{1}{c}{7.05 (0.98, 0.02)} & \multicolumn{1}{c}{-2.09 (0.18, 0.82)} & \multicolumn{1}{c}{-2.17 (0.07, 0.93)} & \multicolumn{1}{c}{-1.76 (0.01, 0.99)} &  &  \\
\multicolumn{1}{c|}{} & \multicolumn{1}{c}{$2^5$} & \multicolumn{1}{c}{15} & \multicolumn{1}{c}{-1.70 (0.99, 0.01)} & \multicolumn{1}{c}{-0.75 (0.21, 0.79)} & \multicolumn{1}{c}{-0.53 (0.10, 0.91)} & \multicolumn{1}{c}{0.92 (0.02, 0.98)} &  &  \\
\multicolumn{1}{c|}{} & \multicolumn{1}{c}{$2^6$} & \multicolumn{1}{c}{16} & \multicolumn{1}{c}{-3.51 (0.99, 0.01)} & \multicolumn{1}{c}{0.10 (0.23, 0.77)} & \multicolumn{1}{c}{0.60 (0.11, 0.89)} & \multicolumn{1}{c}{-0.06 (0.03, 0.97)} &  &  \\
\multicolumn{1}{c|}{} & \multicolumn{1}{c}{$2^7$} & \multicolumn{1}{c}{17} & \multicolumn{1}{c}{-1.95 (0.99, 0.01)} & \multicolumn{1}{c}{-1.22 (0.24, 0.76)} & \multicolumn{1}{c}{-1.18 (0.11, 0.89)} & \multicolumn{1}{c}{-2.15 (0.03, 0.97)} &  &  \\
\multicolumn{1}{c|}{} & \multicolumn{1}{c}{$2^8$} & \multicolumn{1}{c}{18} & \multicolumn{1}{c}{-2.30 (0.99, 0.01)} & \multicolumn{1}{c}{-1.30 (0.25, 0.75)} & \multicolumn{1}{c}{0.47 (0.12, 0.88)} & \multicolumn{1}{c}{2.68 (0.03, 0.97)} &  &  \\
\multicolumn{1}{c|}{} & \multicolumn{1}{c}{$2^9$} & \multicolumn{1}{c}{19} & \multicolumn{1}{c}{nan (0.99, 0.01)} & \multicolumn{1}{c}{-0.85 (0.25, 0.75)} & \multicolumn{1}{c}{3.64 (0.12, 0.88)} & \multicolumn{1}{c}{-0.04 (0.04, 0.97)} &  &  \\
\multicolumn{1}{c|}{} & \multicolumn{1}{c}{$2^{10}$} & \multicolumn{1}{c}{20} & \multicolumn{1}{c}{nan (0.99, 0.01)} & \multicolumn{1}{c}{1.75 (0.25, 0.75)} & \multicolumn{1}{c}{-0.11 (0.12, 0.88)} & \multicolumn{1}{c}{-5.97 (0.04, 0.96)} &  &  \\
\multicolumn{1}{c|}{} & \multicolumn{1}{c}{$2^{11}$} & \multicolumn{1}{c}{21} & \multicolumn{1}{c}{nan (0.99, 0.01)} & \multicolumn{1}{c}{-2.02 (0.25, 0.75)} & \multicolumn{1}{c}{-2.59 (0.12, 0.88)} & \multicolumn{1}{c}{3.26 (0.04, 0.96)} &  &  \\
\multicolumn{1}{c|}{} & \multicolumn{1}{c}{$2^{12}$} & \multicolumn{1}{c}{22} & \multicolumn{1}{c}{nan (0.99, 0.01)} & \multicolumn{1}{c}{-2.75 (0.25, 0.75)} & \multicolumn{1}{c}{-4.15 (0.12, 0.88)} & \multicolumn{1}{c}{-7.93 (0.04, 0.96)} &  &  \\
\multicolumn{1}{c|}{} & \multicolumn{1}{c}{$2^{13}$} & \multicolumn{1}{c}{23} & \multicolumn{1}{c}{nan (0.99, 0.01)} & \multicolumn{1}{c}{-4.07 (0.25, 0.75)} & \multicolumn{1}{c}{0.29 (0.12, 0.88)} & \multicolumn{1}{c}{1.88 (0.04, 0.96)} &  &  \\
\multicolumn{1}{c|}{} & \multicolumn{1}{c}{$2^{14}$} & \multicolumn{1}{c}{24} & \multicolumn{1}{c}{nan (0.99, 0.01)} & \multicolumn{1}{c}{-3.97 (0.25, 0.75)} & \multicolumn{1}{c}{-0.83 (0.12, 0.88)} & \multicolumn{1}{c}{1.30 (0.04, 0.96)} &  &  \\
\multicolumn{1}{c|}{} & \multicolumn{1}{c}{ $2^{15}$} & \multicolumn{1}{c}{25} & \multicolumn{1}{c}{nan (0.99, 0.01)} & \multicolumn{1}{c}{ 7.01 (0.25, 0.75)} & \multicolumn{1}{c}{-1.46 (0.12, 0.88)} & \multicolumn{1}{c}{2.07 (0.04, 0.96)} &  &  \\
\multicolumn{1}{c|}{\multirow{-15}{*}{$10^6$}} & \multicolumn{1}{c}{$2^{16}$} & \multicolumn{1}{c}{26} & \multicolumn{1}{c}{nan 
 (0.99, 0.01)} & \multicolumn{1}{c}{-1.31 (0.25, 0.75)} & \multicolumn{1}{c}{-5.10 (0.12, 0.88)} & \multicolumn{1}{c}{-11.71 (0.04, 0.96)} &  &  \\\hline \hline 
\end{tabular}
\label{tab:a16x16}
\end{table*}

\begin{table*}[]
\caption{Table providing the data for HHL and Psi-HHL algorithms (with $\alpha=60^\circ$, $70^\circ$, $80^\circ$) for $32 \times 32$ matrices.  $A$ is a non-diagonal matrix and $\vec{b}$ is an unequal superposition. Each data point is an average over 10 repetitions. } 
\begin{tabular}{c|cccccccc}
\hline\hline
\multicolumn{1}{c|}{} & \multicolumn{1}{c}{} & \multicolumn{1}{c}{} & \multicolumn{1}{c}{HHL} & \multicolumn{3}{c}{Psi-HHL} &  &  \\
\multicolumn{1}{c|}{} & \multicolumn{1}{c}{} & \multicolumn{1}{c}{} & & \multicolumn{1}{c}{$60^\circ$} & \multicolumn{1}{c}{$70^\circ$} & \multicolumn{1}{c}{$80^\circ$} &  &  \\
\multicolumn{1}{c|}{\multirow{-3}{*}{\# shots}} & \multicolumn{1}{c}{\multirow{-3}{*}{$\approx \mathcal{\kappa}$}} & \multicolumn{1}{c}{\multirow{-3}{*}{\# qubits}} & \multicolumn{1}{c}{PFD ($P(0), P(1)$)} & \multicolumn{1}{c}{PFD ($P'(0), P'(1)$)} & \multicolumn{1}{c}{PFD ($P'(0), P'(1)$)} & PFD ($P'(0), P'(1)$) &  &  \\\hline
\multicolumn{1}{c|}{} & $2^2$ & \multicolumn{1}{c}{14} & \multicolumn{1}{c}{-5.23 (0.17, 0.83)} & \multicolumn{1}{c}{-3.73 (0.11, 0.89)} & \multicolumn{1}{c}{-3.56 (0.08, 0.92)} & -3.42 (0.10, 0.90) &  &  \\
\multicolumn{1}{c|}{} &$2^3$ & \multicolumn{1}{c}{15} & \multicolumn{1}{c}{-2.27 (0.05, 0.95)} & \multicolumn{1}{c}{-1.28 (0.13, 0.87)} & \multicolumn{1}{c}{-1.16 (0.04, 0.96)} & -1.09 (0.02, 0.98) &  &  \\
\multicolumn{1}{c|}{} & $2^4$ & \multicolumn{1}{c}{16} & \multicolumn{1}{c}{-3.27 (0.02, 0.98)} & \multicolumn{1}{c}{-1.90 (0.18, 0.82)} & \multicolumn{1}{c}{-1.70 (0.07, 0.93)} & -1.47 (0.02, 0.98) &  &  \\
\multicolumn{1}{c|}{} & $2^5$ & \multicolumn{1}{c}{17} & \multicolumn{1}{c}{-0.87 (0.01, 0.99)} & \multicolumn{1}{c}{-0.95 (0.21, 0.79)} & \multicolumn{1}{c}{-0.45 (0.10, 0.90)} & -0.49 (0.02, 0.98) &  &  \\
\multicolumn{1}{c|}{} & $2^6$ & \multicolumn{1}{c}{18} & \multicolumn{1}{c}{-2.71 (0.00,1.00)} & \multicolumn{1}{c}{-0.56 (0.23, 0.77)} & \multicolumn{1}{c}{-0.80 (0.11, 0.89)} & 0.49 (0.03, 0.97) &  &  \\
\multicolumn{1}{c|}{} & $2^7$ & \multicolumn{1}{c}{19} & \multicolumn{1}{c}{-0.10 (0.00, 1.00)} & \multicolumn{1}{c}{-0.93 (0.24, 0.76)} & \multicolumn{1}{c}{-0.05 (0.11,0.89)} & -0.61 (0.03, 0.97) &  &  \\
\multicolumn{1}{c|}{} & $2^8$ & \multicolumn{1}{c}{20} & \multicolumn{1}{c}{5.48 (0.00, 1.00)} & \multicolumn{1}{c}{-0.91 (0.25, 0.75)} & \multicolumn{1}{c}{1.97 (0.12, 0.88)} & -4.47 (0.03, 0.97) &  &  \\
\multicolumn{1}{c|}{} &$2^9$ & \multicolumn{1}{c}{21} & \multicolumn{1}{c}{nan (0.00, 1.00)} & \multicolumn{1}{c}{-3.74 (0.25, 0.75)} & \multicolumn{1}{c}{2.07 (0.12, 0.88)} & -2.92 (0.03, 0.97) &  &  \\
\multicolumn{1}{c|}{} &$2^{10}$ & \multicolumn{1}{c}{22} & \multicolumn{1}{c}{nan (0.00, 1.00)} & \multicolumn{1}{c}{-1.17 (0.25, 0.75)} & \multicolumn{1}{c}{2.07 (0.12, 0.88)} & -0.50 (0.03, 0.97) &  &  \\
\multicolumn{1}{c|}{} & $2^{11}$ & \multicolumn{1}{c}{23} & \multicolumn{1}{c}{nan (0.00, 1.00)} & \multicolumn{1}{c}{4.90 (0.25, 0.75)} & \multicolumn{1}{c}{7.32 (0.12, 0.88)} & -3.07 (0.03, 0.97) &  &  \\
\multicolumn{1}{c|}{} & $2^{12}$ & \multicolumn{1}{c}{24} & \multicolumn{1}{c}{nan (0.00, 1.00)} & \multicolumn{1}{c}{6.11 (0.25, 0.75)} & \multicolumn{1}{c}{-1.09 (0.12, 0.88)} & 8.85 (0.03, 0.97) &  &  \\
\multicolumn{1}{c|}{} & $2^{13}$ & \multicolumn{1}{c}{25} & \multicolumn{1}{c}{nan (0.00, 1.00)} & \multicolumn{1}{c}{0.76 (0.25, 0.75)} & \multicolumn{1}{c}{-4.61 (0.12, 0.88)} & -0.63 (0.03, 0.97) &  &  \\
\multicolumn{1}{c|}{\multirow{-13}{*}{$10^6$}} &$2^{14}$ & \multicolumn{1}{c}{26} & \multicolumn{1}{c}{nan (0.00, 1.00)} & \multicolumn{1}{c}{8.74 (0.25, 0.75)} & \multicolumn{1}{c}{7.76 (0.12, 0.88)} & 1.40 (0.03, 0.97) &  &  \\
\hline \hline 
\end{tabular} 
\label{tab:a32x32}
\end{table*}

\begin{table*}[]
\caption{Table providing the data for HHL and Psi-HHL algorithms (with $\alpha=60^\circ$, $70^\circ$, $80^\circ$) for $64 \times 64$ matrices.  $A$ is a non-diagonal matrix and $\vec{b}$ is an unequal superposition. Each data point is an average over 10 repetitions. } 
\begin{tabular}{c|cccccccc}
\hline\hline
 & \multicolumn{1}{c}{} &  & \multicolumn{1}{c}{HHL} & \multicolumn{3}{c}{Psi-HHL} &  &  \\
 & \multicolumn{1}{c}{} &  & & $60^\circ$ & $70^\circ$ & \multicolumn{1}{c}{$80^\circ$} &  &  \\
\multirow{-3}{*}{\# shots} & \multicolumn{1}{c}{\multirow{-3}{*}{$\approx \mathcal{\kappa}$}} & \multirow{-3}{*}{\# qubits} & PFD ($P(0), P(1)$) & PFD ($P'(0), P'(1)$) & PFD ($P'(0), P'(1)$) & PFD ($P'(0), P'(1)$) &  &  \\\hline
 & $2^2$ & 16 &-3.13 (0.82, 0.17)   & 0.01 (0.12, 0.88) & 0.21 (0.09, 0.91) & 0.82 (0.10, 0.90) &  &  \\
 & $2^3$ & 17 & -2.26 (0.94, 0.05)  & -1.06 (0.01, 0.88) & -0.95 (0.04, 0.96) & -1.01 (0.02, 0.98) &  &  \\
 & $2^4$ & 18 & -2.91 (0.98, 0.02) & -1.82 (0.17, 0.83) & 1.47 (0.03, 0.97) & -1.56 (0.02, 0.98) &  &  \\
 &$2^5$  &  19 & -2.16 (0.99, 0.01) & -1.27 (0.21, 0.79) & -1.57 (0.10, 0.91) & -0.86 (0.02, 0.98) &  &  \\
 & $2^6$ & 20 & -1.28 (1.00, 0.00)  & -0.55 (0.23, 0.77) & -1.26 (0.10, 0.90) & -1.06 (0.02, 0.98) &  &  \\
 & $2^7$ & 21 & -6.18 (1.00, 0.00)  & 0.16 (0.24, 0.76) & -0.10 (0.11, 0.89) & -0.10 (0.03, 0.97) &  &  \\
 & $2^8$ & 22 & nan (1.00, 0.00)    & -0.86 (0.25, 0.75) & -0.45 (0.11, 0.89) & 0.18 (0.03, 0.97) &  &  \\
 & $2^9$ & 23 &nan (1.00, 0.00)     & -0.40 (0.25, 0.75) & 0.05 (0.12, 0.88) & -2.46 (0.03, 0.97) &  &  \\
\multirow{-9}{*}{$10^6$} &$2^{10}$ & 24 & nan (1.00, 0.00) & 1.47 (0.25, 0.75) & 1.47 (0.12, 0.88) & 2.09 (0.03, 0.97) &  &  \\\hline \hline 
\end{tabular} 
\label{tab:a64x64}
\end{table*}

\clearpage 

\section{Details on calculations involving quantum chemistry matrices} \label{SM: chemistry-matrices}

\subsection{$A$ matrix and $\vec{b}$ generation} \label{SM: chemistry-matrices-A}

We begin by explaining the procedure for generating an $A$ matrix of size $n \times n$. We use the configuration state function-based Graphical Unitary Group Approach-Configuration Interaction Singles and Doubles (GUGA-CISD) method available in the GAMESS program \cite{GAMESS} to construct the Hamiltonian for a given molecule in a specified single particle basis. GUGA-CI ensures that the Hamiltonian is constructed in an orthogonal basis; hence, it is Hermitian. Also, because it is a spin-adapted CI method, it avoids spin contamination of the wave function, thereby giving accurate results. In order to generate the $A$ matrix, we remove the first row and the first column of the GUGA-CISD Hamiltonian matrix and subtract the diagonal entries of the resulting matrix with the energy obtained from M{\o}ller-Plesset theory to second order in perturbation (also called MP2 in literature). The latter step is carried out for the purpose of normal ordering, which in turn ensures that HHL would output the correlation part of the ground state energy directly instead of the total ground state energy (which can be written as a sum of the HF energy and the correlation energy, with the latter being much smaller than the former). This is desirable when one executes HHL-LCC calculations on NISQ/late-NISQ era quantum hardware, as the error in the total ground state energy can be much greater than the correlation energy itself. We also note that unlike the traditional approach where the HF energy is subtracted from the diagonal entries, we take away the MP2 energy, in order to make sure that $A$ is positive definite. In a sense, subtracting MP2 energy could be thought of as a level shift over and beyond subtracting the HF energy. \\ 

We now focus our attention on the elements of the $A$ matrix. The size of an $A$ matrix, $m \times m$, generated from a quantum chemical calculation, which we can denote as  $A_{mol}$, is not necessarily of the dimension $n \times n$. In such cases, we embed the $A_{mol}$ matrix of dimension ${m\times m}$ as a diagonal block in a larger matrix of dimension $n \times n$, with the other diagonal block being an identity matrix of size $(n-m) \times (n-m)$, denoted as $\mathcal{I}$. In such a case, we solve a system of linear equations given by 

\begin{eqnarray}
\begin{pmatrix}
A_{mol} & 0 \\
0 & \mathcal{I}
\end{pmatrix} 
\begin{pmatrix}
\vec{x} \\
\vec{0}
\end{pmatrix} = \begin{pmatrix}
\vec{b} \\
\vec{0}
\end{pmatrix}. 
\end{eqnarray}

This approach ensures that the eigenvalues of $A_{mol}$ remain unaffected, as long as its maximum eigenvalue of the scaled matrix is $\approx$ 1. We choose our matrices carefully so that this condition is satisfied. In the event that the maximum diagonal entry of $A$ is less than 1, it is recommended to pad with the largest (or the smallest) diagonal entry of $A$ so that the number of shots expended is reduced. As an example, we picked a $3 \times 3$ sparse matrix such that the largest diagonal entry is 0.70: \(
\begin{pmatrix}
0.25 & 0.10 & 0.00 \\
0.10 & 0.45 & 0.20 \\
0.00 & 0.20 & 0.70
\end{pmatrix}
\), and considered two cases: padding with 1 and with 0.70. We found that the condition number, $\mathcal{\kappa}$, is larger in the latter (5.16) than in the former (4.21). \\ 

Since for the problems we consider, $\mathcal{\kappa}$ increases with system size monotonically, and further, going from an unpadded to a padded matrix always right shifts a data point, it is sufficient to carry out $\mathcal{\kappa}$ versus $m$ analysis for the unpadded case, as it upper bounds the complexity of the padded scenario of $\mathcal{\kappa}$ versus $n$. 

\begin{figure}[t]
\centering
\begin{tabular}{cc}
\includegraphics[scale=0.4]{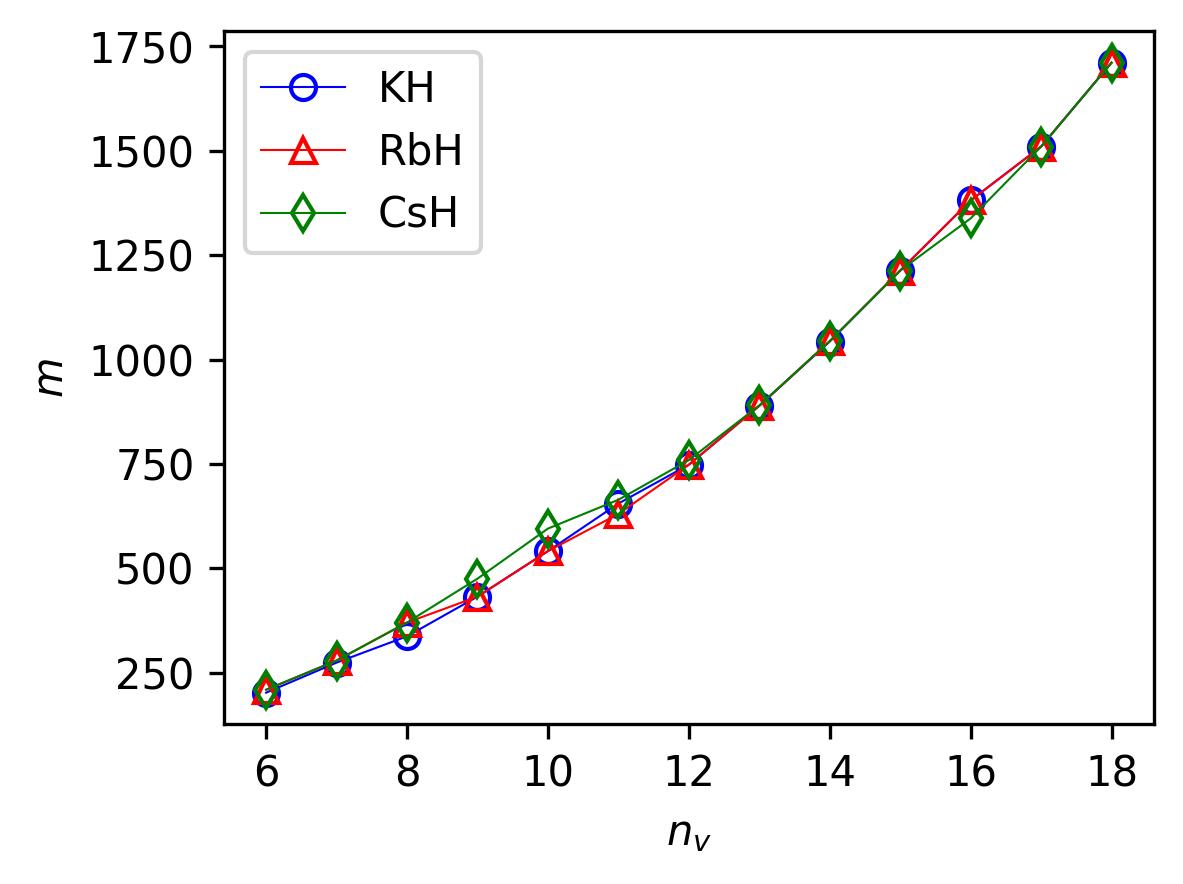}&
\includegraphics[scale=0.4]{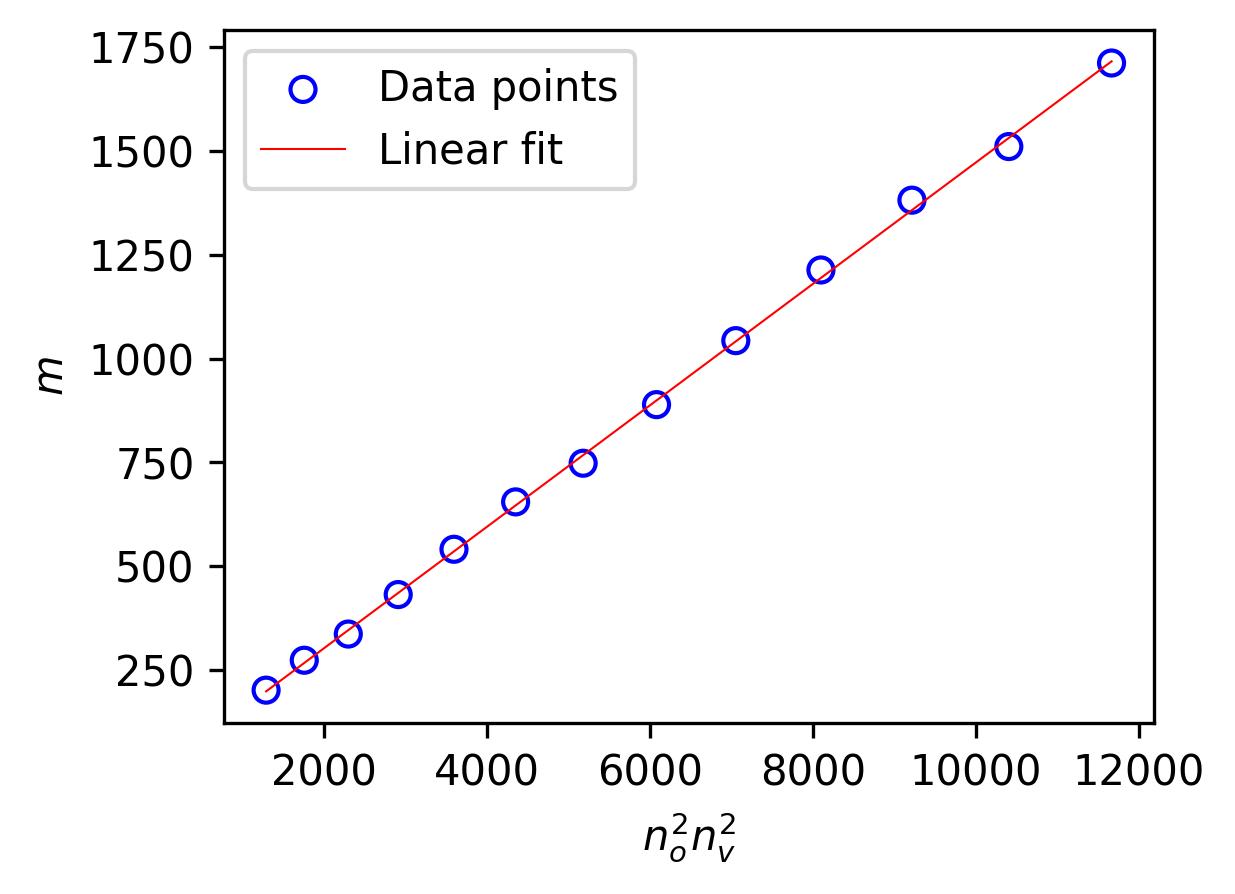}\\
(a)&(b)\\
\includegraphics[scale=0.4]{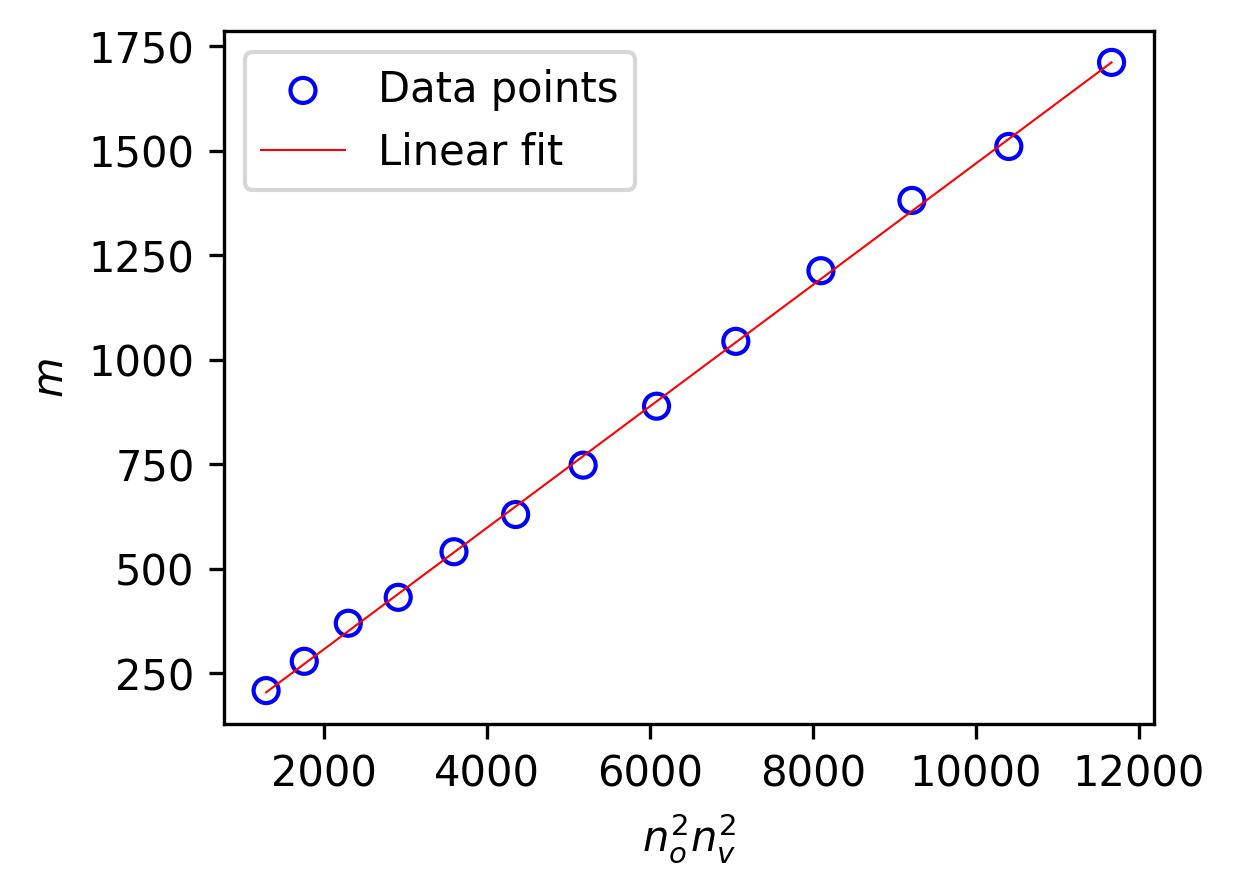}&
\includegraphics[scale=0.4]{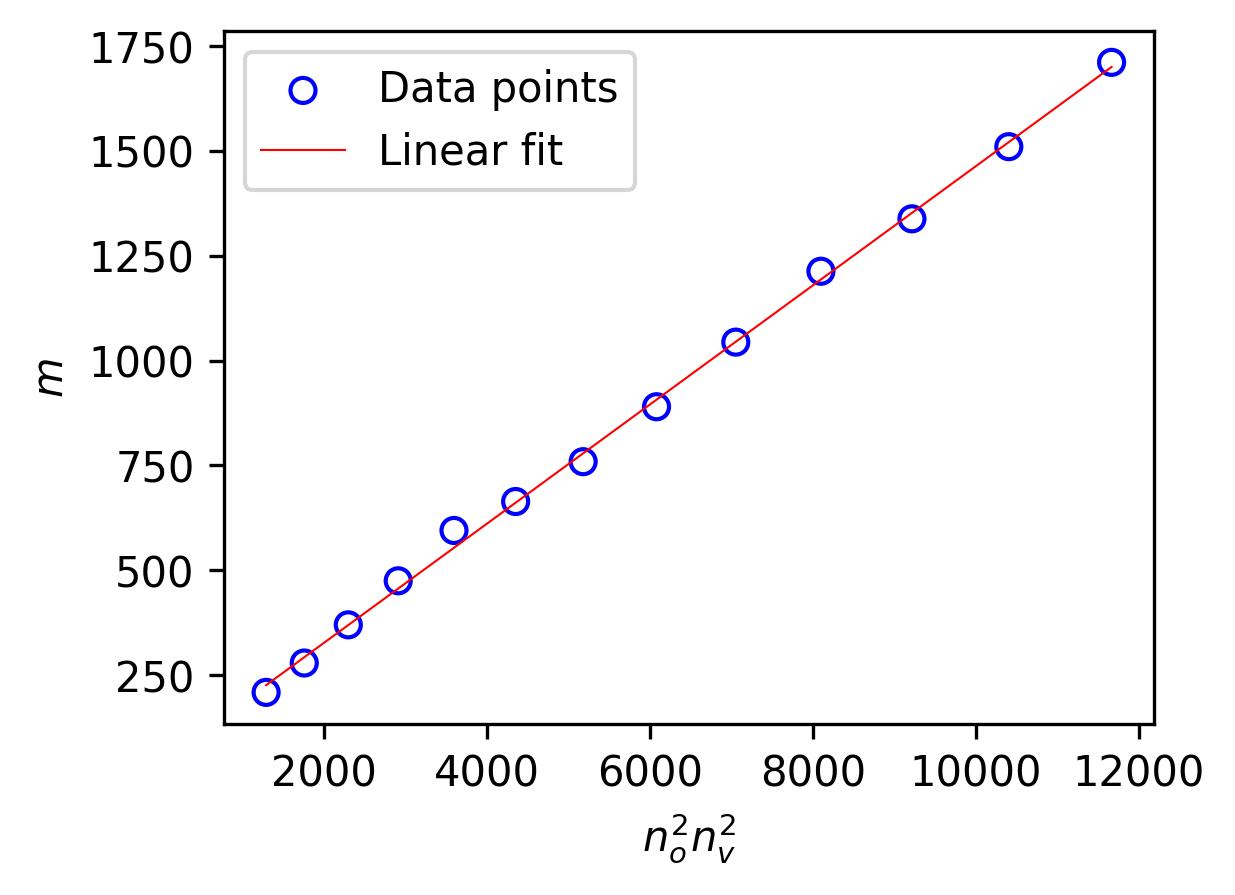}\\(c)&(d)
\end{tabular}
\caption{Sub-figure (a) shows the increase in system size ($m$) with the number of virtuals, $n_v$, for KH, RbH and CsH molecules in the LCCSD approximation, while the sub-figures (b), (c) and (d) show that $m$ increases as $n_o^2n_v^2$ for the three molecules, as one would expect. } 
\label{fig:SM2}
\end{figure}  

\subsection{Orbital character analysis}  \label{SM: chemistry-matrices-B}

We now explain the procedure that we adopted for selecting orbitals, based on orbital character. We begin with RbH, and then proceed to explain our procedure for CsH (with accompanying data provided in Table \ref{tab:A400}). \\ 

We fix the number of occupied orbitals to 6, 8 and 10. We now describe the method adopted in choosing virtual molecular orbitals (MOs). Instead of taking the approach of ordering the virtual MOs by energy in ascending order and then imposing an energy cut-off to define our active space, we pick those MOs that are dominated by important atomic orbitals (AOs). We select the important AOs, and for the purposes of this work, we pick those AOs in the order of the principal quantum numbers. For example, in the case of RbH, we can decide the important AOs from the set that arise from our choice of basis, namely the Sapporo double zeta basis: $5s$ through $7p$. For the purposes of our analysis on condition number versus system size, we pick all the AOs that the basis allows in order to obtain more data points for fitting (details given in Table~\ref{tab:A400}). We now move to some specifics for RbH: for the smallest system size considered (1 virtual MO), we pick the first virtual from our sorted MO list, which is the MO dominated by the Rb $5s-5p_z$ hybrid, and thus our active space that is built out of 6 occupied MOs and 1 virtual one has a system size, $m \times m$, of $209 \times 209$, and subsequently upon padding, an $A$ matrix size, $n \times n$, of $256 \times 256$. For the next system size, we add the next MO in the sorted list, which is dominated by Rb $5s$, leading to a $279 \times 279$ matrix, which, after padding, becomes a matrix of size $512 \times 512$. We continue to construct $A$ matrices in this fashion until we reach $m=2296$, for which $\mathcal{\kappa}$ is about 211.94. \\ 

We do the same for KH and CsH molecules, we fix the number of occupied orbitals to 6, 8 and 10 occupied oritals while progressively increasing the number of valence orbitals. The order in which we add virtual MOs as well as each such virtual MO's details are mentioned in Table \ref{tab:A400}. \\ 

\setlength{\tabcolsep}{0.6em}
\begin{table*}
\caption{Table providing the data for condition number ($\mathcal{\kappa}$) of the $m \times m$ $A_{mol}$ matrix versus system size ($m$), for KH, RbH and CsH molecules, where we vary the virtual orbitals to increase $m$. In the table, $n_{o}$ is the number of active spin orbitals, and $n_v$ is the number of unoccupied spin orbitals. The column labelled `Reordered virtual MOs' represents the atomic orbitals reordered according to the ones that contribute dominantly to the virtual molecular orbital. The column labelled `MO energies' gives the virtual molecular orbital energies in units of milli-Hartree.}
\centering
\begin{tabular}{cccccccccc}
\hline\hline
\multicolumn{1}{l}{\cellcolor[HTML]{FFFFFF}Molecule} &
  Reordered virtual MOs &
  MO energies &
  \multicolumn{2}{c}{\cellcolor[HTML]{FFFFFF} $n_o=6$} &
  \multicolumn{2}{c}{\cellcolor[HTML]{FFFFFF}$n_o=8$} &
  \multicolumn{2}{c}{\cellcolor[HTML]{FFFFFF}$n_o=10$} &
   \\
\multicolumn{1}{l}{\cellcolor[HTML]{FFFFFF}} &
  \multicolumn{1}{l}{\cellcolor[HTML]{FFFFFF}} &
  \multicolumn{1}{l}{\cellcolor[HTML]{FFFFFF}} &
  \multicolumn{1}{c}{\cellcolor[HTML]{FFFFFF}m} &
  \multicolumn{1}{c}{\cellcolor[HTML]{FFFFFF}$\kappa$ $(n_o, n_v)$} &
  \multicolumn{1}{c}{\cellcolor[HTML]{FFFFFF}m} &
  \multicolumn{1}{c}{\cellcolor[HTML]{FFFFFF}$\kappa$ $(n_o, n_v)$} &
  \multicolumn{1}{c}{\cellcolor[HTML]{FFFFFF}m} &
  \multicolumn{1}{c}{\cellcolor[HTML]{FFFFFF}$\kappa$ $(n_o, n_v)$} &
   \\\hline
\cellcolor[HTML]{FFFFFF} &
  K $4s $ &
  1151.70 &
  202 &
  233.65 (6, 6) &
  \multicolumn{1}{c}{\cellcolor[HTML]{FFFFFF}580} &
  41.03 (8, 8) &
  \multicolumn{1}{c}{\cellcolor[HTML]{FFFFFF}1484} &
  2195.80 (10, 10) &
   \\
\cellcolor[HTML]{FFFFFF} &
  K $4p_y$ &
  3609.70 &
  274 &
  234.44 (6, 7) &
  \multicolumn{1}{c}{\cellcolor[HTML]{FFFFFF}745} &
  41.57 (8, 9) &
  \multicolumn{1}{c}{\cellcolor[HTML]{FFFFFF}1761} &
  2206.59 (10, 11) &
   \\
\cellcolor[HTML]{FFFFFF} &
  K $4p_x$ &
  3609.70 &
  337 &
  235.12 (6, 8) &
  \multicolumn{1}{c}{\cellcolor[HTML]{FFFFFF}932} &
  236.57 (8, 10) &
  \multicolumn{1}{c}{\cellcolor[HTML]{FFFFFF}2046} &
  2217.42 (10, 12) &
   \\
\cellcolor[HTML]{FFFFFF} &
  K $4p_z$ &
  3663.70 &
  432 &
  236.80 (6, 9) &
  \multicolumn{1}{c}{\cellcolor[HTML]{FFFFFF}1110} &
  237.74 (8, 11) &
  \multicolumn{1}{c}{\cellcolor[HTML]{FFFFFF}2451} &
  2323.69 (10, 13) &
   \\
\cellcolor[HTML]{FFFFFF} &
  K $3d_{xy}$ &
  743.00 &
  541 &
  238.28 (6, 10) &
  \multicolumn{1}{c}{\cellcolor[HTML]{FFFFFF}1296} &
  238.90 (8, 12) &
  \multicolumn{1}{c}{\cellcolor[HTML]{FFFFFF}2900} &
  2849.98 (10, 14) &
   \\
\cellcolor[HTML]{FFFFFF} &
  K $3dy^2$ &
  743.00 &
  655 &
  244.66 (6, 11) &
  \multicolumn{1}{c}{\cellcolor[HTML]{FFFFFF}1540} &
  250.34 (8, 13) &
  \multicolumn{1}{c}{\cellcolor[HTML]{FFFFFF}3393} &
  2904.40 (10, 15) &
   \\
\cellcolor[HTML]{FFFFFF} &
  K $3d_{yz}$ &
  747.40 &
  748 &
  250.45 (6, 12) &
  \multicolumn{1}{c}{\cellcolor[HTML]{FFFFFF}1808} &
  307.03 (8, 14) &
  \multicolumn{1}{c}{\cellcolor[HTML]{FFFFFF}3794} &
  2933.74 (10, 16) &
   \\
\cellcolor[HTML]{FFFFFF} &
  K $3d_{xz}$ &
  747.40 &
  889 &
  265.31 (6, 13) &
  \multicolumn{1}{c}{\cellcolor[HTML]{FFFFFF}2100} &
  312.83 (8, 15) &
  \multicolumn{1}{c}{\cellcolor[HTML]{FFFFFF}4203} &
  2964.04 (10, 17) &
   \\
\cellcolor[HTML]{FFFFFF} &
  K $3dz^2$ &
  755.30 &
  1043 &
  340.59 (6, 14) &
  \multicolumn{1}{c}{\cellcolor[HTML]{FFFFFF}2358} &
  316.09 (8, 16) &
  \multicolumn{1}{c}{\cellcolor[HTML]{FFFFFF}4788} &
  3610.73 (10, 18) &
   \\
\cellcolor[HTML]{FFFFFF} &
  K $6s$ &
  3.70 &
  1213 &
  348.68 (6, 15) &
  \multicolumn{1}{c}{\cellcolor[HTML]{FFFFFF}2624} &
  319.44 (8, 17) &
  \multicolumn{1}{c}{\cellcolor[HTML]{FFFFFF}5265} &
  3629.57 (10, 19) &
   \\
\cellcolor[HTML]{FFFFFF} &
  K $6p_y$ &
  72.70 &
  1381 &
  371.54 (6, 16) &
  \multicolumn{1}{c}{\cellcolor[HTML]{FFFFFF}2972} &
  389.11 (8, 18) &
  \multicolumn{1}{c}{\cellcolor[HTML]{FFFFFF}5750} &
  3648.68 (10, 20) &
   \\
\cellcolor[HTML]{FFFFFF} &
  K $6p_x$ &
  72.70 &
  1510 &
  374.07 (6, 17) &
  \multicolumn{1}{c}{\cellcolor[HTML]{FFFFFF}3278} &
  391.14 (8, 19) &
  \multicolumn{1}{c}{\cellcolor[HTML]{FFFFFF}6427} &
  3674.83 (10, 21) &
   \\
\cellcolor[HTML]{FFFFFF} &
  K $6p_z$ &
  82.80 &
  1711 &
  380.25 (6, 18) &
  \multicolumn{1}{c}{\cellcolor[HTML]{FFFFFF}3592} &
  393.20 (8, 20) &
   &
  \multicolumn{1}{l}{\cellcolor[HTML]{FFFFFF}} &
   \\
\cellcolor[HTML]{FFFFFF} &
  K $5s$ &
  147.20 &
  1909 &
  385.51 (6, 19) &
  \multicolumn{1}{c}{\cellcolor[HTML]{FFFFFF}3996} &
  395.77 (8, 21) &
   &
  \multicolumn{1}{l}{\cellcolor[HTML]{FFFFFF}} &
   \\
\cellcolor[HTML]{FFFFFF} &
  K $5p_z$ &
  692.70 &
  2062 &
  386.30 (6, 20) &
  \multicolumn{1}{l}{\cellcolor[HTML]{FFFFFF}} &
  \multicolumn{1}{l}{\cellcolor[HTML]{FFFFFF}} &
   &
  \multicolumn{1}{l}{\cellcolor[HTML]{FFFFFF}} &
   \\
\cellcolor[HTML]{FFFFFF} &
  K $5p_y$ &
  702.90 &
  2295 &
  389.76 (6, 21) &
  \multicolumn{1}{l}{\cellcolor[HTML]{FFFFFF}} &
  \multicolumn{1}{l}{\cellcolor[HTML]{FFFFFF}} &
   &
  \multicolumn{1}{l}{\cellcolor[HTML]{FFFFFF}} &
   \\
\cellcolor[HTML]{FFFFFF} &
  K $5p_x$ &
  702.90 &
  \multicolumn{1}{l}{\cellcolor[HTML]{FFFFFF}} &
  \multicolumn{1}{l}{\cellcolor[HTML]{FFFFFF}} &
  \multicolumn{1}{l}{\cellcolor[HTML]{FFFFFF}} &
  \multicolumn{1}{l}{\cellcolor[HTML]{FFFFFF}} &
   &
  \multicolumn{1}{l}{\cellcolor[HTML]{FFFFFF}} &
   \\
\cellcolor[HTML]{FFFFFF} &
  H $2s$ &
  934.00 &
  \multicolumn{1}{l}{\cellcolor[HTML]{FFFFFF}} &
  \multicolumn{1}{l}{\cellcolor[HTML]{FFFFFF}} &
  \multicolumn{1}{l}{\cellcolor[HTML]{FFFFFF}} &
  \multicolumn{1}{l}{\cellcolor[HTML]{FFFFFF}} &
   &
  \multicolumn{1}{l}{\cellcolor[HTML]{FFFFFF}} &
   \\
\cellcolor[HTML]{FFFFFF} &
  H $2p_y$ &
  1668.70 &
  \multicolumn{1}{l}{\cellcolor[HTML]{FFFFFF}} &
  \multicolumn{1}{l}{\cellcolor[HTML]{FFFFFF}} &
  \multicolumn{1}{l}{\cellcolor[HTML]{FFFFFF}} &
  \multicolumn{1}{l}{\cellcolor[HTML]{FFFFFF}} &
   &
  \multicolumn{1}{l}{\cellcolor[HTML]{FFFFFF}} &
   \\
\cellcolor[HTML]{FFFFFF} &
  H $2p_x$ &
  1668.70 &
  \multicolumn{1}{l}{\cellcolor[HTML]{FFFFFF}} &
  \multicolumn{1}{l}{\cellcolor[HTML]{FFFFFF}} &
  \multicolumn{1}{l}{\cellcolor[HTML]{FFFFFF}} &
  \multicolumn{1}{l}{\cellcolor[HTML]{FFFFFF}} &
   &
  \multicolumn{1}{l}{\cellcolor[HTML]{FFFFFF}} &
   \\
\multirow{-21}{*}{\cellcolor[HTML]{FFFFFF}KH} &
  H $2p_z$ &
  1914.10 &
  \multicolumn{1}{l}{\cellcolor[HTML]{FFFFFF}} &
  \multicolumn{1}{l}{\cellcolor[HTML]{FFFFFF}} &
  \multicolumn{1}{l}{\cellcolor[HTML]{FFFFFF}} &
  \multicolumn{1}{l}{\cellcolor[HTML]{FFFFFF}} &
   &
  \multicolumn{1}{l}{\cellcolor[HTML]{FFFFFF}} &
   \\\hline
\cellcolor[HTML]{FFFFFF} &
  Rb $5s-5p_z$ hybrid&
  940.00 &
  209 &
  102.73 (6, 6) &
  632 &
  104.88 (8, 8) &
  \multicolumn{1}{r}{\cellcolor[HTML]{FFFFFF}1401} &
  108.16 (10, 10) &
   \\
\cellcolor[HTML]{FFFFFF} &
  Rb $5s$ &
  925.80 &
  279 &
  103.00 (6, 7) &
  744 &
  105.81 (8, 9) &
  \multicolumn{1}{r}{\cellcolor[HTML]{FFFFFF}1673} &
  111.85 (10, 11) &
   \\
\cellcolor[HTML]{FFFFFF} &
  Rb $5p_x$ &
  2283.00 &
  370 &
  103.74 (6, 8) &
  932 &
  106.63 (8, 10) &
  \multicolumn{1}{r}{\cellcolor[HTML]{FFFFFF}1963} &
  115.27 (10, 12) &
   \\
\cellcolor[HTML]{FFFFFF} &
  Rb $5p_y$ &
  2283.00 &
  432 &
  104.65 (6, 9) &
  1110 &
  110.27 (8, 11) &
  \multicolumn{1}{r}{\cellcolor[HTML]{FFFFFF}2321} &
  125.33 (10, 13) &
   \\
\cellcolor[HTML]{FFFFFF} &
  Rb $5p_z$ &
  2305.80 &
  541 &
  105.46 (6, 10) &
  1296 &
  113.64 (8, 12) &
  \multicolumn{1}{r}{\cellcolor[HTML]{FFFFFF}2713} &
  156.32 (10, 14) &
   \\
\cellcolor[HTML]{FFFFFF} &
  Rb $4d_{xy}$ &
  635.40 &
  630 &
  109.05 (6, 11) &
  1540 &
  123.51 (8, 13) &
  \multicolumn{1}{r}{\cellcolor[HTML]{FFFFFF}3141} &
  182.23 (10, 15) &
   \\
\cellcolor[HTML]{FFFFFF} &
  Rb $4dx^2$ &
  635.40 &
  748 &
  112.38 (6, 12) &
  1808 &
  154.29 (8, 14) &
  \multicolumn{1}{r}{\cellcolor[HTML]{FFFFFF}3537} &
  206.22 (10, 16) &
   \\
\cellcolor[HTML]{FFFFFF} &
  Rb $4d_{xz}$ &
  641.20 &
  889 &
  122.12 (6, 13) &
  2100 &
  179.90 (8, 15) &
  \multicolumn{1}{r}{\cellcolor[HTML]{FFFFFF}3951} &
  207.93 (10, 17) &
   \\
\cellcolor[HTML]{FFFFFF} &
  Rb $4d_{yz}$ &
  641.20 &
  1044 &
  152.70 (6, 14) &
  2358 &
  203.86 (8, 16) &
  \multicolumn{1}{r}{\cellcolor[HTML]{FFFFFF}4460} &
  213.88 (10, 18) &
   \\
\cellcolor[HTML]{FFFFFF} &
  Rb $4dz^2$ &
  666.90 &
  1213 &
  178.09 (6, 15) &
  2624 &
  205.55 (8, 17) &
  \multicolumn{1}{r}{\cellcolor[HTML]{FFFFFF}4932} &
  214.42 (10, 19) &
   \\
\cellcolor[HTML]{FFFFFF} &
  Rb $7s $ &
  3.20 &
  1381 &
  202.02 (6, 16) &
  2972 &
  211.53 (8, 18) &
  \multicolumn{1}{r}{\cellcolor[HTML]{FFFFFF}5422} &
  214.96 (10, 20) &
   \\
\cellcolor[HTML]{FFFFFF} &
  Rb $7p_x$ &
  64.20 &
  1510 &
  203.70 (6, 17) &
  3278 &
  212.06 (8, 19) &
  \multicolumn{1}{r}{\cellcolor[HTML]{FFFFFF}6013} &
  216.27 (10, 21) &
   \\
\cellcolor[HTML]{FFFFFF} &
  Rb $7p_y$ &
  64.20 &
  1711 &
  209.67 (6, 18) &
  3592 &
  212.60 (8, 20) &
   &
  \multicolumn{1}{l}{\cellcolor[HTML]{FFFFFF}} &
   \\
\cellcolor[HTML]{FFFFFF} &
  Rb $6s$ &
  78.40 &
  1860 &
  210.19 (6, 19) &
  3996 &
  213.87 (8, 21) &
   &
  \multicolumn{1}{l}{\cellcolor[HTML]{FFFFFF}} &
   \\
\cellcolor[HTML]{FFFFFF} &
  Rb $6s$ &
  130.80 &
  2062 &
  210.72 (6, 20) &
  \multicolumn{1}{l}{\cellcolor[HTML]{FFFFFF}} &
  \multicolumn{1}{l}{\cellcolor[HTML]{FFFFFF}} &
   &
  \multicolumn{1}{l}{\cellcolor[HTML]{FFFFFF}} &
   \\
\cellcolor[HTML]{FFFFFF} &
  Rb $6p_x$ &
  477.70 &
  2295 &
  211.94 (6, 21) &
  \multicolumn{1}{l}{\cellcolor[HTML]{FFFFFF}} &
  \multicolumn{1}{l}{\cellcolor[HTML]{FFFFFF}} &
   &
  \multicolumn{1}{l}{\cellcolor[HTML]{FFFFFF}} &
   \\
\cellcolor[HTML]{FFFFFF} &
  Rb $6p_y$ &
  477.70 &
  \multicolumn{1}{l}{\cellcolor[HTML]{FFFFFF}} &
  \multicolumn{1}{l}{\cellcolor[HTML]{FFFFFF}} &
  \multicolumn{1}{l}{\cellcolor[HTML]{FFFFFF}} &
  \multicolumn{1}{l}{\cellcolor[HTML]{FFFFFF}} &
   &
  \multicolumn{1}{l}{\cellcolor[HTML]{FFFFFF}} &
   \\
\cellcolor[HTML]{FFFFFF} &
  Rb $6p_z$ &
  487.50 &
  \multicolumn{1}{l}{\cellcolor[HTML]{FFFFFF}} &
  \multicolumn{1}{l}{\cellcolor[HTML]{FFFFFF}} &
  \multicolumn{1}{l}{\cellcolor[HTML]{FFFFFF}} &
  \multicolumn{1}{l}{\cellcolor[HTML]{FFFFFF}} &
   &
  \multicolumn{1}{l}{\cellcolor[HTML]{FFFFFF}} &
   \\
\cellcolor[HTML]{FFFFFF} &
  H $2p_x$ &
  1678.20 &
  \multicolumn{1}{l}{\cellcolor[HTML]{FFFFFF}} &
  \multicolumn{1}{l}{\cellcolor[HTML]{FFFFFF}} &
  \multicolumn{1}{l}{\cellcolor[HTML]{FFFFFF}} &
  \multicolumn{1}{l}{\cellcolor[HTML]{FFFFFF}} &
   &
  \multicolumn{1}{l}{\cellcolor[HTML]{FFFFFF}} &
   \\
\cellcolor[HTML]{FFFFFF} &
  H$2p_y$ &
  1678.20 &
  \multicolumn{1}{l}{\cellcolor[HTML]{FFFFFF}} &
  \multicolumn{1}{l}{\cellcolor[HTML]{FFFFFF}} &
  \multicolumn{1}{l}{\cellcolor[HTML]{FFFFFF}} &
  \multicolumn{1}{l}{\cellcolor[HTML]{FFFFFF}} &
   &
  \multicolumn{1}{l}{\cellcolor[HTML]{FFFFFF}} &
   \\
\multirow{-21}{*}{\cellcolor[HTML]{FFFFFF}{RbH}}
&
  H$2p_z$ &
  1903.50 &
  \multicolumn{1}{l}{\cellcolor[HTML]{FFFFFF}} &
  \multicolumn{1}{l}{\cellcolor[HTML]{FFFFFF}} &
  \multicolumn{1}{l}{\cellcolor[HTML]{FFFFFF}} &
  \multicolumn{1}{l}{\cellcolor[HTML]{FFFFFF}} &
   &
  \multicolumn{1}{l}{\cellcolor[HTML]{FFFFFF}} &
   \\\hline
\end{tabular}\label{tab:A400}
\end{table*}
\setlength{\tabcolsep}{0.6em}
\begin{table*}
\ContinuedFloat
\centering
\caption{(\textit{Continued.})}
\begin{tabular}{cccccccccc}
\hline\hline
\multicolumn{1}{l}{\cellcolor[HTML]{FFFFFF}Molecule} &
  Reordered virtual MOs &
  MO energies &
  \multicolumn{2}{c}{\cellcolor[HTML]{FFFFFF} $n_o=6$} &
  \multicolumn{2}{c}{\cellcolor[HTML]{FFFFFF}$n_o=8$} &
  \multicolumn{2}{c}{\cellcolor[HTML]{FFFFFF}$n_o=10$} &
   \\
\multicolumn{1}{l}{\cellcolor[HTML]{FFFFFF}} &
  \multicolumn{1}{l}{\cellcolor[HTML]{FFFFFF}} &
  \multicolumn{1}{l}{\cellcolor[HTML]{FFFFFF}} &
  \multicolumn{1}{c}{\cellcolor[HTML]{FFFFFF}m} &
  \multicolumn{1}{c}{\cellcolor[HTML]{FFFFFF}$\kappa$ $(n_o, n_v)$} &
  \multicolumn{1}{c}{\cellcolor[HTML]{FFFFFF}m} &
  \multicolumn{1}{c}{\cellcolor[HTML]{FFFFFF}$\kappa$ $(n_o, n_v)$} &
  \multicolumn{1}{c}{\cellcolor[HTML]{FFFFFF}m} &
  \multicolumn{1}{c}{\cellcolor[HTML]{FFFFFF}$\kappa$ $(n_o, n_v)$} &
   \\\hline
   \cellcolor[HTML]{FFFFFF}&
  Cs $6p_x$ &
  364.80 &
  209 &
  77.05 (6, 6) &
  632 &
  81.29 (8, 8) &
  \multicolumn{1}{r}{\cellcolor[HTML]{FFFFFF}1484} &
  94.54 (10, 10) &
   \\
\cellcolor[HTML]{FFFFFF} &
  Cs $6p_y$ &
  364.80 &
  279 &
  77.18 (6, 7) &
  812 &
  92.18 (8, 9) &
  \multicolumn{1}{r}{\cellcolor[HTML]{FFFFFF}1711} &
  95.68 (10, 11) &
   \\
\cellcolor[HTML]{FFFFFF} &
  Cs $6p_z$ &
  382.00 &
  370 &
  80.54 (6, 8) &
  1016 &
  93.40 (8, 10) &
  \multicolumn{1}{r}{\cellcolor[HTML]{FFFFFF}2007} &
  100.02 (10, 12) &
   \\
\cellcolor[HTML]{FFFFFF} &
  Cs $6s$ &
  1455.30 &
  475 &
  91.32 (6, 9) &
  1044 &
  94.51 (8, 11) &
  \multicolumn{1}{r}{\cellcolor[HTML]{FFFFFF}2321} &
  104.14 (10, 13) &
   \\
\cellcolor[HTML]{FFFFFF} &
  Cs $8s$ &
  172.00 &
  595 &
  92.54 (6, 10) &
  1339 &
  98.79 (8, 12) &
  \multicolumn{1}{r}{\cellcolor[HTML]{FFFFFF}2714} &
  114.07 (10, 14) &
   \\
\cellcolor[HTML]{FFFFFF} &
  Cs$7p_y$ &
  25.50 &
  664 &
  93.63 (6, 11) &
  1540 &
  102.85 (8, 13) &
  \multicolumn{1}{r}{\cellcolor[HTML]{FFFFFF}3141} &
  157.90 (10, 15) &
   \\
\cellcolor[HTML]{FFFFFF} &
  Cs $7p_x$ &
  25.50 &
  759 &
  97.83 (6, 12) &
  1808 &
  112.65 (8, 14) &
  \multicolumn{1}{r}{\cellcolor[HTML]{FFFFFF}3537} &
  184.86 (10, 16) &
   \\
\cellcolor[HTML]{FFFFFF} &
  Cs $7s-7p_y$ hybrid &
  75.60 &
  890 &
  101.83 (6, 13) &
  2100 &
  156.01 (8, 15) &
  \multicolumn{1}{r}{\cellcolor[HTML]{FFFFFF}3951} &
  186.03 (10, 17) &
   \\
\cellcolor[HTML]{FFFFFF} &
  Cs $7s$ &
  85.50 &
  1044 &
  111.49 (6, 14) &
  2358 &
  182.96 (8, 16) &
  \multicolumn{1}{r}{\cellcolor[HTML]{FFFFFF}4461} &
  192.15 (10, 18) &
   \\
\cellcolor[HTML]{FFFFFF} &
  Cs $5dy^2$ &
  307.30 &
  1213 &
  154.45 (6, 15) &
  2624 &
  184.12 (8, 17) &
   &
  \multicolumn{1}{l}{\cellcolor[HTML]{FFFFFF}} &
   \\
\cellcolor[HTML]{FFFFFF} &
  Cs $5d_{xy}$ &
  307.30 &
  1338 &
  181.34 (6, 16) &
  2973 &
  190.17 (8, 18) &
   &
  \multicolumn{1}{l}{\cellcolor[HTML]{FFFFFF}} &
   \\
\cellcolor[HTML]{FFFFFF} &
  Cs $5d_{xz}$ &
  309.70 &
  1510 &
  182.49 (6, 17) &
  \multicolumn{1}{l}{\cellcolor[HTML]{FFFFFF}} &
  \multicolumn{1}{l}{\cellcolor[HTML]{FFFFFF}} &
   &
  \multicolumn{1}{l}{\cellcolor[HTML]{FFFFFF}} &
   \\
\cellcolor[HTML]{FFFFFF} &
  Cs $5d_{yz}$ &
  309.70 &
  1711 &
  188.52 (6, 18) &
  \multicolumn{1}{l}{\cellcolor[HTML]{FFFFFF}} &
  \multicolumn{1}{l}{\cellcolor[HTML]{FFFFFF}} &
   &
  \multicolumn{1}{l}{\cellcolor[HTML]{FFFFFF}} &
   \\
\cellcolor[HTML]{FFFFFF} &
  Cs $5dz^2$ &
  310.90 &
  \multicolumn{1}{l}{\cellcolor[HTML]{FFFFFF}} &
  \multicolumn{1}{l}{\cellcolor[HTML]{FFFFFF}} &
  \multicolumn{1}{l}{\cellcolor[HTML]{FFFFFF}} &
  \multicolumn{1}{l}{\cellcolor[HTML]{FFFFFF}} &
   &
  \multicolumn{1}{l}{\cellcolor[HTML]{FFFFFF}} &
   \\
\cellcolor[HTML]{FFFFFF} &
  H $2s$ &
  882.50 &
  \multicolumn{1}{l}{\cellcolor[HTML]{FFFFFF}} &
  \multicolumn{1}{l}{\cellcolor[HTML]{FFFFFF}} &
  \multicolumn{1}{l}{\cellcolor[HTML]{FFFFFF}} &
  \multicolumn{1}{l}{\cellcolor[HTML]{FFFFFF}} &
   &
  \multicolumn{1}{l}{\cellcolor[HTML]{FFFFFF}} &
   \\
\cellcolor[HTML]{FFFFFF} &
  H $2p_x$ &
  1716.80 &
  \multicolumn{1}{l}{\cellcolor[HTML]{FFFFFF}} &
  \multicolumn{1}{l}{\cellcolor[HTML]{FFFFFF}} &
  \multicolumn{1}{l}{\cellcolor[HTML]{FFFFFF}} &
  \multicolumn{1}{l}{\cellcolor[HTML]{FFFFFF}} &
   &
  \multicolumn{1}{l}{\cellcolor[HTML]{FFFFFF}} &
   \\
\cellcolor[HTML]{FFFFFF} &
  H $2p_y$ &
  1716.80 &
  \multicolumn{1}{l}{\cellcolor[HTML]{FFFFFF}} &
  \multicolumn{1}{l}{\cellcolor[HTML]{FFFFFF}} &
  \multicolumn{1}{l}{\cellcolor[HTML]{FFFFFF}} &
  \multicolumn{1}{l}{\cellcolor[HTML]{FFFFFF}} &
   &
  \multicolumn{1}{l}{\cellcolor[HTML]{FFFFFF}} &
   \\
\multirow{-18}{*}{\cellcolor[HTML]{FFFFFF}CsH} &
  H $2p_z$ &
  1716.80 &
  \multicolumn{1}{l}{\cellcolor[HTML]{FFFFFF}} &
  \multicolumn{1}{l}{\cellcolor[HTML]{FFFFFF}} &
  \multicolumn{1}{l}{\cellcolor[HTML]{FFFFFF}} &
  \multicolumn{1}{l}{\cellcolor[HTML]{FFFFFF}} &
   &
  \multicolumn{1}{l}{} &\\\hline\hline
\end{tabular}
\label{tab:A300}
\end{table*}

\begin{table*}[]
\centering
 \caption{Table providing the minimum and maximum eigenvalues of the matrix $A_{mol}$ of size $m \times m$, denoted in the table as $\lambda_{\mathrm{min}}^{A_{mol}}$ and $\lambda_{\mathrm{max}}^{A_{mol}}$ respectively, for KH, RbH and CsH systems. $n_o$ and $n_v$ denote the number of occupied and virtual orbitals, respectively. $\lambda_{\mathrm{min}}^{A_{mol}}$  and $\lambda_{\mathrm{max}}^{A_{mol}}$  are in units of milli-Hartree. }
\begin{tabular}{ccccccccccccc}
\hline\hline
Molecule &
  \multicolumn{4}{c}{$n_o=6$} &
  \multicolumn{4}{c}{$n_o=8$} &
  \multicolumn{4}{c}{\cellcolor[HTML]{FFFFFF}$n_o=10$} \\
 &
  \multicolumn{1}{c}{m} &
  \multicolumn{1}{c}{$n_v$} &
  \multicolumn{1}{c}{$\lambda_{\mathrm{min}}^{A_{mol}}$} &
  \multicolumn{1}{c}{$\lambda_{\mathrm{max}}^{A_{mol}}$} &
  \multicolumn{1}{c}{m} &
  \multicolumn{1}{c}{$n_v$} &
  \multicolumn{1}{c}{$\lambda_{\mathrm{min}}^{A_{mol}}$}&
  \multicolumn{1}{c}{$\lambda_{\mathrm{max}}^{A_{mol}}$} &
  \multicolumn{1}{c}{m} &
  \multicolumn{1}{c}{$n_v$} &
  \multicolumn{1}{c}{$\lambda_{\mathrm{min}}^{A_{mol}}$} &
  \multicolumn{1}{c}{$\lambda_{\mathrm{max}}^{A_{mol}}$} \\\hline
 &
  202 &
  6 &
  733.73 &
  25918.07 &
  580 &
  8 &
  730.05 &
  29960.23 &
  1484 &
  10 &
  127.91 &
  280908.57 \\
 &
  274 &
  7 &
  731.91 &
  29570.15 &
  745 &
  9 &
  728.20 &
  30271.74 &
  1761 &
  11 &
  127.29 &
  280909.73 \\
 &
  337 &
  8 &
  730.17 &
  29604.25 &
  932 &
  10 &
  127.95 &
  30272.05 &
  2046 &
  12 &
  126.64 &
  280910.95 \\
 &
  432 &
  9 &
  728.35 &
  29727.55 &
  1110 &
  11 &
  127.33 &
  30272.33 &
  2451 &
  13 &
  120.90 &
  280914.12 \\
 &
  541 &
  10 &
  128.01 &
  29729.71 &
  1296 &
  12 &
  126.71 &
  30272.59 &
  2900 &
  14 &
  98.56 &
  280914.37 \\
 &
  655 &
  11 &
  127.38 &
  29730.03 &
  1540 &
  13 &
  120.92 &
  30272.79 &
  3393 &
  15 &
  96.78 &
  281150.00 \\
 &
  748 &
  12 &
  126.76 &
  29730.35 &
  1808 &
  14 &
  98.59 &
  30272.81 &
  3794 &
  16 &
  95.80 &
  281161.10 \\
 &
  889 &
  13 &
  120.98 &
  29731.63 &
  2100 &
  15 &
  96.83 &
  30293.88 &
  4203 &
  17 &
  94.84 &
  281171.57 \\
 &
  1043 &
  14 &
  98.65 &
  29731.69 &
  2358 &
  16 &
  95.87 &
  30304.25 &
  4788 &
  18 &
  77.86 &
  281200.60 \\
 &
  1213 &
  15 &
  96.89 &
  29830.59 &
  2624 &
  17 &
  94.89 &
  30314.00 &
  5265 &
  19 &
  77.45 &
  281200.93 \\
 &
  1381 &
  16 &
  95.93 &
  29834.52 &
  2972 &
  18 &
  77.91 &
  30317.70 &
  5750 &
  20 &
  76.99 &
  281201.42 \\
 &
  1510 &
  17 &
  94.96 &
  29839.08&
  3278 &
  19 &
  77.50 &
  30317.87 &
  6427 &
  21 &
  76.56 &
  281469.15 \\
&
  1711 &
  18 &
  77.98 &
  29854.11 &
  3592 &
  20 &
  77.10 &
  30318.05 &
  \multicolumn{1}{c}{} &
  \multicolumn{1}{c}{} &
  \multicolumn{1}{c}{} &
  \multicolumn{1}{c}{} \\
 &
  1909 &
  19 &
  77.57 &
   29854.20  &
  3996 &
  21 &
  76.62 &
  30328.12 &
  \multicolumn{1}{c}{} &
  \multicolumn{1}{c}{} &
  \multicolumn{1}{c}{} &
  \multicolumn{1}{c}{} \\
 &
  2062 &
  20 &
  77.17 &
  29854.30 &
  \multicolumn{1}{c}{} &
  \multicolumn{1}{c}{} &
  \multicolumn{1}{c}{} &
  \multicolumn{1}{c}{} &
  \multicolumn{1}{c}{} &
  \multicolumn{1}{c}{} &
  \multicolumn{1}{c}{} &
  \multicolumn{1}{c}{} \\
\multirow{-17}{*}{KH} &
  2295 &
  21 &
  76.69 &
  29894.66 &
  \multicolumn{1}{c}{} &
  \multicolumn{1}{c}{} &
  \multicolumn{1}{c}{} &
  \multicolumn{1}{c}{} &
  \multicolumn{1}{c}{} &
  \multicolumn{1}{c}{} &
  \multicolumn{1}{c}{} &
  \multicolumn{1}{c}{} \\\hline
 &
  209 &
  6 &
  108.95 &
   11193.43 &
  632 &
  8 &
  108.02 &
  11330.03 &
  1401 &
  10 &
  106.22 &
  11489.85 \\
 &
  279 &
  7 &
  108.73 &
  11200.46&
  744 &
  9 &
  107.07 &
  11330.11 &
  1673 &
  11 &
  102.72 &
  11490.43 \\
 &
  370 &
  8 &
  108.02 &
  11206.98 &
  932 &
  10 &
  106.25 &
  11330.19 &
  1963 &
  12 &
  99.68 &
  11491.02 \\
 &
  432 &
  9 &
  107.08 &
  11207.05 &
  1110 &
  11 &
  102.75 &
  11330.84 &
  2321 &
  13 &
  91.77 &
  11503.07 \\
 &
  541 &
  10 &
  106.26 &
  11207.13 &
  1296 &
  12 &
  99.70 &
  11331.49 &
  2713 &
  14 &
  81.20 &
  12694.32 \\
&
  630 &
  11 &
  102.77 &
  11207.95 &
  1540 &
  13 &
  91.80 &
  11339.56 &
  3141 &
  15 &
  72.02 &
  13125.74 \\
 &
  748 &
  12 &
  99.73 &
  11208.41 &
  1808 &
  14 &
  81.23 &
  12533.58 &
  3537 &
  16 &
  71.65 &
  14777.36 \\
 &
  889 &
  13 &
  91.84 &
  11216.40 &
  2100 &
  15 &
  72.05 &
  12963.58 &
  3951 &
  17 &
  71.28 &
  14822.69 \\
 &
  1044 &
  14 &
  81.27 &
  12410.75 &
  2358 &
  16 &
  71.68 &
  14614.13 &
  4460 &
  18 &
  70.82 &
  15148.61 \\
 &
  1213 &
  15 &
  72.10 &
  12840.73 &
  2624 &
  17 &
  71.31 &
  14659.47 &
  4932 &
  19 &
  70.64 &
  15148.86 \\
&
  1381 &
  16 &
  71.72 &
  14490.85 &
  2972 &
  18 &
  70.85 &
  14988.70 &
  5422 &
  20 &
  70.46 &
  15149.12 \\
 &
  1510 &
  17 &
  71.35 &
  14536.14 &
  3278 &
  19 &
  70.67 &
  14988.95 &
  6013 &
  21 &
  70.05 &
  15151.35 \\
 &
  1711 &
  18 &
  70.90 &
  14865.97 &
  3592 &
  20 &
  70.50 &
  14989.21 &
  \multicolumn{1}{l}{} &
  \multicolumn{1}{l}{} &
  \multicolumn{1}{l}{} &
  \multicolumn{1}{l}{} \\
&
  1860 &
  19 &
  70.72 &
  14866.22 &
  3996 &
  21 &
  70.09 &
  14991.44 &
  \multicolumn{1}{l}{} &
  \multicolumn{1}{l}{} &
  \multicolumn{1}{l}{} &
  \multicolumn{1}{l}{} \\
 &
  2062 &
  20 &
  70.55 &
  14866.47 & 
  \multicolumn{1}{l}{} &
  \multicolumn{1}{l}{} &
  \multicolumn{1}{l}{} &
  \multicolumn{1}{l}{} &
  \multicolumn{1}{l}{} &
  \multicolumn{1}{l}{} &
  \multicolumn{1}{l}{} &
  \multicolumn{1}{l}{} \\
\multirow{-16}{*}{RbH} &
  2295 &
  21 &
  70.15 &
  14868.69 &
  \multicolumn{1}{c}{} &
  \multicolumn{1}{c}{} &
  \multicolumn{1}{c}{} &
  \multicolumn{1}{c}{} &
  \multicolumn{1}{c}{} &
  \multicolumn{1}{c}{} &
  \multicolumn{1}{c}{} &
  \multicolumn{1}{c}{} \\\hline
 &
  209 &
  6 &
  119.60 &
  9216.72 &
  632 &
  8 &
  114.40 &
  9306.70 &
  1484 &
  10 &
  99.94 &
  9447.99 \\
&
  279 &
  7 &
  119.41 &
  9216.73 &
  812 &
  9 &
  101.30 &
  9338.37 &
  1711 &
  11 &
  98.74 &
  9448.39 \\
 &
  370 &
  8 &
  114.52 &
  9224.27 &
  1016 &
  10 &
  99.97 &
  9338.71 &
  2007 &
  12 &
  94.46 &
  9448.85 \\
 &
  475 &
  9 &
  101.35 &
  9255.84 &
  1044 &
  11 &
  98.81 &
  9339.05 &
  2321 &
  13 &
  90.73 &
  9449.29 \\
 &
  595 &
  10 &
  100.02 &
  9256.34 &
  1339 &
  12 &
  94.53 &
  9339.23 &
  2714 &
  14 &
  82.84 &
  9449.94 \\
 &
  664 &
  11 &
  98.86 &
  9256.39 &
  1540 &
  13 &
  90.80 &
  9339.41 &
  3141 &
  15 &
  63.54 &
  10033.15 \\
 &
  759 &
  12 &
  94.61 &
  9256.47 &
  1808 &
  14 &
  82.91 &
  9340.27 &
  3537 &
  16 &
  63.38 &
  11718.57 \\
 &
  890 &
  13 &
  90.89 &
  9256.59 &
  2100 &
  15 &
  63.60 &
  9923.04 &
  3951 &
  17 &
  63.23 &
  11763.91 \\
 &
  1044 &
  14 &
  83.03 &
  9257.34 &
  2358 &
  16 &
  63.44 &
  11609.02 &
  4461 &
  18 &
  62.86 &
  12079.72 \\
 &
  1213 &
  15 &
  63.71 &
  9840.84 &
  2624 &
  17 &
  63.29 &
  11654.36 &
  &
 &
 &
  \\
 &
  1338 &
  16 &
  63.56 &
  11526.12 &
  2973 &
  18 &
  62.92 &
  11966.65 &
 &
 &
&
   \\
 &
  1510 &
  17 &
  63.40 &
  11571.52 &
  \multicolumn{1}{c}{} &
  \multicolumn{1}{c}{} &
  \multicolumn{1}{c}{} &
  \multicolumn{1}{c}{} &
  \multicolumn{1}{c}{} &
  \multicolumn{1}{c}{} &
  \multicolumn{1}{c}{} &
  \multicolumn{1}{c}{} \\
\multirow{-13}{*}{CsH} &
  1711 &
  18 &
  63.04 &
  11884.73 &
  \multicolumn{1}{c}{} &
  \multicolumn{1}{c}{} &
  \multicolumn{1}{c}{} &
  \multicolumn{1}{c}{} &
  \multicolumn{1}{c}{} &
  \multicolumn{1}{c}{} &
  \multicolumn{1}{c}{} &
  \multicolumn{1}{c}{}\\\hline\hline
\end{tabular}
\label{tab:A200}
\end{table*}

\begin{table*}[!h]
\caption{Average PFDs and range of PFDs of HHL and Psi-HHL for Li$_2$ and KH molecules, both in Sapporo-DKH3-DZP-2012  basis set. The average PFDs are calculated over a range of 30 repetitions. }
\begin{tabular}{
>{\columncolor[HTML]{FFFFFF}}c 
>{\columncolor[HTML]{FFFFFF}}c 
>{\columncolor[HTML]{FFFFFF}}c 
>{\columncolor[HTML]{FFFFFF}}c 
>{\columncolor[HTML]{FFFFFF}}c }
\hline\hline
\cellcolor[HTML]{FFFFFF} &
  \multicolumn{2}{c}{\cellcolor[HTML]{FFFFFF}Li$_2$} &
  \multicolumn{2}{c}{\cellcolor[HTML]{FFFFFF}KH} \\ 
\multirow{-3}{*}{\cellcolor[HTML]{FFFFFF}\# shots} &
  HHL (PFD, range) &
  Psi-HHL (PFD, range) &
  HHL (PFD, range) &
  Psi-HHL (PFD, range) \\\hline
$5\times 10^2$ &
  (-7.68, 40.22) &
  (-2.79, 29.58) &
  \multicolumn{1}{l}{\cellcolor[HTML]{FFFFFF}} &
  \multicolumn{1}{l}{\cellcolor[HTML]{FFFFFF}} \\
$10^3$ &
  (-4.08, 19.80) &
  (-1.46, 8.34) &
  (nan, nan) &
  (-11.98, 106.87) \\
$5 \times 10^3$ &
  (-2.94, 8.89) &
  (-2.11, 7.56) &
  \multicolumn{1}{l}{\cellcolor[HTML]{FFFFFF}} &
  \multicolumn{1}{l}{\cellcolor[HTML]{FFFFFF}} \\
$10^4$ &
  (-0.58, 6.00) &
  (1.07, 4.47) &
  (0.34, 68.36) &
  (1.29, 44.70) \\
$5\times 10^4$ &
  \multicolumn{1}{l}{\cellcolor[HTML]{FFFFFF}} &
  \multicolumn{1}{l}{\cellcolor[HTML]{FFFFFF}} &
  (0.73, 32.51) &
  (0.68, 16.78) \\
$10^5$ &
  \multicolumn{1}{l}{\cellcolor[HTML]{FFFFFF}} &
  \multicolumn{1}{l}{\cellcolor[HTML]{FFFFFF}} &
  (0.73, 17.00) &
  (0.72, 9.02) \\
$5\times 10^5$ &
  \multicolumn{1}{l}{\cellcolor[HTML]{FFFFFF}} &
  \multicolumn{1}{l}{\cellcolor[HTML]{FFFFFF}} &
  (2.65, 10.59) &
  (1.11, 5.65) \\
$10^6$ &
  \multicolumn{1}{l}{} &
  \multicolumn{1}{l}{} &
  (3.99, 5.09) &
  (0.22, 4.05) \\\hline\hline
\end{tabular}
\label{tab:A100}
\end{table*}

\begin{table*}[]
\caption{Table providing the details on the condition numbers ($\kappa$), the average total energy, the energy difference ($\Delta E$) between the total HHL (Psi-HHL) energy and classical value of the LCCSD total energy and the average PFDs over 30 repetitions with respect to the bond length of KH molecule. All the energies are in units of milli-Hartree. `Eq' denotes the equilibrium geometry. }  
\begin{tabular}{cccccccc}
\hline\hline
Bond length (Bohr) & $\kappa$ &  \multicolumn{3}{c}{HHL}       & \multicolumn{3}{c}{Psi-HHL}  \\
\multicolumn{1}{c}{} &
  \multicolumn{1}{c}{} &
  \multicolumn{1}{c}{Total energy} &
  \multicolumn{1}{c}{$\Delta E$} &
  \multicolumn{1}{c}{PFD} &
  \multicolumn{1}{c}{Total energy} &
  \multicolumn{1}{c}{$\Delta E$} &
  \multicolumn{1}{c}{PFD} 
   \\ \hline 
2.82              & 393.21        & -598878.025 & -0.124 & -3.92  & -598877.968 & -0.068 & -2.48 \\
3.29              & 374.93       & -598940.871 & -0.086 & -3.84 & -598940.844 & -0.059 & 2.31  \\
3.76              & 367.32      & -598967.447 & -0.003 & -1.96 & -598967.447 & -0.004 & -0.26  \\
4.23 (Eq)         & 365.40         & -598975.800   & 0.007  & 1.46  & -598975.802 & 0.004  & 0.34  \\
4.71              & 371.38        & -598974.517 & -0.010  & -10.00    & -598974.509 & -0.007 & 1.39 \\
5.18 &
  385.16 &
  \multicolumn{1}{c}{-598968.421} & -0.001 & -0.44 & -598968.419 & 0.000 & -0.08\\
  5.65  & 373.56    &  -598960.199&  -0.001  & -0.41   & -598960.196 & 0.000 & 0.26 \\\hline\hline
\end{tabular}
\label{tab:A500}
\end{table*} 

\end{appendices}
\end{document}